\documentclass[11pt]{report}
\usepackage{longtable}
\usepackage{graphicx}
\usepackage{latexsym}
\makeglossary
\makeindex
\newcounter{mytab}
\newcommand{\mycap}[3]{\refstepcounter{mytab}
   \label{#1} \\ \vspace{#2} Table \Alph{chapter}.\arabic{mytab}: #3}
\newcommand{\mycapw}[3]{\refstepcounter{mytab}
   \label{#1} \\ \vspace{#2} \parbox{6in}{Table \Alph{chapter}.\arabic{mytab}: #3}}
\newtheorem{thm}{Theorem}[chapter]
\newtheorem{defn}{Definition}[chapter]
\newtheorem{eg}{Example}[chapter]
\newcommand{\egbox}[1]{
\begin{eg}
\rm
#1 \hfill $\Box$
\end{eg}
}
\newcommand{\bibann}[1]{}
\newcommand{\mynote}[1]{}
\newcommand{\myfoot}[1]{}
\newcommand{\xxx}[1]{}
\newcommand{\git}[1]{\glossary{#1} {\it #1}}
\newcommand{\beq}{\begin{equation}}
\newcommand{\eeq}{\end{equation}}
\newcommand{\beqa}{\begin{eqnarray}}
\newcommand{\eeqa}{\end{eqnarray}}
\newcommand{\ben}{\begin{enumerate}}
\newcommand{\een}{\end{enumerate}}
\newlength{\dummysp}
\newcommand{\spc}{\hbox{\hspace{\dummysp}}}
\newcommand{\gappeq}{\mathrel{\rlap {\raise.5ex\hbox{$>$}}
{\lower.5ex\hbox{$\sim$}}}}
\newcommand{\lappeq}{\mathrel{\rlap{\raise.5ex\hbox{$<$}}
{\lower.5ex\hbox{$\sim$}}}}
\newcommand{\myref}[1]{(\ref{#1})}
\newcommand{\nnn}{ \nonumber \\ }
\newcommand{\mod}{{\; \mtxt{mod} \; }}

\newcommand{\tr}{\mathop{{\hbox{tr} \, }}\nolimits}
\newcommand{\mtxt}[1]{{\mathop{\hbox{{\small #1}}}\nolimits}}
\newcommand{\stxt}[1]{{\mathop{\hbox{{\scriptsize #1}}}\nolimits}}
\newcommand{\ttxt}[1]{{\mathop{\hbox{{\tiny #1}}}\nolimits}}
\newcommand{\bbar}[1]{{\overline{#1}}}

\newcommand{\diag}{\mathop{{\hbox{diag} \, }}\nolimits}
\newcommand{\Imag}{\mathop{{\hbox{Im} \, }}\nolimits}
\newcommand{\Real}{\mathop{{\hbox{Re} \, }}\nolimits}

\newcommand{\vev}[1]{{\langle #1 \rangle}}
\newcommand{\bigvev}[1]{{\left\langle #1 \right\rangle}}
\newcommand{\ord}[1]{{{\cal O}(10^{#1})}}
\newcommand{\ordnt}[1]{{{\cal O}(#1)}}

\newcommand{\ket}[1]{{ | #1 \rangle }}
\newcommand{\bra}[1]{{ \langle #1 | }}

\newcommand{\Db}{{\bar {\cal D}}}
\newcommand{\W}{{\cal W}}

\newcommand{\Zbf}{{{\bf Z}}}
\newcommand{\zbf}{{{\bf z}}}

\newcommand{\Rbf}{{{\bf R}}}
\newcommand{\Pbf}{{{\bf P}}}
\newcommand{\Tbf}{{{\bf T}}}
\newcommand{\Wbf}{{{\bf W}}}
\newcommand{\bbf}{{{\bf b}}}

\newcommand{\half}{{1 \over 2}}

\newcommand{\third}{{1 \over 3}}
\newcommand{\twthird}{{2 \over 3}}

\newcommand{\p}{{\partial}}
\newcommand{\e}{{\epsilon}}
\newcommand{\s}{{\sigma}}

\newcommand{\eelat}{{\Lambda_{E_8 \times E_8}}}
\newcommand{\elat}{{\Lambda_{E_8}}}
\newcommand{\ux}{$U(1)_X$}
\newcommand{\eetee}{$E_8 \times E_8$}

\newcommand{\gsmc}{$G_{SM} \times G_C$}

\newcommand{\LamX}{\Lambda_X}
\newcommand{\LamC}{\Lambda_C}

\newcommand{\uone}{$U(1)$}
\newcommand{\bz}{{\bar z}}
\newcommand{\Aem}{$ \{ V_A, a_{1A}, a_{3A} \} $}
\newcommand{\Bem}{$ \{ V_B, a_{1B}, a_{3B} \} $}
\newcommand{\opI}{{\bf (I)}}
\newcommand{\opII}{{\bf (II)}}
\newcommand{\adot}{{{\dot a}}}
\newcommand{\bdot}{{{\dot b}}}
\newcommand{\ccdot}{{{\dot c}}}
\newcommand{\dddot}{{{\dot d}}}

\newcommand{\LamH}{\Lambda_H}
\newcommand{\tK}{{\tilde K}}
\newcommand{\emset}{$\{ V, a_1, a_3, a_5 \}$}
\newcommand{\GNA}{{G_{\stxt{NA}}}}
\newcommand{\GUO}{{G_{\stxt{UO}}}}
\newcommand{\GSM}{{G_{\stxt{SM}}}}
\newcommand{\GGUT}{{G_{\stxt{GUT}}}}
\newcommand{\LamHsq}{\Lambda_H^2}
\newcommand{\LamSB}{\Lambda_{\ttxt{SUSY}}}
\newcommand{\LamU}{\Lambda_U}
\newcommand{\LamTh}{{\Lambda_{SU(3)^3}}}
\newcommand{\tra}[1]{\mathop{{\hbox{tr}_{#1} \, }}\nolimits}
\newcommand{\tbeta}[1]{{b_{#1}^{\stxt{tot}}}}
\newcommand{\icoup}[1]{{\alpha_{#1}^{-1}}}

\newcommand{\dbtw}{{\delta b_2}}
\newcommand{\dbth}{{\delta b_3}}
\newcommand{\dbY}{{\delta b_Y}}
\newcommand{\dbYp}{{\delta b_Y'}}

\newcommand{\dkY}{{\delta k_Y}}
\newcommand{\dkYm}{{\delta k_Y^{\stxt{min}}}}

\newcommand{\nasm}{$SU(3)_C \times SU(2)_L$}
\newcommand{\GSME}{{SU(3) \times SU(2) \times U(1)}}

\newcommand{\bGS}{{b_{\stxt{GS}}}}
\newcommand{\ipr}[2]{{\langle #1 | #2 \rangle }}
\newcommand{\bfe}[1]{{\bf #1 \hspace{3pt}}}

\newcommand{\bsa}{${\rm BSL}_{\rm A}$}
\newcommand{\MSB}{M_{\ttxt{SUSY}}}
\newcommand{\swZ}{\sin^2 \theta_W (m_Z)}
\newcommand{\smi}{\s_-}
\newcommand{\spl}{\s_+}
\newcommand{\Hcal}{{\cal H}}
\newcommand{\tw}{{(\theta,\ell)}}
\def\[{\left [}
\def\]{\right ]}
\def\({\left (}
\def\){\right )}

\def\|{ \hbox{\Large $|$}\; }
\begin{document}

\bibliographystyle{unsrt}

\def\thefootnote{\fnsymbol{footnote}}

\begin{titlepage} 

\hfill    LBNL-50169

\hfill    UCB-PTH-02/19

\hfill    hep-ph/0204315

\hfill    April 26, 2002

\begin{center}

\vspace{30pt}

{ \bf \Huge Heterotic Orbifolds}

\end{center}

\vspace{5pt}

\begin{center}
{\sl Joel Giedt\footnote{E-Mail: {\tt JTGiedt@lbl.gov}}}

\end{center}

\vspace{5pt}

\begin{center}

{\it Department of Physics, University of California, \\
and Theoretical Physics Group, 50A-5101, \\
Lawrence Berkeley National Laboratory, Berkeley,
CA 94720 USA.}\footnote{This work was supported in part by the
Director, Office of Science, Office of High Energy and Nuclear
Physics, Division of High Energy Physics of the U.S. Department of
Energy under Contract DE-AC03-76SF00098 and in part by the National
Science Foundation under grant PHY-0098840.}

\end{center}

\vspace{5pt}

\begin{center}

{\bf Abstract}

\end{center}

\vspace{5pt}

A review of orbifold geometry is given, followed
by a review of the construction of four-dimensional
heterotic string models by compactification on
a six-dimensional $Z_3$ orbifold.  Particular attention
is given to the details of the transition from a classical
theory to a first-quantized theory.  Subsequently,
a discussion is given of the systematic
enumeration of all standard-like three generation
models subject to certain limiting conditions.
It is found that the complete set is described
by 192 models, with only five possibilities for
the hidden sector gauge group.  It is argued
that only four of the hidden sector gauge groups
are viable for dynamical supersymmetry breaking,
leaving only 175 promising models in the class.
General features of the spectra of matter states in
all 175 models are discussed.  Twenty patterns of representations
are found to occur.  Accomodation of the Minimal
Supersymmetric Standard Model (MSSM) spectrum is
addressed.  States beyond those
contained in the MSSM and nonstandard hypercharge
normalization are shown to be generic, though some
models do allow for the usual hypercharge normalization
found in $SU(5)$ embeddings of the Standard Model
gauge group.  Only one of the twenty patterns of
representations, comprising seven of the 175 models,
is found to be without an anomalous $U(1)$.
Various quantities of interest in effective supergravity
model building are tabulated for the set of 175 models.
String scale gauge coupling unification is shown
to be possible, albeit contrived, in an example model.

\end{titlepage}

\renewcommand{\thepage}{\roman{page}}
\setcounter{page}{1}

\tableofcontents

%

\chapter*{Acknowledgements}
\typeout{Acknowledgements}
\addcontentsline{toc}{chapter}{Acknowledgements}
This work would have been impossible without the
assistance, guidance and support of many people.
My wife has been constant in her encouragement,
and has made sacrifices so that I could pursue
a career in physics.  For her selfless support
I will be forever grateful.
I heartily thank my research advisor,
Mary K.~Gaillard, who has been nothing but encouraging
during my time at Berkeley, and who taught me many
things, including how to be a productive researcher
in the face of complex questions.
I am extremely grateful
to my undergraduate research advisor, Jeffrey P.~Greensite.
His immeasurable help made it possible for me
to study physics at Berkeley, and I learned
a great deal about how to do quality research
by working with him.  I would also like to
thank Alan Weinstein and Ori Ganor for serving
on my dissertation committee and taking the
time to read this thesis.  Alan has made many
helpful comments which resulted in important
improvements.

My family has been entirely supportive of my
goals.  My mother, father and grandparents all nurtured
my interests in the sciences at an early age.
My mother found ways to expose me to research scientists
during my childhood.  These experiences
founded my desire to become a professional
scientist.  I am sincerely appreciative of all her
efforts to provide for my unusual needs.
I have a debt of gratitude to a number of excellent
science teachers in the California public
primary, secondary and higher educational systems.
In particular I would like to thank Misters
Atchison, Doyle and Mauney.  I
would like to thank the taxpayers of California
for supporting the Cal State University and University
of California systems.  Further,
I would like to thank U.S.~taxpayers and
our national leaders
for their support of the Department of Education,
the National Science Foundation and
the Department of Energy.  I have
received significant financial support from these
agencies throughout my education.
I have benefitted greatly from
the opportunities which are made available
through the programs these agencies administer.

A number of distinguished physicists have helped
me along the way.  I would like to thank Ron Adler,
Nima Arkani-Hamed, Korkut Bardacki,
Pierre Bin\'etruy, John Burke, Bob Cahn, Gene Commins,
Emilian Dudas, Alon Faraggi, Christophe Grojean,
Lawrence Hall, Dave Jackson,
Oliver Johns, Susan Lea, 
Geoff Marcy, Brent Nelson, 
Bob Rogers and Bruno Zumino.
Each of these persons has in one way or another
contributed to my development as a young
physicist.

Finally, I would like to acknowledge my Good Fortune.
To be born in an era, in a society, when and where
the common person can receive well over twenty years
of public education is a rare circumstance not
shared by most of our ancestors, nor by the majority
of the people alive on earth today.
I strive to remain cognizant of this fact and try
to bear my education humbly.  My health has been
excellent throughout my life, and I feel
most fortunate to be granted the faculties
required to develop a deep appreciation of
the ideas and discoveries which have been
the focus of my studies.  Reflecting on how I got
to this point, I realize that Fate has been
kind to me---so it only seems right I that
acknowledge its role.

\chapter*{Prefatory Remarks}
\typeout{Prefatory Remarks}
\addcontentsline{toc}{chapter}{Prefatory Remarks}
What follows is my doctoral thesis.  Its purpose is
two-fold.  First, I review ``well-known'' aspects
of orbifold geometry and its application to the
weakly-coupled heterotic string.  These are the
contents of Chapters \ref{oge}-\ref{mss}. Second, I describe
my own research, the emphasis of which has been
the construction of semi-realistic $Z_3$ heterotic
orbifold models within a restricted class.
This material is contained in Chapters \ref{slz}-\ref{app}
and is based on my two recent articles \cite{Gie01b,Gie01c}.

I have written Chapter \ref{oge} to be elementary and
accessible to a wide audience.  It is my hope that it
might prove useful to those who are just beginning a
study of orbifolds.  I was able to do this because the
topic is fairly self-contained and does not require
a large amount of preparatory knowledge beyond that
already possessed by most graduate students in theoretical
physics or mathematics.

Chapter \ref{mss} is best supplemented with standard
texts on string theory, say, the first volume of Green, Schwarz
and Witten \cite{GSW87a}.  Moreover, Chapter
\ref{mss} assumes familiarity with the theory of
Lie algebras and groups, especially the Cartan
system of roots and weights.  The
review of string theoretic aspects of heterotic
orbifold theory is somewhat heuristic.  I would have preferred
to have given a complete and self-contained discussion
rather than what the reader will find in Chapter \ref{mss}.
However, after attempting this project, I gradually
began to understand that such an endeavor would---if
properly done---encompass a rather large book!  Therefore,
I have resorted to the more ``poetic'' presentation
of Chapter \ref{mss}, supplementing it with adequate
references to the several very good reviews and texts
which are widely available.  Thus, the intent of
Chapter \ref{mss} is to provide the reader with
a general impression of the approach taken to building
semi-realistic string theories, and to introduce
crucial terms within a context which provides,
hopefully, an intuitive sense of their meanings.

Does string theory have anything to do with the
material universe?  It is questionable whether or
not we will ever have {\it compelling} evidence which
would indicate the answer to this question.
I do not expect string theory to ever stand on the
same experimental footing as, say, the Standard Model
of elementary particle physics.

If an affirmative answer is forthcoming, I believe it
will probably come from the application of string
theory to strongly coupled Quantum Chromodynamics (QCD),
where string-like behavior has already been ``observed''
in quark confinement, the Regge trajectories
of the hadron spectrum, and in Lattice Gauge Theory
simulations of the confining phase of QCD.

However, the application of string theory to QCD is not the
topic of this thesis.  Rather, the line of research
taken up here envisions string theory as an underlying
theory behind {\it all} of the fundamental interactions between
elementary particles observed in the laboratory.
The energy scale where the ``stringy'' nature of the
underlying theory really becomes apparent is many
orders of magnitude beyond the reach of particle
accelerators, at least in the case of the weakly-coupled
heterotic string.  Optimistically, a few distinctive
``stringy'' remanants might possibly be observed in,
say, searches for fractionally charged particles, or
(very optimistically) ultra-high energy cosmic
ray experiments.

The real advantage to string theory is not that it
provides hard and fast predictions for experimental observations
which are just around the corner.  (It does however
{\it constrain} what might be observed.)  Rather, its chief
strength is in its (perhaps unique) ability to provide
resolutions to troubling theoretical difficulties.  I
will discuss these in the Motivations section which
follows.  Given that it apparently resolves these
difficulties, it is important to determine whether it
can simultaneously accomodate our view of the material
universe.  Can it consistently account for what is
observed?  This has been the main focus of my research.
Another important question to answer is the following.
If string theory {\it is} the underlying theory of
the interactions of elementary particles, in what ways
does it limit effective theories describing what
{\it will} be observed in the coming years?  This
is the topic of ``stringy'' constraints on physics
beyond the Standard Model of elementary particle physics.
My research also touches on this question.

At this point I remark that the universe
is larger than the material universe!  For instance, there
is the universe of ideas, abstract structures and
mathematics without any apparent applications.
Regardless of the suitability of string theory
for the description of aspects of the material universe,
it {\it is} a description of fascinating mathematical
structures.  As such, I would like to find a vehicle
to communicate my studies to the mathematical community.
There are numerous difficulties, however, in doing
so.  Foremost among these is a difference in language.
For instance, physicists talk about ``fields'' while
mathematicians talk about ``sections of fiber bundles.''
Bridging this gap is no small task, due to the years of
specialized training in separate departments of the
academy.  I had originally thought to provide
enough elementary discussions, footnotes and appendices
to render this thesis accessible to non-physicists.
I eventually determined that the goal was not terribly
realistic, as I would have to learn quite a bit
more mathematics than I had time to do in the final
year of my doctoral program and because the amount
of theoretical physics material that I would have
to review was going to be a great quantity.

Consequently, I am afraid that I have had to resort
to an appendix which outlines a few good references which
a person could read to become sufficiently
familiar with the terminology, techniques and models
of modern particle theory.  These references review
these topics better than I ever could, and I
recommend that readers interested in
better understanding this thesis consider
reading them concurrently, at least poetically.


\chapter*{Motivations${}^{\hbox{{\rm \normalsize *}}}$}
\typeout{Motivations}
\addcontentsline{toc}{chapter}{Motivations}
\def\thefootnote{\fnsymbol{footnote}}
\footnotetext[1]{This section summarizes reasons why
theoretical physicists are interested in string theory.
It is directed toward a non-specialist audience
and is therefore somewhat elementary in its
discussion.}
\def\thefootnote{\arabic{footnote}}
\setcounter{footnote}{0}
The quantum theory of fields
provides an adequate
description of the electromagnetic, weak and strong
interactions;
to these gauge interactions, one may
add the mass interactions of quarks and leptons,
including ``mass-mixing'' interactions such as
those which appear in the charged current.
For a variety of reasons,
it is widely believed that new particles and
interactions beyond those now known
may soon be detected in precision 
low-energy experiments and ``next generation''
high-energy colliders.\footnote{See for example
\cite{GGS99} for a survey of some the reasons
why particles and interactions beyond
the Standard Model are anticipated.}
Moreover, quantized field theory with its point-like
interactions runs into difficulties in two
rather significant respects.  First, it allows
for the spontaneous emission and reabsorption
of particles which are ``off mass-shell,''
$E^2 \not= {\bf p}^2 + m^2$,
where $m$ is the measured mass.
These ``virtual'' particles yield quantum
corrections to the interactions and
field strengths (the normalization
of fields) of quarks and leptons 
which are often infinite.  The usual response
to this is to point out that it is only the
quantum corrected field strengths and
interaction strengths (measured by
coupling parameters) which are physically
meaningful, since in an experimental
process we always measure quantities
which have all quantum effects included.
Thus, the original, uncorrected ``bare''
field strengths and coupling parameters
are not fundamental, and should be adjusted
such that the infinities which arise from
quantum corrections are canceled when
we compute (infrared safe) rates, lifetimes and other
physical processes.  The systematic
implementation of this philosophy goes
by the name {\it renormalization}.  In
order to cancel the infinities, they
must be rendered finite by imposing a
cutoff of some kind; this taming
of infinities is known as {\it regularization.}

The game of regularization and renormalization
is perfectly capable of rendering quantum
field theory a successful calculation tool
for describing and predicting observed
processes.  Indeed, the program has proven
quite a triumph in the cases of quantum electrodynamics,
the electroweak theory and weakly coupled QCD,
as well as a mixed success in the description of
hadrons \glossary{hadrons} and their interactions.
However, many physicists are
unsatisfied with this state of affairs and
seek to understand the infinities which 
arise and how they might be avoided.  When one
studies the effects of the virtual particles,
it is found that the infinites may be traced
back to the contribution of virtual
particles with very large energies or momenta;
these are the \git{ultraviolet} divergences
of quantized field theories.\footnote{
The {\it infrared} \glossary{infrared} divergences which occur from
virtual particles with extremely low
energies or momenta are well-understood.}  The most
naive way to regularize a typical quantum
field theory is to impose an artificial
cutoff on the energies and momenta of
virtual particles.  One may imagine that
the particle content and interactions of
a theory has a limited range of
validity, depending on energy and momentum
scales.  One can further posit that the
quantum field theory used to calculate the
effects of virtual particles is beyond its
range of validity in precisely that region
where the effects become disturbingly large.
This would explain why bizarre results
are obtained in the calculation of
quantum corrections:  we have gone beyond
the limitations of our theory.  The question
then becomes, what {\it is} the correct
theory at these higher energies and momenta?
I will refer to this as the {\it underlying
theory.}

The underlying theory must satisfy certain
requirements.  First, it must have as its
low energy limit a quantized theory of fields
which includes quarks and leptons and their
interactions.  Second, it must be finite;
that is, we want a theory where artificial
cutoffs are no longer necessary and where
ultraviolet behavior is ``softened'' in a natural
way.  Third, it must provide
a quantum description of gravity.
Remarkably, a theory
exists which is believed to possess all
of these properties!
Actually, a few exist, all string
theories of one brand or another.  However,
they are all thought to be limits of
a more fundamental theory---the mysterious
{\it M-theory.}  In this work, I shall
chiefly be concerned with the
\git{weakly coupled} \git{heterotic}
string theory \cite{GHMR85}.
In its original construction,
this is a theory with ten space-time
dimensions.  The non-observation of
spatial dimensions other than the three
of the everyday world suggests that
the six extra dimensions, should they
really exist, be hidden from us in
some way.

The oldest way to accomplish
this goes back to \git{Kaluza-Klein theory} \cite{KKth},
where extra dimensions are made very
small and compact.  Difficulties arise in
obtaining \git{chiral fermions} in the effective
four-dimensional theory.  These difficulties
were surmounted at the field theory level
by Chapline et al.~\cite{CGS}
through compactification on \git{quotient
manifolds.}  Shortly after the invention
of the heterotic string, Candelas et al.~illustrated
how compactification of this type of string
on a Calabi-Yau manifold could
produce an effective theory with
{\it local} ${\it N=1}$ {\it supersymmetry}
and chiral fermions in four dimensions \cite{CHSW85}.
However, Calabi-Yau manifolds pose technical
difficulties for the explicit calculation of
many important quantities in the effective theory.
Concurrent to the intiation of the Calabi-Yau studies,
Dixon, Harvey, Vafa and Witten showed how to use of a certain
class of quotient spaces---orbifolds---to build
more calculable four-dimensional string
models \cite{DHVW85,DHVW86}.
Orbifolds are Euclidean except at a finite number of points.
This is a great simplification over Calabi-Yau manifolds.
Since the six-dimensional
orbifolds studied in this context may
be viewed as singular limits of certain Calabi-Yau
manifolds, we can presume that many of the more
{\it important features} of these manifolds can be
understood in this orbifold limit.  A comparison
of quantities which can be calculated in
both theories bears out the validity this presumption.

Orbifold compactifications
of heterotic string theory
are of interest because many aspects
of the theory are tractable
and because they can generate realistic models.
Like Calabi-Yau manifolds, orbifolds can generate
{\it local} ${\it N=1}$ {\it supersymmetry}
in four dimensions.
Supersymmetry protects the hierarchy between the
\git{Planck scale}
($m_P = 1/\sqrt{8 \pi G} = 2.44 \times 10^{18}$ GeV,\footnote{
A GeV is approximately the rest mass energy of a
proton or neutron.}
where $G$ is Newton's gravitational coupling)
and the {\it electoweak scale} ($m_Z = 91.19$ GeV,
where this is the mass of the {\it Z boson});
in a non-supersymmetric theory we
would expect this hierarchy to
be destabilized by the effects of virtual particles.
$N=1$ supersymmetry is desirable
to accomodate the standard model, a chiral theory with particles which
lie in complex representations
of gauge groups.
Local supersymmetry is required in order to have
a realistic mass spectrum for the states beyond
those contained in the Standard Model (SM),
the {\it superpartners} to the quarks, leptons
and gauge bosons.  This is because a realistic
spectrum may only be obtained by so-called
\git{soft supersymmetry breaking} in the low
energy effective theory, which is achieved by the
spontaneous breaking of local supersymmetry in an
effective supergravity theory valid at higher energies.

In this thesis, I will restrict my attention
to the $Z_3$ orbifold, which is often referred
to simply as ``the $Z$ orbifold.''  It is the
canonical example and has been studied extensively,
yet many of the more detailed
issues of this construction of a four-dimensional
heterotic theory remain
uninvestigated.  Several aspects of the $Z_3$ orbifold make
it simpler than other orbifold constructions.
To explain these would require jargon which
will be defined below;
I will not enumerate
them here but will note these simplifying features
as they arise in the discussion which follows.

\newpage

\renewcommand{\thepage}{\arabic{page}}
\setcounter{page}{1}
\def\thefootnote{\arabic{footnote}}
\setcounter{footnote}{0}

%

\chapter{Orbifold geometry}
\label{oge}
In this chapter I discuss the geometry of
\git{orbifolds.}  I begin in Section \ref{1do} with a very simple (and
currently popular) example, the one-dimensional orbifold.
Next, I look at the two-dimensional case in Section \ref{2do}.
Finally, I arrive in Section \ref{6do}
at the orbifold which will concern us throughout
the remainder of this work, the $Z_3$ orbifold.
Thus I build gradually to the six-dimensional construction, so that
key concepts are introduced in simpler one- and two-dimensional
examples.

\section{One-dimensional Orbifold}
\label{1do}
I begin with the simplest orbifold
that can be constructed.  It provides an
introduction to some of the concepts, 
terminology and notation common to 
the discussion of orbifolds.

Let $\Rbf$ be the real number line.
Define a one-dimensional lattice $\Lambda$ with
lattice spacing $a \in \Rbf$:
\beq
\Lambda = \left\{ \; n a \; | \; n \in \Zbf \; \right\},
\label{1dlt}
\eeq
where $\Zbf$ is the set of integers.
The elements $\ell \in \Lambda$ will
be referred to as \git{lattice vectors.}
We construct a \git{torus} by identifying 
points on the line with each other
if they are related under addition
by a lattice vector:
\beq
x \simeq x + \ell \qquad 
\forall \; x \in \Rbf, \, \ell \in \Lambda.
\label{1dte}
\eeq
This generates a torus whose \git{fundamental domain} can
be chosen
as $[0,a)$.  That is, any other point in
$\Rbf$ maps into this domain by the identification
made in \myref{1dte}; $\Rbf$ is 
the \git{covering space} for the torus.

For example, the points labeled ``x'' in Figure \ref{1dtf}
are equivalent on the torus.  
The one-dimensional torus is topologically equivalent
(more precisely, \git{homeomorphic}) to a circle.
We notate this construction
\beq
\Tbf = \Rbf / \Lambda .
\label{jstg}
\eeq
The torus $\Tbf$ is \git{compact} while the
real number line $\Rbf$ is \git{non-compact.}

An ``active attitude'' may be taken:
Eq.~\myref{1dte} states that points which
are related to each other
by a lattice vector translation
are equivalent to each other.
The discrete group of translations
defined by the lattice is referred
to as the \git{lattice group}, which
is often also denoted $\Lambda$.
The lattice group is
an invariance or \git{isometry group} of $\Rbf$.
This terminology leads to a brief
description of the torus construction:
we ``divide out'' or ``mod out'' the lattice
group $\Lambda$ from the space $\Rbf$.  
The lattice group affords an
equivalence relation $r_\Lambda$,
defined by \myref{1dte}, which
partitions the real number line into a set
of \git{equivalence classes;} an equivalence
class is a set of elements which are
all equivalent to each other.  The set of
equivalence classes is called
the \git{quotient set} or \git{quotient space}
determined by $r_\Lambda$ and is denoted by
${\bf R}/r_\Lambda$.  However, most
people use the shorthand ${\bf R}/\Lambda$,
as in \myref{jstg}.  Since
$\Rbf$ and $\Lambda$ are groups, it is also
correct to refer to ${\bf R}/\Lambda$ as
a \git{coset space}.  Each element
in the fundamental domain $[0,a)$ is in one-to-one
correspondence with an equivalence
class contained in ${\bf R}/\Lambda$.  Given
$x \in [0,a)$ we can reach (generate) every element
in the equivalence class corresponding to $x$ by the
action of the lattice group $\Lambda$ 
on $x$.

\begin{figure}[h]
\begin{center}
\unitlength=1mm
\begin{picture}(100,30)
\put(10,20){\line(1,0){80}}
\put(20,15){\makebox(0,0){$-a$}}
\put(50,15){\makebox(0,0){$0$}}
\put(80,15){\makebox(0,0){$a$}}
\put(20,19){\line(0,1){2}}
\put(50,19){\line(0,1){2}}
\put(80,19){\line(0,1){2}}
\put(23,20){\makebox(0,0){x}}
\put(53,20){\makebox(0,0){x}}
\put(83,20){\makebox(0,0){x}}
\end{picture}
\end{center}
\caption{The torus $\Tbf$ embedded into the
covering space $\Rbf$.}
\label{1dtf}
\end{figure}

A \git{toroidal orbifold} may be constructed from a torus by
supplementing \myref{1dte} with
other equivalence relations.
A \git{twist operator} is used to define
each equivalence relation.  For the
toroidal orbifolds considered here,
the twist must be an \git{automorphism}
of the lattice used in constructing
the torus.  An automorphism $\theta$
of a lattice $\Lambda$ is a transformation which
maps lattice vectors into lattice vectors:
\beq
\theta \ell \in \Lambda, \qquad \forall \; \ell \in \Lambda.
\label{azyy}
\eeq
Twist operators are most often generators of
a discrete rotation group on the compact \git{manifold}
which is to be ``twisted'' into an orbifold,
typically a torus.

In the simplified one-dimensional case that I present here,
rotation is not well defined.  
Instead, I take the twist 
operator to be the \git{parity} operator ${\bf P}$:
\beq
\Pbf x = -x, \qquad \forall \; x \in \Rbf.
\eeq
The parity operation
is an automorphism of the lattice $\Lambda$
since it satisfies \myref{azyy}.
It is very simple to see that this is the
case.  Any lattice vector may be written
as $n a$, where $n$ is an integer and $a$
is the lattice spacing.  Applying the parity
operation, 
\beq
\Pbf (n a) = -(n a) = (-n) a.
\label{rsim}
\eeq
But since $-n$ is also an integer, it can
be seen that the right-hand side of \myref{rsim}
is a lattice vector of $\Lambda$ too.  

The equivalence relation generated
by ${\bf P}$ is
\beq
x \simeq {\bf P} x,  \, \qquad \forall \; x \in {\bf R} .
\label{1doe}
\eeq
Notice that ${\bf P}^2=1$.  Thus, ${\bf P}$ realizes
the cyclic group of order two,
commonly referred to as $Z_2$.  The group generated
by the twist operators is known as the \git{point group}
of the orbifold.
We then have as the point group for our
simple one-dimensional orbifold
\beq
Z_2 = \{ 1, \Pbf \} .
\eeq
The orbifold we have constructed,
denoted $\Omega(1,Z_2)$, is the
quotient space $\Tbf /Z_2$.  

It is very common to define operators which
combine the action of the point group $Z_2$ with
the lattice group $\Lambda$:
\beq
(\omega,\ell) x = \omega x + \ell,
\qquad \forall \; x \in \Rbf,
\eeq
where $\omega \in Z_2, \; \ell \in \Lambda$.
The collection of all such operators
forms a group $S$ known as the \git{space group}
of the orbifold.
It is not difficult to check that these
operators have the multiplication
rule
\beq
(\omega_1 , \ell_1)  (\omega_2, \ell_2)
= (\omega_1 \omega_2, \ell_1 + \omega_1 \ell_2).
\label{yati}
\eeq
An isomorphism of the point group is a bijective map $g$
which satisfies\myfoot{Double-check defn.}
\beq
g(\omega_1)  g(\omega_2) = g(\omega_1 \omega_2).
\eeq
Similarly an isomorphism of the lattice group is a bijective
map $h$ which satisfies
\beq
h(\ell_1)  h(\ell_2) = h(\ell_1 + \ell_2).
\eeq
The projection of the space group onto the point
group defined by
\beq
\pi_1(\omega,\ell) \cdot x = \omega x,
\qquad \forall \; x \in \Rbf,
\eeq
is a homomorphism to the point group, as can be seen
from \myref{yati}.  On the other hand the projection
of the space group onto the lattice group defined by
\beq
\pi_2(\omega,\ell) \cdot x = x + \ell,
\qquad \forall \; x \in \Rbf,
\eeq
is not a homorphism to the lattice group.  From \myref{yati}
we have
\beq
\pi_2((\omega_1,\ell_1)(\omega_2,\ell_2)) \cdot x = x +
\ell_1 + \omega_1 \ell_2, \qquad \forall \; x \in \Rbf,
\eeq
whereas
\beq
\( \pi_2(\omega_1,\ell_1) \circ \pi_2(\omega_2,\ell_2)\) \cdot x
= x + \ell_1 + \ell_2,
\qquad \forall \; x \in \Rbf.
\eeq
For this reason, the space group
is the \git{semi-direct product} of
the point group and the lattice group.

We have so far constructed the one-dimensional
orbifold in a two step process, imposing
the equivalence \myref{1dte} and 
then equivalence \myref{1doe}.  The
space group $S$ affords a more compact
description:
\beq
x \simeq g x, \qquad \forall \; x \in \Rbf,
\; g \in S .
\eeq
The orbifold may be denoted $\Omega(1,Z_2)=\Rbf / S$.

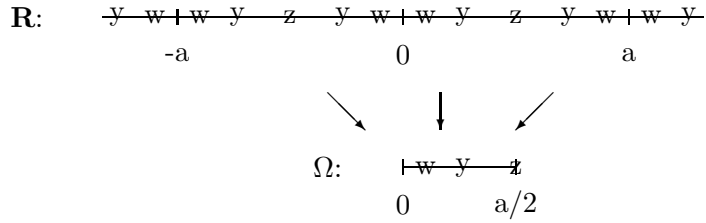
\begin{figure}[h]
\begin{center}
\unitlength=1mm
\begin{picture}(100,40)
\put(0,30){\makebox(0,0){$\Rbf$:}}
\put(10,30){\line(1,0){80}}
\put(20,25){\makebox(0,0){-a}}
\put(50,25){\makebox(0,0){0}}
\put(80,25){\makebox(0,0){a}}
\put(20,29){\line(0,1){2}}
\put(50,29){\line(0,1){2}}
\put(80,29){\line(0,1){2}}
\put(23,30){\makebox(0,0){w}}
\put(53,30){\makebox(0,0){w}}
\put(83,30){\makebox(0,0){w}}
\put(17,30){\makebox(0,0){w}}
\put(47,30){\makebox(0,0){w}}
\put(77,30){\makebox(0,0){w}}
\put(28,30){\makebox(0,0){y}}
\put(58,30){\makebox(0,0){y}}
\put(88,30){\makebox(0,0){y}}
\put(12,30){\makebox(0,0){y}}
\put(42,30){\makebox(0,0){y}}
\put(72,30){\makebox(0,0){y}}
\put(35,30){\makebox(0,0){z}}
\put(65,30){\makebox(0,0){z}}
\put(55,20){\vector(0,-1){5}}
\put(70,20){\vector(-1,-1){5}}
\put(40,20){\vector(1,-1){5}}
\put(40,10){\makebox(0,0){$\Omega$:}}
\put(50,10){\line(1,0){15}}
\put(50,9){\line(0,1){2}}
\put(65,9){\line(0,1){2}}
\put(50,5){\makebox(0,0){0}}
\put(65,5){\makebox(0,0){a/2}}
\put(53,10){\makebox(0,0){w}}
\put(58,10){\makebox(0,0){y}}
\put(65,10){\makebox(0,0){z}}
\end{picture}
\end{center}
\caption{The orbifold ${\bf T}/Z_2$
and its embedding into the covering space $\Rbf$.
Points marked with the
same letter are equivalent.}
\label{1dof}
\end{figure}

In Figure \myref{1dof} I illustrate
the orbifold.  All points marked with
the same letter are equivalent.  Note that
in the fundamental domain $[0,a)$ of the
torus we now have pairs of points which
are equivalent, except for the \git{fixed points}
$x=0$ and $x=a/2$.  (A proper definition of
fixed points will
be given below.)  On the other hand,
the fundamental domain of
the orbifold is $[0,a/2]$, for every other
point in $\Rbf$ may be mapped into this interval.

Note that we can ``fit'' a local
one-dimensional coordinate system,
commonly referred to as the \git{tangent space}, at points
$w$ and $y$.  However, things become confused at the fixed
points, such as $z$.  On the covering space,
both directions leaving $z$ are equivalent;
we cannot fit a well-defined
tangent space in the neighborhood of $z$.
This ``difficulty'' is a general feature
of orbifold fixed points; it distinguishes
orbifolds from manifolds.

\begin{figure}[h]
\begin{center}
\unitlength=1mm
\begin{picture}(100,25)
\put(30,5){\line(-1,0){25}}
\put(30,5){\line(-2,1){23}}
\put(35,5){\makebox(0,0){a/2}}
\put(0,5){\makebox(0,0){0}}
\put(0,17){\makebox(0,0){a}}
\put(7,15){\vector(-1,-4){2}}
\put(43,5){\vector(1,0){10}}
\put(65,5){\line(1,0){25}}
\put(95,5){\makebox(0,0){a/2}}
\put(62,5){\makebox(0,0){0}}
\end{picture}
\end{center}
\caption{Fold the fundamental domain
of the one-dimensional torus to form
the one-dimensional orbifold.  Points
which overlap are equivalent.}
\label{1dfold}
\end{figure}
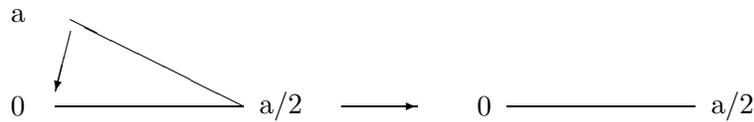

One may imagine folding the fundamental
domain $[0,a)$ of the torus at $x=a/2$
and bending it back over so the two
half-segments overlap and the ends of
the fundamental domain touch, as shown
in Figure \ref{1dfold}.  Points which
overlap are equivalent.

\begin{figure}[h]
\begin{center}
\includegraphics[height=5.0in,width=1.0in,
angle= -90, bb = 275 100 330 700, clip]{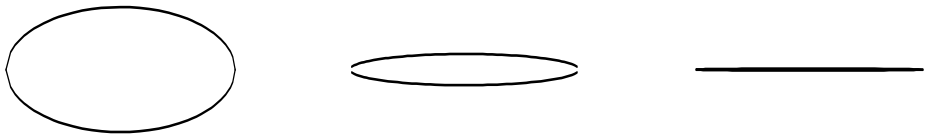}
\end{center}
\caption{Ellipse collapsing to segment under identification.}
\label{ellf}
\end{figure}

The torus may be pictured as an ellipse
in the two-dimensional plane.  
We may think of the orbifold construction
as an ellipse whose eccentricity $\e$
is taken to the limit $\e \to 1$ (the minor
axis approaches vanishing length relative
to the length of the major axis) while
the length of the major axis is held fixed;
see Figure \ref{ellf}.
Thus, we can no longer resolve the
``top'' of the ellipse from the ``bottom.''  The
points $x=0$ and $x=a/2$ coincide with the ends
of the major axis in this picture, and one
can see intuitively that they are in a 
certain sense singular.

\begin{figure}[h]
\begin{center}
\includegraphics[height=5.0in,width=2.0in,
angle= -90, bb = 250 200 350 650, clip]{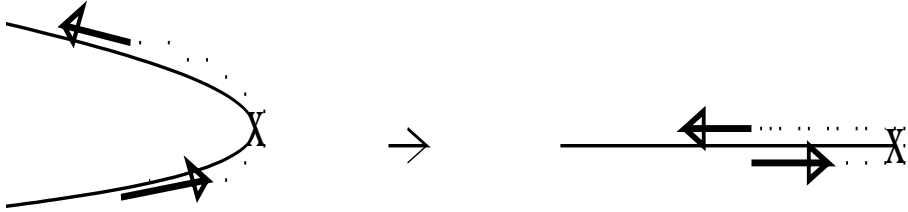}
\end{center}
\caption{Parallel transport about a fixed point in
the one-dimensional orbifold leads to a rotation
by $\pi$, regardless of how small the loop becomes.
Curvature at the fixed point is infinite.}
\label{1dsin}
\end{figure}

Specifically, let us consider motion on the ellipse which
covers the orbifold.  In the vicinity of the fixed
points, the corresponding motion on the orbifold
has an interesting behavior.
Keeping the ellipse with shrinking minor axis
in mind, we imagine parallel transporting
a vector about the fixed point, which has
been marked by an ``x'' in the Figure \ref{1dsin}.
The starting and ending orientations are indicated.
On the orbifold, the starting point and the ending
point are equivalent.  Thus, the path is a closed ``loop''
on our singular space.
It can be seen that from the orbifold perspective,
the vector has undergone a rotation $\Delta \theta = \pi$ as
one passes ``around'' the fixed point.  This
is independent of how small a loop we take.
The curvature $\kappa(s)$ of a curve\footnote{See
for example \cite{MP77}.}
is the magnitude of the
rate of change in the unit tangent
vector ${\bf T}(s)$ with respect to path length $s$:
\beq
\kappa(s) = \left| {d {\bf T}(s) \over ds} \right|.
\eeq
About the fixed point we have
\beq
\kappa \propto \lim_{\Delta s \to 0} {\Delta \theta
\over \Delta s} =
\lim_{\Delta s \to 0} {\pi
\over \Delta s} = \infty .
\eeq
We see that the fixed point is a point of infinite
curvature.  An orbifold is not a manifold because
of the existence of curvature singularities
at the fixed points.

Let us examine these points in greater detail
and see in what sense they are ``fixed.''
Note first that $\Pbf x = x$ if $x=0$.  That is,
$x=0$ is neutral under the action of the point
group.  Next note that 
\beq
\Pbf (a/2) = -a/2 = a/2 - a,
\eeq
so that $\Pbf (a/2) \simeq a/2$ under the
action of the lattice group, \myref{1dte}.
This leads us to the general definition
of a fixed point.
Let $\omega$ be an element of the point group
of an orbifold.  Then a \git{fixed point}
of the operator $\omega$ is a solution $x_f$ to
\beq
\omega x_f + \ell = x_f,
\label{pgfp}
\eeq
with $\ell$ some element of the lattice $\Lambda$.

\egbox{
The points $x=0$ and
$x=a/2$ are the fixed points contained
in the fundamental domain of our orbifold.
\label{aaa2}
}

All other fixed points (integral and
half-integral multiples of $a$) are
equivalent to one of these two fixed
points under the action of the lattice
group.  In this sense, our orbifold has
``two'' fixed points; the correct
statment is that the orbifold possesses
precisely two inequivalent fixed points.
Note that \myref{pgfp}
may be written more succinctly as neutrality
with respect to a space group element:
\beq
(\omega,\ell) \cdot x_f = x_f.
\label{aaa1}
\eeq
Through \myref{aaa1}
each pair $(\omega,\ell)$
can be put into one-to-one correspondence
with an element $x_f$ of the covering
space $\Rbf$.  Each of these elements is a fixed point of
the twist operator $\omega$ on
the orbifold.  That is to say, given a
pair $(\omega,\ell)$, the solution
$x_f \in \Rbf$ to \myref{aaa1} is unique
and always exists.

\egbox{
In Example \ref{aaa2}, we saw that the fixed point
$x_f = 0$ corresponded to the lattice
vector $\ell = 0$ while the fixed point
$x_f = a/2$ corresponded to the lattice
vector $\ell = a$.
}

More generally,
\beq
(\Pbf, na) {n \over 2} a = {n \over 2} a,
\qquad \forall \; n \in \Zbf .
\eeq
Thus, the fixed point $na/2$ corresponds
to the lattice vector $na$.  Since all
fixed points with $n$ even are equivalent
to $x_f=0$, this fixed point corresponds
to the sublattice spanned by lattice 
vectors $na$ with $n$ even.  Similarly,
the fixed point $x_f=a/2$ corresponds to
the sublattice spanned by lattice
vectors $na$ with $n$ odd.

\section{Two-dimensional Orbifold}
\label{2do}
The next example is a generalization of the last.
We extend to the two-dimensional real manifold
$\Rbf^2$ and mod out by a lattice generated
by linearly independent elements $e_1,e_2 \in \Rbf$:
\beq
\Lambda = \left\{ \; \sum_{i=1}^2 m^i e_i \; \left|
\; m^1, m^2 \in {\bf Z} \; \right. \right\}
\label{2dtl}
\eeq
The basis vectors characterize the shape
and size of the lattice {\it via}
\beq
\sqrt{e_1 \cdot e_1} = a_1, \qquad
\sqrt{e_2 \cdot e_2} = a_2, \qquad
e_1 \cdot e_2 = a_1 a_2 \cos \alpha.
\eeq
The torus described by $\Tbf^2 = \Rbf^2/\Lambda$
is obtained by imposing equivalence relations
\beq
x \simeq x + \ell, \qquad \forall \; x \in \Rbf^2,
\, \ell \in \Lambda.
\label{2dte}
\eeq
We again define the twist operator to be the parity
operation $\Pbf \cdot x = -x$ and impose the identification
\beq
x \simeq {\bf P} \cdot x \, \qquad \forall \; x \in \Tbf^2
\label{1aaa}
\eeq
to construct the orbifold $\Omega(2,Z_2)=\Tbf^2/Z_2$.  
We note that $\Pbf$ is equivalent to a
rotation by angle $\pi$.  Thus, the point
group is a discrete subgroup of the full rotation
group $O(2)$ of the real manifold $\Rbf^2$; this
is the usual circumstance in toroidal orbifolds,
and will be true in the six-dimensional $Z_3$
case considered in Section \ref{6do}.
It is easy to check that \myref{1aaa} is an automorphism
of the lattice \myref{2dtl}.  As discussed in
the one-dimensional example of
Section \ref{1do} this is a necessary condition
for the consistency of the orbifold construction.

\begin{figure}[hp!]
\begin{center}
\unitlength=1mm
\begin{picture}(100,100)
\thicklines
\put(0,20){\line(1,0){100}}
\put(0,50){\line(1,0){100}}
\put(0,80){\line(1,0){100}}
\put(20,0){\line(0,1){100}}
\put(50,0){\line(0,1){100}}
\put(80,0){\line(0,1){100}}
\thinlines
\put(20,20){\line(1,1){60}}
\put(35,20){\line(1,2){30}}
\put(65,50){\line(1,2){15}}
\put(50,50){\circle*{3.0}}
\put(50,65){\circle*{3.0}}
\put(65,50){\circle*{3.0}}
\put(65,65){\circle*{3.0}}
\put(15,50){\makebox(0,0){x}}
\put(25,50){\makebox(0,0){x}}
\put(45,50){\makebox(0,0){x}}
\put(55,50){\makebox(0,0){x}}
\put(75,50){\makebox(0,0){x}}
\put(85,50){\makebox(0,0){x}}
\put(15,80){\makebox(0,0){x}}
\put(25,80){\makebox(0,0){x}}
\put(45,80){\makebox(0,0){x}}
\put(55,80){\makebox(0,0){x}}
\put(75,80){\makebox(0,0){x}}
\put(85,80){\makebox(0,0){x}}
\put(10,50){\makebox(0,0){y}}
\put(30,50){\makebox(0,0){y}}
\put(40,50){\makebox(0,0){y}}
\put(60,50){\makebox(0,0){y}}
\put(70,50){\makebox(0,0){y}}
\put(90,50){\makebox(0,0){y}}
\put(10,80){\makebox(0,0){y}}
\put(30,80){\makebox(0,0){y}}
\put(40,80){\makebox(0,0){y}}
\put(60,80){\makebox(0,0){y}}
\put(70,80){\makebox(0,0){y}}
\put(90,80){\makebox(0,0){y}}
\put(50,15){\makebox(0,0){d}}
\put(50,25){\makebox(0,0){d}}
\put(50,45){\makebox(0,0){d}}
\put(50,55){\makebox(0,0){d}}
\put(50,75){\makebox(0,0){d}}
\put(50,85){\makebox(0,0){d}}
\put(50,10){\makebox(0,0){e}}
\put(50,30){\makebox(0,0){e}}
\put(50,40){\makebox(0,0){e}}
\put(50,60){\makebox(0,0){e}}
\put(50,70){\makebox(0,0){e}}
\put(50,90){\makebox(0,0){e}}
\put(55,55){\makebox(0,0){b}}
\put(75,75){\makebox(0,0){b}}
\put(45,45){\makebox(0,0){b}}
\put(60,60){\makebox(0,0){c}}
\put(70,70){\makebox(0,0){c}}
\put(40,40){\makebox(0,0){c}}
\put(54,58){\makebox(0,0){r}}
\put(58,66){\makebox(0,0){s}}
\put(62,74){\makebox(0,0){t}}
\put(46,42){\makebox(0,0){r}}
\put(42,34){\makebox(0,0){s}}
\put(38,26){\makebox(0,0){t}}
\put(76,72){\makebox(0,0){r}}
\put(72,64){\makebox(0,0){s}}
\put(68,56){\makebox(0,0){t}}
\end{picture}
\end{center}
\caption{The orbifold ${\bf T}^2 / Z_2$,
for the special case of $a_1 = a_2, \, \alpha = 0$.
Points marked
with the same letter are equivalent.  The fundamental
region of the torus is $[0,1) \times [0,1)$.  Four fixed points
exist in this region, marked by solid dots.}
\label{t2z2fig}
\end{figure}
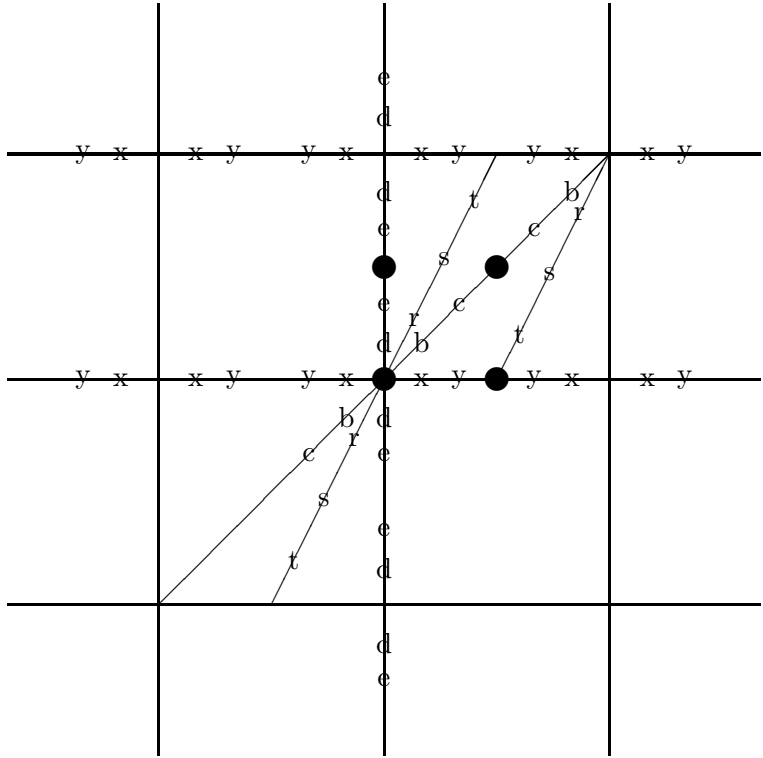

\begin{figure}[hp!]
\begin{center}
\includegraphics[height=7.0in,width=6.0in,angle=0]{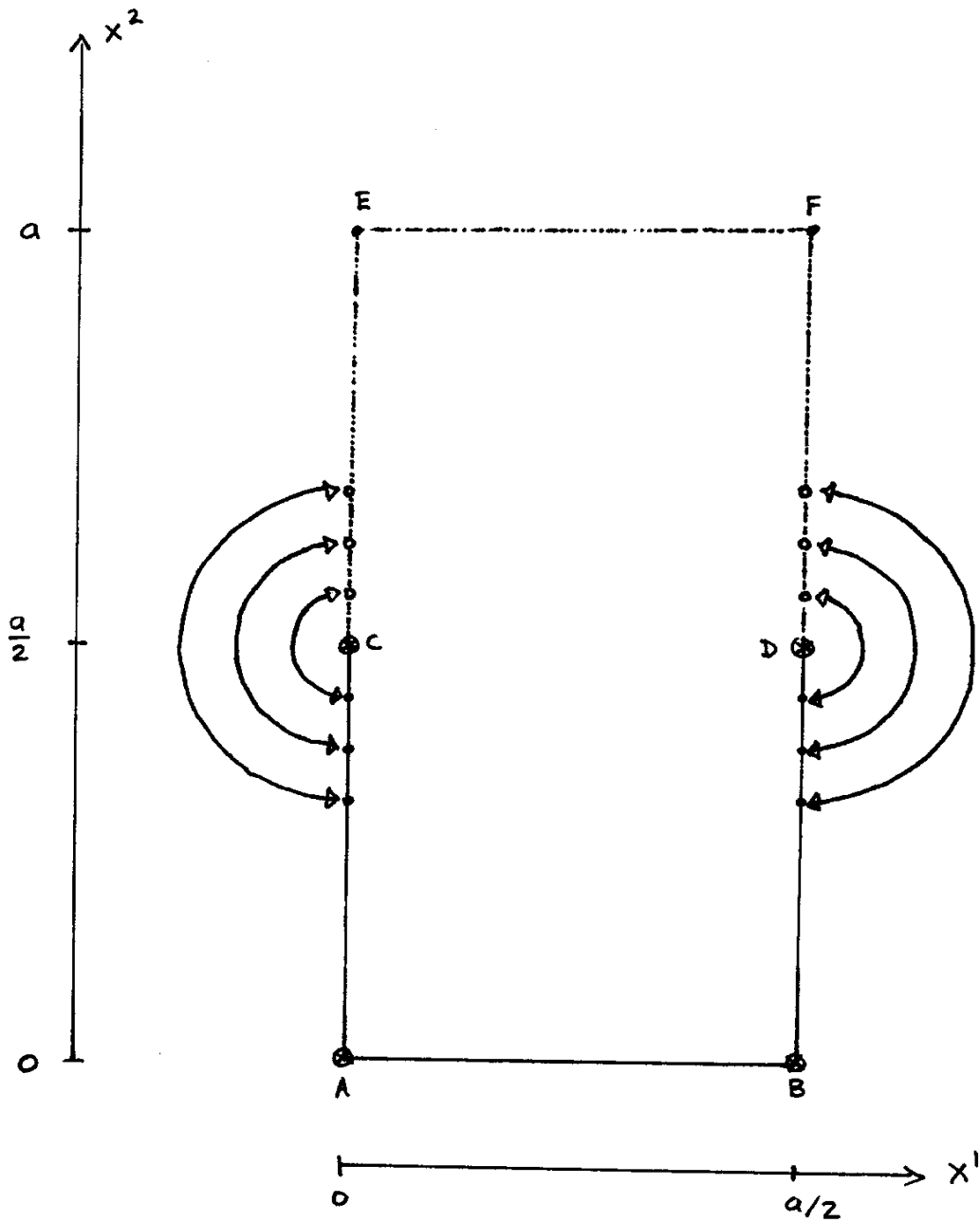}
\end{center}
\caption{${\cal F}$ for the $Z_2$ orbifold.  Open points
along the open boundary (dotted) are identified (examples shown
by arrows) with closed points
along the closed boundary (solid), forming a ``pillow.''}
\label{z2f2}
\end{figure}

The orbifold, embedded into the covering
space $\Rbf^2$, is depicted in Figure \ref{t2z2fig}
for the special case of $a_1 = a_2 \equiv a, \, \alpha = 0$.
Locating equivalent points is now a bit
more complicated, but with a brief study one
can convince oneself that points labeled
with the same letter are related---by a combination
of operations (\ref{2dte},\ref{1aaa}).
With a bit more effort, one should be able to convince
oneself that a valid choice for the fundamental domain
${\cal F} \subset \Rbf^2$ of the orbifold is
\beq
{\cal F} = [0,a/2] \times [0,a/2]
+ (0,a/2) \times (a/2,a) .
\eeq
The notation here specifies a sum of two squares $M_1 \times M_2$,
where $M_1$ is an interval parallel to the $x^1$-axis
and $M_2$ is and interval parallel to the $x^2$-axis.
For greater clarity,
this choice of ${\cal F}$ is shown in Figure \ref{z2f2}.
The limit points along the open boundary are identified
with points on the closed boundary as indicated by
arrows.
That is: $\bbar{AC}$ is sewn to $\bbar{CE}$; $\bbar{AB}$
is sewn to $\bbar{EF}$; and, $\bbar{BD}$ is sewn to $\bbar{DF}$.
This is done such that $A$ and $E$ connect, as do $B$ and $F$.
Thus on the orbifold the edges are sewn together
about $x^2 = a/2$ to form a four-cornered ``pillow.''
The space is clearly compact, and has four ``corners'' where the
space looks locally like a cone; at each one of these
corners there is a point where the curvature is singular.  These
four conical singularities are the fixed points of
the orbifold:  $A, \, B, \, C$ and $D$.

These four inequivalent fixed points can
also be found by the algebraic method.
It is easy to check that (using the space
group notation)
\beq
(\Pbf, m^i e_i) {m^j e_j \over 2}
= {m^j e_j \over 2} 
\eeq
for all pairs of integers $(m^1,m^2)$.
Thus, the fixed points of the orbifold
are given by $m^i e_i /2$.
The lattice vectors given by
\beq
(m^1,m^2) \in \{ (0,0), \; (0,1), \; (1,0), \; (1,1) \}
\label{fppa}
\eeq
give fixed points $m^i e_i /2$ which
are inequivalent to each other and which are
related to all other fixed points 
by a lattice group equivalence \myref{2dte}.

\section[The Six-dimensional $Z_3$ Orbifold]
{The Six-dimensional ${\bf Z_3}$ Orbifold}
\label{6do}
Toroidal orbifolds of dimension larger than those
so far considered are mere generalizations
of the two-dimensional construction just described.  However,
the increase in dimensionality allows for
many more possibilities.  We will be concerned
with six-dimensional orbifolds in the applications
considered in subsequent chapters.  This is because
the heterotic string theory, as originally formulated,
has nine spatial dimensions.  To construct an
effective theory with only the three spatial
dimensions we observe, it is necessary to
somehow hide the extra six.  The standard
approach to this is to make the six extra
dimensions compact and very small---a
characteristic length on the order of
$8 \times 10^{-33}$ centimeters!
For reasons which will be explained in later
chapters, promising models follow from the
assumption that the six-dimensional
compact space is an orbifold.  In this
thesis I concentrate on the possibility that
it is a $Z_3$ orbifold.

\subsection{Construction}
\label{abas}
The six-dimensional $Z_3$ orbifold
may be constructed from a six-dimensional Euclidean
space $\Rbf^6$.  One defines basis vectors
$e_1, \ldots, e_6$ satisfying
\beq
e_i^2 = e_{i+1}^2 = 2 R_i^2, 
\qquad e_i \cdot e_{i+1} = -1 R_i^2,
\qquad i=1,3,5,
\label{2f}
\eeq
with a vector $x \in \Rbf^6$
having real-valued components:
\beq
x = \sum_{i=1}^6 x^i e_i ,
\qquad x^i \in \Rbf \quad \forall \;
i = 1, \ldots, 6.
\label{comp}
\eeq
Note that $x^i \not= x \cdot e_i$ since
the root basis \myref{2f} is a skew basis consisting of
elements which do not have unit norm.
Each of the three pairs $e_i,e_{i+1} \; (i=1,3,5)$
define a two-dimensional subspace which
is referred to below as the ``$i$th complex plane.''
The $i$th such pair also defines
a two-dimensional $SU(3)$ root lattice, obtained from
the set of all linear combinations of the
form $n_i e_i + n_{i+1} e_{i+1}$ with
$n_i,n_{i+1}$ both integers.
Taking together all six basis vectors $e_1,\ldots,e_6$,
we obtain the $SU(3)^3$ root lattice $\LamTh$,
formed from all linear combinations of the
basis vectors $e_1, \ldots, e_6$ with integer
coefficients:
\beq
\LamTh = \left\{ \left. \; \sum_{i=1}^6 \ell^i e_i \; \right|
\; \ell^i \in \Zbf \; \right\} .
\eeq
Note that the \git{radii} $R_i$ in \myref{2f} are
not fixed; neither are angles not appearing in
\myref{2f}, such as $e_1 \cdot e_3$.  These free
parameters determine the size and shape of the
unit cell of the lattice $\LamTh$, and are
encoded in {\it K\"ahler-} \glossary{K\"ahler-moduli} or
\git{T-moduli} $T^{ij}$.
These moduli depend on
the metric $G_{ij}=e_i \cdot e_j$ ($i,j=1,\ldots,6$)
of the six-dimensional compact
space, as well as an antisymmetric two-form
$B_{ij}$.  Of particular interest are the
\git{diagonal} T-moduli $T^i \equiv T^{ii}$.
Up to normalization conventions on the
$T^i$ and $B_{ij}$,
the diagonal T-moduli are defined by
\beq
T^i = \sqrt{\det G^{(i)}} + i B_{i,i+1}, \quad i=1,3,5.
\eeq
Here, $G^{(i)}$ is the metric of the $i$th
complex plane:
\beq
G^{(i)} = \pmatrix{ e_i \cdot e_i & e_i \cdot e_{i+1} \cr
e_{i+1} \cdot e_i & e_{i+1} \cdot e_{i+1} \cr}
= R_i^2 \pmatrix{2 & -1 \cr -1 & 2 \cr}.
\eeq

\begin{figure}[h!]
\begin{center}
\includegraphics[height=4.0in,width=6.0in]{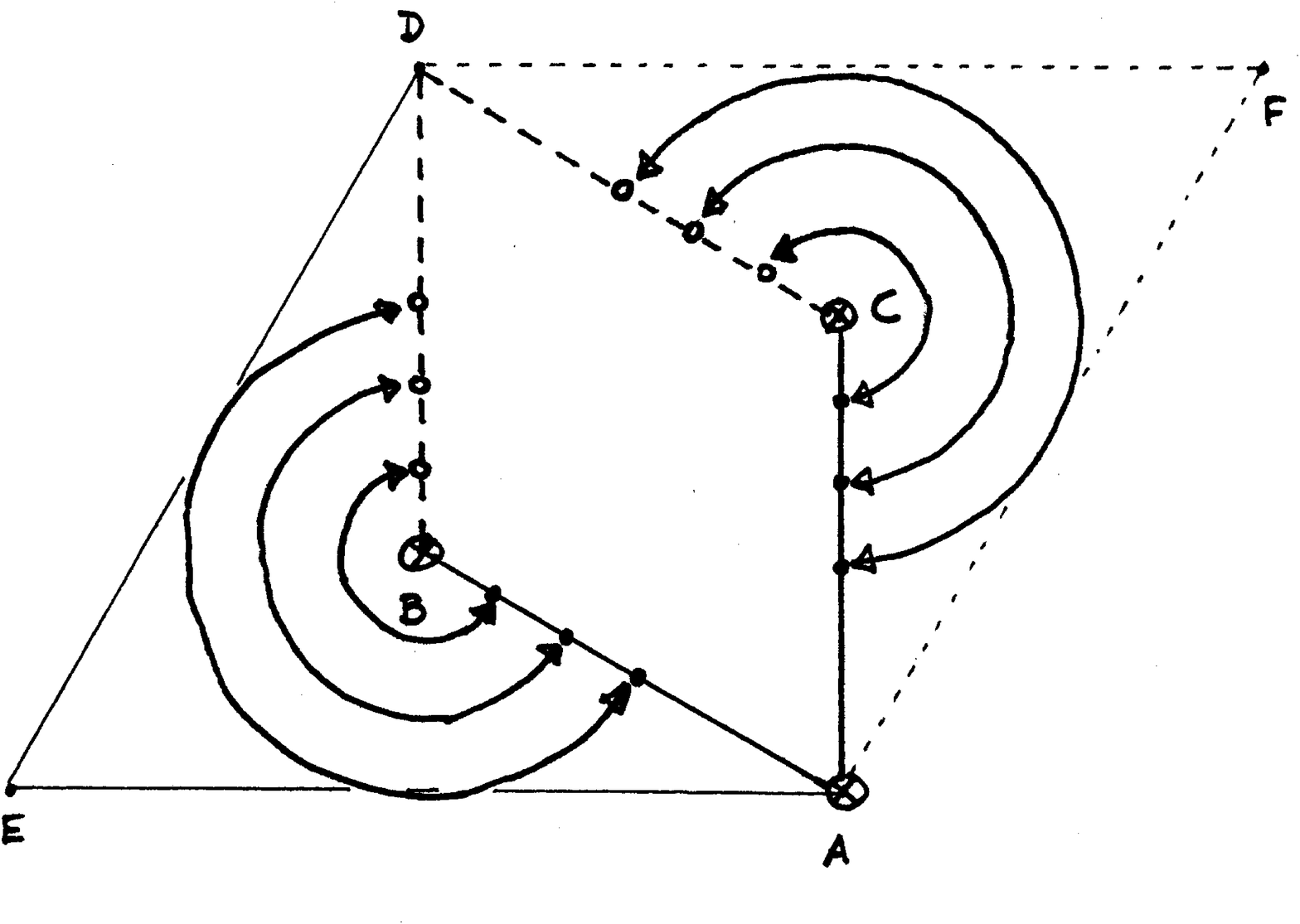}
\end{center}
\caption{The parallelogram $\bbar{BACDB}$ depicts
the two-dimensional $Z_3$ orbifold.
Also shown is how it sits within
the $SU(3)$ root torus (the larger parallelogram).
Here I take $E=0$, $A=1$, $D= e^{i\pi/3}$,
$F=A+D$, $B=F/3$ and $C=2F/3$.
Open points
along the open boundary (dashed) are identified (examples shown
by arrows) with closed points
along the closed boundary (solid), forming a
three-cornered ``pillow.''
The fixed points are at $A$, $B$ and $C$.
We can also view
the parallelogram $\bbar{BACDB}$ as the projection
of the six-dimensional $Z_3$ orbifold
into one of the three complex planes.
Similarly, the larger parallelogram can be viewed
as the projection of the $SU(3)^3$ root torus
in one of the three complex planes.}
\label{z3f1}
\end{figure}

Translations in $\Rbf^6$ by elements of
$\LamTh$,
\beq
x \to x + \ell, \qquad \ell \in \LamTh,
\qquad \forall \; x \in \Rbf^6,
\eeq
form the lattice group; thus we obtain the
six-dimensional torus $\Tbf^6 = \Rbf^6/\LamTh$.
A suitable choice for the fundamental
domain of this torus in any one of the three
complex planes is given by the parallelogram
$\bbar{DEAFD}$ of Figure \ref{z3f1}.  (The
interpretation of the figure is more transparent
in terms of the {\it complex basis} which
will be introduced in Section \ref{cbss} below.)

The twist operator $\theta$ is a simultaneous $2\pi/3$ rotation
of each of the three complex planes.
Its action on the basis vectors is
\beq
\theta \cdot e_i = e_{i+1}, \qquad \theta \cdot e_{i+1} = - e_i - e_{i+1},
\qquad i=1,3,5.
\label{pnh}
\eeq
It is easy to check that $\theta^3 =1$.
The twist operator $\theta$ generates the orbifold point group,
\beq
Z_3 = \{ 1, \theta, \theta^2 \}.
\label{pni}
\eeq
It can be seen from \myref{pnh} that the
twist operator maps any element of $\LamTh$ into $\LamTh$.
Consequently, we can define the product group
generated by the combined action of the
point group and the lattice group---
the space group $S$.  As in the one- and two-dimensional
examples considered above,
a generic element is written $(\omega,\ell)$,
with $\omega \in Z_3$ and $\ell \in \LamTh$.
The space group has four generators:
$(\theta,0), \; (1,e_1), \; (1,e_3)$ and $(1,e_5)$.

\egbox{
Using \myref{pnh} and the space
group multiplication rule \myref{yati} one can write
\beq
(1,e_2) = (\theta,0) \cdot (1,e_1) \cdot (\theta,0) \cdot(\theta,0).
\eeq
}

Acting on any element $x \in \Rbf^6$,
\beq
(\omega,\ell) \cdot x = \omega \cdot x + \ell
= \sum_{i=1}^6 [x^i (\omega \cdot e_i) + \ell^i e_i],
\eeq
where $\omega \cdot e_i$ can be
obtained by (repeated) application of \myref{pnh}.

\egbox{
The pure twist element $(\theta,0)$
transforms $x \in \Rbf^6$ according to
\beq
\theta \cdot x = \sum_{i=1,3,5}
\left[ x^i e_{i+1} - x^{i+1} (e_i + e_{i+1}) \right]
= \sum_{i=1,3,5}
\left[ - x^{i+1} e_i + (x^i - x^{i+1}) e_{i+1} \right].
\eeq
We can express the action of $\theta$
in terms of components by $x \to \theta \cdot x = x'$
with
\beq
x^i \to (x')^i = - x^{i+1},
\qquad
x^{i+1} \to (x')^{i+1} = x^i - x^{i+1},
\qquad
i=1,3,5.
\label{cmpt}
\eeq
The difference between the coefficients in
\myref{pnh} versus \myref{cmpt} is due to the
fact that the root basis is a skew basis.
Eq.~\ref{cmpt} leads to a matrix for the action of $\theta$
on the components:
\beq
M(\theta) = \diag [ m(\theta), m(\theta), m(\theta) ],
\qquad
m(\theta) = \pmatrix{0 & -1 \cr 1 & -1 \cr} .
\label{tmat}
\eeq
}

Having described the space group, we define the
six-dimensional $Z_3$ orbifold.
\begin{defn} \label{abat}
The orbifold $\Omega(6,Z_3) = \Rbf^6 / S$ is the
quotient space constructed when we deem
the points $x,x' \in \Rbf^6$ equivalent
if they are related to each other under the
action of the space group $S$:
$x' \simeq x$ if and only if
there exists an element $(\omega,\ell) \in S$ such that
$x' = (\omega,\ell) \cdot x$.
\end{defn}
A suitable fundamental domain for
the orbifold, projected into any one of the three complex
planes, is depicted in Figure \ref{z3f1}.
Sewing of open boundaries to closed boundaries
is suggested by the arrows.  Whereas in the
two-dimensional $Z_2$ orbifold of Section \ref{2do}
could be pictured as a four-cornered ``pillow,''
we now obtain for the $Z_3$ orbifold a three-cornered ``pillow.''
Since the orbifold is six-dimensional,
we actually have three such pillow spaces
associated with the projection of the orbifold
into each of the three complex planes.

\subsection{Conjugacy Classes and Fixed Points}
\label{ccfp}
The \git{conjugacy class} associated with the
space group element $(\omega,\ell)$ is the set
\beq
\left\{ \; (\omega',\ell') \cdot (\omega,\ell)
 \cdot (\omega',\ell')^{-1}
\quad \left| \quad (\omega',\ell') \in S \; \right. \right\},
\eeq
This partition of the space group is of prime
importance in the application of the six-dimensional
$Z_3$ orbifold to the heterotic string, as
will be seen in Section \ref{obo} below.
It will be useful to have a full determination
of the conjugacy classes of space group.
This is provided by the following examples.

\egbox{
\rm
Let us consider the conjugacy class associated with
a space group element which is a pure lattice
translation $(1,\ell_0)$.  First note that
the space group multiplication rule \myref{yati} implies
\beq
(\omega,\ell)^{-1} = (\omega^{-1}, -\omega^{-1}  \ell).
\eeq
Again using \myref{yati} it is easy to check that
\beq
(\omega,\ell)  (1,\ell_0)  (\omega,\ell)^{-1}
= (1,\omega  \ell_0).
\eeq
Thus, in the case of the $Z_3$ orbifold where only
three choices of $\omega$ are available (cf.~Eq.~\myref{pni}),
the conjugacy class associated with $(1,\ell_0)$
is given by
\beq
\{ (1,\ell_0), (1,\theta  \ell_0), (1,\theta^2  \ell_0) \} .
\eeq
}

\egbox{
\rm
Now I determine the conjugacy class associated with
a twisted space group element $(\theta,\ell_0)$.
\beqa
(\omega,\ell)(\theta,\ell_0)(\omega,\ell)^{-1}
& = & (\omega \theta,\omega \ell_0 + \ell)(\omega^{-1},-\omega^{-1} \ell) \nnn
& = & (\omega \theta \omega^{-1},
- \omega \theta \omega^{-1} \ell + \omega \ell_0 + \ell) \nnn
& = & (\theta,(1-\theta)\ell+\omega \ell_0).
\label{cct}
\eeqa
In the last step I have used the fact that the point
group $Z_3$ is abelian.

To begin with, consider $\omega = 1$.  It is not hard to
check that
\beqa
& & \ell \equiv \sum_{i=1}^6 \ell^i e_i \quad \Rightarrow \quad
(1-\theta)\ell = \sum_{i=1}^6 \ell^i v_i, \nnn
&& v_i \equiv e_i - e_{i+1}, \quad v_{i+1} \equiv e_i + 2 e_{i+1},
\quad i=1,3,5.
\eeqa
Then within the conjugacy class of $(\theta,\ell_0)$ are
all elements of the form
\beq
(\theta,\ell_0 + \sum_{i=1}^6 m^i v_i)
\label{1caa}
\eeq
with $m^i \in \Zbf$.  Note that I have relabeled $\ell^i \to m^i$.

For $\omega = \theta$ we have
\beq
(\theta,(1-\theta)\ell + \omega \ell_0) =
(\theta,(1-\theta)\ell + \theta \ell_0) =
(\theta,(1-\theta)(\ell-\ell_0) + \ell_0).
\eeq
Without loss of generality we can take $\ell' \equiv \ell - \ell_0$.
Then we again obtain the same set of elements as
for the $\omega=1$ case above, with $(\ell')^i \equiv m^i$
in \myref{1caa}.

For $\omega = \theta^2$ we have
\beq
(\theta,(1-\theta)\ell + \omega \ell_0) =
(\theta,(1-\theta)\ell + \theta^2 \ell_0) =
(\theta,(1-\theta)(\ell-\theta \ell_0- \ell_0) + \ell_0).
\eeq
Without loss of generality we can take $\ell' \equiv
\ell - \theta \ell_0- \ell_0$.
We obtain the same set of elements, with
$(\ell')^i \equiv m^i$ in \myref{1caa}.

Therefore the conjugacy class of $(\theta,\ell_0)$ is
given by
\beq
\{ (\theta, \ell_0 + \sum_{i=1}^6 m^i v_i) \}_{m^i \in \Zbf} .
\label{tcc}
\eeq
\label{tccl}
}

I now show that there are 27 inequivalent
classes and that each class is labeled by
a \git{class leader} of the form
\beq
\ell(n_1,n_3,n_5) = n_1 e_1 + n_3 e_3 + n_5 e_5,
\qquad n_i = 0, \pm 1 .
\label{ccld}
\eeq
First suppose given the lattice vector
$k= \sum_{i=1}^6 k^i e_i$ and look
for a solution to
\beq
k = \ell(n_1,n_3,n_5) + \sum_{i=1}^6 m^i v_i
= \sum_{i=1,3,5} [n_i e_i + m^i (e_i - e_{i+1})
+ m^{i+1} (e_i + 2 e_{i+1}) ].
\label{lfrm}
\eeq
It is not hard to check that this leads to
constraints equivalent to $(i=1,3,5)$:
\beqa
3 m^{i+1} & = & k^i + k^{i+1} - n_i \label{fjat} \\
m^i &=& k^i - n_i - m^{i+1}
\eeqa
So first pick $n_i = 0, \pm 1$ such that a
solution $m^{i+1} \in \Zbf$ exists for the first
equation---one of these three values of
$n_i$ will always work.  Then plug these values into the
second equation to determine $m^i$.  This
shows that all lattice vectors can be
written in the form \myref{lfrm}.  Now I
show that the leaders belong to different
classes.

Suppose $\ell(n_1,n_3,n_5)$ and $\ell(n_1',n_3',n_5')$
are in the same class.  Then we can take $k \to \ell(n_1',n_3',n_5')$
in \myref{lfrm} above.  Eq.~\myref{fjat} gives $(i=1,3,5)$:
\beq
3 m^{i+1} = n_i' - n_i.
\eeq
Because $n_i$ and $n_i'$ take only values
$0, \pm 1$, the only solution is $m^{i+1}=0$
and $n_i' = n_i$.  Therefore if $\ell(n_1,n_3,n_5)$
and $\ell(n_1',n_3',n_5')$ belong to the same
class, they are identical to each other.  Q.E.D.

As was the case for the one- and two-dimensional
orbifolds discussed in Sections \ref{1do} and \ref{2do},
certain points $x_f \in \Rbf^6$ are fixed under the action
of space group elements with $\omega=\theta$:
\beq
(\theta,\ell) \cdot x_f = \theta \cdot x_f + \ell = x_f.
\eeq
It is not hard to solve this equation; one finds
that the fixed points are in one-to-one correspondence
with elements of $\LamTh$:
\beq
x_f(\ell) = (1-\theta)^{-1} \cdot \ell.
\label{2.3}
\eeq

Because of the identification of points under $S$
on the orbifold, most of the fixed points \myref{2.3}
are equivalent to each other.  In fact, only 27
inequivalent fixed points exist, and they
are in one-to-one correspondence with the
conjugacy class leaders \myref{ccld}, by way of \myref{2.3}:
\beq
f_{n_1,n_3,n_5} \equiv x_f(\ell(n_1,n_3,n_5))
= (1-\theta)^{-1} \cdot \ell(n_1,n_3,n_5).
\label{iefp}
\eeq
It is not difficult to prove these statements.
The key steps of the proof are outlined
in the following examples.

\egbox{
Consider two fixed points corresponding to lattice
vectors $\ell_a$ and $\ell_b$
in the same conjugacy class.  Then from the
considerations of Example \ref{tccl} we know
that there exist $m^i \in \Zbf$ such that
\beq
\ell_b = \ell_a + \sum_{i=1}^6 m^i v_i,
\eeq
where the notation is as in \myref{tcc}.
Furthermore, we found in Example \ref{tccl}
that the lattice vector $\ell$ defined by
$\ell^i \equiv m^i$ satisfies
\beq
\sum_{i=1}^6 m^i v_i = (1-\theta) \ell.
\eeq
Then the corresponding fixed points, given
by \myref{2.3}, are related by
\beq
x_f(\ell_b) = (1-\theta)^{-1}[\ell_a + (1-\theta) \ell]
= x_f(\ell_a) + \ell.
\eeq
Thus, fixed points corresponding to lattice vectors
in the same twisted conjugacy class differ by a lattice
vector; they are therefore equivalent on the orbifold.
}

\egbox{
In this example I show that the fixed points
$x_f(0)=0$ and $x_f(e_1) = (1-\theta)^{-1} e_1$ are not related by
a lattice vector.  By similar arguments it is
easy to check that each of the 27 fixed points
given in \myref{iefp} are inequivalent.

Suppose $x_f(e_1) \simeq x_f(0)$.  Then there
exists a lattice vector $\ell= \ell^1 e_1 + \ell^2 e_2$ such that
\beq
(1-\theta)^{-1} e_1 = \ell \quad \Rightarrow \quad
e_1 = (1-\theta)\ell = (\ell^1 + \ell^2)e_1 +
(2\ell^2-\ell^1)e_2.
\eeq
It is easy to check that linear independence of
$e_1$ and $e_2$ then requires $3 \ell^2 = 1$,
which cannot be satisfied for $\ell^2 \in \Zbf$.
Thus we arive at a contradiction.
}

\subsection{The Complex Basis}
\label{cbss}
For the applications in the following chapter,
the root basis employed in \myref{comp} is inconvenient.
Rather, we use a \git{complex basis} which is defined
in terms of the components $x^i$ appearing in \myref{comp}
according to
\beq
z^i \equiv x^i + e^{2 \pi i / 3} x^{i+1},
\qquad
\bz^i \equiv x^i + e^{-2 \pi i / 3} x^{i+1},
\qquad
i=1,3,5.
\label{ccmp}
\eeq
This is motivated by supposing that in the $i$th complex
plane $e_i$ lies along the real axis while from \myref{2f}
we see that $e_{i+1}$ lies at 120 degrees counterclockwise
from the real axis; i.e., along $e^{2 \pi i / 3}$.  This
picture is of course the origin of the usage of ``complex
plane'' for each of the three pairs $e_i,e_{i+1}$ ($i=1,3,5$).

From \myref{cmpt} it is not hard to show that the
twist operator acts on $z^i$ as a pure phase rotation
($i=1,3,5$):
\beq
z^i \to (z')^i = (x')^i + e^{2 \pi i / 3} (x')^{i+1}
= e^{2 \pi i/3} \( x^i + e^{2 \pi i / 3} x^{i+1} \)
= e^{2 \pi i/3} z^i.
\label{twac}
\eeq
Similarly,
\beq
\bz^i \to (\bz')^i = e^{-2 \pi i/3} \bz^i.
\eeq
In the complex basis, the matrix realization
\myref{tmat} is given instead by
\beq
M_c(\theta) = \diag ( e^{2 \pi i/3}, e^{2 \pi i/3}, e^{2 \pi i/3} )
\label{cmat}
\eeq
when acting on $(z^1,z^3,z^3)$ and is the complex
conjugate $[M_c(\theta)]^*$ when acting on vectors in
the conjugate representation space $(\bz^1,\bz^3,\bz^3)$.
It is this decomposition into irreducible representations---no
mixing between $z^i$ and $\bz^i$, in contrast to the
mixing between $x^i$ and $x^{i+1}$ in \myref{tmat}---which
eases manipulations when we come to applications
below.  The $Z_3$ nature of the point group is
obvious from \myref{cmat}.  It is the generator of
the center of $SU(3)$ in the fundamental representation.

In an abuse of notation I shall often write
$\theta = e^{2 \pi i/3}$, so that \myref{cmat}
becomes
\beq
M_c(\theta) = \diag ( \theta, \theta, \theta ).
\eeq
Furthermore, I collect the components \myref{ccmp}
into a three-tuple $\zbf = (z^1,z^3,z^5)$ and write
the twist action \myref{twac} as
\beq
\zbf \to \zbf' = \theta \zbf,
\eeq
where $\theta$ here simply means scalar multiplication
by $e^{2 \pi i/3}$.  It is also clear from \myref{ccmp}
that the complex parameterization of a lattice vector $\ell$
takes the form
\beq
b^i(\ell) \equiv \ell^i + e^{2 \pi i / 3} \ell^{i+1},
\qquad
\bar b^i(\ell) \equiv \ell^i + e^{-2 \pi i / 3} \ell^{i+1},
\qquad
i=1,3,5.
\label{lcmp}
\eeq
Thus the action of a general space group element
$(\omega,\ell)$ with $\omega=\theta^n$ ($n=0,\pm 1$)
is given by
\beq
\zbf \to \zbf' = (\theta^n,\ell) \zbf = \theta^n \zbf + \bbf(\ell)
= e^{2n\pi i/3} \zbf + \bbf(\ell),
\label{cspc}
\eeq
where $\bbf = (b^1,b^3,b^5)$.
The correspondence \myref{2.3} between fixed points
and lattice group elements finds its expression in
the complex basis through
\beq
\zbf_f(\bbf) = (1-\theta)^{-1} \bbf
= {1 \over \sqrt{3}} e^{i \pi / 6} \bbf.
\label{uiwt}
\eeq

%

\chapter{Heterotic String}
\label{mss}
The \git{heterotic string}, introduced in \cite{GHMR85},
is a ten-dimensional theory.  \, One path to a four-dimen\-sional
analogue is to associate six of the spatial dimensions
instead with a very small six-dimensional orbifold.
In this chapter my principle intent is
to describe this application of orbifold
geometry.\footnote{
It is worth noting that quotient
space constructions for extra dimensions
were applied in a field theory context
some years prior to the construction of
four-dimensional strings on orbifolds,
with important consequences such as chiral
fermions \cite{CGS}.}

The material contained in this chapter is
not new, nor is it the result of research that I have
contributed to.  It represents entirely the work of others
and it is well-known to most string theorists---certainly
the older generation.
Readers not familiar with string theory at the
level of, say, the first volume of Green, Schwarz and
Witten \cite{GSW87a} would do well to simultaneously tackle the
suggested reading described in Appendix \ref{mysg},
as I cannot possibly provide an adequate introduction
to this large topic in so short a work.  However,
I do my best to keep the discussion accessible
to a wider audience.

I begin in Section \ref{rot} with a brief
reminder of the original ten-dimensional theory.
In Section \ref{obo} I discuss the four-dimensional
heterotic string obtained from the $Z_3$ orbifold.
The closed string boundary conditions are
of chief importance when the \git{target space} is an orbifold.
This leads to a significant modification in the Hilbert space
of \git{physical states.}
In Section \ref{eep} I stress aspects
which result from using the \eetee\ heterotic string
as a starting point.

Finally, in Section \ref{rec}
I give a set of heuristic rules which allow one
to calculate the spectrum of massless states for the
orbifold constructions studied here.  I have chosen
to avoid a complete description of the string
theoretic details which lead to these rules.
My first reason is that these aspects have been
adequately reviewed elsewhere.  (References to these
reviews will be given in the discussion below.)
Moreover, to discuss these matters in a self-contained
way would require me to review much more of
string theory than there is space for here and
to discuss string theory beyond the leading
order in perturbation theory (see below).
My second reason is that an understanding of these details is not
important to the original work that
I performed, application of heterotic string
theory to the construction of \git{semi-realistic} models.
The description of what I actually did in my
research program is the topic of Chapters \ref{slz}-\ref{app}.
To follow the discussion of these chapters,
a detailed understanding of all the string theoretic
details is not necessary.

Throughout, I work in the context of perturbative string
theory, and for the most part only to leading order.
The perturbation series corresponds to string
\git{world-sheet} (the two-dimensional surface
swept out by the string) diagrams of increasing complexity.
These are labeled by the \git{genus} of the diagram,
starting at genus zero---often referred to as ``tree level''
in string theory.  The next order, genus one, is
often referred to as the ``one loop level'' in string
theory, because the world-sheet diagram is a two-dimensional
torus.  An analysis of the one loop consistency of
the theory leads one to impose various projections
on the Hilbert space of \git{physical states.}  The projections
in the original ten-dimensional theory
are referred to as \git{GSO projections}
after Gliozzi, Scherk and Olive \cite{GSO}.
In the four-dimensional constructions they
are referred to as \git{generalized GSO projections.}
Although I use the projections which follow
from such considerations, I will not discuss
the one loop analysis here.

\section{Ten-dimensional Construction}
\label{rot}
\subsection{Classical String}
\subsubsection{The World-Sheet}
The heterotic string can be regarded as a two-dimensional
field theory.  Our \git{base space} is parameterized by a
time-like coordinate $\tau$ and space-like coordinate
$\s$; this space is the \git{world-sheet}
of the string.  More precisely, denote the two-dimensional
world-sheet as $M_2$, a two-dimensional
space-time with Lorentzian metric.
Invariances of the classical string action
allow one to transform to
a Minkowski world-sheet frame, where the
action takes the form
\beq
S = - {1 \over 2 \pi} \int d^2\s \, \[ \p_\alpha X_\mu \p^\alpha X^\mu
+ 4i \psi^\mu (\p_\tau-\p_\s) \psi_\mu
+ \sum_{I=1}^{16} \p_\alpha X^I \p^\alpha X^I \].
\eeq
In this frame the invariant arclength is given by:
\beq
ds^2 = -d\tau^2 + d\s^2.
\eeq
The parameter $\tau$ labels proper time in
the frame of the string.
The world-sheet coordinate $\s$ labels points along the string
in its proper frame, with $\s \to \s + \pi$ as one goes once around
the string.
It is convenient to define \git{right-moving} and \git{left-moving}
world-sheet coordinates
\beq
\s_- = \tau - \s, \qquad \s_+ = \tau + \s .
\label{baab}
\eeq

\subsubsection{Covariant Description}
The \git{fields} in the theory give a map
of this base space into a \git{target space}
which is a \git{Riemannian supermanifold.}\footnote{See
Ref.~\cite{DeW92} for an extensive discussion
of supermanifolds.}
That is, the target space is the Cartesian product of
a real manifold and a Grassmannian manifold (points
labeled by anticommuting c-numbers).
The heterotic string is a map described by
\beq
(\s_+,\s_-) \to (X^\mu(\s_+,\s_-), \psi^\mu(\s_-), X^I(\s_+)).
\label{tma}
\eeq
The right-hand side of \myref{tma} belongs to the space
$M_{10} \times G_{10}^- \times \Tbf_{16}$.
Here, $M_{10}$ is a ten-dimensional
Minkowski space-time. $X^\mu(\s_+,\s_-)$
transform in the vector representation
of the corresponding rotation group $O(1,9)$.
$G_{10}^-$ is a ten-dimensional {\it odd vector space;}
i.e., the functions on $G_{10}^-$ form a Grassmann algebra.
$O(1,9)$ target space symmetry is also imposed
on these coordinates, and they are in the
vector representation.  However, each
component $\psi^\mu(\s_-)$
is given by a \git{Majorana-Weyl} \,
$O(1,1)$ spinor of negative chirality.
That is, there exists a basis for the
two-dimensional Dirac matrices where
\beq
(\psi^\mu(\tau,\s))^* = \psi^\mu(\tau,\s)
\qquad
\gamma^{3} \psi^\mu(\tau,\s)
= - \psi^\mu(\tau,\s).
\eeq
Here, $\gamma^{3}=\gamma^0 \gamma^1$ is the chirality matrix
in two dimensions, which anticommutes with the two-dimensional
Dirac matrices:  $\{\gamma^{3},\gamma^m \} = 0$ ($m=0,1$).
Coordinates $X^I(\spl) \; (I=1,\ldots,16)$
are the image of the string on a torus
$\Tbf^{16} = \Rbf^{16}/\Lambda$, with $\Lambda$ a sixteen-dimensional
lattice.  In the quantized heterotic theory,
one finds that internal consistency imposes strong
restrictions on what we may choose for $\Lambda$.
The only consistent choices are the \eetee\ root
lattice or the ${\rm spin}(32)/Z_2$ lattice.
I will not discuss these aspects here, but refer
the reader to \cite{GSW87b}.  
In this thesis I will only consider the case
where $\Lambda = \eelat$.  This lattice is described
in detail in Section \ref{eer} below.

Because the string is closed, periodic boundary conditions
must be satisfied for the map \myref{tma} to be single-valued
(or possibly double-valued in the case of fermions
of odd world-sheet parity) on the target space.
The coordinates $X^I(\s_+)$ are required
to be periodic under $\s \to \s + \pi$, up to
a lattice vector (a factor of $\pi$ is
conventionally included):
\beq
X^I(\s_+ + \pi) = X^I(\s_+) + \pi L^I, \qquad
L^I \in \eelat.
\label{bbc}
\eeq
The other coordinates must satisfy boundary conditions
\beq
X^\mu(\s + \pi,\tau)  =  X^\mu(\s,\tau), \qquad
\psi^\mu(\s_- - \pi)  =  \pm \psi^\mu(\s_-).
\label{fbc}
\eeq

Thus, the set of maps $M_2 \to M_{10} \times \Tbf_{16} \times G_{10}^-$
that describe a string configuration is restricted by:
(i) equations of motion; (ii) periodic boundary conditions;
(iii) the constraint that some coordinates in \myref{tma}
depend on only $\s_-$ or $\s_+$.  (Another constraint
appears in the quantized theory, whose classical
analogue is not clear to me---the GSO projection, described below.)
These restrictions account for the action formulation of the theory,
the closed string interpretation, and the fact that
the theory is actually obtained as a hybridization\footnote{It
is this feature which is the source of the name ``heterotic.''}
of ``parent'' theories---the closed bosonic string and the
closed superstring---which are subjected to
projections on the allowed classical trajectories.
(For more details consult the references provided
in Appendix~\ref{mysg}.)

\subsubsection{Light-cone Description}
When one applies the canonical formalism\myfoot{It would be
nice if I could either use a well-defined terminology
here or explain what I mean by this term.
Maybe sit down with Goldstein and straighten
out my terminology.  Maybe reference it.
Also, ``the above system'' is poorly defined.} to the above
system, one finds that not all of the canonical momenta
are independent.  The equations which relate them\myfoot{Where
do these equations come from?}
form a system of constraints ($\p_\pm = \p/\p \s_\pm$)
\beq
\psi^\mu \p_- X_\mu = 0, \qquad
\p_+ X^\mu \p_+ X_\mu = 0, \qquad
\p_- X^\mu \p_- X_\mu + {i \over 2} \psi^\mu \p_- \psi_{\mu} = 0,
\label{ceq}
\eeq
which must be satisfied
by solutions to the Euler-Lagrange equations of motion
\beq
\p_+ \psi^\mu = 0, \qquad
\p_+ \p_- X^\mu = 0.
\label{baaa}
\eeq

There exists some arbitrariness in how the constraint
equations \myref{ceq} may be
satisfied, corresponding to a reparameterization
invariance in the action.\myfoot{Would be nice
to have a discussion of action formulation which
I could refer to here.}  We can exploit this
invariance to ``gauge-fix'' the system and eliminate
dependent degrees of freedom.  The gauge which has
proven most useful\footnote{A good discussion of light-cone
gauge is can be found in Sections 2.3.1 and 4.3.1 of
\cite{GSW87a}.} is referred to as \git{light-cone gauge.}
In it, the time-like direction $\mu=0$ and
one space-like direction $\mu=9$ are singled
out for special treatment.  We define
\beq
X^\pm(\s,\tau) = X^0(\s,\tau) \pm X^9(\s,\tau),
\qquad
\psi^\pm(\s_-) = \psi^0(\s_-) \pm \psi^9(\s_-).
\eeq
Light-cone gauge uses residual invariance\myfoot{What
residual invariance? Reparameterization invariance
of the action?} to set
\beq
X^+(\s,\tau) \equiv x^+ + p^+ \tau,
\qquad \psi^+(\s,\tau) \equiv 0,
\label{gfx}
\eeq
for all $\s,\tau$.  Here, $x^+,p^+$ are constants.
The constraint equations are then satisfied by making
$X^-(\s,\tau)$ and $\psi^-(\s_-)$ functions\myfoot{Why
not just give the explicit functions here instead of F and G?}
of the \git{transverse} coordinates $X^i(\s,\tau)$, $\psi^i(\s-)$,
$i=1,2,\ldots,8$:
\beqa
X^-(\s,\tau) &=& F[X^i(\s,\tau),\psi^i(\s-)] \\
\psi^-(\s_-) &=& G[X^i(\s,\tau),\psi^i(\s-)]
\eeqa
Thus, the transverse coordinates
$X^i(\s,\tau), \psi^i(\smi) \; (i=1,\ldots,8)$
carry\myfoot{I might
want to spell out what exactly I mean by this.} the string dynamics.

\subsection{Mode Expansions}
Quantization of the string is much like
quantization of the Klein-Gordan and Dirac fields.
The Hilbert space is most easily constructed
in terms of Fourier modes.  Thus as a preliminary
step toward quantization I will give mode
expansions.  I only describe the transverse
coordinates in light-cone gauge, as the other
coordinates are then determined\myfoot{Ordering
ambiguities may exist here which would be
interesting to study.} by \myref{ceq} and \myref{gfx}.

We first solve the equations of motion
\myref{baaa} classically
by Fourier mode expansion.  For the transverse bosons
we have
\beq
X^i(\s,\tau)  =  x^i + p^i \tau + {i \over 2} \sum_{n \not= 0}
{1 \over n} \alpha_n^i e^{-2 i n \smi} 
+ {i \over 2} \sum_{n \not= 0}
{1 \over n} \tilde \alpha_n^i e^{-2 i n \spl}.
\label{fmep}
\eeq
The portion $x^i + p^i \tau$ is called the \git{zero modes}
contribution while the remainder is referred to as
the \git{oscillator modes} contribution.
It is customary to break up the left- and right-moving modes
(recall definition~\myref{baab}):
\beqa
X^i(\s,\tau) & = & X_R^i(\smi) + X_L^i(\spl) , \\
X_R^i(\s_-) & = & \half x^i + \half p^i \s_- + {i \over 2} \sum_{n \not= 0}
{1 \over n} \alpha_n^i e^{-2 i n \s_-} ,
\label{mder} \\
X_L^i(\spl) & = & \half x^i + \half p^i \s_+ + {i \over 2} \sum_{n \not= 0}
{1 \over n} \tilde \alpha_n^i e^{-2 i n \s_+} .
\label{mdel}
\eeqa
For the zero modes contribution
$x^i + p^i \tau$, I split it up in
a symmetric way among the left- and right-movers, which
turns out to be the right thing to do when it comes
to the quantization of the theory.

For the world-sheet fermions we must distinguish
between the two types of boundary conditions \myref{fbc}.
The \git{Neveu-Schwarz (NS)} boundary conditions (odd) lead to
\beq
\psi^i = \sum_{r \in \Zbf + \half} b_r^i e^{-2ir \smi} ,
\label{baag}
\eeq
while for the \git{Ramond (R)} boundary conditions (even) we have
\beq
\psi^i = \sum_{n \in \Zbf} d_n^i e^{-2in \smi}.
\label{baah}
\eeq
Note that the mode coefficients $b_r^i$ and $d_n^i$
are Grassmann numbers.

The case of the internal bosons $X^I$ is slightly more
complicated because of \myref{bbc}.  We begin by writing
the solution to the equations of motion without the restriction to
left-movers:
\beq
X^I(\s,\tau) = x^I + p^I \tau + L^I \s
+ {i \over 2} \sum_{n \not= 0}{1 \over n} \alpha_n^I e^{-2 i n \s_-}
+ {i \over 2} \sum_{n \not= 0}{1 \over n} \tilde \alpha_n^I e^{-2 i n \s_+}.
\label{zmcb}
\eeq
The third term provides
the non-trivial boundary condition \myref{bbc} to the
left-mover as $\s \to \s + \pi$, provided $L \in \eelat$,
as we will see.
Now we decompose into left- and right-movers to get
\beqa
X_R^I(\smi) &=& \half x^I + \half (p^I - L^I) \smi
+ {i \over 2} \sum_{n \not= 0}{1 \over n} \alpha_n^I e^{-2 i n \s_-},
\label{prte}
\\
X_L^I(\spl) &\equiv& X^I(\spl) = \half x^I + \half (p^I + L^I) \spl
+ {i \over 2} \sum_{n \not= 0}{1 \over n} \tilde \alpha_n^I e^{-2 i n \s_+}.
\label{plte}
\eeqa
Then we \git{project out} the right-movers with
the constraint that coefficients contributing
$\smi$ dependence vanish.  That is,
we require $\p_- X_R^I \equiv 0$ for all $\smi$, which
in turn implies that the mode coefficients
$\half(p^I - L^I)$ and $\alpha_n^I$ vanish.\footnote{Note that
I do not require $\half x^I$ to vanish in \myref{prte}
because we would like it to also be non-vanishing in
\myref{plte}.  In the quantized theory we will find
that a way around this ``difficulty'' exists.}
In \myref{prte} this requires $p^I - L^I = 0$, or
\beq
p^I \equiv L^I .
\label{ouuw}
\eeq
Consequently in \myref{plte} and elsewhere we may substitute
$\half (p^I + L^I) = L^I$.
Then the boundary condition \myref{bbc}
is satisfied since
\beq
\half (p^I + L^I) (\spl + \pi) =
\half (p^I + L^I) \spl + L^I \pi.
\label{babr}
\eeq

\subsection{Quantum Mechanical String}
\label{bsac}
As in more elementary field theories with singular
Lagrangians (such as Quantum Electrodynamics),
quantization of the string is most straightforward if
fixed to a gauge where unphysical degrees of freedom
have been removed \cite{GT90}.  Thus, we quantize in light-cone gauge.

To quantize the theory, we promote the mode coefficients
to operators on a Hilbert space.  For the oscillator
modes of \myref{mder} and \myref{mdel}
we have an infinite tensor product of simple harmonic
oscillator Hilbert spaces, one for each pair
$(\alpha_n^i, \alpha_{-n}^i)$ or
$(\tilde \alpha_n^i, \tilde \alpha_{-n}^i)$, where $n>0$.
This follows from the canonical commutation
relations which are imposed on the oscillator modes:
\beq
[\alpha_m^j, \alpha_n^k] = m \delta^{jk} \delta_{m+n,0}, \qquad
[\tilde \alpha_m^j, \tilde \alpha_n^k] = m \delta^{jk} \delta_{m+n,0},
\qquad
[\alpha_m^j, \tilde \alpha_n^k] = 0.
\eeq
Thus the construction is straightforward and
I will not discuss it further in this brief review.
Some care, however,
must be taken with the zero modes in the bosonic
expansions \myref{fmep} and \myref{zmcb}---as I now
describe in some detail.  My reason for reviewing
these aspects carefully has to do with the importance
of zero modes in the four-dimensional theory to be
described in Section \ref{obo}.  The discussion
which follows also serves to illustrate
the sort of projections of product spaces which
are typical in the construction of the full Hilbert
space of a consistent first-quantized string theory.

For the right-movers \myref{mder} we make a
(classical) $\to$ (quantum) transition
to operators on a (bosonic) Hilbert space $\Hcal_R^B$:
\beq
\half x^i \to x_R^i, \qquad \half p^i \to p_R^i.
\label{cprm}
\eeq
For the left-movers \myref{mdel}, operators on a {\it different}
Hilbert space $\Hcal_L^B$ are identified:
\beq
\half x^i \to x_L^i, \qquad \half p^i \to p_L^i.
\label{cplm}
\eeq
The space-time boson portion of the Hilbert space of the
heterotic theory is
a subspace of the tensor product of the spaces
$\Hcal_L^B$ and $\Hcal_R^B$:
\beq
\Hcal^B \subset \Hcal_L^B \otimes \Hcal_R^B.
\label{baad}
\eeq
{\it Which} subspace will be made clear in the discussion
which follows.

More precisely, the total position $x^i$
and total momentum $p^i$ which appear in \myref{fmep}
are promoted to operators $\hat x^i$ and $\hat p^i$
on the space $\Hcal_L^B \otimes \Hcal_R^B$, of the forms
\beq
\hat x^i = x_L^i \otimes 1_R + 1_L \otimes x_R^i, \qquad
\hat p^i = p_L^i \otimes 1_R + 1_L \otimes p_R^i.
\label{ptdf}
\eeq
$1_R$ and $1_L$ are identity operators on
$\Hcal_R^B$ and $\Hcal_L^B$ respectively.
Semi-canonical commutation
relations are imposed on the
left- and right-moving zero mode operators:
\beq
[x_L^j, p_L^k] = {i \over 2} \delta^{jk}, \qquad
[x_R^j, p_R^k] = {i \over 2} \delta^{jk}.
\eeq
These yield canonical commutation relations for the
zero mode operators on $\Hcal_L^B \otimes \Hcal_R^B$:
\beq
[\hat x^j, \hat p^k] = i \delta^{jk}.
\label{baac}
\eeq
To check that this is true, note that
\beq
[x_L^j \otimes 1_R, 1_L \otimes p_R^k] =
[1_L \otimes x_R^j,p_L^k \otimes 1_R] = 0.
\eeq
Consequently
\beqa
[\hat x^j, \hat p^k] &=&
[x_L^j \otimes 1_R,p_L^k \otimes 1_R] +
[1_L \otimes x_R^j,1_L \otimes p_R^k] \nnn
&=& 
[x_L^j,p_L^k] \otimes 1_R +
1_L \otimes [x_R^j,p_R^k] ,
\eeqa
immediately leading to \myref{baac}.

Comparing \myref{fmep} and \myref{ptdf} we have
the (classical) $\to$ (quantum) correspondence
\beqa
x^i + p^i \tau & \to & \hat x^i + \hat p^i \tau \nnn
& = & \hat x^i + (p_L^i \otimes 1_R) \spl + (1_L \otimes p_R^i) \smi
- (p_L^i \otimes 1_R - 1_L \otimes p_R^i) \s .
\eeqa
For this to be consistent with the replacements
\myref{cprm} and \myref{cplm}---in
the expressions \myref{mder} and \myref{mdel}---we require
\beq
p_L^i \otimes 1_R - 1_L \otimes p_R^i \equiv 0
\label{pimt}
\eeq
on the subspace $\Hcal^B$ alluded to in Eq.~\myref{baad}
above.  Thus $\Hcal^B$ is not
the tensor product of left- and right-moving Hilbert
spaces, but is the maximal \git{projective Hilbert space}
spanned by states\footnote{It is common practice in physics
to refer to the vectors in a Hilbert space as \git{states.}}
contained in $\Hcal_L^B \otimes \Hcal_R^B$
satisfying \myref{pimt}.
This defines the subspace indicated by \myref{baad}.

The Hilbert space projection is best described
in \git{momentum space.}  The Hilbert spaces $\Hcal_R^B$
and $\Hcal_L^B$ are each spanned by infinite orthonormal sequences
of eigenvectors of $p_R^i$ and $p_L^i$ respectively
($i=1,2,\ldots,8$):
\beqa
p_L^i \ket{\chi_{n}}_L^B &=& k_{Ln}^i \ket{\chi_{n}}_L^B
\qquad n=1,2,\ldots, \\
p_R^i \ket{\eta_{p}}_R^B &=& k_{Rp}^i \ket{\eta_{p}}_R^B
\qquad p=1,2,\ldots, 
\eeqa
eigenvalue multiplicities counted.
From these bases, we
construct the infinite sequence obtained by tensor
products:
\beq
\ket{\chi_{n}}_L^B \otimes \ket{\eta_{p}}_R^B
\qquad n,p=1,2,\ldots
\label{baae}
\eeq
This sequence is \git{complete} on $\Hcal_L^B \times \Hcal_R^B$;
i.e., the sequence is not contained in a larger orthonormal
system.\footnote{See for example \cite{AG93}.}
To have \myref{pimt} we ``project out'' (i.e., delete)
all vectors in the sequence \myref{baae} which do
not vanish under the action of
$p_L^i \otimes 1_R - 1_L \otimes p_R^i$; that is,
we keep only vectors which are in the \git{nullity} of
this operator.  This leads to \git{level matching,}
since the only states which ``survive'' projection
are those for which
\beq
(p_L^i \otimes 1_R) \cdot
(\ket{\chi_{n}}_L^B \otimes \ket{\eta_{p}}_R^B)
= (1_L \otimes p_R^i) \cdot
(\ket{\chi_{n}}_L^B \otimes \ket{\eta_{p}}_R^B),
\qquad i=1,2,\ldots,8.
\label{baaf}
\eeq
I.e., we keep the states with matching left- and right-moving
momentum eigenvalues: $k_{Ln}^i \equiv k_{Rp}^i \; (i=1,\ldots,8)$.

\begin{defn}
The infinite subsequence of vectors (states)
$\{ \ket{\psi_1}^B, \ket{\psi_2}^B, \ldots \}$
belonging to \myref{baae}
which also satisfy \myref{baaf} is the orthonormal momentum eigenbasis
of $\Hcal^B$.  These states are said to ``survive'' the projection
$\Hcal_L^B \otimes \Hcal_R^B \to \Hcal^B$.
The Hilbert space $\Hcal^B$ is
the \git{closed linear envelope} of
the orthonormal momentum eigenbasis.\footnote{
That is, consider the set of all finite linear combinations
\beq
c_1 \ket{\psi_1}^B + c_2 \ket{\psi_2}^B + \cdots
+ c_n \ket{\psi_n}^B,
\eeq
with $c_1, \ldots, c_n$ complex numbers.
This is the \git{linear envelope} of the set $\{ \ket{\psi_m}^B \}$.
With the addition of all \git{limit points} of the linear envelope,
we obtain the \git{closed linear envelope.}
}
\end{defn}

The correspondence between elements of \myref{baae}
satisfying \myref{baaf} defines a map
$\Zbf_+ \to \Zbf_+ \times \Zbf_+$ between
labels.  We write this as $m \to (n(m),p(m))$,
defined by the identification
\beq
\ket{\psi_{m}}^B =
\ket{\chi_{n(m)}}_L^B \otimes \ket{\eta_{p(m)}}_R^B 
\eeq
for those values of $(n,p)$ such that \myref{baaf}
is satisfied.  It is of interest to
note that for the surviving sequence each member has {\it total} eigenvalues
$k_{m}^i = k_{Ln(m)}^i + k_{Rp(m)}^i \; (i=1,\ldots,8)$ with
respect to the total momentum operator $\hat p^i$ in
\myref{ptdf}.  Because of the level matching, we have
($i=1,2,\ldots,8$)
\beq
k_{Ln(m)}^i = k_{Rp(m)}^i = \half k_{m}^i,
\eeq
which is the precise description of what is
meant by the (classical) $\to$ (quantum)
correspondences $p^i/2 \to p_{L,R}^i$ stated
in \myref{cprm} and \myref{cplm}.

We next consider the quantization of the zero
mode part of \myref{zmcb}.  We make the correspondence
\beq
x^I + p^I \tau + L^I \s \to
\hat x^I + \hat p^I \tau + \hat L^I \s ,
\label{tyur}
\eeq
where on the right we have operators on
a Hilbert space corresponding to the
sixteen-dimensional \eetee\ torus subspace of the
target space.
The winding mode operator $\hat L^I$ is taken
to commute with the other mode operators.\footnote{
Classically, $L^I$ labels a countable multiplicity in
the solutions to the equations of motion. 
Thus the Poisson brackets $\{x^I,L^J \}$ and
$\{ p^I, L^J \}$ vanish.  Dirac's prescription
for quantization
instructs us to extend this to the operator algebra
of the quantum theory.}
Thus the zero mode operators have canonical
commutation relations
\beq
[\hat x^I , \hat p^J] = i \delta^{IJ}
\eeq
with all others vanishing.
Now we seek a realization of these operators
on a tensor product of Hilbert spaces $\Hcal_L^T \otimes \Hcal_R^T$.
Following what was done for the transverse bosons $X^i(\s,\tau)$
above, we assume
\beq
\hat x^I = x_L^I \otimes 1_R + 1_L \otimes x_R^I, \quad
\hat p^I = p_L^I \otimes 1_R + 1_L \otimes p_R^I, \quad
\hat L^I = L_L^I \otimes 1_R + 1_L \otimes L_R^I.
\label{babk}
\eeq
The commutation relations for $\hat x^I, \hat p^I, \hat L^I$
are satisfied provided
\beq
[x_L^I,p_L^J]= [x_R^I,p_R^J] ={i \over 2} \delta^{IJ},
\eeq
with all others vanishing.
Substitution of \myref{babk}
into \myref{tyur} yields\myfoot{Check algebra.}
\beqa
\hat p^I \tau + \hat L^I \s & = &
\[ (p_L^I + L_L^I) \otimes 1_R \] \spl
+ \[ 1_L \otimes (p_R^I-L_R^I) \] \smi \nnn
& & + \( 1_L \otimes p_R^I - p_L^I \otimes 1_R \)  \s
+ \( 1_L \otimes L_R^I - L_L^I \otimes 1_R \)  \tau .
\label{uyqy}
\eeqa
By the same reasoning which led to \myref{pimt},
we require a projection
$\Hcal_L^T \otimes \Hcal_R^T \to \Hcal^T$
such that on $\Hcal^T$:
\beq
1_L \otimes p_R^I - p_L^I \otimes 1_R \equiv 0, \qquad
1_L \otimes L_R^I - L_L^I \otimes 1_R \equiv 0.
\label{babm}
\eeq
Furthermore, in the case of the heterotic string we require a
projection such that the right-moving modes in \myref{uyqy}
vanish on $\Hcal^T$:\footnote{A similar relation is imposed
on the right-moving oscillator modes $\alpha_n^I$.}
\beq
1_L \otimes (p_R^I - L_R^I) \equiv 0.
\label{babl}
\eeq
This is accomplished as above.  We define orthonormal
bases for $\Hcal_R^T$ and $\Hcal_L^T$ respectively
$(I=1,\ldots,16)$:
\beqa
p_R^I \ket{\eta_{p}}_R^T &=& k_{Rp}^I \ket{\eta_{p}}_R^T
\qquad p=1,2,\ldots \nnn
p_L^I \ket{\chi_{n}}_L^T &=& k_{Ln}^I \ket{\chi_{n}}_L^T
\qquad n=1,2,\ldots, \nnn
L_R^I \ket{\eta_{p}}_R^T &=& w_{Rp}^I \ket{\eta_{p}}_R^T
\qquad p=1,2,\ldots, \nnn
L_L^I \ket{\chi_{n}}_L^T &=& w_{Ln}^I \ket{\chi_{n}}_L^T
\qquad n=1,2,\ldots
\eeqa
From these bases we construct the infinite orthonormal
sequence of tensor products
\beq
\ket{\chi_{n}}_L^T \otimes \ket{\eta_{p}}_R^T
\qquad n,p=1,2,\ldots
\label{baai}
\eeq
In the projection,
we retain only those vectors in the sequence
\myref{baai} which satisfy
the constraints \myref{babm} and \myref{babl} $(I=1,\ldots,16)$:
\beqa
(1_L \otimes p_R^I) \cdot (\ket{\chi_{n}}_L^T \otimes \ket{\eta_{p}}_R^T)
&=&
(p_L^I \otimes 1_R) \cdot (\ket{\chi_{n}}_L^T \otimes \ket{\eta_{p}}_R^T),
\label{baaj} \\
(1_L \otimes L_R^I) \cdot (\ket{\chi_{n}}_L^T \otimes \ket{\eta_{p}}_R^T)
&=&
(L_L^I \otimes 1_R) \cdot (\ket{\chi_{n}}_L^T \otimes \ket{\eta_{p}}_R^T),
\label{baak} \\
{[} 1_L \otimes (p_R^I - L_R^I)] \cdot
(\ket{\chi_{n}}_L^T \otimes \ket{\eta_{p}}_R^T) &=& 0.
\label{baal}
\eeqa
This projection defines a map $m \to (n(m),p(m))$
defined by the identification
\beq
\ket{\psi_{m}}^T =
\ket{\chi_{n(m)}}_L^T \otimes \ket{\eta_{p(m)}}_R^T
\label{babn}
\eeq
for those values of $(n,p)$ such that \myref{baaj}-\myref{baal}
are satisfied.

\begin{defn}
The infinite subsequence of vectors (states)
$\{ \ket{\psi_1}^T, \ket{\psi_2}^T, \ldots \}$
belonging to \myref{baai}
which also satisfy \myref{baaj}-\myref{baal} is the
orthonormal momentum eigenbasis
of $\Hcal^T$.  These states
survive the projection
$\Hcal_L^T \otimes \Hcal_R^T \to \Hcal^T$.
The Hilbert space $\Hcal^T$ is
the closed linear envelope of this basis.
\end{defn}

The level-matching conditions \myref{baaj} and \myref{baak}
imply that the total eigenvalue $k_m^I$ of $\hat p^I$
and the left- and right-moving momentum eigenvalues
are related by $k_{Ln(m)}^I=k_{Rp(m)}^I=k_m^I/2$
for vectors in the sequence \myref{babn}, and that
similarly for the eigenvalues $w_m^I$ of $\hat L^I$ we have
$w_{Ln(m)}^I=w_{Rp(m)}^I=w_m^I/2$.  On the other hand
\myref{baal} implies $k_{Rp}^I = w_{Rp}^I$,
which in turn yields $k_m^I = w_m^I$, the quantum
analogue of the classical constraint \myref{ouuw}.
From these facts we also find that the left-moving operator
$(p_L^I + L_L^I) \otimes 1_R$
appearing in \myref{uyqy} has eigenvalues $k_m^I$:
\beq
\[ (p_L^I + L_L^I) \otimes 1_R \] \ket{\psi_m}^T
= k_m^I \ket{\psi_m}^T .
\label{babp}
\eeq
It is convenient to define
\beq
H^I \equiv (p_L^I + L_L^I) \otimes 1_R.
\label{babq}
\eeq
The classical boundary condition \myref{bbc}, which was
satisfied because of \myref{babr}, is now satisfied at
the operator level on $\Hcal^T$ because of
\beq
\[ H^I \spl \] \ket{\psi_m}^T \to
\[ H^I (\spl + \pi) \] \ket{\psi_m}^T =
\[ H^I \spl + k_m^I \pi \] \ket{\psi_m}^T ,
\label{babs}
\eeq
provided $(k_m \equiv (k_m^1,\ldots,k_m^{16}))$
\beq
k_m \in \eelat .
\label{babt}
\eeq

In addition to the world-sheet boson
factors $\Hcal^B$ and $\Hcal^T$ of the full
Hilbert space, which have just been described, we have a world-sheet
fermion
factor which is the sum of the \git{Neveu-Schwarz (NS) sector}\
and the \git{Ramond (R) sector.}  The
two sectors correspond to the two choices of boundary
conditions in \myref{fbc}.  One imposes
canonical anticommutation relations on the modes appearing
in \myref{baag} and \myref{baah}:
\beq
\{ b_r^i, b_s^j \} = \delta^{ij} \delta_{r+s,0}, \qquad
\{ d_m^i, d_n^j \} = \delta^{ij} \delta_{m+n,0}.
\eeq

For the NS sector we define a vacuum state which is
annihilated by all positive modes:
\beq
b_{r>0}^i \ket{0}^{(NS)} = 0 \qquad i=1,\ldots,8.
\eeq
On the other hand
the zero mode algebra for the R sector implies that
the vacuum state is in a spinor representation
of $SO(8)$, which we write as $\ket{\alpha}^{(R)}$.
It too is defined to be annihilated by all positive
modes:
\beq
d_{m>0}^i \ket{\alpha}^{(R)} = 0 \qquad i=1,\ldots,8.
\eeq
By acting on $\ket{0}^{(NS)}$ with negative modes $b_{r<0}^i$
we generate an infinite sequence of vectors:
\beq
\ket{0}^{(NS)}, b_{r<0}^i \ket{0}^{(NS)},
b_{r<0}^i b_{s<0}^j \ket{0}^{(NS)}, \ldots
\label{baam}
\eeq

\begin{defn}
The closed linear envelope of the sequence \myref{baam} is the
Neveu-Schwarz Hilbert space $\Hcal^{(NS)}$.
\end{defn}

By acting on $\ket{\alpha}^{(R)}$ with modes $d_{m \leq 0}^i$
we generate another infinite sequence of vectors:
\beq
\ket{\alpha}^{(R)}, d_{m \leq 0}^i \ket{\alpha}^{(R)},
d_{m \leq 0}^i d_{n \leq 0}^j \ket{\alpha}^{(R)}, \ldots
\label{baan}
\eeq

\begin{defn}
The closed linear envelope of the sequence \myref{baan} is the
Ramond Hilbert space $\Hcal^{(R)}$.
\end{defn}

The next step is to define \git{fermion number operators}
\beq
{\bf F}(NS) = \sum_{r>0} \sum_{i=1}^8 b_{-r}^i b_r^i, \qquad
{\bf F}(R) = \sum_{m>0} \sum_{i=1}^8 d_{-m}^i d_m^i,
\eeq
and corresponding \git{G-parity operators}
\beq
G(NS) = - (-1)^{{\bf F}(NS)}, \qquad
G(R) = (-1)^{{\bf F}(R)}.
\eeq
Note that each element in the sequence \myref{baam}
is either odd or even with respect to $G(NS)$. 
We can decompose $\Hcal^{(NS)}$ into the direct
sum of a subspace $\Hcal^{(NS)}_+$ which is even
with respect to $G(NS)$ and a subspace $\Hcal^{(NS)}_-$
which is odd with respect to $G(NS)$;  thus
we write $\Hcal^{(NS)}=\Hcal^{(NS)}_+ \oplus \Hcal^{(NS)}_-$.
A similar decomposition occurs for $\Hcal^{(R)}$ with
respect to $G(R)$.

\begin{defn}
The G-parity even subspace $\Hcal^{(NS)}_+$ of
$\Hcal^{(NS)}$ is the closed linear envelope of the
infinite subsequence of elements of \myref{baam}
which are neutral with respect to $G(NS)$.
Similarly
the G-parity even subspace $\Hcal^{(R)}_+$ of
$\Hcal^{(R)}$ is the closed linear envelope of the
infinite subsequence of elements of \myref{baan}
which are neutral with respect to $G(R)$.
The two projective Hilbert spaces so obtained
are said to be those which ``survive'' the
{\bf GSO projection} \cite{GSO}.
\end{defn}

\begin{defn}
The Hilbert space $\Hcal^F$ is the direct sum of
the projective Hilbert spaces
$\Hcal^{(NS)}_+$ and $\Hcal^{(R)}_+$:
\beq
\Hcal^F \equiv \Hcal^{(NS)}_+ \oplus \Hcal^{(R)}_+
\eeq
\end{defn}

Finally, we assemble the various factors to
give the complete description of the Hilbert
space of the ten-dimensional heterotic theory.
\begin{defn}
The Hilbert space $\Hcal$ of the ten-dimensional
heterotic theory is given by
\beq
\Hcal \equiv \Hcal^B \otimes \Hcal^T \otimes \Hcal^F .
\eeq
\end{defn}
The Hilbert space may also be written as the direct sum
of the {\it overall} NS and R sectors:
\beq
\Hcal = (\Hcal^B \otimes \Hcal^T \otimes \Hcal^{(NS)}_+)
\oplus (\Hcal^B \otimes \Hcal^T \otimes \Hcal^{(R)}_+).
\eeq
In the four-dimensional construction which we next
consider, we will find a proliferation of sectors,
corresponding to a greater variety of ways in which
closed string boundary conditions can be satisfied
on an orbifold.

\section[Four-dimensional $Z_3$ Construction]
{Four-dimensional ${\bf Z_3}$ Construction}
\label{obo}
I now describe how the ten-dimensional construction
may be modified to yield a theory with
a four-dimensional space-time.  The essential
idea is to suppose that six of the spatial
dimensions of the original theory have
as their target space the six-dimensional
$Z_3$ orbifold, of size so small that the
extra dimensions cannot be resolved (by mere mortals).
Consistency of the theory
requires other modifications, as we will see.
The result is a much more realistic theory:
three noncompact spatial dimensions; a gauge
symmetry group of dimension smaller than
\eetee; much more variety in the possibilities
for the low-energy spectrum and effective
couplings.

\subsection{Classical String}
\label{4dcs}
At the classical level, the image of the motion
of the string in the six-dimensional
compact space is specified by a two parameter
map $X(\s,\tau)$ which has
a component expression of the form \myref{comp}:
\beq
X(\s,\tau)=
\sum_{i=1}^6 X^i(\s,\tau) \, e_i.
\label{baap}
\eeq

Recall that the parameter $\s$ labels points along the string,
and that $\s \to \s + \pi$ as one goes once around
the string.  Since the heterotic theory is a theory of closed strings,
$X(\s,\tau)$ and $X(\s + \pi,\tau)$
should be equivalent points on the orbifold.
As a consequence of Definition~\ref{abat}, $X(\s,\tau)$
need only be closed up to a space group element.\myfoot{Alan
says that the mathematical description is in terms
of the \git{homotopy class} of the strings.}
For the {\it $(\omega,\ell)$ sector,}
\beq
X(\s + \pi,\tau)
= (\omega,\ell) \cdot X(\s,\tau).
\label{2.1}
\eeq

If we apply some other space group element $(\omega',\ell')$ to
\myref{2.1}, we find
\beq
(\omega',\ell') \cdot X(\s + \pi,\tau) =
\left[ (\omega',\ell') \cdot (\omega,\ell) 
\cdot (\omega',\ell')^{-1} \right] \cdot
(\omega',\ell') \cdot X(\s,\tau).
\eeq
Because $(\omega',\ell') \cdot X(\s,\tau)$ and $X(\s,\tau)$
are equivalent world-sheet maps into the orbifold, the boundary
condition
\beq
X(\s + \pi,\tau) =
(\omega',\ell') \cdot (\omega,\ell)
\cdot (\omega',\ell')^{-1} \cdot X(\s,\tau)
\label{2.2}
\eeq
must be treated as equivalent to \myref{2.1}.
That is, boundary conditions in the same conjugacy class
as $(\omega,\ell)$
are equivalent\myfoot{Alan says these are described
by the \git{free homotopy classes.}} because they are
related to each other under the action of
the space group~\cite{DHVW86}.

Recall from Section \ref{ccfp}
that there are 27 such conjugacy classes
associated with sectors twisted by $\theta$
and that each conjugacy class corresponds
to the 27 inequivalent fixed points of
the $Z_3$ orbifold.  Thus, the inequivalent fixed points provide
a labeling system for the conjugacy classes.
Since these sectors do not mix with
each other under the action of the space group,
I regard them as 27 {\it different} twisted sectors.

We have seen in Section \ref{rot}
that in the heterotic string there exist
internal degrees of freedom:  right-moving
world-sheet fermions $\psi^i(\s_-)$ and
left-moving internal world-sheet bosons
$X^I(\s_+)$.  I denote these
collectively by $\Psi(\s,\tau)$.
Nontrivial boundary conditions are typically
extended to these other fields $\Psi(\s,\tau)$
in each sector (untwisted, 27 twisted, 27 antitwisted).
For the $(\omega,\ell)$ sector,
defined by \myref{2.1}, the extension
may be written schematically as
\beq
\Psi(\s+\pi,\tau) = U(\omega,\ell) \cdot \Psi(\s,\tau) .
\label{2r}
\eeq
Here, $U$ is a map from the space group $S$ to
a transformation group ${\cal T}$ acting on
the target space of $\Psi(\s,\tau)$.
Consistency requires this map to
be a \git{homomorphism} of the space group:
\beq
U(\omega,\ell) \cdot U(\omega',\ell')
\simeq U(\omega \omega',\omega \ell' + \ell),
\label{2p}
\eeq
where ``$\simeq$'' denotes equivalence,\myfoot{
Alan had various comments here.  He thought I
should identify the larger manifold that
we quotient here.  E.g., quotient of $\Rbf^6 \times \Rbf^{16}$.
He also mentioned the phrase ``bundle over the orbifold''
and \git{orbibundle.} }
the precise meaning of which depends
on the nature of $\Psi(\s,\tau)$, as I will illustrate
in explicit examples below.

As mentioned in Section~\ref{abas}, the space group has
four generators, which we may choose as
$(\theta,0)$, $(1,e_1)$, $(1,e_3)$ and $(1,e_5)$.
Now suppose we specify the map $U : S \to {\cal T}$
for these operators.  That is, we fix
$U(\theta,0),U(1,e_1),U(1,e_3)$ and $U(1,e_5)$.
Any element of $(\omega,\ell) \in S$ may be obtained from a product
of the four generators
$(\theta,0),(1,e_1),(1,e_3)$ and $(1,e_5)$, so the homomorphism
requirement \myref{2p} implies $U(\omega,\ell)$
by taking the corresponding product of
$U(\theta,0),U(1,e_1),U(1,e_3)$ and $U(1,e_5)$.

\egbox{
\rm
Consider the sixteen internal bosonic degrees
of freedom $X^I(\s_+)$.  In the twisted sectors,
the $X^I(\s_+)$ are typically
assigned nontrivial boundary conditions
according to a homomorphism $U$.
As described above, we may define $U$ through
a map of the space group generators
into transformations on the $X^I(\s_+)$.
In the \git{shift embedding} construction studied here, this consists of
a set of translations on the target space
for the $X^I(\s_+)$:
\beqa
\[ U( \theta,0) \cdot X(\s_+) \]^I & = & X^I(\s_+) + \pi V^I, \nnn
\[ U( 1,e_i) \cdot X(\s_+) \]^I & = &
   X^I(\s_+) + \pi a_i^I, \qquad \forall \; i=1,3,5.
\label{edf}
\eeqa
The vector $V$ is referred to as the \git{shift embedding}
of the space group generator $(\theta,0)$;
equivalently, $V$ embeds the twist operator $\theta$.
Likewise, the vectors $a_i$ embed the
other three space group generators $(1,e_i)$,
$i=1,3,5$ respectively.  They
are referred to as \git{Wilson lines}
because of their interpretation as background
gauge fields in the compact space.\myfoot{Alan
suggests I relate the conventional notion
of a Wilson loop.}
(It is worth noting that nontrivial
gauge field configurations in an extra-dimensional
compact space were used by Hosotani in a field theory
context to achieve gauge symmetry breaking \cite{Hos83b};
the nontrivial $a_1,a_3$ in the \bsa\ models represent
a ``stringy'' version of the Hosotani mechanism,
allowing one to obtain \git{standard-like} $G$;
that is, $G$ is a product group with factors $SU(3) \times
SU(2)$ from the start.)

I now construct a general twisted sector
embedding by applying the homomorphism constraint,
making use of the embeddings \myref{edf}.
First note the space group multiplication
($n_1$, $n_3$ and $n_5$ are integral powers $0,\pm 1$)
\beq
(1,e_1)^{n_1} \cdot (1,e_3)^{n_3} \cdot (1,e_5)^{n_5} \cdot
(\theta,0) =
(\theta,n_1 e_1 + n_3 e_3 + n_5 e_5).
\label{hty}
\eeq
This is the space group element
$(\omega,\ell) = (\theta,n_1 e_1 + n_3 e_3 + n_5 e_5)$
labeling one of
the 27 twisted conjugacy classes.

We build up the corresponding embedding
$U(\theta,n_1 e_1 + n_3 e_3 + n_5 e_5)$
with products of $U$s defined in \myref{edf} according
to the multiplication of space group elements
in \myref{hty}:
\beq
U(1,e_1)^{n_1} \cdot U(1,e_3)^{n_3} \cdot U(1,e_5)^{n_5} \cdot
U(\theta,0) =
U(\theta,n_1 e_1 + n_3 e_3 + n_5 e_5).
\eeq
Then from \myref{edf} it is easy to see that
the embedding of the 
boundary condition \myref{2.1} for 
twisted sector labeled by $(n_1,n_3,n_5)$ is
given by:
\beqa
X^I(\s_+ + \pi) & = & U(\theta,n_1 e_1 + n_3 e_3 + n_5 e_5)^I_J
\, X^J(\s_+) \nnn
& = & X^I(\s_+) + \pi E^I(n_1,n_3,n_5),
\label{2q} \\
E(n_1,n_3,n_5) & \equiv & V + n_1 a_1 + n_3 a_3 + n_5 a_5.
\label{evd}
\eeqa
Note that the total shift 
is described by a sixteen-dimensional
embedding vector $E(n_1,n_3,n_5)$.

Consistency conditions \cite{IMNQ88,BLT88} for \emset\ following
from the homomorphism condition \myref{2p}
have been accounted for in
the embeddings enumerated\footnote{The origin and significance
of these embedding vectors will be discussed in Chapter~\ref{slz}.}
in Appendix \ref{emt}.
For example, $(\theta,n_1 e_1 + n_3 e_3 + n_5 e_5)^3=(1,0)$
implies that we must have
\beq
[U(\theta,n_1 e_1 + n_3 e_3 + n_5 e_5)^3]^I_J
\, X^J(\s_+) = X^I(\s_+) + 3 \pi E^I(n_1,n_3,n_5)
\simeq X^I(\s_+).
\eeq
This last step is true because the $X^I(\s_+)$
have as their target space the
the \eetee\ root torus where
\beq
X^I(\s_+) \simeq X^I(\s_+) + \pi L^I,
\qquad \forall \; L \in \eelat,
\eeq
{\it and} the embedding vectors are constrained
to satisfy $3 E(n_1,n_3,n_5) \in \eelat$.
The results of a detailed study of these aspects
of the underlying string theory 
\cite{IMNQ88,BLT88} have been built into
the embeddings given in Appendix \ref{emt}
and the recipes given below.
}

\egbox{ \label{baas}
The NSR fermions $\psi^i(\s_-)$, $i=1,\ldots,6$,
associated with the compact space are also assigned
modified boundary conditions for each conjugacy class.
In fact, the modification must mirror what occurs
for $X_R^i(\s_-)$ for world-sheet supersymmetry
to be preserved in the right-moving sectors,
an important ingredient for the absence of \git{tachyons}
in the physical spectrum.  Because the world-sheet
fermions enter into the action through a term of
the form $\psi^i \p_+ \psi_i$, shifts are not an
invariance.  However, the point group action is,
and it is this which we ``lift'' to the fermions.

I denote the six components in the root lattice basis:
\beq
\psi(\s_-) \equiv \sum_{i=1}^6 \psi^i(\s_-) e_i.
\label{baau}
\eeq
For the NS sector associated with the
conjugacy class of space group element $(\omega,\ell)$,
we have
\beq
\psi(\s_- - \pi) = - \omega \cdot \psi(\s_-)
= - \sum_{i=1}^6 \psi^i(\s_-) \omega \cdot e_i.
\label{baav}
\eeq
As for the bosonic fields of the compact space,
we obtain the action of the point group element
$\omega$ on the lattice basis vectors $e_i$ through
\myref{pnh}.
For the R sector associated with the
conjugacy class of space group element $(\omega,\ell)$,
we have
\beq
\psi(\s_- - \pi) = \omega \cdot \psi(\s_-)
= \sum_{i=1}^6 \psi^i(\s_-) \omega \cdot e_i.
\label{baaw}
\eeq
}

As noted in Section~\ref{ccfp}, the boundary conditions
are labeled by the conjugacy classes of the
space group; it is clear that in the general
case, the extension $U$ in \myref{2r}---and
more specifically the embedding
$E(n_1,n_3,n_5)$---will be different
for each conjugacy class.
In the description of string states,
it is therefore convenient to decompose the
Hilbert space into sectors, with each sector
corresponding to a particular conjugacy class.\myfoot{I
might actually want to discuss first-quantization
in a separate section.  I am pretty certain that
each classical conjugacy class is, by a sort of
correspondence principle which is really an assumption,
promoted to its own Hilbert space.  After all, a
string state cannot have definite $\hat p_L$ and $\hat
p_R$ eigenvalues {\it and} be at a fixed point.  This
would violate the uncertainty principle.}
For the $Z_3$ orbifold, one
has an \git{untwisted} sector, 27 \git{twisted} sectors
corresponding to fixed point (conjugacy class)
labels $(n_1,n_3,n_5)$, $n_i = 0,\pm 1$,
and 27 \git{antitwisted} sectors with similar labeling.
The 27 (anti)twisted sectors are often lumped together
and regarded as a single (anti)twisted sector, since the
(anti)twist (i.e., the point group element)
is identical among them; I prefer
not to do this here.
The antitwisted sector of the $Z_3$ orbifold
merely contains the antiparticle
states of the twisted sector, so
we need not discuss it below.

\subsection{Mode Expansions}
Rather than \myref{baap}, it is best to use the
complex parameterization.  That is, we define
complex string coordinates according to \myref{ccmp}:
\beq
Z^i(\s,\tau) \equiv X^i(\s,\tau) + e^{2 \pi i / 3} X^{i+1}(\s,\tau),
\quad
\bbar{Z}^i(\s,\tau) \equiv X^i(\s,\tau) + e^{-2 \pi i / 3} X^{i+1}(\s,\tau),
\quad
i=1,3,5.
\label{scmp}
\eeq
In this basis, \myref{2.1} for $\omega = \theta^n$
takes the form corresponding to \myref{cspc},
\beq
\Zbf(\s+\pi,\tau) = e^{2n\pi i/3} \Zbf(\s,\tau) + {\bf b},
\label{uiat}
\eeq
where ${\bf b} = (b^1,b^3,b^5)$ and
$b^i$ are given by \myref{lcmp}.

From \myref{baaa}, the equations of motion for these string coordinates
are
\beq
\p_+ \p_- Z^i = (\p_\tau^2-\p_\s^2) Z^i = 0.
\eeq
The general solution can be written in the form
\beq
Z^i(\s,\tau) = z^i + q^i \tau + M^i \s + f^i(\smi) + \tilde f^i(\spl) ,
\label{baaq}
\eeq
where $\p_+ f^i(\smi) \equiv 0$ and $\p_- \tilde f^i(\spl) \equiv 0$.
For the $\tw$ sector we must have
\beq
f^i(\smi - \pi) = e^{2\pi i/3} f^i(\smi), \qquad
\tilde f^i(\spl + \pi) = e^{2\pi i/3} \tilde f^i(\spl),
\eeq
according to \myref{uiat}.  Thus \myref{baaq} leads to the
Fourier mode expansion
\beqa
Z^i(\s,\tau) & = & z^i + q^i \tau + M^i \s
+ {i \over 2} \sum_{n \in \Zbf}
{1 \over n + 1/3} \alpha_{n+1/3}^i e^{-2 i (n+1/3) \smi} \nnn
&& + {i \over 2} \sum_{n \in \Zbf}
{1 \over n - 1/3} \tilde \alpha_{n-1/3}^i e^{-2 i (n-1/3) \spl},
\label{iyat}
\eeqa
where the normalization of mode coefficients is
a matter of convention, chosen for future convenience.
Similarly for the $\bbar{Z}^i$ we find
\beqa
\bbar{Z}^i(\s,\tau) & = & \bz^i + \bar q^i \tau + \bbar{M}^i \s
+ {i \over 2} \sum_{n \in \Zbf}
{1 \over n - 1/3} \alpha_{n-1/3}^i e^{-2 i (n-1/3) \smi} \nnn
&& + {i \over 2} \sum_{n \in \Zbf}
{1 \over n + 1/3} \tilde \alpha_{n+1/3}^i e^{-2 i (n+1/3) \spl}.
\eeqa
From $\bbar{Z}^i(\s,\tau)=[Z^i(\s,\tau)]^*$ it follows that
\beq
\bz^i = (z^i)^*, \qquad
\bar q^i = (q^i)^*, \qquad
\bbar{M}^i = (M^i)^*,
\eeq
for the zero modes, while for the \git{oscillator} modes
\beq
\alpha_{n-1/3}^i = (\alpha_{-n+1/3}^i)^*, \qquad
\tilde \alpha_{n+1/3}^i = (\tilde \alpha_{-n-1/3}^i)^*.
\eeq

Eq.~\myref{uiat} together with the mode
expansion \myref{iyat} implies for the zero
modes
\beq
\theta z^i + b^i = z^i + \pi M^i, \quad
\theta q^i = q^i, \quad \theta M^i = M^i,
\eeq
since \myref{uiat} must hold for all $\s,\tau$.
This has solution
\beq
M^i \equiv q^i \equiv 0, \quad
z^i \equiv (1-\theta)^{-1} b^i.
\label{uits}
\eeq
That is, the \git{center of mass} coordinates
$z^i$ are fixed points, in correspondence with
the lattice group element $b^i$ according
to Eq.~\myref{uiwt} of Section \ref{cbss}.

Corresponding to this complex basis is a Cartesian
basis $(i=1,3,5)$:
\beq
Z^i(\s,\tau) \equiv U^i(\s,\tau) + i V^i(\s,\tau), \qquad
\bbar{Z}^i(\s,\tau) \equiv U^i(\s,\tau) - i V^i(\s,\tau).
\label{baat}
\eeq
Zero modes $u^i,p_u^i,L_u^i$ of $U^i$ and
zero modes $v^i,p_v^i,L_v^i$ of $V^i$ and are
given by
\beqa
z^i &=& u^i + i v^i, \qquad \bz^i = u^i - i v^i,
\\
q^i &=& p_u^i + i p_v^i, \qquad \bar q^i = p_u^i - i p_v^i,
\\
M^i &=& L_u^i + i L_v^i, \qquad \bbar{M}^i = L_u^i - i L_v^i.
\eeqa
Eq.~\myref{uits} implies
\beq
p_u^i = p_v^i = L_u^i = L_v^i \equiv 0.
\eeq

It is not difficult to decompose \myref{iyat} into left-
and right-moving coordinates, $Z^i(\s,\tau) \equiv Z^i_L(\spl)
+ Z^i_R(\smi)$.  The oscillator modes have already been
split up, so we need only note that
\beq
z^i + q^i \tau + M^i \s =
\[\half z^i + \half ( q^i + M^i) \spl\]
+\[\half z^i + \half ( q^i - M^i) \smi \].
\label{baar}
\eeq
As in the ten-dimensional case, we have split the
center of mass coordinate $z^i$ evenly between
the left-moving part (first term in brackets) and
the right-moving part (second term in brackets).

Next I examine the world-sheet fermions with indices
corresponding to the six-dimensional compact space.
These were already discussed to some extent in
Example \ref{baas} above.  Starting from the root
basis \myref{baau} we define a complex basis $(i=1,3,5)$:
\beq
\chi^i(\smi) \equiv \psi^i(\smi)
+ e^{2 \pi i / 3} \psi^{i+1}(\smi), \qquad
\bar \chi^i(\smi) \equiv \psi^i(\smi)
+ e^{-2 \pi i / 3} \psi^{i+1}(\smi).
\eeq
In analogy to \myref{baat} we may also define a
Cartesian basis:
\beq
\chi^i(\smi) \equiv \psi_u^i + i \psi_v^i, \qquad
\bar \chi^i(\smi) \equiv \psi_u^i - i \psi_v^i.
\label{baba}
\eeq

The embedding of the orbifold
action in the Neveu-Schwarz (NS) and Ramond (R) subsectors
of the $\tw$ sector---already
described in Eqs.~\myref{baav} and \myref{baaw} respectively---is
a simple phase rotation in the complex basis:
\beq
\chi^i(\smi-\pi) = \left\{ \begin{array}{lc} 
-e^{2\pi i/3} \chi^i(\smi) & {\rm(NS)} \\
e^{2\pi i/3} \chi^i(\smi) & {\rm(R)}
\end{array} \right. 
\label{baax}
\eeq
The conjugate fields have boundary conditions
corresponding to the $\tw^{-1}$ sector:
\beq
\bar \chi^i(\smi-\pi) =  \left\{ \begin{array}{lc}
-e^{-2\pi i/3} \bar \chi^i(\smi) & {\rm(NS)} \\
e^{-2\pi i/3} \bar \chi^i(\smi) & {\rm(R)}
\end{array} \right. 
\label{baay}
\eeq

The equations of motion \myref{baaa} now read
\beq
\p_+ \chi^i = \p_+ \bar \chi^i = 0.
\label{baaz}
\eeq
Eqs.~\myref{baax}-\myref{baaz} together imply mode
expansions
\beqa
\chi^i(\smi) &=&
\sum_{r \in \Zbf + 1/2} b_{r+1/3}^i e^{-2i(r+1/3)\smi} \qquad {\rm(NS)}, \nnn
\chi^i(\smi) &=&
\sum_{m \in \Zbf} d_{m+1/3}^i e^{-2i(m+1/3)\smi} \qquad {\rm(R)}, \nnn
\bar \chi^i(\smi) &=&
\sum_{r \in \Zbf + 1/2} \bar b_{r-1/3}^i e^{-2i(r-1/3)\smi}
\qquad {\rm(NS)}, \nnn
\bar \chi^i(\smi) &=&
\sum_{m \in \Zbf} \bar d_{m-1/3}^i e^{-2i(m-1/3)\smi} \qquad {\rm(R)}.
\eeqa

The world-sheet fermions $\psi_u^i,\psi_v^i$ in
the Cartesian basis \myref{baba} are Majorana-Weyl.
A basis therefore exists where these fields
are real, and it follows that in this basis the
Fourier mode coefficients are related by:
\beqa
\bar b_{-r-1/3}^i &=& \( b_{r+1/3}^i\)^* \qquad {\rm(NS)}, \nnn
\bar d_{-m-1/3}^i &=& \( d_{r+1/3}^i\)^* \qquad {\rm(R)}.
\label{babi}
\eeqa

The world-sheet fermions $\psi^a(\smi) \; (a=7,8)$ corresponding
to the four-dimensional Minkowski space $M_4$
in light-cone gauge (two physical polarizations) are unaffected
by the orbifold action; that is, the space group
homomorphism \myref{2r} for these fields is the identity map.
Thus they have the same ``modings'' (i.e., \git{monodromy})
as in the ten-dimensional
case, Eqs.~\myref{baag} and \myref{baah}, for the
NS and R sectors respectively.  The shift embedding
\myref{2q} for the bosonic gauge degrees of freedom
$X^I(\spl) \; (I=1,\ldots,16)$ does not affect the
moding for these coordinates; the expansions still
take the ten-dimensional form \myref{zmcb}.  What {\it is}
changed is the set of allowed values for the winding vector $L^I$.

\subsection{Quantum Mechanical String}
Quantization of the classical theory just described
is quite similar to what was done in the ten-dimensional case,
described in Section \ref{bsac}.  However, the new
features---twisted boundary conditions, fractional
modings, etc.---must be realized at the operator level.
In particular, the requirement \myref{2.1} is extended
to the quantized theory, where $X(\s,\tau)$ is promoted
to a quantum operator.  For each conjugacy class
$c,c',\ldots$ (cf. Section \ref{ccfp}) we construct
a Hilbert space $\Hcal_c, \Hcal_{c'}, \ldots$,
respectively, of \git{physical states.}
Various projections which generalize the projective Hilbert
space constructions of the ten-dimensional theory
are typically required.  The full Hilbert space
of the heterotic orbifold theory is then taken to be the
direct sum of the various sectors:
\beq
\Hcal \equiv \Hcal_c \oplus \Hcal_{c'} \oplus \cdots
\label{babb}
\eeq

As we saw in Section \ref{ccfp}, each conjugacy class is
labeled by a space group element $(\omega,\ell)$,
the class leader.  Thus we may also write \myref{babb}
\beq
\Hcal \equiv \Hcal_{(\omega,\ell)} \oplus
\Hcal_{(\omega',\ell')} \oplus \cdots
\label{babc}
\eeq
Each of the Hilbert spaces
$\Hcal \equiv \Hcal_{(\omega,\ell)},
\Hcal_{(\omega',\ell')}, \ldots$
is referred to
as a \git{sector} of the theory.
Each sector is distinguished by an
inequivalent set of boundary conditions
in the corresponding classical description.
This includes internal degrees of freedom such
as the world-sheet fermions.  Thus we have NS and R
\git{subsectors} for each class leader $(\omega,\ell)$:
\beq
\Hcal_{(\omega,\ell)} = \Hcal_{(\omega,\ell)}^{(NS)} \oplus 
\Hcal_{(\omega,\ell)}^{(R)}.
\eeq

I now illustrate aspects of the modified construction
in the orbifold case, in order to emphasize how
the quantum theory is different from that of the
ten-dimensional theory.  Quantization of the untwisted sector
$(\omega,\ell)=(1,\ell)$---grouping together
subsectors labeled by different
winding vectors---is essentially the same as
quantization in the ten-dimensional theory, since no
fractional modings occur.  Rather, a projection
onto states invariant with respect to the orbifold
action and its embedding is enforced; the projection
thus defines the untwisted sector Hilbert space as a subspace
of the ten-dimensional Hilbert space.  The
phenomenologial consequences
of this projection are summarized in Section \ref{rec}
below.  The more interesting case is that of a
$\tw$ sector; I will concentrate on this in the
following discussion.

We construct $\Hcal_{\tw L}^B \otimes \Hcal_{\tw R}^B$,
the product of left- and right-moving spaces.  Zero
mode operators are defined (for the six-dimensional
compact space coordinates):
\beqa
\hat z^i &=& z_L^i \otimes 1_R + 1_L \otimes z_R^i, \qquad
\hat {\bar z}^i = \bar z_L^i \otimes 1_R + 1_L \otimes \bar z_R^i, \nnn
\hat q^i &=& q_L^i \otimes 1_R + 1_L \otimes q_R^i, \qquad
\hat {\bar q}^i = \bar q_L^i \otimes 1_R + 1_L \otimes \bar q_R^i, \nnn
\hat M^i &=& M_L^i \otimes 1_R + 1_L \otimes M_R^i, \qquad
\hat {\bbar{M}}^i = \bbar{M}_L^i \otimes 1_R + 1_L \otimes \bbar{M}_R^i.
\label{babd}
\eeqa
Mirroring the classical theory, we can also write
these in terms of a Cartesian basis, defined by:
\beqa
z_L^i &=& u_L^i + i v_L^i, \qquad
\bz_L^i = u_L^i - i v_L^i, \nnn
q_L^i &=& p_{uL}^i + i p_{vL}^i, \qquad
\bar q_L^i = p_{uL}^i - i p_{vL}^i, \nnn
M_L^i &=& L_{uL}^i + i L_{vL}^i, \qquad
\bbar{M}_L^i = L_{uL}^i - i L_{vL}^i,
\eeqa
with similar equations for the right-movers.
The Cartesian basis is then treated in the same manner as the
Cartesian zero modes in \myref{ptdf}.  We impose
semi-canonical commutation relations:
\beq
[u_L^i,p_{uL}^j]= [u_R^i,p_{uR}^j] =
[v_L^i,p_{vL}^j]= [v_R^i,p_{vR}^j] =
{i \over 2} \delta^{ij} ,
\eeq
with all others vanishing.  This implies for the
complex operators
\beq
[z_L^i,\bar q_L^j]= [\bar z_R^i,q_R^j] = i \delta^{ij},
\eeq
with all others vanishing.  For the tensored
operators we thereby obtain
\beq
[\hat z^i,\hat {\bar q}^j]= [\hat {\bar z}^i,\hat q^j] = 2i \delta^{ij}.
\label{fcmz}
\eeq

Recall the classical decomposition to left- and right-movers
\myref{baar}.  In the quantum theory instead, the $1/2$ factor
(heuristically) indicates that we act only on half the
Hilbert space; more precisely, we make the
(classical) $\to$ (quantum) correspondence $(i=1,3,5)$
\beq
\half (q^i+M^i)\spl \to [(q_L^i + M_L^i) \otimes 1_R] \spl, \qquad
\half (q^i-M^i)\smi \to [1_L \otimes (q_R^i - M_R^i)] \smi .
\eeq
On the other hand
\beq
q^i \tau + M^i \s \to \hat q \tau + \hat M \s.
\eeq
Thus the correspondence principle applied to \myref{baar}
yields the constraint
\beq
\hat q \tau + \hat M \s \equiv
[(q_L^i + M_L^i) \otimes 1_R] \spl +
[1_L \otimes (q_R^i - M_R^i)] \smi .
\label{babe}
\eeq
It is easy to check---by substitution of the expressions
for $\hat q,\hat M$ in \myref{babd}---that Eq.~\myref{babe}
holds if and only if we work on a subspace of
$\Hcal_{\tw L}^B \otimes \Hcal_{\tw R}^B$ where
\beq
q_L \otimes 1_R - 1_L \otimes q_R \equiv 0, \qquad
M_L \otimes 1_R - 1_L \otimes M_R \equiv 0.
\label{babf}
\eeq
This defines a projective Hilbert space
$\Hcal_{\tw L}^B \otimes \Hcal_{\tw R}^B \to \Hcal_{\tw}^B$,
as I now describe.

We define complete orthonormal sequences on
$\Hcal_{\tw L}^B$ and $\Hcal_{\tw R}^B$
respectively, satisfying
\beqa
q_L^i \ket{\chi_n}_{\tw L}^B &=& \kappa_{Ln}^i \ket{\chi_n}_{\tw L}^B,
\qquad n=1,2,\ldots \nnn
q_R^i \ket{\eta_p}_{\tw R}^B &=& \kappa_{Rp}^i \ket{\eta_p}_{\tw R}^B,
\qquad p=1,2,\ldots \nnn
M_L^i \ket{\chi_n}_{\tw L}^B &=& w_{Ln}^i \ket{\chi_n}_{\tw L}^B,
\qquad n=1,2,\ldots \nnn
M_R^i \ket{\eta_p}_{\tw R}^B &=& w_{Rp}^i \ket{\eta_p}_{\tw R}^B,
\qquad p=1,2,\ldots
\eeqa
eigenvalue multiplicities counted.  Then we construct the
corresponding complete orthonormal sequence on
the tensor product space 
$\Hcal_{\tw L}^B \otimes \Hcal_{\tw R}^B$:
\beq
\ket{\chi_n}_{\tw L}^B \otimes \ket{\eta_p}_{\tw R}^B,
\qquad
n,p=1,2,\ldots
\label{babh}
\eeq
From this sequence we select the subsequence of elements
which are in the nullity of both operators in \myref{babf}.
We write these as:
\beq
\ket{\psi_{m}}_{\tw}^B \equiv
\ket{\chi_n(m)}_{\tw L}^B \otimes \ket{\eta_p(m)}_{\tw R}^B,
\qquad m=1,2,\ldots
\label{babg}
\eeq
with $n,p$ ranging all possible choices for which
\beq
\kappa_{Ln}^i = \kappa_{Rp}^i, \qquad w_{Ln}^i = w_{Rn}^i .
\eeq

\begin{defn}
The Hilbert space $\Hcal_{\tw}^B$ is the closed
linear envelope of the set \myref{babg}.  That is,
$\Hcal_{\tw}^B$ is the space spanned by elements
of \myref{babh} which satisfy the level-matching
conditions \myref{babf}.
\end{defn}

Quantization of the world-sheet fermions associated
with the six-dimensional compact space proceeds
in the usual way:  we impose semi-canonical
anticommutation relations between mode operators
and their hermitian conjugates:
\beqa
\{ b_{r+1/3}^i, (b_{r+1/3}^i)^\dagger \} &=& 2 \qquad {\rm(NS)}, \nnn
\{ d_{m+1/3}^i, (d_{m+1/3}^i)^\dagger \} &=& 2 \qquad {\rm(R)},
\eeqa
with all others vanishing.  The factor of 2 is due to the
normalization in the classical definitions \myref{baba}.
In the quantum theory \myref{babi} become
\beqa
\bar b_{-r-1/3}^i &=& \( b_{r+1/3}^i\)^\dagger \qquad {\rm(NS)}, \nnn
\bar d_{-m-1/3}^i &=& \( d_{r+1/3}^i\)^\dagger \qquad {\rm(R)}.
\label{babj}
\eeqa
Then we can rephrase the above anticommutation relations as
\beqa
\{ b_{r+1/3}^i, \bar b_{s-1/3}^j \}
&=& 2 \delta^{ij} \delta_{r+s,0} \qquad {\rm(NS)}, \nnn
\{ d_{m+1/3}^i, \bar d_{n-1/3}^j \}
&=& 2 \delta^{ij} \delta_{m+n,0} \qquad {\rm(R)},
\eeqa
with all others vanishing.  To these we append the modes of
the world-sheet fermions $\psi^a(\smi) \; (a=7,8)$ associated
with $M_4$ in light-cone gauge.  These have the modings and
anticommutation relations of the ten-dimensional theory.

We define $\Hcal_\tw^{(NS)}$ to be the space on which the operators
$b_{r+1/3}^i,\bar b_{r-1/3}^i,b_r^a$ act.\footnote{It is worth
noting that both twisted modes $b_{r+1/3}^i$
and antitwisted modes $\bar b_{r-1/3}^i$ act on
$\Hcal_\tw^{(NS)}$.}  The vacuum is unique and
defined by $(r\in \Zbf+1/2; i=1,3,5;a=7,8)$:
\beq
b_{r+1/3>0}^i \ket{0}_\tw^{(NS)}
=\bar b_{r-1/3>0}^i \ket{0}_\tw^{(NS)}
=b_{r>0}^a \ket{0}_\tw^{(NS)} \equiv 0.
\eeq

\begin{defn}
The space $\Hcal_\tw^{(NS)}$ is the closed linear
envelope of the infinite sequence of vectors
obtained from the repeated application of operators $b_{r+1/3<0}^i,
\bar b_{r-1/3<0}^i,b_{r<0}^a$ to $\ket{0}_\tw^{(NS)}$,
including $\ket{0}_\tw^{(NS)}$ itself (zero applications).
\end{defn}

In the case of $\Hcal_\tw^{(R)}$, the fermions
$\psi^a(\smi) \; (a=7,8)$ have zero modes which
satisfy the $SO(2)$ Clifford algebra
\beq
\{ d_0^a, d_0^b \} = \delta^{ab} .
\eeq
This $SO(2)$ is a subalgebra of the light-cone decomposition
of an $SO(1,3)$ spinor representation associated with
the four-dimensional target space $M_4$.  The vacuum
thus fills out a nontrivial representation of this
$SO(2)$ subalgebra, and we denote this vacuum degeneracy
by a label $\alpha$.  The vacuum conditions are now
\beq
d_{m+1/3>0}^i \ket{\alpha}_\tw^{(R)}
=\bar d_{m-1/3>0}^i \ket{\alpha}_\tw^{(R)}
=b_{m>0}^a \ket{\alpha}_\tw^{(R)} \equiv 0.
\eeq

\begin{defn}
The space $\Hcal_\tw^{(R)}$ is the closed linear
envelope of the infinite sequence of vectors
obtained from the repeated application of operators $d_{m+1/3<0}^i,
\bar d_{m-1/3<0}^i,d_{m \leq 0}^a$ to $\ket{\alpha}_\tw^{(R)}$,
including $\ket{\alpha}_\tw^{(R)}$ itself (zero applications).
\end{defn}

Finally, we define G-parity operators $G(NS)$ and $G(R)$.
Then we make a decomposition of the Hilbert spaces
into G-parity even and odd subspaces, as in the
ten-dimensional case:\footnote{A small modification
to the G-parity operators is required due to the fractional
moding.  See for example \cite{DHVW86}.}
\beqa
\Hcal_\tw^{(NS)} &=& \Hcal_{\tw+}^{(NS)} \oplus \Hcal_{\tw-}^{(NS)}, \nnn
\Hcal_\tw^{(R)} &=& \Hcal_{\tw+}^{(R)} \oplus \Hcal_{\tw-}^{(R)}.
\eeqa

\begin{defn}
The projective Hilbert space $\Hcal_\tw^{F}$ is the
direct sum of G-parity even NS and R sectors.  That is,
\beq
\Hcal_\tw^{F} \equiv \Hcal_{\tw+}^{(NS)} \oplus \Hcal_{\tw+}^{(R)}.
\eeq
\end{defn}

Beside the Hilbert spaces just described, we also
have the space $\Hcal_\tw^T$ associated with the
internal gauge degrees of freedom $X^I(\spl)$.
Quantization here is essentially the same as
in the ten-dimensional case, except for some
restrictions on eigenvalues of the zero mode
operators $p_{L,R}^I, L_{L,R}^I$ to be
addressed in Section \ref{rec}.

\begin{defn}
The full $\tw$ sector Hilbert space is
\beq
\Hcal_\tw \equiv \Hcal_\tw^B \otimes \Hcal_\tw^T \otimes
\Hcal_\tw^F.
\eeq
\end{defn}

\section[$E_8 \times E_8$ and $G$]{${\bf E_8 \times E_8}$ and ${\bf G}$}
\label{eep}
The tree level\footnote{Cf.~the introductory
comments to this chapter.}
gauge group $G$ obtained in the
four-dimensional string construction
described in Section \ref{obo} is a rank sixteen subgroup of \eetee.
The theory on the orbifold involves
``twisting'' the \eetee\ heterotic string.
Even though $G$ is a subgroup of \eetee, its
description on the string side reflects
the \eetee\ symmetry of the original theory.  That is,
$G$ is ``embedded into \eetee.''  To clarify what
is meant by this phrase, I will rehearse a well-known
physical example in Section \ref{su3f}.
Then in Section \ref{eer} I describe the
\eetee\ root system and clarify how we
use it in the description of representations
of $G$.  In Section \ref{rec} explicit examples
of the discussion given here will be presented.

The mathematical theory of Lie algebras and groups
plays a prominant role in the discussion which follows,
as well as in subsequent chapters.
A reader not familiar with this topic
will have great difficulty understanding the
remainder of this thesis,
and should first study some of the numerous
texts and reviews available.  For example,
the basic facts and better known physical
applications of Lie algebras and groups can be found
in Refs.~\cite{Cor84,GM94,Gou82,Geo99,Cah84,Sla81}.

\subsection[$SU(3)_F$ Example]{${\bf SU(3)_F}$ Example}
\label{su3f}
Recall that each irrep of a Lie group
can be identified with
a weight diagram; points on the weight diagram
are labeled by weight vectors.  Well-known
examples are the flavor $SU(3)_F$ weight diagrams of
hadrons containing only $u,d,s$ valence quarks.\myfoot{Show
some diagrams.  Describe them.}
In this case, the weight vectors are two-dimensional,
$(\lambda_1,\lambda_2)$,
with entries corresponding\myfoot{Concrete example...} to eigenvalues
of two basis elements $H^1,H^2$ of a Cartan
subalgebra of $SU(3)_F$.  If we work in the
limit $m_u=m_d \ll m_s$, then $SU(3)_F$ is
not a good symmetry, but the flavor isospin
subgroup $SU(2)_F$ is.  In a well-chosen
basis for $SU(3)_F$,
the weight diagrams of
$SU(2)_F$ are one-dimensional subdiagrams
of the $SU(3)_F$ weight diagrams.  The points
of the one-dimensional $SU(2)_F$ weight
diagrams are labeled by eigenvalues of the 
basis element $I_3$
of a Cartan subalgebra of $SU(2)_F$.  However,
we could just as well continue to label states
by the $SU(3)_F$ weight vectors; the isospin
quantum numbers would be determined by an
appropriate linear combination
\beq
I_3 = \alpha^1 H^1 + \alpha^2 H^2
\label{2.5}
\eeq 
of $SU(3)_F$ Cartan generators.  The additional
information contained in the two-dimensional
$SU(3)_F$ weight vectors, strangeness, determines
quantum numbers under a global $U(1)_S$
symmetry group which commutes with
$SU(2)_F$.  The generator of $U(1)_S$
is given by
\beq
S = s^1 H^1 + s^2 H^2.
\label{2.6}
\eeq
Consistency of this decomposition requires
that for any irrep $R$ of $SU(3)_F$,
\beq
\tra{R} (I_3 S) = 0 \quad
\Rightarrow \sum_{i,j=1}^2 \kappa^{ij} \alpha^i s^j = 0, 
\label{2s}
\eeq
where $\kappa^{ij}$ is defined by
\beq
\tra{R} (H^i H^j) = X(R) \, \kappa^{ij}.
\eeq
To summarize, the symmetry
group is $G_F = SU(2)_F \times U(1)_S$; states are
conveniently labeled by $SU(3)_F$ weight vectors, which allow
one to determine the quantum numbers with
respect to $G_F$; the weight diagrams of $SU(2)_F$ are
best recognized as subdiagrams of $SU(3)_F$ weight diagrams.
We say that $G_F$ is embedded into $SU(3)_F$.

In complete analogy, an irrep of
the gauge symmetry group $G$ of a given orbifold
model will be described by a set of basis states labeled by
weight vectors of \eetee.
The weights with respect to nonabelian
factors of $G$ as well \uone\ charges of the irrep
are determined by these \eetee\ weight
vectors, just as was the case in the $SU(3)_F$
example above.

\subsection[The $E_8 \times E_8$ Root System]
{The ${\bf E_8 \times E_8}$ Root System}
\label{eer}
In this subsection I briefly review some salient
aspects of the $E_8$ and $E_8 \times E_8$
root systems.  Other discussions of the topics
described below can be found in standard textbooks on string theory,
such as \cite{GSW87a,BL94,Pol98b}, as well as texts on
Lie algebras and groups, such as \cite{Cor84}.

Much of the analysis necessary for determining
the spectrum of states in an orbifold model
has to do with the eigenvalues or \git{weights}
of states\footnote{A given ``state''
corresponds to a vector in the \git{representation space}
(equivalently, \git{carrier space} or \git{module}) of the gauge
symmetry group $G$, which is a rank sixteen
subgroup of \eetee.  It is conventional to
work with an eigenbasis with respect to the
Cartan generators.} $\ket{W}$ under
basis elements $H^I$ $(I=1,\ldots,16)$
of the Cartan subalgebra of \eetee:\footnote{In the
underlying theory, the $H^I$ are nothing but the
operators defined in \myref{babq} above.  We saw there
that periodic boundary conditions for the internal bosons
$X^I(\spl)$ imply that the eigenvalues of $H^I$ belong to
the \eetee\ root lattice (cf.~Eq.~\myref{babt}.}
\beq
H^I \ket{W} = W^I \ket{W}.
\eeq
The weights of the adjoint representation
are referred to as \git{roots.}
For \eetee, the adjoint
representation {\it is} the fundamental representation
and higher dimensional representations are obtained
from tensor products of the adjoint representation
with itself.
Weight vectors add when the tensor products
are taken to form higher dimensional
representations; consequently, the weight diagrams of
the higher dimensional representations fill
out a weight lattice, spanned by the basis vectors
of the adjoint representation weight diagram.
In the case of \eetee, this is the
root lattice $\eelat$.
A basis in the root space may be chosen such
that the $E_8$ root lattice can be written
as the (infinite) set of eight-dimensional
vectors
\beq
\elat  =  \left\{ (n_1, \ldots, n_8), \; (n_1 + \half, \ldots, n_8 + \half)
\quad \left| \quad n_1, \ldots, n_8 \in \Zbf, \; \;
\sum_{i=1}^8 n_i = 0 \mod 2 \right. \right\}.
\label{eld}
\eeq
Note that the components of a given $E_8$ root lattice
vector are either all integral or all half-integral.
Lattice vectors $\ell \in \elat$ which satisfy
$\ell \cdot \ell = 2$ (where the ordinary eight-dimensional
``dot product'' is implied) yield the 240
nonzero $E_8$ roots, which we denote $e_1, \ldots, e_{240}$.
By convention, we take as \git{positive roots} those $e_i$
whose first nonzero entry (counting left to right) is
positive.\footnote{Other \git{lexicographic ordering}
conventions for the determination of positivity
for roots will be considered in Chapter \ref{hyc}
below.}
A \git{simple root} is a positive root which
cannot be obtained from the sum of two positive roots.
Eight simple roots exist for $E_8$, which we denote by
$\alpha_1, \ldots, \alpha_8$.  These form a basis for
the $E_8$ root lattice given in \myref{eld}, which may
alternatively be written as
\beq
\elat = \left\{ \left. \; \sum_{i=1}^8 m_i \alpha_i \quad
\right| \quad m_i \in \Zbf \; \right\}.
\label{kitr}
\eeq
The $E_8 \times E_8$ root lattice is constructed
by taking the direct sum of two copies of $\elat$,
which we distinguish by labels $(A)$ and $(B)$:
\beq
\eelat = \Lambda_{E_8}^{(A)} \oplus \Lambda_{E_8}^{(B)}.
\eeq
Thus, an $E_8 \times E_8$ root lattice vector $\ell$ is
a sixteen-dimensional vector satisfying
\beq
\ell = ( \ell_A ; \ell_B ), \qquad
\ell_A \in \Lambda_{E_8}^{(A)}, \qquad 
\ell_B \in \Lambda_{E_8}^{(B)},
\eeq
where we have denoted the first eight entries of $\ell$
by $\ell_A$ and the last eight entries of $\ell$
by $\ell_B$, as in the main text.  The 480 nonzero
roots of $E_8 \times E_8$ are given in this
notation by $(e_i;0)$ and $(0;e_i)$, where $e_i$
is one of the 240 nonzero $E_8$ roots.  Similarly,
the sixteen simple roots of $E_8 \times E_8$ are given
by $(\alpha_i;0)$ and $(0;\alpha_i)$, where
$\alpha_i$ is one of the eight $E_8$ simple roots.
We denote the sixteen roots by $\alpha_1, \ldots,
\alpha_{16}$.  Corresponding to \myref{kitr},
by taking all linear combinations of the sixteen
simple roots with integer-valued coefficients,
one recovers the root lattice $\eelat$.  That is,
\beq
\eelat = \left\{ \left. \; \sum_{i=1}^{16} 
m^i \alpha_i \quad \right| 
\quad m^i \in \Zbf \; \right\} .
\label{2l}
\eeq

The sixteen entries
of a root lattice vector $(n_1, \ldots, n_8; n_9, \ldots, n_{16})$
correspond to eigenvalues with respect to
a basis of the \eetee\ Cartan subalgebra,
which we write as $H^I \; (I = 1, \ldots, 16)$
and which is Cartesian:
\beq
\tra{R} (H^I H^J) = X(R) \; \delta^{IJ} ,
\label{2a}
\eeq
where the trace is taken over an
\eetee\ irrep $R$.  In particular,
the adjoint representation (A) corresponds to
the 480 roots described above.
It is not hard to check
from \myref{eld} that $X(A)=60$, which
is twice the value typically used by phenomenologists.
Thus, the $H^I$ in \myref{2a} and the
eigenvalues in \myref{eld} are larger
by a factor of $\sqrt{2}$ than the
phenomenological normalization.

Of particular importance is the map
of roots $\alpha_i$ into the Cartan
subalgebra defined by
\beq
H(\alpha_i) = \sum_{I=1}^{16} \alpha_i^I H^I.
\label{rtm}
\eeq
From this, one defines an inner product on
the root space:
\beq
\langle \alpha_i | \alpha_j \rangle
\equiv \tra{A} \left[ H(\alpha_i) \cdot H(\alpha_j) \right].
\label{ird}
\eeq
Using \myref{2a}, it is not hard to see that
\beq
\langle \alpha_i | \alpha_j \rangle
= X(A) \; \alpha_i \cdot \alpha_j .
\label{2b}
\eeq
It can be seen that the Dynkin index
$X(A)'$ of the basis \myref{rtm}
is related to the index of \myref{2a}
by $X(A)'=2 X(A)$.  Thus, the generators
\myref{rtm} are larger by a factor
of 2 than the phenomenological
normalization; we return to this point
in Chapter \ref{hyc} below.
The Cartan matrix of a Lie algebra is
defined by
\beq
A_{ij} = {2 \; \langle \alpha_i | \alpha_j \rangle
\over \langle \alpha_j | \alpha_j \rangle },
\label{dcm}
\eeq
where $i,j$ run over the simple roots.
Using \myref{2b} and $\alpha_i^2 = 2$,
it is easy to see that \myref{dcm}
is simply expressed in terms of the
sixteen-dimensional simple root vectors:
\beq
A_{ij} = {\alpha_i \cdot \alpha_j}.
\label{2c}
\eeq
In the orbifold constructions studied here,
a subset of the \eetee\ simple roots
survive, and by computing the submatrices
according to \myref{2c}, we can identify
the nonabelian factors in the surviving
gauge group $G$, using widely available
tables for the Cartan matrices of Lie
algebras (e.g., Ref.~\cite{Cor84}).

\section{Recipes}
\label{rec}
We next write down without proof recipes
for the generation of the spectrum of pseudo-massless
states.  Where possible, I have attempted to
motivate the rules in a heuristic fashion,
avoiding a detailed discussion
of the underlying string theory.  For further
details, see the reviews \cite{Har86,orbrvw,BL99},
texts \cite{GSW87a,GSW87b,BL94,Pol98b}, and references therein.

\bfe{Nonzero root gauge states.}  We write these
states as $\ket{\alpha}$ where $\alpha$ satisfies:
\beq
\alpha^2 = 2 , \qquad \alpha \in \eelat ,
\label{2d}
\eeq
\beq
\alpha \cdot a_i \in \Zbf, \quad \forall \; i=1,3,5,
\label{2e}
\eeq
\beq
\alpha \cdot V \in \Zbf .
\label{nzr}
\eeq
Eq.~\myref{2d} merely states that $\alpha$ is
an \eetee\ root.  For nontrivial \emset, 
several roots of \eetee\ will
not satisfy (\ref{2e},\ref{nzr}).
Consequently, the nonzero roots of $G$ will be a subset of the
\eetee\ roots.  The states 
$\ket{\alpha}$ are eigenstates of the
generators $H^I$ of the \eetee\ Cartan subalgebra:
\beq
H^I \ket{\alpha} = \alpha^I \ket{\alpha}, 
\qquad I=1,\ldots,16.
\label{2j}
\eeq
To determine $G$, one first (fully)
decomposes the solutions of (\ref{2d}-\ref{nzr})
into orthogonal subsets.  That is, for $a \not= b$
the subset $\{ \alpha_{a1}, \ldots, \alpha_{a n_a} \}$
is orthogonal to the subset $\{ \alpha_{b1}, \ldots, \alpha_{b n_b} \}$
provided 
\beq
\alpha_{ai} \cdot \alpha_{bj} = 0, \qquad
\forall \quad i=1,\ldots,n_a,
\quad j=1,\ldots,n_b .
\eeq
The $a$th such subset corresponds to a nonabelian 
simple subgroup $G_a$ of $G$,
and the solutions $\alpha_{a1}, \ldots, \alpha_{a n_a}$ 
belonging to this subset are the nonzero roots
of $G_a$.  One next determines which of the
$\alpha_{a1}, \ldots, \alpha_{a n_a}$ are simple
roots.  From the
simple roots one can compute the Cartan 
matrix for $G_a$ using \myref{2c}
and thereby determine the group $G_a$.

\egbox{
In the special class of embeddings to be discussed
in Chapter~\ref{slz}, there are precisely eight solutions to
(\ref{2d}-\ref{nzr}) which {\it do not} have all first eight
entries vanishing:
\beq
\alpha_{1,1} \, , \, \alpha_{1,2} =
(\underline{1,-1},0,0,0,0,0,0;0,\ldots,0), \quad
\alpha_{2,1}\, , \, \ldots \, , \, \alpha_{2,6} =
(0,0,\underline{1,-1,0},0,0,0;0,\ldots,0).
\label{nz32}
\eeq
Here (and elsewhere below),
all permutations of underlined entries
should be taken.  The vectors in \myref{nz32} are the nonzero
roots of the observable sector gauge
group $G_O$.  The first set 
in \myref{nz32} is orthogonal to
all vectors in the second set; therefore, these
two sets correspond to different simple factors,
one with two nonzero roots and the other with six;
the two groups must be $SU(2)$ and $SU(3)$.
It is easy to check that the simple roots are
\beq
\alpha_{1,1} = (1,-1,0,0,0,0,0,0;0,\ldots,0),
\label{2m}
\eeq
\beq
\alpha_{2,1} = (0,0,1,-1,0,0,0,0;0,\ldots,0), \qquad
\alpha_{2,2} = (0,0,0,1,-1,0,0,0;0,\ldots,0).
\label{2.39}
\eeq
The simple roots \myref{2.39} give the Cartan
matrix for $SU(3)$, using \myref{2c}.
}

\bfe{Zero root gauge states.}  We write these states
in an orthonormal basis
$\ket{I}$, where $I=1, \ldots, 16$.  These correspond
to gauge states for the Cartan subalgebra of $G$,
in the Cartesian basis $H^I$ discussed above.
They of course have vanishing \eetee\ weights:
\beq
H^I \ket{J} = 0, \quad \forall \; I,J=1,\ldots,16.
\eeq
The group $G$ typically has a nonabelian part $\GNA$ which
is a product of $m$ simple factors, and a \uone\ part
$\GUO$ which is a product of $n$ {\uone}s:
\beq
G = \GNA \times \GUO, \qquad
\GNA = G_1 \times G_2 \times \cdots \times G_m ,
\qquad \GUO = U(1)_1 \times U(1)_2 \times \cdots
   \times U(1)_n .
\label{gde}
\eeq
For the orbifold models studied in the following
chapters,
the simple factors $G_a$ ($a=1,\ldots,m$)
are either $SU(N)$ or $SO(2N)$ groups.  Each $G_a$
has its own Cartan subalgebra with
a corresponding basis $H_a^1, \ldots, H_a^{r_a}$,
where $r_a$ is the rank of $G_a$.
Each basis element $H_a^i$ is a linear combination
of the \eetee\ Cartan basis elements $H^I$:
\beq
H_a^i = \sum_{I=1}^{16} h_a^{iI} H^I .
\label{cgc}
\eeq
This is the analogue of \myref{2.5}.
It should not be too surprising that 
corresponding linear
combinations of the \eetee\ Cartan gauge states
$\ket{I}$ are taken to obtain Cartan gauge states
of $G_a$:
\beq
\ket{a;i} = \sum_{I=1}^{16} h_a^{iI} \ket{I} .
\eeq
Similarly, the generator $Q_a$ of the factor $U(1)_a$
may be written as
\beq
Q_a = \sum_{I=1}^{16} q_a^I H^I
\label{Qdf}
\eeq
(this is the analogue of \myref{2.6})
and the corresponding gauge state
\beq
\ket{a} = \sum_{I=1}^{16} q_a^I \ket{I} .
\eeq
It is convenient to choose the states
$\ket{a}$ to be orthogonal (I discuss normalization
below):
\beq
\ipr{a}{b} = q_a \cdot q_b = 0 \quad
\mtxt{if} \quad a \not= b.
\label{u1o}
\eeq
For the Cartan states
$\ket{a;i}$, it is more convenient that their inner
product reproduce the Cartan matrix $A^a$ for the
group $G_a$:
\beq
\ipr{a;i}{b;j} = h_a^i \cdot h_b^j = \delta_{ab} A^a_{ij}.
\eeq
It is hopefully apparent from \myref{2c}
that this equation is satisfied
if we take $h_a^i$ to be the
sixteen-dimensional simple root
vectors for $G_a$: $h_a^i \equiv \alpha_{ai}$.
We therefore rewrite \myref{cgc} as
\beq
H_a^i = H(\alpha_{ai}) = \sum_{I=1}^{16} \alpha_{ai}^I H^I ,
\label{2g}
\eeq
where we use the notation of \myref{rtm};
as mentioned there, these generators are
larger by a factor of two than the
phenomenological normalization.

Naturally, we want the $\GNA$ Cartan states orthogonal to
the $\GUO$ states:
\beq
\ipr{a}{b;j} = q_a \cdot \alpha_{bj} = 0,
\qquad \forall \; a,b,j.
\label{qao}
\eeq
The $q_a$ are therefore chosen to be orthogonal to
the simple roots and to each other.  With $n$
\uone\ factors, as in \myref{gde}, the choice of $q_a$
is determined only up to reparameterizations
which preserve the orthogonality conditions 
(\ref{u1o},\ref{qao}).
In practice, most choices for the \uone\
generators lead to several of them being
\git{anomalous;} i.e., $\tr Q_a \not= 0$, with
the trace taken over the pseudo-massless
spectrum of matter states.  It is then useful to make
redefinitions such that only one \uone\
is anomalous.  Let
\beq
t_a = \tr Q_a, \quad t_b = \tr Q_b, \quad
s_a = q_a^2, \quad s_b = q_b^2,
\eeq
with $t_a, t_b$ both nonzero.  
Then define generators $Q_a' = \sum_I (q_a')^I H^I$
and $Q_b' = \sum_I (q_b')^I H^I$ via
\beq
q_a'  =  t_b q_a - t_a q_b , \qquad
q_b'  =  t_a s_b q_a + t_b s_a q_b .
\eeq
It is easy to see that 
$\tr Q_a' = t_b t_a - t_a t_b = 0$,
so that the anomaly is isolated to $Q_b'$.
Furthermore, orthogonality is maintained:
\beq
q_a' \cdot q_b' = t_a t_b (s_b q_a^2 - s_a q_b^2)
= t_a t_b (s_b s_a - s_a s_b) = 0.
\eeq
By repeating this process,
one can easily isolate the anomaly
to a single factor, \ux.

\bfe{Untwisted matter states.}  We denote these states
as $\ket{K;i}$, $i=1,3,5$.  Here, $K$ is a sixteen-vector,
denoting weights under the \eetee\ Cartan generators
$H^I$:
\beq
H^I \ket{K;i} = K^I \ket{K;i},
\qquad I=1,\ldots,16.
\label{2k}
\eeq
Furthermore, $K$ must satisfy
\beq
K^2 = 2 , \qquad K \in \eelat ,
\label{2h}
\eeq
\beq
K \cdot a_i \in \Zbf, \quad \forall \; i=1,3,5.
\label{2i}
\eeq
\beq
K \cdot V = \third \mod 1 ,
\label{ums}
\eeq
It can be seen from comparison
to (\ref{2d}-\ref{nzr}) 
that the weights $K$ of untwisted matter states
differ from the weights of nonzero
root gauge states
only in the last condition,
\myref{nzr} versus \myref{ums}:  untwisted matter
states correspond to a different 
subset of the nonzero \eetee\ roots
which satisfy \myref{2e}.
(The remaining subset corresponds 
to untwisted antimatter states.)  The
multiplicity of three carried by the index $i$ 
in $\ket{K;i}$ corresponds
to a ground state degeneracy in 
the underlying theory \cite{DHVW85},
which I will not discuss here.  It is one of the nice
features of the $Z_3$ orbifold which aids in easily
obtaining three generation constructions.  However, it
also means that for fixed $K$,
the three generations $i=1,3,5$ have
identical \uone\ charges and are
in identical irreps,
as can easily be checked using
(\ref{Qdf},\ref{2g},\ref{2k}):
\beqa
H_a^j \; \ket{K;i} & = & \alpha_{aj} \cdot K \; \ket{K;i},
\label{wtd} \\
Q_a \; \ket{K;i} & = & q_a \cdot K \; \ket{K;i}.
\label{2n}
\eeqa
That is, the weight $\lambda_{aj}^K = \alpha_{aj} \cdot K$
is independent of $i$ and similarly for the charge
$q_a^K = q_a \cdot K$.

In order to determine the matter
spectrum, we need more than just
the weights \myref{wtd};
we need to be able to
group the basis states 
$\ket{K_1;i}, \ldots, \ket{K_{d(R)};i}$
which make up a given irrep 
$R$ of dimension $d(R)$.
Suppose an incoming matter state $\ket{K;i}$ 
interacts with a gauge supermultiplet
state corresponding to
a nonzero root $\alpha_{aj}$ of $G_a$.  This interaction
is described by inserting a current $J(\alpha_{aj})$,
which acts like a raising or lowering operator
with respect to some $SU(2)$ subgroup of $G_a$:
\beq
\bra{K';i} J(\alpha_{aj}) \ket{K;i}
= \ipr{K';i}{K+\alpha_{aj};i}
= \delta_{K',K+\alpha_{aj}} .
\eeq
For fixed family index $i$, vectors
$K'$ related to $K$ by the addition of
one of the nonzero roots
of $G_a$ are in the same irrep.
Collecting all vectors $K'$ related to
$K$ in this way (and satisfying
(\ref{2h}-\ref{ums})), we fill
out the vertices of a weight diagram 
of an irrep of $G_a$.  
Due to \myref{qao},
$K'$ and $K$ give the same \uone\ charges
(as they must):
\beq
q_b \cdot K' = q_b \cdot \alpha_{aj} + q_b \cdot K
= q_b \cdot K .
\eeq

\bfe{Twisted non-oscillator matter states.}  We denote these
as $\ket{\tK;n_1,n_3,n_5}$, where $n_i =0,\pm 1$
specify which of the 27 fixed points (conjugacy
classes) the state
corresponds to and $\tK$ is a sixteen-vector
giving the weights with respect to the \eetee\
Cartan generators $H^I$, similar to
Eqs.~(\ref{2j},\ref{2k}) above.  However, the $\tK$
{\it do not} correspond to points on $\eelat$.
Rather (cf. \myref{evd}),
\beq
\tK^2 = 4 / 3,
\qquad
\tK = K +  E(n_1,n_3,n_5),
\qquad 
K \in \eelat.
\label{tps}
\eeq
The condition
$\tK^2 = 4/3$ guarantees $\tK \not\in \eelat$
since all elements $L \in \eelat$ have $L^2 = 0 \mod 2$,
as can be checked by inspection of \myref{2l}.
Weights and charges under $G$ are calculated as
for the untwisted states, only now the
shifted weights $\tK$ are used.
In particular,
\beqa
Q_a \; \ket{\tK;n_1,n_3,n_5} 
& = & q_a \cdot \tK \; \ket{\tK;n_1,n_3,n_5} \nnn
& = & \left[ q_a \cdot K + q_a 
\cdot E(n_1,n_3,n_5) \right] \; \ket{\tK;n_1,n_3,n_5}.
\label{2t}
\eeqa
Thus, the twisted matter states have 
charges shifted by
\beq
\delta_a(n_1,n_3,n_5) = q_a \cdot E(n_1,n_3,n_5)
\eeq
from what would occur
in the decomposition of
\eetee\ representations onto
a subgroup with \uone\ factors.
The quantity $\delta_a(n_1,n_3,n_5)$ is
the \git{Wen-Witten defect} \cite{WW85},
a problematic contribution
which is uniform for a given twisted
sector.
It is precisely this feature
which is responsible for difficulties accomodating
the hypercharges of the MSSM 
spectrum and the generic appearance of
states with fractional electric charge,
as will be discussed below.  
Comparison
to \myref{evd} shows that with $a_5 \equiv 0$,
the embedding vector $E(n_1,n_3,n_5)$ is
independent of $n_5$.  
It follows that
states which differ only by the value of $n_5$
have identical \uone\ charges and are
in identical irreps of the gauge group $G$.  This is
how three generations in twisted
sectors are naturally generated in
the class of models considered here.
Filling out irreps of $G_b$
is accomplished by collecting all $\tK'$
which are related to $\tK$ through
$\tK' = \tK + \alpha_{bj}$, similar
to what was done for untwisted states.
Of course, the other quantum numbers
$n_1,n_3,n_5$ must match.

It was stated above that higher dimensional
irreps of \eetee\ are, in a way, relevant to
massless states in the twisted sectors.  We
are now in a position to address this comment.
In Chapter \ref{app} we will discuss a model
with an embedding (cf.~\myref{evd}) such that
\beq
3E(1,1,n_5) = (0,0,-1,-1,-1,5,2,2;3,1,1,0,1,0,0,0).
\eeq
It is easy to check that a solution to \myref{tps}
is obtained if
\beq
K = (0,0,0,0,0,-2,-1,-1;-1,-1,0,0,0,0,0,0).
\eeq
However, $K^2=8$, so this is not a root of \eetee,
but the weight of a higher dimensional \eetee\ irrep.
Of course, the weight of the state
$\ket{\tK;n_1,n_3,n_5}$ is $\tK$ and not $K$,
so it seems unimportant that $K^2>2$.
However, $q_a \cdot K$ in \myref{2s}
would be the ``conventional''
charge while $q_a \cdot E(1,1,n_5)$ is the
Wen-Witten defect; in this interpretation
the charge $q_a \cdot K$ which would
occur if the defect were absent is
that of the decomposition a higher dimensional 
\eetee\ irrep.
If nothing else, it creates the illusion that
some massive states of the uncompactified
\eetee\ heterotic string are shifted down
into the massless spectrum when compactified
on the six-dimensional orbifold.

Finally, we note that projections
analogous to (\ref{2i},\ref{ums}) are not
required in the twisted sectors of
a $Z_3$ orbifold \cite{IMNQ88,BLT88}.  As a result,
study of this orbifold is significantly
simpler than most other orbifold constructions,
where projections in the twisted
sectors are rather complicated.

\bfe{Twisted oscillator matter states.}  We denote these
as $\ket{\tK;n_1,n_3,n_5;i}$, where $i =1,3,5$
conveys an additional multiplicity of three,
due to different ways to excite
the vacuum in the underlying string theory
with the analogue of harmonic oscillator
raising operators.  Three types of
oscillators---corresponding to the three
complex planes of the six-dimensional
compact space---excite the vacuum to
generate a massless state.  These are
the operators $\tilde \alpha_{-1/3}^i$
which appear in \myref{iyat}.
The $\tK$
are again shifted \eetee\ weights, but they
have a smaller norm (to compensate for
energy associated with the excited vacuum):
\beq
{\tK}^2 = 2 / 3,
\qquad
\tK = K +  E(n_1,n_3,n_5),
\qquad 
K \in \eelat.
\eeq
The determination of weights, irreps and
charges is identical to that for the other matter states
discussed above.

%

\chapter{Standard-like ${\bf Z_3}$ Orbifold Models}
\label{slz}
With this chapter we begin the application of
formal details described above to situations
of phenomenological interest.  The intent is to
study and describe generic features of a large
number of models.  By looking at a wide breadth of
semi-realistic heterotic orbifold constructions,
it is hoped that generic features\myfoot{Repetitive
use of ``generic features.''} will become
apparent, and that to a certain extent we will
gain an intuitive picture of what the ``predictions''
of this sort of string theory really are.

Section \ref{ceft} is a review of well-known facts.
First I draw the broad outlines
of how string theory is connected to an effective
field theory of particles and their interactions.
Then the states always present in $Z_3$ orbifold
models are summarized.  In Section \ref{appr} I
discuss different approaches to compactification
of the extra six dimensions of the heterotic
string.  I then restrict the scope of investigations
taken up here by defining what I call the
$BSL_A$ {\it class,} a set of heterotic
$Z_3$ orbifold models meeting certain limiting
conditions.  As an example, they are {\it standard-like,}
meaning that their gauge symmetry group $G$ already
has factors $SU(3) \times SU(2) \times U(1)$ from
the start, without any spontaneous symmetry breaking
by a Higgs effect.

In Section \ref{iem}, I report the
first part of original research conducted by
myself into the \bsa\ class.  I determine {\it all}
of the possible models within this class.
The number of such models would by first appearances
seem to be quite large; however, a number of
equivalence relations can be exploited to
greatly reduce the list.  This was already
carried out in part by Casas, Mondragon
and Mu\~noz in \cite{CMM89}; I complete
the analysis and give a full enumeration of
the possibilities.  Section \ref{iem} is based
on a recent article of mine \cite{Gie01b}.

The next two sections review well-known aspects
of the effective supergravity description of
superstrings.  In Section \ref{ssb} I discuss
the breaking of supersymmetry in these theories
through an effect known as
{\it gaugino condensation.}  This is a very
important topic, as one of the first signs
of supersymmetry which would be experimentally
observed is the existence of new particles which
could plausibly be interpreted as supersymmetric
partners ({\it superpartners}) to the Standard Model particles.
By studying the properties of any candidate
superpartners, we could make inferences about
the nature of supersymmetry breaking, should
this indeed be their origin.  How supersymmetry
is broken in the low energy effective theory
constrains the description of physics at much
higher energy, and in particular the effective
supergravity description of superstrings.  This
will in turn have implications for heterotic
orbifold models.  As explained in Section \ref{ssb}
however, we do not need to wait for the discovery
of superpartners to place constraints on string
models.  The nonobservation of superpartners, to
date, already limits the \bsa\ models in significant
ways.

In Section \ref{au1} I address the
implications of an anomalous \ux\ in the
theory.  A \ux\ factor has already been discussed
briefly above in Section \ref{rec}.  However in
Section \ref{au1} the \ux\ is discussed in the context
of effective supergravity.

Finally, in Section
\ref{mds} I discuss results of
my calculation of spectra of matter states
for models in the \bsa\ class---previously
reported in \cite{Gie01c}.
The calculations are based on the application of
the recipes given in Section \ref{rec},
to the models enumerated in Section \ref{iem}.
A number of conclusions are drawn based on these
results.

This chapter relies on common knowledge in modern
theoretical particle physics.  It is best if the
reader understands the perturbation theory
of quantized gauge field theories,
is familiar with globally supersymmetric field
theories, especially {\it Super-Yang-Mills,}
and has a practical knowledge of supergravity
as it relates to the soft terms of the MSSM.
Readers not sufficiently prepared in these
topics would do well to concurrently
consult relevant references suggested
in Appendix \ref{mysg}.

\section{Constructing the Effective Field Theory}
\label{ceft}
To make contact with the world of particle physics,
one is interested in the effective theory produced
by heterotic string theory at energy scales far below
the string scale $\LamH \sim 10^{17}$ GeV.
The current upper limits on direct experimental probes of
fundamental interactions are in the neighborhood of
a few hundred GeV.  Many experiments (proton decay,
electric dipole moments, etc.) are sensitive to
underlying physical processes characterized by {\it much} higher energies
(e.g., roughly $10^{15}$ to $10^{16}$ GeV in the case
of proton decay).
However, the observable processes are best
described by effective theories valid
at experimentally accessible energy scales,
to be compared with theoretical predictions
of the effective theory derived from the
high scale theory.
The first
step in constructing the low energy
effective theory is to determine
the string states with masses much less than $\LamH$.
Secondly, one must derive the interactions between
these states and an appropriate description for these
interactions.

In the context of perturbative string
theory, systematic methods for the
accomplishment of these tasks exist, subject to certain
technical difficulties which we will not discuss
here, since for the most part we work only at leading order in
string perturbation theory.  Interactions are described
by scattering amplitudes between string states.  In
particular, these amplitudes can be studied in the
limit where external momenta are taken to be much
less than the string scale, often referred to as
the {\it zero-slope limit} \cite{Sch71}.

One then matches
the results onto a field theory.  That
is, one constructs a local field theory lagrangian
which, when quantized, would have single particle
states with the same properties (mass, spin, charge,
etc.)~as the low-lying string
states.  Moreover, the field theory
scattering amplitudes are required to match the
string scattering amplitudes at low external momenta.
Thus, one can talk about the ``particle'' states
which arise from the ``field theory limit'' of the
string.

A study of the heterotic string at tree level shows that
the string states are organized into a tower of
mass levels, with the lowest level
massless.  For the four-dimensional heterotic
string, subject to certain qualifications which
will not trouble us here,\footnote{E.g., the large radius
limit of the extra dimensions---where massive string states
can drop far below $\LamH$.} the only string states
with masses significantly below $\LamH$ are those which lie
at the massless level of the string.
However, genus one corrections can be significant
if, for example, an anomalous \ux\ is present.\footnote{Recall
from Section \ref{rec} that this is a factor \ux\
of the gauge symmetry group $G$ for which $\tr Q_X \not= 0$.}
As will be discussed in Section \ref{au1} below,
the cancelation of the corresponding anomaly
leads to a {\it Fayet-Illiopoulos (FI) term.}
The tree level spectrum of masses can be dramatically altered
when this one loop effect is taken into account.
For this reason, I hereafter refer to the states which
are massless at tree level as {\it pseudo-massless.}
Furthermore, the gauge symmetry group $G$ is
broken to a subgroup at an energy scale $\LamX$ very
near the string scale $\LamH$.

As a matter of fact, many of the pseudo-massless states
have masses near $\LamH$ once the
one loop corrections are accounted for!
This is because the scalar fields which
I will refer to as {\it Xiggs fields} acquire $\ordnt{\LamX}$
vacuum expectation values ({\it vevs});
explicit calculations detailed below
show that $\LamH/1.73 \leq \LamX \leq \LamH$
in the 175 models studied here, indicating that
$\LamX$ is more or less the string scale $\LamH$.
The Xiggs vevs cause several
chiral (matter) superfields to
get effective ``vector'' superpotential couplings
\beq
W \ni {1 \over m_P^{n-1}} \vev{\phi^1 \cdots \phi^n} A A^c .
\label{ems}
\eeq
Here, $A$ and $A^c$ are conjugate with respect to the
gauge group which survives after spontaneous symmetry
breaking caused by the \ux\ FI term.
The right-hand side of \myref{ems} is an effective
supersymmetric mass term, which generally results in masses
\beq
m_{eff} \sim \ordnt{\LamX^n / m_P^{n-1}} 
\approx \ordnt{\LamH^n / m_P^{n-1}} .
\label{mef}
\eeq
With $n=1$ in \myref{mef}, the effective masses
are near the string scale.  Due to the numerous
gauge symmetries present in the models considered
here, as well as discrete symmetries
known as {\it orbifold selection rules}
(see for example \cite{HV87,FIQS90,BL99}), not all operators
of the form $A A^c$ will have couplings with $n=1$
in \myref{ems}.  Because of this, a hierarchy of mass scales
is a general prediction of models with a \ux\ factor
(all but seven of the 175 models studied here).
I return to this point in Chapter \ref{app},
where I briefly discuss gauge coupling unification.

By construction, the spectrum is that of an $N=1$ four-dimensional
locally supersymmetric theory.  Furthermore,
the compact space is a six-dimensional $Z_3$ orbifold.
Certain parts of the spectrum are well-known to
be present by virtue of these facts alone \cite{DHVW85};
I have not discussed these states
in Chapter \ref{mss}; They are:  the supergravity
multiplet, the {\it dilaton} supermultiplet and
nine chiral multiplets $T^{ij}$ whose scalar
components correspond to the K\"ahler-
or T-moduli of the compact space.  (See for example
\cite{Gre97} for a discussion of toric
moduli.)

The remainder of the spectrum depends on the
choice of {\it embedding,} and it is this part
of the spectrum which we must calculate separately for
each of the 175 models.  The embedding-dependent spectrum
consists of massless chiral multiplets of
matter states and massless vector multiplets of gauge states.
Once the vacuum shifts to cancel the FI
term, some gauge symmetries are spontaneously broken
and chiral matter multiplets (which are linear
combinations of Xiggses) get ``eaten'' by some of the vector
multiplets to form massive vector multiplets.
Examples of the ``degree of freedom balance sheet''
may be found for example in~\cite{GG00}.

\section[The ${\rm BSL}_{A}$ Class]{The ${\rm {\bf BSL}}_{\bf A}$ Class}
\label{appr}
As described in Chapter \ref{mss},
the heterotic string theory as originally
formulated \cite{GHMR85} has a ten-dimensional space-time.
To construct a four-dimensional theory, one
typically associates six of the spatial dimensions
of the original theory with a very small compact
space.  One route to ``compactifying'' the six
extra dimensions, which has been the subject
of intense research for several years now,
is to take the six-dimensional space
to be an orbifold \cite{DHVW85,DHVW86},
such as the six-dimensional
$Z_3$ orbifold described above.

Four-dimensional heterotic string theories
obtained by orbifold compactification
take two broad paths to the treatment of internal string
degrees of freedom not associated with four-dimensional
space-time.  On the one hand, these degrees of freedom
are associated with two-dimensional {\it free fermionic}
fields \cite{ABK87};
on the other, some are associated with
two-dimensional {\it bosonic} fields propagating
in a constant background, as was described in
Chapter \ref{mss}.
\index{free fermionic heterotic string models}
\index{bosonic heterotic string models}
\glossary{bosonic heterotic string models}
\glossary{free fermionic heterotic string models}

Remarkable progress in the construction of
realistic four-dimensional {\it free fermionic}
heterotic string models \cite{AEHN88}
has been made in the last several years:
a high standard has been established
recently by Cleaver, Faraggi, Nanopoulos
and Walker in their construction 
and analysis \cite{CFN99,CFNW00,CFNW01a,CFNW01b}
of a ``Minimal Superstring
Standard Model'' based on the free
fermionic model of Ref.~\cite{FNY90}.
The Minimal Superstring Standard Model
has only the matter content of the Minimal
Supersymmetric Standard Model\footnote{For a review of the MSSM, 
see for example Refs.~\cite{mssmr}.}
(MSSM)
at scales significantly
below the string scale $\LamH \sim 10^{17}$ GeV.
Furthermore, the hypercharge normalization
\glossary{hypercharge normalization}
is conventional.\footnote{Hypercharge normalization
and what is meant by ``conventional'' will
be discussed in detail in Chapter \ref{hyc}.}

Similarly realistic
four-dimensional {\it bosonic} heterotic string models
have not yet been engineered,
though the foundations of such an effort
were laid some time ago
\cite{DHVW85,DHVW86,INQ87,IMNQ88}.  Some of the most
promising models were of the $Z_3$ orbifold variety,
\glossary{discrete Wilson lines}
\glossary{continuous Wilson lines}
with nonvanishing {\it Wilson lines} (the embedding
vectors $a_1,a_3,a_5$ discussed in Section \ref{obo} above)
chosen such that
the matter spectrum naturally had three generations.
\glossary{generations}
One such model was introduced
by Ib\'a\~nez, Kim, Nilles and Quevedo
in Ref.~\cite{IKNQ87}, which I will refer
to as the Bosonic Standard-Like-I (BSL-I) model.
\glossary{matter} \glossary{vacuum expectation value}
\glossary{effective field theory}
\glossary{field theory limit}
The model was subsequently
studied in great detail by two
groups: Ib\'a\~nez, Nilles, Quevedo et al.~in
Refs.~\cite{FINQ88b,FIQS90}; Casas and Mu\~noz in
Refs.~\cite{CM88}.  As is often the case in
supersymmetric models, the vacuum (i.e., the
configuration of scalar vevs in the effective supergravity
theory) of the BSL-I model
is not unique; different choices lead to different
low energy \glossary{low energy} effective theories.  A particularly
encouraging vacuum was the one chosen
by Font, Ib\'a\~nez, Quevedo and Sierra (FIQS)
in Section~4.2 of Ref.~\cite{FIQS90}; in what
follows, I will refer to this effective
string-derived theory as the FIQS model.
Departures from realism in the FIQS model were pointed
out recently in \cite{GG00} and \cite{Gie01a}.  
In the latter article,
I suggested that a scan over three generation
constructions analogous to the BSL-I model be
conducted, in the search for a more realistic model.
Ultimately, I would like to attempt models
with realism comparable to that of
the free fermionic Minimal Superstring Standard Model.
In large part, the research summarized
in this chapter is aimed at this goal.

This research has consisted mostly of a
model {\it dependent} study of bosonic standard-like
$Z_3$ orbifolds.  Model {\it independent}
analyses are appealing because they paint a wide swath
and highlight general predictions of a class of
theories.  Too often, however, one is left wondering
whether the limiting assumptions made in
such analyses really reflect the properties
of some class of explicit, consistent
underlying theories.  At some point it is necessary
to ``get dirt on oneself'' and investigate whether or
not the broad assumptions made in model
independent analyses are ever valid.  This is
one of the motivations for model dependent
studies such as the one taken up here.
Another reason to study explicit string constructions
is that certain peculiarities are
more readily apparent under close examination.
One well-known example, which will be discussed
in Chapter \ref{hyc}, is the generic presence of
\glossary{exotic states}
\glossary{Grand Unified Theory}
exotic states with hypercharges which do not
occur in typical Grand Unified Theories\footnote{ 
For a review of non-supersymmetric GUTs see  
Refs.~\cite{gutrvw,Sla81} and for supersymmetric
extensions see Refs.~\cite{sgtrvw}.} (GUTs).
\glossary{model independent}
\glossary{model dependent}
\glossary{standard-like}

One objection to model dependent studies in
four-dimensional string theories is that the
number of possible constructions is enormous.
However, in at least one respect the enormity
is not as great as it would appear.
Already in the second of the two seminal papers by
Dixon, Harvey, Vafa and Witten \cite{DHVW86}, it was realized
that many ``different'' orbifold models are
in fact equivalent.  Two models are
equivalent if their Hilbert spaces are isomorphic
and the induced map between operators preserves
physical interpretations.  That is, suppose two
Hilbert spaces ${\cal H}$ and ${\cal H}'$.  The
two spaces are isomorphic if and only if there
exists a bijective map $\phi: {\cal H} \to {\cal H}'$ such
that $\ipr{\phi(f)}{\phi(g)} = \ipr{f}{g}$ for
any $f,g \in {\cal H}$.  Furthermore, suppose
$T$ is an operator in ${\cal H}$ with a given
physical interpretation (e.g., the Hamiltonian).
The isomorphism $\phi$ induces a map from $T$
to an operator on ${\cal H}'$, the
composition $T' \equiv \phi \circ T \circ \phi^{-1}$.  We demand
that $T'$ have the same physical interpretation
in the theory associated with ${\cal H}'$.
Two models {\it not} related in this way
are said to be ``physically distinct.''

Casas, Mondragon and Mu\~noz
(CMM) have shown in detail how equivalence
relations among orbifold compactifications can be
used to greatly reduce the number of embeddings
(in the present context the set \emset\
introduced in Section \ref{obo})
which must be studied in order to produce all
physically distinct models within a given
class of constructions \cite{CMM89}.
In particular, they applied these techniques to a special
class of bosonic standard-like heterotic string models;
for convenience, I will refer to this as the
$BSL_A$ {\it class.}  For completeness,
I give its technical definition.\footnote{
In simpler terms, the definition given
here implies that
we follow the construction outlined in \cite{INQ87},
with three generations by the method
suggested in \cite{IKNQ87},
subject to the additional restrictions
imposed by CMM (items (iii) and (iv) below).}
\begin{defn}
The ${\rm {\bf BSL}}_{\bf A}$ {\bf class}
consists of all bosonic \eetee\ heterotic $Z_3$ orbifold 
models with the following properties:
\ben
\item[(i)]
symmetric treatment of left- and right-movers
and a shift embedding $V$ of the twist operator $\theta$;
\item[(ii)] two nonvanishing Wilson lines $a_1,a_3$
and one vanishing Wilson line $a_5=0$;
\item[(iii)] observable sector gauge group
\beq
G_O = SU(3) \times SU(2) \times U(1)^5 ;
\label{CMMgo}
\eeq
\item[(iv)] a quark doublet representation
$(3,2)$ in the untwisted sector.
\een
\label{bsad}
\end{defn}
CMM found that models satisfying
(i-iv) may be described (in part) by
one of just nine observable sector
embeddings; here, ``observable'' refers to the first
eight entries of each of the nonvanishing embedding vectors,
$V, a_1, a_3$; it is this which determines
properties (iii) and (iv) listed above.
In Ref.~\cite{Gie01b} I showed that
these nine observable sector embeddings are equivalent
to a smaller set of six embeddings;
the calculations are given
here in Appendix \ref{erp}.
To fully specify a model, the observable sector embedding
must be completed with a hidden sector embedding---the
last eight entries of each of the nonvanishing
embedding vectors, $V, a_1, a_3$.
In Ref.~\cite{Gie01b} I enumerated all possible ways
to complete the embeddings in the hidden sector,
using equivalence relations to reduce this set to
a ``mere'' 192 models.  The calculation is
reviewed in the following section;
the CMM observable sector embeddings are
given in Table \ref{tabcmm}
and my results for the hidden sector embeddings
can be found in Tables \ref{tab1}-\ref{tab9b}
of Appendix \ref{emt}.

\section{Completion of Standard-like Embeddings}
\label{iem}
As discussed in Chapter \ref{mss},
for heterotic $Z_3$ orbifold models with
discrete Wilson lines, the embedding is expressed in terms
of four sixteen-dimensional vectors:  the twist embedding $V$
and three Wilson lines $a_1, a_3$ and $a_5$;
each of the four vectors is given
by one-third of a vector belonging to
the $E_8 \times E_8$ root lattice:
\beq
3V \in \eelat, \qquad 3a_i \in \eelat, \quad \forall
\; i = 1,3,5.
\label{lvc}
\eeq
It is convenient to denote the vector formed
from the first eight entries of $V$ by $V_A$
and the vector formed from the last eight
entries of $V$ by $V_B$, so that the twist
embedding $V$ may be written as $V = (V_A;V_B)$.
Eq.~\myref{lvc} then implies 
\beq
3 V_A \in \Lambda_{E_8}^{(A)},
\qquad 3V_B \in \Lambda_{E_8}^{(B)},
\label{vab}
\eeq
where $\Lambda_{E_8}^{(A)}$
and $\Lambda_{E_8}^{(B)}$ are the two
copies of the $E_8$ root lattice used
to construct $\eelat$.
Similarly, we write $a_i= (a_{iA};a_{iB})$ for
each $i=1,3,5$.  In addition to
\myref{vab}, constraint \myref{lvc} becomes
\beq
3 a_{iA} \in \Lambda_{E_8}^{(A)},
\qquad 3 a_{iB} \in \Lambda_{E_8}^{(B)},
\qquad \forall \; i = 1,3,5.
\label{aabd}
\eeq
The set $\{ V_A, a_{1A}, a_{3A}, a_{5A} \}$ dictates the
space group transformation properties
of the underlying string degrees of
freedom corresponding to the first $E_8$ factor
of the gauge group; i.e, the
set ``embeds the first $E_8$.''  Similarly, the set
$\{ V_B, a_{1B}, a_{3B}, a_{5B} \}$ embeds the
second $E_8$.
As described in Chapter \ref{mss},
for discrete Wilson lines constructions,
the embedding of the gauge degrees of freedom
has the effect of breaking each $E_8$ down to
a rank eight subgroup:
\beq
E_8(A) \to G_O, \qquad E_8(B) \to G_H,
\eeq
where $G_O$ and $G_H$ are usually coined
the ``observable'' and ``hidden''
sector gauge groups.

In Chapter \ref{mss} we saw that
models with three generations of
matter can be obtained by choosing
the third Wilson line $a_5$ to
vanish, as exploited for example in Refs.~\cite{IKNQ87,FIQS90}.
Consequently, three generation models
of this ilk are specified by the
set of embedding vectors $ \{ V, a_1, a_3 \} $.
For this reason, I will ignore
$a_5$ in the remainder of this section.
The observable sector gauge group $G_O$
is determined entirely by the set of
observable sector embedding vectors \Aem.
Many such sets lead to a standard-like 
observable sector gauge group $G_O$ of the form \myref{CMMgo}.
CMM have determined observable
sector embeddings of this type, with the additional
requirement of quark doublets---$(3,2)$ 
irreps under the $SU(3) \times SU(2)$ subgroup
of \myref{CMMgo}---in the untwisted sector
(item (iv) in Definition \ref{bsad} above).
As noted in the previous section, CMM have found that any
observable sector embedding satisfying these two
conditions is equivalent to some one of only nine \Aem;
they are displayed in Table \ref{tabcmm}.
Although they argue that these nine observable sector
embeddings are inequivalent, in Appendix \ref{erp} I show that
three more equivalences exist:
\beq
\mtxt{CMM 3} \simeq \mtxt{CMM 1}, \qquad
\mtxt{CMM 5} \simeq \mtxt{CMM 4}, \qquad
\mtxt{CMM 7} \simeq \mtxt{CMM 6}.
\label{cmmiss}
\eeq
Thus, the number of inequivalent observable
sector embeddings satisfying the CMM conditions
is presumably six; I take CMM observable sector embeddings
1, 2, 4, 6, 8 and 9 as representatives of these six.
These are the six inequivalent observable sector
embeddings of the \bsa\ class.
This does not mean that only six {\it models}
of this type exist.  For each choice of the six inequivalent
\Aem\ there will be many
possible hidden sector embeddings \Bem,
not all of which are equivalent.
CMM have left the hidden sector embedding unspecified
and the purpose of my recent paper \cite{Gie01b} was to enumerate
the allowed ways (up to equivalences) of
embedding the hidden sector.  The details of
that calculation are reviewed in this section.

One might wonder whether or not the hidden
sector embedding has any phenomenological
relevance from the ``low energy'' ($\lappeq$ 100 TeV)
point of view.
I now point out three ways in which
the hidden sector embedding is crucial to understanding
the low energy physics predicted by a given model.
Firstly, the conditions \myref{tps} for twisted sector
states depend on the full embedding $ \{ V, a_{1}, a_{3} \} $.
Thus, the hidden sector embedding
is important because the spectrum of
twisted sector states, including those
charged under the observable
sector gauge group $G_O$, depends on \Bem.
Secondly, twisted sector fields in nontrivial
irreps of $G_O$ are typically charged under $U(1)$
factors contained in the hidden sector gauge
group $G_H$; the spectrum of hidden $U(1)$ charges
will also depend on the hidden sector embedding,
since nonabelian factors in $G_H$
constrain the \uone\ generators by \myref{qao}.
Finally, the hidden sector embedding is
relevant to model building because $G_H$
and the nontrivial matter irreps under
nonabelian factors of $G_H$ play a crucial role in models
of dynamical supersymmetry breaking,
as discussed briefly in Section \ref{ssb} below.

\begin{table}[ht!]
\begin{center}
\begin{tabular}{clll}
CMM No. & \hspace{25pt} $3V_A$ & \hspace{25pt} $3a_{1A}$
& \hspace{25pt} $3a_{3A}$ \\ \hline 
1 & (-1,-1,0,0,0,2,0,0) & (1,1,-1,-1,2,0,0,0) & (0,0,0,0,0,0,2,0) \\ 
2 & (-1,-1,0,0,0,2,0,0) & (1,1,-1,-1,-1,-1,0,0) & (0,0,0,0,0,0,2,0) \\
3 & (-1,-1,0,0,0,2,0,0) & (1,1,-1,-1,-1,0,1,0) & (0,0,0,0,0,2,1,1) \\ 
4 & (-1,-1,0,0,0,2,0,0) & (1,1,-1,-1,-1,0,1,0) & (0,0,0,0,0,0,2,0) \\ 
5 & (-1,-1,0,0,0,2,0,0) & (1,1,-1,-1,-1,-1,1,-1) & (0,0,0,0,0,2,1,1) \\
6 & (-1,-1,0,0,0,2,0,0) & (1,1,-1,-1,-1,2,1,0) & (0,0,0,0,0,1,1,2) \\ 
7 & (-1,-1,0,0,0,2,0,0) & (1,1,-1,-1,-1,2,1,0) & (0,0,0,0,0,0,2,0) \\ 
8 & (-1,-1,0,0,0,1,1,0) & (1,1,-1,-1,-1,1,1,1) & (0,0,0,0,0,1,2,1) \\ 
9 & (-1,-1,0,0,0,1,1,0) & (1,1,-1,-1,-1,-2,0,1) & (0,0,0,0,0,0,0,2) \\
\hline
\end{tabular}
\end{center}
\caption{Observable sector embeddings.}
\label{tabcmm}
\end{table}

The allowed ways of completing the embeddings
of Table \ref{tabcmm} may be determined from
the consistency conditions (which ensure
{\it world sheet modular invariance}---a
property which is necessary for the
absence of quantum anomalies---of the
underlying string theory) 
presented in Refs.~\cite{IMNQ88,BLT88}:
\beq
3 V_B \in \elat, \qquad 3 a_{iB} \in \elat,
\label{vai}
\eeq
\beq
3 V \cdot V \in \Zbf, \qquad
3 a_i \cdot a_j \in \Zbf, \qquad
3 V \cdot a_i \in \Zbf.
\label{pp2}
\eeq
(The consistency conditions \myref{vai} were already
given in \myref{vab} and \myref{aabd} above; the
last two equations in \myref{pp2} must hold
for all choices of $i$ and $j$.)
For example, the first embedding in Table \ref{tabcmm}
has $9 V_A \cdot a_{1A} = -2$.  
Then the hidden sector embeddings which complete
CMM 1 must satisfy $9 V_B \cdot a_{1B} = 2 \mod 3$ since
\beq
V \cdot a_1 = V_A \cdot a_{1A} + V_B \cdot a_{1B}
\eeq
and from \myref{pp2} we see that $9 V \cdot a_1$ must
be a multiple of three.

An infinite number of solutions to \myref{vai} and
\myref{pp2} exist, even after the CMM conditions
of \myref{CMMgo} and untwisted $(3,2)$
irreps are imposed (conditions (iii-iv) of Definition 1).
This does not imply an infinite
number of {\it physically distinct} models.  For
example, trivial permutation redundancies such as
\beq
\pmatrix{ V_B^I \cr a_{1B}^I \cr a_{3B}^I}
\leftrightarrow 
\pmatrix{ V_B^J \cr a_{1B}^J \cr a_{3B}^J},
\qquad \forall \; I,J = 1, \ldots, 8
\label{triv}
\eeq
allow for different embeddings which give identical
physics.  Redundancies related to the signs of
entries also exist (to be addressed later).
Moreover, we will see below that an upper
bound may be placed on the magnitude of the entries
of the embedding vectors; that is, any embedding
with an entry whose magnitude is 
greater than the bound is equivalent to
another embedding which respects the bound.
Once these redundacies are eliminated 
the number of consistent hidden sector 
embeddings is large ($10^4 \sim 10^5$),
though no longer infinite.
However, just as with the observable sector
embeddings, the equivalence relations
exploited by CMM allow for a dramatic
reduction when one determines the physically
distinct models.  

I have carried out an automated reduction
using the equivalence relations enumerated by CMM,
which they have denoted ``(i)'' through ``(vi)''.
Their operations ``(ii)'' through ``(v)'' would affect
the observable embedding and are thus irrelevant
to our analysis.  This leaves two
equivalence relations,
presented here for ease of reference.
\begin{enumerate}
\item[\opI] The addition of a root lattice
vector $\ell \in \elat$ 
to any one of the vectors 
$V_B, a_{1B}$ or $a_{3B}$;
it is important to stress that any one
of these embedding vectors may be shifted
{\it independently}:
\beq
V_B \to V_B + \ell \qquad \mtxt{or} \qquad
a_{iB} \to a_{iB} + \ell, \quad i = 1 \, \mtxt{or} \, 3.
\label{otI}
\eeq
\item[\opII] A {\it Weyl reflection} performed
{\it simultaneously} on each of the embedding vectors
in the set $\{ V_B, a_{1B}, a_{3B} \}$:
\beq
V_B \to V_B - (V_B \cdot e_j) e_j,
\qquad
a_{iB} \to a_{iB} - (a_{iB} \cdot e_j) e_j,
\quad i = 1 \, \mtxt{and} \, 3.
\label{otII}
\eeq
\end{enumerate}
In keeping with the notation of Appendix~\ref{eer},
$e_j$ is one of the 240 nonzero roots of $E_8$.
In what follows I will refer to these as
operations \opI\ and \opII.

Operation \opI\ corresponds to an invariance
under translations by elements of the
$E_8$ root lattice $\elat$.  This transformation
group is nothing but the lattice group
associated with $\elat$; I will denote
this group as $\Tbf$.  Since operation \opI\
allows each vector $V_B$, $a_{1B}$ and $a_{3B}$
to be shifted by a different $E_8$ root lattice
vector, it is actually $\Tbf^3 = \Tbf \times \Tbf
\times \Tbf$ which is the corresponding invariance
group.  Operation \opII\
corresponds to an invariance under
the $E_8$ Weyl group, which I denote $\Wbf$.  To systematically
analyze possible equivalences between
different hidden sector embeddings under
operations \opI\ and \opII, it
is therefore vital to have a rudimentary
understanding of these two groups and their
combined action on the representation space
$\Rbf^8$; i.e., real-valued eight-dimensional
vectors such as $V_B, a_{1B}$ and $a_{3B}$.  
It is also helpful to develop a
concise notation for certain essential features
of $\Tbf$ and $\Wbf$.  For these 
purposes we now embark on a minor study
of these two groups.

It is convenient to notate the elements of $\Tbf$
as $T_\ell$, where $\ell$ is the lattice vector
by which the translation is performed:
\beq
T_\ell P = P + \ell, \qquad \ell \in \elat,
\qquad \forall \; P \in \Rbf^8.
\label{trn}
\eeq
Weyl reflections by any of the 240 nonzero $E_8$ roots
belong to $\Wbf$; I write these as $W_i$ with the
subscript corresponding to the $E_8$ root $e_i$
used in the reflection:
\beq
W_i : P \to W_i P = P - (P \cdot e_i ) e_i,
\qquad \forall \; i = 1, \ldots 240 ,
\qquad \forall \; P \in \Rbf^8.
\label{Wdf}
\eeq
It is not difficult to check that for each
of these operators $W_i^2 = 1$, so that each
is its own inverse; thus, the Weyl group
$\Wbf$ can be built up by taking all possible
products of the 240 $W_i$:
\beq
\Wbf = \{1, W_i, W_i W_j, \ldots \}.
\label{Wgd}
\eeq
The $E_8$ Weyl group is a nonabelian
finite group of {\it order} (the number
of elements) $696 \, 729 \, 600$.  On the
other hand, there are only 240 Weyl
reflections $W_i$.  Thus, the generic element
of $\Wbf$ is not a simple reflection \myref{Wdf},
but is a product of several such reflections.
In what follows, I write generic elements
of the Weyl group in calligraphic type:
${\cal W}_I \in \Wbf$, with
$I = 1, \ldots, 696 \, 729 \, 600$.
Thus, for each element ${\cal W}_I$
of $\Wbf$, Weyl reflections $W_j, W_k, \ldots, W_m$
exist such that
\beq
{\cal W}_I = W_j W_k \cdots W_m .
\label{stw}
\eeq
I point out one more property of the Weyl
group $\Wbf$, which we will have occasion to
appeal to below:  an $E_8$ root lattice
vector, when subjected to a Weyl
group transformation,
yields back an $E_8$ root lattice vector.
Explicitly, if $\ell \in \elat$ and
${\cal W}_I \in \Wbf$, then there exists
a $k \in \elat$ such that
\beq
{\cal W}_I \ell = k .
\label{fnk}
\eeq
In mathematical parlance, ${\cal W}_I$ is
an {\it automorphism} of $\elat$.

With these tools in hand, we can prove a useful theorem.
\begin{thm}
\label{thma}
If ${\cal W}_I \in \Wbf$ and $T_\ell \in \Tbf$,
then there exists a $T_k \in \Tbf$
such that ${\cal W}_I T_\ell = T_k {\cal W}_I$.
\end{thm}
To see this, let $P \in \Rbf^8$ and compute
\beq
{\cal W}_I T_\ell P = {\cal W}_I (P + \ell)
= {\cal W}_I P + {\cal W}_I \ell.
\label{lch}
\eeq
The last step follows from the fact that
${\cal W}_I$ is a linear operator---a property which
is evident from \myref{Wdf} and \myref{stw}.
Using \myref{fnk}, the right-handed side of
\myref{lch} can be rewritten
\beq
{\cal W}_I P + {\cal W}_I \ell = {\cal W}_I P + k = T_k {\cal W}_I P .
\eeq
I.e., ${\cal W}_I T_\ell = T_k {\cal W}_I$, as was to be
shown.

A sequence of operations \opI\ and \opII\
has the form of a product of various elements of
$\Tbf$ and $\Wbf$.  Theorem \ref{thma} allows one to rewrite
any sequence of operations \opI\ and
\opII, whatever the order and number of
operations of each type, in the form
\beq
{\cal O} = T_\ell {\cal W}_I,
\qquad T_\ell \in \Tbf,
\qquad {\cal W}_I \in \Wbf .
\eeq
I stress that the element $T_\ell$ may be different
for each of the embedding vectors $V_B$, $a_{1B}$
and $a_{3B}$, but that the Weyl group element ${\cal W}_I$
acting on these vectors must be the same.
Typically, ${\cal W}_I$ will
be a generic element of the Weyl group
taking the form \myref{stw},
corresponding to a string of operations of type
\opII.  Thus, we arrive at the following rather useful
conclusion:  any sequence of operations
\opI\ and \opII, whatever the order
and number of operations of each type, is equal
in effect to a sequence of operations
of type \opII, followed by a {\it single}
operation of type \opI, allowing for
different shifts for each of the three
embedding vectors.  Symbolically,
we need only consider equivalences of the form
\beq
{\cal O} = T_\ell W_j W_k \cdots W_m .
\eeq

Suppose two embeddings $\{ V_B, a_{1B}, a_{3B} \}$
and $\{ V_B', a_{1B}', a_{3B}' \}$.  We want to
determine whether these two embeddings are 
equivalent.  Based on the results of the last
paragraph, we see that it is sufficient to
first tabulate all points in the {\it orbit}
of $\{ V_B, a_{1B}, a_{3B} \}$
under $\Wbf$, and then to check whether any of these
points are related to $\{ V_B', a_{1B}', a_{3B}' \}$
by operation \opI;
the orbit of $\{ V_B, a_{1B}, a_{3B} \}$
under $\Wbf$ is tabulated by computing the
transformations
$\{ {\cal W}_I V_B, {\cal W}_I a_{1B}, {\cal W}_I a_{3B} \}$
for all $696 \, 729 \, 600$ elements ${\cal W}_I$ of
the $E_8$ Weyl group.  If the two embeddings
are related in this way, then they are equivalent.

As mentioned above, for a given \Aem, the number
of consistent \Bem\ is infinite; the following
definition exploits operation \opI\ to immediately
and efficiently eliminate enough redundancy
to obtain a finite set.
\begin{defn}
\label{defa}
An embedding $\{ V_B, a_{1B}, a_{3B} \}$ is in
{\bf minimal} form provided:
\begin{enumerate}
\item[{\rm (a)}] $3 V_B^I \in \Zbf$, $3 a_{1B}^I \in \Zbf$
and $3 a_{3B}^I \in \Zbf$ for each choice $I = 1, \ldots, 8$;
\item[{\rm (b)}] $|3 V_B^I| \leq 2$, $|3 a_{1B}^I| \leq 2$ 
and $|3 a_{3B}^I| \leq 2$ for each choice $I = 1, \ldots, 8$;
\item[{\rm (c)}] no more than one entry of each vector
$3V_B$, $3 a_{1B}$ and $3 a_{3B}$ has absolute value two,
and any such entry is the left-most nonzero entry.
\end{enumerate}
\end{defn}
Any embedding may be reduced to minimal form by means of
operation \opI.  I will demonstrate the veracity
of this statement by considering $V_B$ which are
not minimal.  It will be understood that similar
statements hold for $a_{1B}$ and $a_{3B}$ which
are not minimal, since operations of type \opI\
are allowed to act independently on $V_B$, $a_{1B}$ and
$a_{3B}$.

From \myref{vai} one sees that $3 V_B$ is an $E_8$
root lattice vector.  As explained in Section~\ref{eer},
the entries of an $E_8$ root lattice vector are either
all integral or all half-integral.  In the latter case,
part (a) of Definition \ref{defa} will not be satisfied.  However,
operation \opI\ allows us to shift
\beq
3 V_B \to 3 V_B + 3 \ell,
\qquad \ell \in \elat .
\label{13i}
\eeq
If we take $\ell$ to be any lattice vector with half-integral
entries, then \myref{13i} transforms $3 V_B$ to a lattice
vector with integral entries.
Now suppose $3 V_B$ satisfies part (a) of Definition \ref{defa}
but $| 3 V_B^I| > 2$ for one or more choices of $I$.
It is in all cases possible to find a lattice
vector $\ell$ such that \myref{13i} generates an
equivalent $3 V_B$ which satisfies part (b)
of Definition \ref{defa}.  To see this, first note
that repeated shifts \myref{13i} by vectors
\beq
3 \ell \in \left\{ \pm (\underline{ 3, 3, 0,0,0,0,0,0}), \;
(\underline{3, -3, 0,0,0,0,0,0})
\right\}
\label{13iii}
\eeq
(underlining indicates that any permutation
of entries may be taken)
allows $3 V_B$ to be translated to a form
where no entry has absolute value greater
than three.  If the original $3 V_B$ satisfied
\myref{vai}, then the translated one will as
well, since the sum of two lattice vectors
is also a lattice vector.  As explained in Section~\ref{eer},
an $E_8$ root lattice vector must have its
entries sum to an even number (the final
condition in \myref{eld}).  Then from \myref{vai}
we know that
\beq
\sum_{I=1}^8 3 V_B^I = 0 \mod 2 .
\label{13ii}
\eeq
If for any $I$ the translated vector
has $3 V_B^I = \pm 3$, then \myref{13ii}
implies that there must be a $J \not= I$ such
that $3 V_B^J$ is an odd integer.  If
$3 V_B^J = \pm 3$, then a final shift by
one of the vectors in \myref{13iii} allows
us to set $V_B^I \to 0$ and $V_B^J \to 0$.
For example:
$$
3 V_B = (\ldots,3,\ldots,3,\ldots) \qquad \mtxt{and} \qquad
3 \ell = (\ldots,-3,\ldots,-3,\ldots)
$$
\beq
\mtxt{gives} \qquad
3 V_B \to 3 V_B + 3 \ell = (\ldots,0,\ldots,0,\ldots).
\eeq
On the other hand, if
$3 V_B^J = \pm 1$, then a final shift by
one of the vectors in \myref{13iii} allows
us to set $V_B^I \to 0$ and $V_B^J \to \mp 2$.
From the above manipulations, it should be clear
that a shift \myref{13i} by an appropriate vector
\myref{13iii} will eliminate any pair of $\pm 2$s
appearing in $3 V_B$ in favor of a pair of $\pm 1$s.
Similarly, if a $\pm 1$ precedes a $\pm 2$ (reading
left to right), the order may be reversed---possibly
altering signs---by a shift \myref{13i} by an
appropriate vector \myref{13iii}.
In this way, we are always able to transform any
$V_B$ satisfying parts (a) and (b) of Definition \ref{defa}
into an equivalent form which also satisfies part (c)
of Definition \ref{defa}.

It is a simple excercise to verify that
Weyl reflections \myref{Wdf} using $E_8$ roots of the form
$e_i = (\underline{1,-1,0,\ldots,0})$ exchange
two entries; it is also easy to check that
Weyl reflections using roots of the form
$e_i = (\underline{1,1,0,\ldots,0})$
exchange two entries and flip both signs.
I will refer to these as ``integral'' Weyl
reflections.  The second type uses
$E_8$ roots of the form $e_i = (\pm 1/2, \ldots, \pm 1/2)$
with an even number of positive entries,
and I will refer to these as ``half-integral''
Weyl reflections.  These tend to have more
dramatic effects; for example, $3 V_B = (1,\ldots,1)$
can be reflected to $3 V_B = (2,2,0,\ldots,0)$
using $e_i = (1/2,1/2,-1/2,\ldots,-1/2)$.
By such manipulations, together with operation \opI,
it is well-known that only five inequivalent
twist embeddings $V=(V_A;V_B)$ exist (including $V=0$).
Consistency with a given CMM $V_A$
restricts $V_B$ to one or two
choices.  We can eliminate 
remaining redundancies related
to integral Weyl reflections by enforcing
ordering and sign conventions on $a_{1B}$ and $a_{3B}$.
With this in mind, I make the following definition.
\begin{defn}
An embedding $\{ V_B, a_{1B}, a_{3B} \}$
is in {\bf canonical} form if
$$3 V_B = (2,1,1,0,0,0,0,0)$$ for CMM 1 through 7,
$$3 V_B = (1,1,0,0,0,0,0,0)$$ or
$$3 V_B = (2,1,1,1,1,0,0,0)$$ for CMM 8 and 9; and,
$a_{1B}$ and $a_{3B}$ are {\bf first} fixed to minimal form,
and {\bf then} subjected to whatever integral Weyl reflections
are required such that they satisfy the following conditions:
\begin{enumerate}
\item[{\rm (a)}]
	$V_B^I = V_B^{I+1} \; \Rightarrow \; a_{1B}^I \geq a_{1B}^{I+1}, \;
	I=1,\ldots,7$;
\item[{\rm (b)}]
	$V_B^I=0 \; \Rightarrow \; a_{1B}^I \geq 0, \; I = 3,\ldots,7$;
\item[{\rm (c)}]
	$a_{1B}^7 = 0 \; \Rightarrow \; a_{1B}^8 \geq 0$ while
	$a_{1B}^7 \not= 0 \; \Rightarrow \; 3 a_{1B}^8 \geq -1$;
\item[{\rm (d)}]
	$V_B^I = V_B^{I+1}$ and $a_{1B}^I = a_{1B}^{I+1} \;
	\Rightarrow \; a_{3B}^I \geq a_{3B}^{I+1}, \;
	I=1,\ldots,7$;
\item[{\rm (e)}]
   $V_B^I=a_{1B}^I=0 \; \Rightarrow \;
   a_{3B}^I \geq 0,\; I = 3,\ldots,6$;
\item[{\rm (f)}]
   $a_{1B}^7 = a_{1B}^8 =0$ or
   $a_{1B}^6 = a_{1B}^7 = a_{3B}^6 = 0 \; \Rightarrow \; a_{3B}^7 \geq 0$;
\item[{\rm (g)}]
   $a_{1B}^6 = a_{1B}^7 = 0$ and $a_{3B}^6 \not= 0$ and 
	$a_{1B}^8 \not= 0 \; \Rightarrow \; 3 a_{3B}^7 \geq -1$;
\item[{\rm (h)}]
	$a_{1B}^7 = a_{1B}^8 = a_{3B}^7 = 0 \; \Rightarrow \;
	a_{3B}^8 \geq 0$;
\item[{\rm (i)}]
	$a_{1B}^7 = a_{1B}^8 =0$ and $a_{3B}^7 \not= 0
	\; \Rightarrow \; 3 a_{3B}^8 \geq -1$;
\end{enumerate}
\end{defn}
It is straightforward, though tedious, to
verify that any $a_{1B}$ and $a_{3B}$ of minimal
form can be transformed to satisfy the conditions
listed above using the integral Weyl reflections;
I do not present a proof here as the manipulations
are lengthy and elementary.
Transforming all embeddings $\{ V_B, a_{1B}, a_{3B} \}$
to canonical form, we arrive at a set for which
no two are related purely by integral Weyl reflections.

With the definition \myref{Wdf},
it is not difficult to check
\beq
W_i W_j W_i = W_k, \qquad
e_k = e_j - (e_j \cdot e_i) e_i.
\label{trf}
\eeq
Recall that the entries of $E_8$ roots $e_i$ 
are either all integral or all half-integral.
I denote integral roots with undotted subscripts from the
beginning of the alphabet, $e_a, e_b, \ldots$ and
half-integral roots with dotted subscripts from
the beginning of the alphabet, $e_\adot, e_\bdot, \ldots$.
It should be clear that $e_\adot - (e_\adot \cdot e_a) e_a$
is a half-integral root since $e_\adot \cdot e_a \in \Zbf$.
Thus we can specialize \myref{trf} to obtain, for example,
\beq
W_a W_\adot W_a = W_\ccdot, \qquad
e_\ccdot = e_\adot - (e_\adot \cdot e_a) e_a.
\label{poi}
\eeq
We can then perform manipulations such as
\beq
W_\adot W_a = W_a W_a W_\adot W_a = W_a W_\ccdot,
\label{pf1}
\eeq
\beq
W_\adot W_\bdot W_a = W_a W_a W_\adot W_a W_a W_\bdot W_a
= W_a W_\ccdot W_\dddot,
\label{pf2}
\eeq
where $W_\ccdot$ is defined explicitly in \myref{poi} and
$W_\dddot = W_a W_\bdot W_a$ is defined analogously.
This illustrates how \myref{poi} allows
us to write a generic element \myref{stw} of the Weyl
group $\Wbf$ in the form
\beq
{\cal W}_I = W_a \cdots W_c W_\adot \cdots W_\ccdot .
\eeq
Equivalences related to the string of 
integral Weyl reflections $W_a \cdots W_c$
are eliminated by going to canonical form.
From these considerations
we find that, given a set of canonical embeddings,
equivalences may be identified by the following
procedure:  
\begin{enumerate}
\item[(i)] compute the orbit of 
\Bem\ under strings of half-integral
Weyl reflections;
\item[(ii)] fix the results of (i) to minimal form
by operations of type \opI;
\item[(iii)] fix the results of (ii) to canonical form by
integral Weyl reflections; 
\item[(iv)] check
whether the results of (iii) 
are related by operation \opI\ to any
other embedding in the original set.
\end{enumerate}
The last
step is simply a matter of checking whether
the differences 
$V_B - V_B'$, $a_{1B}-a_{1B}'$
and $a_{3B}-a_{3B}'$ each give
lattice vectors, where \Bem\ is a result
of step (iii) and $\{ V_B', a_{1B}', a_{3B}' \}$
is an element of the original set of
canonical embeddings.

In my automated analysis, I first generated
a list of all possible consistent embeddings
of the hidden sector, constraining them to
be of canonical form.  Since all embeddings can
be reduced to canonical form by way of operations
\opI\ and \opII, we are assured that this list is complete.
The number of ``initial'' embeddings was
at this point already reduced to roughly $10^4$.
Using the procedure outlined in the
previous paragraph, I removed as many of the
redundant embeddings as performing
only 1, 2 and 3 half-integral Weyl reflections
in step (i) would allow.
Because the $E_8$ Weyl group is so large, it proved
to be impractical to act on the initial
embeddings with each of its elements.  It also
proved impractical to perform four or more
half-integral Weyl reflections.\footnote{The number
of positive half-integral roots is 64
(negative roots generate the same
Weyl reflections); four Weyl reflections would have
required roughly $10^7$ different operations
for each embedding.}
The initial list was thereby
reduced to a mere 192 embeddings.  
This list is guaranteed
to be complete, but entries of the list
are not necessarily inequivalent.
However, already in
going from 2 half-integral Weyl reflections
to 3 half-integral Weyl reflections, the
list did not shrink by much.  It would
appear that though there may be some
equivalences remaining, there should not
be very many.
(It is worth pointing out that application
of an analogous procedure to the observable
sector embeddings turned up equivalences
overlooked by CMM, already at the level
of one half-integral Weyl reflection.)

I have, in addition, determined the hidden
sector gauge group $G_H$ for each of the 192
embeddings.  Only five $G_H$ were found to
be possible, displayed above in Table \ref{gcs}.
This is remarkable, considering that
one might naively expect a large subset of the
112 breakings \cite{HM88} of $E_8$ to be present.
Apparently, the CMM requirements of \myref{CMMgo} 
and untwisted quark doublets significantly 
affect what is possible in the hidden sector.

\begin{table}[ht!]
\begin{center}
\begin{tabular}{cc}
Case & $G_H$ \\ \hline 
1 & $SO(10) \times U(1)^3$ \\
2 & $SU(5) \times SU(2) \times U(1)^3$ \\
3 & $SU(4) \times SU(2)^2 \times U(1)^3$ \\
4 & $SU(3) \times SU(2)^2 \times U(1)^4$ \\
5 & $SU(2)^2 \times U(1)^6$ \\ \hline
\end{tabular}
\caption{Allowed hidden sector gauge groups $G_H$. \label{gcs}}
\end{center}
\end{table}

In Appendix \ref{emt}, I present lists of
the hidden sector embeddings which complete the CMM
analysis.  I have not displayed\footnote{
They are, however, available from the author upon
request.} Case 5 $G_H$ models,
since I do not regard them as affording viable scenarios
of hidden sector dynamical supersymmetry breaking, as
discussed in the following section.
Eliminating the Case 5 $G_H$
models from the total of 192, we are left
with 175 models.
The spectrum of massless matter for these models
and a summary of general features is discussed in
Section \ref{mds} below.

\section{Supersymmetry Breaking}
\label{ssb}
The $Z_3$ orbifold models studied here have
$N=1$ local
supersymmetry (supergravity) at the string scale.  In my
analysis, I assume that this supersymmetry
is broken dynamically via gaugino condensation
of an asymptotically free condensing group $G_C$
in the hidden sector.\myfoot{ORs.}  That is, the vacuum expectation
value ({\it vev}) of the gaugino bilinear $\vev{\lambda \lambda}$
acquires a nonvanishing value.  This operator has
mass dimension three; I therefore
define the dynamically generated {\it condensation
scale} $\LamC$ by
\beq
\vev{ \lambda \lambda } = \LamC^3.
\eeq
To estimate the value of $\LamC$, consider
the one loop evolution of the
running gauge coupling $g_C(\mu)$ of $G_C$:
\beq
{d g_C \over d \ln \mu} = \beta(g_C) = {b_C g_C^3 \over  16 \pi^2}.
\label{bfdef}
\eeq
The $\beta$ function coefficient $b_C$ is given by
\beq
b_C = -3 C(G_C) + \sum_R X_C(R).
\label{bcd}
\eeq
Here, $C(G_C)$ is the eigenvalue
of the quadratic Casimir operator for the adjoint
representation of the group $G_C$ while
$X_C(R)$ is the {\it Dynkin index} for the representation $R$,
given by $\tra{R}(T^a)^2 = X_C(R)$ in a Cartesian
basis for the generators $T^a$;
I adhere to a normalization where $X_C=1/2$
for an $SU(N)$ fundamental representation.
The sum runs over chiral supermultiplet
representations.
Provided $b_C$ is negative, the coupling turns strong
at low energies and the dynamical scale
$\LamC$ is generated, in analogy to
$\Lambda_{\stxt{QCD}}$.  
The running of gauge couplings
from an initial unified value $g_H \sim 1$ at a
unification scale,
which in our case is the
string scale $\LamH \sim 10^{17}$ GeV, gives
\beq
\LamC \sim \LamH \exp (8 \pi^2/ b_C g_H^2),
\label{lces}
\eeq
where I have identified $\LamC$ with the
Laundau pole of the running coupling.

Soft mass terms in the low energy effective
lagrangian split the masses of supersymmetry multiplets,
and thereby break supersymmetry;
partners to Standard Model (SM) particles are
generically heavier by the soft mass scale $\MSB$.
The soft terms arise from nonrenormalizable interactions
in the supergravity lagrangian, with masses
proportional to the gaugino condensate
$\vev{ \lambda \lambda }$,
suppressed by inverse powers of the (reduced) Planck mass,
$m_P \equiv 1 / \sqrt{8 \pi G} = 2.44 \times 10^{18}$ GeV.
On dimensional grounds, one expects that the observable
sector supersymmetry breaking scale $\MSB$ is given by
\beq
\MSB \approx \zeta \cdot \vev{ \lambda \lambda} / m_P^2
= \zeta \cdot \LamC^3 / m_P^2,
\label{sbs}
\eeq
with (naively) $\zeta \sim \ordnt{1}$.
For supersymmetry to protect the gauge hierarchy $m_Z \ll m_P$
between the electroweak scale and the fundamental scale,
one requires, say, $\MSB \lappeq 10$ TeV.  Then
\myref{sbs} with $\zeta \sim \ordnt{1}$ implies
$\LamC \lappeq 4 \times 10^{13}$ GeV.  On the other hand,
direct search limits \cite{PDG00} on charged superpartners
require, say, $\MSB \gappeq 50$ GeV, which translates
into $\LamC \gappeq 7 \times 10^{12}$ GeV.
More precise results may be obtained, for instance,
with the detailed supersymmetry breaking models\myfoot{Maybe
I should cite some other groups that have
worked on this?  Bailin and Love, de Carlos et al., etc.}
of Bin\'etruy, Gaillard and Wu (BGW) \cite{BGW} as well
as subsequent ellaborations by Gaillard and
Nelson \cite{GN00}.  These calculations confirm
the naive expectation \myref{sbs}, except that
\beq
\ord{-2} \lappeq \zeta \lappeq \ord{-1},
\label{1a}
\eeq
which tends to increase $\LamC$.  For example,
the lower bound implied by $\MSB \gappeq 50$ GeV
changes to $\LamC \gappeq 9 \times 10^{12}$ GeV
if $\zeta \approx 0.4$, near the upper end of
the range \myref{1a}.
The result is that
\beq
\ord{13} \lappeq {\LamC \over \mtxt{GeV}} \lappeq \ord{14}
\label{1b}
\eeq
is a reasonably firm estimate.

For $G_C = SU(2)$ with no matter,
one has $b_C = -6$.  Substituting
into \myref{lces}, one finds $\LamC \sim 10^{11}$ GeV.
On the other hand, \myref{lces} is a crude estimate;
studies of the BGW effective theory show that
the naive estimate \myref{lces}
can receive significant corrections
due to a variety of effects, and deviations
by an order of magnitude are certainly possible.
Thus, a more reliable bound is $\LamC \lappeq 10^{12}$ GeV.
Since $b_C > -6$ when $G_C$ charged matter
is present, the limit $\LamC \lappeq 10^{12}$ GeV is saturated by the
case with no matter.  In the models considered here, as will be seen
below, $SU(2)$ groups always have many, many matter
representations, and it is unlikely that {\it all} of them
would acquire effective mass couplings
{\it at the unification scale} $\LamH$
so that $b_C = -6$ and 
$\LamC \sim 10^{12}$ GeV could be achieved.
In any case, $10^{12}$ GeV is below the lower bound
in \myref{1b}, set by $\MSB \gappeq 50$ GeV,
the firmer of the soft scale requirements,
so having $b_C = -6$ is marginal at best.
Case 5 of Table \ref{gcs} was therefore considered
to be an unviable hidden sector gauge group.
Certainly, Cases 1 to 4 appear more
promising.
Eliminating the models with the Case 5 gauge group,
only 175 models remain.  The matter spectra of these
models are the topic of Section \ref{mds}.

\section[The Anomalous $U(1)$]{The Anomalous ${\bf U(1)}$}
\label{au1}
Quite commonly in the models considered here,
some of the \uone\ factors contained in the
gauge group $G = G_O \times G_H$ are apparently anomalous:
$\tr Q_a \not= 0$.
As discussed in Section \ref{rec},
redefinitions of the charge generators
allow one to isolate
this anomaly such that only one \uone\
has an apparent trace anomaly.
I denote this factor of $G$ as \ux.
The associated anomaly is canceled by the Green-Schwarz
mechanism \cite{GS84}:  tree level couplings between
the \ux\ vector multiplet and the two-form
field strength (dual to the
universal axion)
are added to the effective action
in such a way that the one loop \ux\ anomaly is
canceled \cite{DIS87,DSW87,ADS87}; the \ux\ only appears to
be anomalous.
When the cancelation is done in
a supersymmetric fashion, a Fayet-Illiopoulos (FI)
term $\xi$ for \ux\ is induced; I 
provide details in Appendix \ref{gsx}.  The result
is an effective D-term for \ux\
of the form:
\beq
D_X = \sum_i {\p K \over \p \phi^i} {\hat q}^X_i \phi^i + \xi,
\qquad \xi = {g_H^2 \tr \hat Q_X \over 192 \pi^2} m_P^2.
\label{1.2}
\eeq
The \ux\ generator $\hat Q_X$ has a normalization consistent
with unification (discussed further below),
${\hat q}^X_i$ is the charge of
the scalar $\phi^i$ with respect to $\hat Q_X$,
$K$ is the K\"ahler potential and
$g_H$ is the unified coupling mentioned briefly
in Section \ref{ssb} above.
Since the scalar potential of the effective supergravity
theory at the string scale $\LamH$ contains
the term $g_H^2 D_X^2 / 2$,
some scalar fields generically shift to cancel the FI
term (i.e., $\vev{D_X}=0$ to leading order)
and get vevs of order $\sqrt{|\xi|}$.
Adopting the terminology
of \cite{Gie01a}, I will refer to these
as {\it Xiggs} fields, 
since they are associated with the breaking
of \ux\ (and typically other factors of $G$)
via the Higgs mechanism.
Generally, the
way in which the FI term may be canceled is
not unique and continuously connected vacua
result.  Pseudo-Goldstone modes, {\it D-moduli} \cite{GG00},
parameterize the flat directions;
dynamical supersymmetry
breaking and loop effects are required to select
the true vacuum and render these
scalar fields massive \cite{GG00,Gai01}.
(Moduli parameterizing flat directions of the
scalar potential are a generic feature of
supersymmetric field theories \cite{BDFS82}.
An example of D-moduli was noted previously in the study
of D-flat directions in \cite{CM88},
parameterized there by the quantity
``$\lambda$,'' which interpolated between
various vacuua.  Such moduli have also been
noted in the study of
flat directions in free fermionic string models,
for instance in Ref.~\cite{CCF01}.)
The FI term $\xi$ has mass dimension two and its
square root therefore gives the approximate scale of
\ux\ breaking, which I hereafter denote
\beq
\LamX \equiv \sqrt{|\xi|}
= { \sqrt{|\tr \hat Q_X |} \over 4 \pi \sqrt{12} }
\times g_H m_P.
\label{1c}
\eeq
In the examples below we will find by explicit calculation
of $\tr \hat Q_X$ for each of the 175 models
that $\LamX \approx \LamH \sim 0.2 \, m_P$.

\section{Discussion of Spectra}
\label{mds}
Automating the matter spectrum recipes
given in Section \ref{rec}, I have determined
the spectra for all 175 models.
I now make some general observations
based on the results of this analysis.
Ignoring the various $U(1)$ charges,
only 20 patterns of irreps were found 
to exist in the 175 models.
These are summarized
in Tables \ref{pt1}-\ref{pt4}.
In all 175 models, twisted oscillator matter states
are singlets of $\GNA$ (cf.~\myref{gde}).
Singlets notated $(1,\dots,1)_0$ are either
untwisted matter states or twisted
non-oscillator matter states
while singlets notated $(1,\dots,1)_1$
are twisted oscillator matter states.  
Only Patterns 2.6, 4.5, 4.7 and 4.8 have no
twisted oscillator states.  In Table~\ref{unpat}
I show the irreps in the untwisted
sector for each of the twenty patterns.
Comparing to Tables~\ref{pt1}-\ref{pt4},
it can be seen that the majority of
states in any given pattern are twisted
non-oscillator states.

In Table~\ref{tb1} I have cross-referenced
the models enumerated in Section \ref{iem} with
the twenty patterns given here.  The
observable sector embeddings are given
in Table \ref{tabcmm} and the
hidden sector embeddings can be found
in Appendix \ref{emt}.
Models are labeled in the format ``$i.j$'' where:
\ben
\item[(a)]
for $i=1,2,4$ or $6$, $i$ is the CMM observable 
sector embedding according to the labeling 
of Table \ref{tabcmm} and $j$ is the
hidden sector embedding label
as per the corresponding choice
of table from the set Tables \ref{tab1}-\ref{tab6};
\item[(b)]
$i=8$ corresponds to the CMM observable
sector embedding 8 according to the labeling
of Table \ref{tabcmm} and $j$ is the
hidden sector embedding
according to the labeling of Table \ref{tab8a};
\item[(c)]
$i=10$ also corresponds to the CMM observable
sector embedding 8 according to the labeling
of Table \ref{tabcmm}, but now $j$ is the
hidden sector embedding
according to the labeling of Table \ref{tab8b};
\item[(d)]
$i=9$ corresponds to the CMM observable
sector embedding 9 according to the labeling
of Table \ref{tabcmm} and $j$ is the
hidden sector embedding
according to the labeling of Table \ref{tab9a};
\item[(e)]
$i=11$ also corresponds to the CMM observable
sector embedding 9 according to the labeling
of Table \ref{tabcmm}, but now $j$ is the
hidden sector embedding
according to the labeling of Table \ref{tab9b}.
\een
I remind the reader that CMM observable
sector embeddings 3, 5 and 7 do not appear
because they are equivalent to 1, 4 and 6
respectively.

All patterns except Pattern 1.1 have
an anomalous \ux\ factor.
I have determined the FI term for each of the
models in the other 19 patterns.
I find that all models
within a particular pattern have the
same FI term; the
corresponding values
of $\LamX$ are displayed
in Table \ref{fir}.
As will be discussed in greater detail
in Section~\ref{gcu}, Kaplunovsky \cite{Kap88} has
estimated the string scale to be
\beq
\LamH \approx g_H \times 5.27 \times 10^{17} \mtxt{GeV}
= 0.216 \times g_H m_P.
\label{3d}
\eeq
Using the values in Table~\ref{fir},
it is easy to check that
\beq
\LamH/1.73 \leq \LamX \leq \LamH .
\label{3e}
\eeq
The effective supergravity lagrangian
describing the field theory limit of the string
is nonrenormalizable.  In principle, all superpotential
and K\"ahler potential operators
allowed by symmetries of the underlying theory
should be present.  As discussed in Appendix~\ref{cma},
there exist field reparameterization invariances
in the effective theory.  These invariances
relate different classical field configurations,
or vacua.  Expansion about a particular vacuum
leads\myfoot{Read more on NL$\s$ models;
try to improve explanation.} to a nonlinear $\s$ model.
For instance, this is reflected in the
presence of superpotential operators such as \myref{ems}
above, with ever increasing numbers $n$ of Xiggses.
For the nonlinear $\s$ model to be perturbative,
it must be possible to truncate the sequence of operators
at some order $n_{\stxt{max}}$ and obtain a
reasonable approximation to the full theory.\myfoot{This
is not the best explanation; the T-moduli
enter in a nonperturbative way, for example;
really it is the Xiggs that we truncate.}
Since the relevant expansion parameter for
nonrenormalizable operators is roughly
$\LamX/m_P$, which from Table~\ref{fir}
lies in the range
\beq
g_H / 8.00 \leq \LamX / m_P \leq  g_H / 4.63 ,
\label{3c}
\eeq
the nonlinear $\s$ model
has a reasonable chance to be perturbative,
provided the unified coupling satisfies
$g_H \lappeq 1$ and the number of operators
contributing to an effective coupling
(such as the $A A^c$ coupling in \myref{ems})
is not too large.  (Generically, the number
of such operators increases with dimension.)

\begin{table}
\begin{center}
\begin{tabular}{ccccc}
Pattern & $\LamX/(g_H m_P)$ & \hspace{50pt} &
Pattern & $\LamX/(g_H m_P)$ \\ \hline 
1.2 & 0.216 & & 2.6, 3.3, 4.6 & 0.170 \\
2.1, 4.2 & 0.125 & & 3.1, 4.3 & 0.148 \\
2.2, 2.3, 4.1 & 0.138 & & 3.2, 4.4, 4.8 & 0.176 \\
2.4 & 0.186 & & 3.4 & 0.181 \\
2.5 & 0.191 & & 4.5, 4.7 & 0.157 \\
\hline
\end{tabular}
\caption{The \ux\ symmetry breaking scale
$\LamX$ for each of the irrep patterns. \label{fir}}
\end{center}
\end{table}

Given the importance of nonvanishing vevs
to the perturbative expansion of the
nonlinear $\s$ model, I next
estimate the range of Xiggs vevs.
I will assume that $g_H \approx 1$ in
\myref{3c}, as suggested by analyses of
the running gauge couplings; for
example, see Section~\ref{gcu} below.
Then from \myref{3c} we have
\beq
\LamX \sim \ord{-1} \; m_P.
\label{3f}
\eeq
Furthermore, I assume that Xiggs fields
have a nearly diagonal K\"ahler potential
at leading order in an expansion about the
vacuum:
\beq
K_{\stxt{Xiggs}} = \sum_i
\bigvev{ {\p^2 K \over \p \phi^i \p \bar \phi^i} }
|\phi^i|^2 + \cdots,
\eeq
with the terms represented by ``$\cdots$'' negligible
in comparison to the explicit terms.  This
assumption is justified by the known
form for the terms in $K$ quadratic in
matter fields for $Z_3$ orbifolds
with nonstandard embedding \cite{twk}, such
as the cases considered here. 
In the limit of vanishing
off-diagonal T-moduli (i.e., $\vev{T^{ij}}=0, \; \forall \; i \not= j$),
\beq
K_{\stxt{quad.-matter}}
= \sum_i {|\phi^i|^2 \over \prod_{j=1,3,5} (T^j + \bar T^j)^{q_j^i}} .
\label{kqm}
\eeq
Here, $q_j^i$ are the {\it modular weights} of the
matter field $\phi^i$:  untwisted states $\ket{K;i}$
have modular weights $q_j^i = \delta_j^i$, while twisted
non-oscillator states $\ket{\tK;n_1,n_3,n_5}$
have modular weights $q_j^i=2/3$ and
twisted oscillator states $\ket{\tK;n_1,n_3,n_5;i}$
have modular weights $q_j^i=2/3+ \delta^i_j$.
Moduli stabilization in the BGW model
gives $\vev{T^j}=1$ or $e^{i\pi/6}$ $\, \forall \; j$.  
Assuming the former value and applying \myref{kqm},
we find
\beq
\bigvev{ {\p^2 K \over \p \phi^i \p \bar \phi^i} }_{\stxt{BGW}} =
\left\{ \begin{array}{lc}
1/2 & \mtxt{untwisted}, \\
1/2^2 & \mtxt{twisted non-oscillator}, \\
1/2^3 & \mtxt{twisted oscillator}.
\end{array} \right.
\eeq
This ignores the possible contribution of terms
$K \ni (c/m_P^2) \; f(T) \; |\phi^i|^2 |\phi^j|^2$, with both
fields $\phi^i,\phi^j$
Xiggses and $f(T)$ a function of the T-moduli.
If we assume $\vev{\phi^{i}} \sim \vev{\phi^{j}}
\sim \LamX$, these quartic terms (which include
$i$-$j$ mixing) are suppressed by $\ordnt{\LamX^2/m_P^2}$
relative to the leading terms.  However,
we still have to estimate
$\vev{\phi^{i}}$ and $\vev{\phi^{j}}$,
so at the end of our analysis we will
have to check whether or not it was
consistent to neglect these quartic terms.
It is also unclear what the moduli-dependent
function $f(T)$ is, and whether or not
the dimensionless coefficient $c$ is $\ordnt{1}$;
an explicit calculation of such higher order
K\"ahler potential terms from the underlying
string theory apparently remains to be accomplished.

In large radius (LR) stabilization schemes 
such as in Refs.~\cite{Cas90,ILR91},
T-moduli vevs as large as 
$13 \lappeq \vev{T^j} \lappeq 16$
are envisioned.
This greatly affects our estimates
for the Xiggs vevs, since we now have
(for the larger value of $\vev{T^j} \approx 16$)
\beq
\bigvev{ {\p^2 K \over \p \phi^i \p \bar \phi^i} }_{\stxt{LR}} =
\left\{ \begin{array}{lc}
1/32 & \mtxt{untwisted}, \\
1/32^2 & \mtxt{twisted non-oscillator}, \\
1/32^3 & \mtxt{twisted oscillator}.
\end{array} \right.
\eeq

Let $N$ be the number of Xiggses, $q^X$ be the
average Xiggs \ux\ charge magnitude, $K''$ be the average value
for the Xiggs metric
$\vev{ \p^2 K / \p \phi^i \p \bar \phi^i }$
and $\phi$
be the average value for $|\vev{\phi^i}|$,
where ``average'' is used loosely.
Then from (\ref{1.2},\ref{1c}) we see that
$\vev{D_X} =0$ implies
\beq
\phi \sim \left( N q^X K'' \right)^{-1/2} \LamX .
\label{3a}
\eeq
In Chapter~\ref{app}
we will see in an explicit
example that the (properly normalized) 
nonvanishing \ux\
charges vary between $1/\sqrt{84}\approx 0.11$
to $6/\sqrt{84}\approx 0.65$.
We take this as an indication that
$1/10 \lappeq q^X \lappeq 2/3$ is reasonable.
In a typical model
there are $3 \times \ordnt{50}$ chiral
matter multiplets.  The number $N$
which may acquire vevs to cancel the FI term
varies from one flat direction to another.
A reasonable range is $1 \lappeq N \lappeq 50$,
given the enormous number of
\gsmc\ singlets in any of the models.

If a single twisted oscillator field $\phi^i$
of charge $1/10$ dominates the FI cancelation
(i.e., $\phi^i$ is the only Xiggs
or all of the other Xiggses have much
smaller vevs so that effectively $N=1$ in \myref{3a}),
then with the BGW T-moduli stabilization
\beq
\phi \sim \sqrt{10 \times 2^3} \; \LamX \sim \ordnt{1} \; m_P,
\eeq
where we have used \myref{3f}.
Such a large vev is certainly troubling.
If the large radius
value $\vev{T^j} \approx 16$ is assumed,
the result is a hundred times worse:
\beq
\phi \sim \sqrt{10 \times 32^3} 
\; \LamX \sim \ord{2} \; m_P.
\eeq
On the other hand, if we had, say, 50 Xiggs fields $\phi^i$
with more average charges of roughly $1/2$
contributing equally to cancel the FI term,
with the typical field a twisted nonoscillator
field, and the BGW stabilization of T-moduli,
\beq
\phi \sim \sqrt{2 \times 2^2 / 50} \; \LamX
\sim \ord{-2} \; m_P.
\eeq
However, for the large radius case,
\beq
\phi \sim \sqrt{2 \times 32^2  / 50} \; \LamX
\sim \ordnt{1} \; m_P.
\eeq
This examination of
\myref{1.2} indicates that
for the BGW stabilization,
Xiggs vevs are naturally
$\ord{-1 \pm 1} \; m_P$.
At the upper end, the $\s$ model would
seem to be in trouble.  The large
radius case appears to be complete catastrophe,
however we arrange cancelation of the FI
term.  To be fair, the quadratic
terms $K \ni (c/m_P^2) \; f(T) \; |\phi^i|^2 |\phi^j|^2$
mentioned above now need to be included
in the estimation of Xiggs vevs, since
they are not of sub-leading order
in the large Xiggs vev limit.

It should be noted, however, that
the principal motivation for the large
radius assumption is to produce appreciable
string scale threshold corrections to
the running gauge couplings, such as
was studied in \cite{ILR91,BL92}; there, the aim was
to achieve gauge
coupling unification at the conventional\footnote{By ``conventional''
I refer to the unification scenario favored by a majority
of high energy theorists,
the renormalization group evolution of gauge couplings
within the MSSM.}
value of approximately $2 \times 10^{16}$ GeV.
In a $Z_3$ orbifold compactification, these
large T-moduli dependent threshold corrections coming from
heavy string states are absent \cite{DKL91,ANT91}.
Nevertheless, it should be clear from
the above analysis that orbifolds which {\it do} have
the T-moduli dependent string threshold corrections
{\it and} a \ux\ factor are likely to also
suffer from a problem of too large Xiggs
vevs in the large radius limit,
because of the noncanonical K\"ahler
potential.

Moderately large, yet perturbative, vevs
such as $\phi \approx m_P/5$
would require large $n$ in
\myref{ems} to generate
significant hierarchies.  This may be a virtue:
in many cases orbifold selection rules and $G$
symmetries require that leading operators
contributing to a given effective low energy
superpotential term have significantly
higher dimension than might be guessed
from \gsmc\ alone.
For example, in the FIQS model (mentioned
in Section \ref{ceft}) the leading down-type quark masses
come from dimension eleven operators.
(I.e., the effective Yukawa matrix elements
are sums of vevs of seventh degree monomials
of Xiggs fields.)

The sum in \myref{1.2} allows for some terms to
be very small if others are $\ordnt{\LamX}$;
I exploited this possibility in a recent study of 
effective quark Yukawa couplings 
induced by Xiggs vevs of
rather different scales \cite{Gie01a}.
Such hierarchies in Xiggs vevs
remain to be (dynamically) motivated from a detailed
study of an explicit scalar potential
which lifts the D-moduli flat directions
\cite{GG00} mentioned in Section \ref{au1}.
The existence of these flat directions
means that the upper bound estimates
made here for Xiggs vevs are not at all robust.  Xiggs of
opposite \ux\ charge may be ``turned on''
along a particular flat direction (as in
the FIQS model).  In that case their contributions
partially cancel each other; it is technically
possible for the Xiggs vevs to be made arbitrarily
large as a result.  Of course, this would quickly
spoil the nonlinear $\s$ model expansion.

The BSL-I model mentioned in Section \ref{ceft}
belongs to Pattern~1.2
and is equivalent to one\myfoot{Try to determine.} of the
models 6.1-3 listed under that pattern in Table~\ref{tb1}.
(CMM found that the BSL-I model observable
sector embedding was equivalent to CMM 7,
and in Appendix \ref{erp} I show that CMM 7
is equivalent to CMM 6.)
In \cite{GG00} it was noted that the FIQS
model suffers from a problem of light
diagonal T-moduli masses; the conclusions
made there do not depend on the choice
of (hidden $SO(10)$ preserving)
flat direction, and therefore hold
for other vacua of the BSL-I model,
such as those studied by Casas and Mu\~noz \cite{CM88}.
As will shortly be
explained, the light mass
problem is a consequence of
having $G_C=SO(10)$ charged matter fields
only in the untwisted sector.  This observation
extends to {\it all}\ models of Pattern 1.2,
as well as to the models of Pattern 1.1.
Because BGW stabilize the diagonal T-moduli
with nonperturbative effects in the hidden
sector (i.e., gaugino condensation),
they simultaneously derive an effective
(soft) mass term for these fields \cite{BGW}.
If the effective moduli masses are much
larger than the gravitino mass, the 
{\it cosmological moduli problem}
\cite{CFKRR83} can be avoided.  In the
BGW effective theory, one finds for the
diagonal T-moduli
\beq
m_T \approx 2 {|\bGS - b_C| \over |b_C|} m_{\tilde G},
\label{mtm}
\eeq
where $b_C$ is the beta function coefficient
for the condensing group $G_C$, $m_{\tilde G}$
is the gravitino mass and $\bGS$ is the {\it Green-Schwarz
counterterm coefficient,} a quantity whose origin is not
important to the present discussion, but which
is briefly explained in Appendix \ref{cma}.  If $\bGS/b_C \approx 10$,
then $m_T \approx 20 m_{\tilde G}$; it was
argued by BGW, and others \cite{GLM99}, that this
may be heavy enough to resolve the
cosmological moduli problem.

However,
as pointed out in Ref.~\cite{GG00},
if $G_C$ has only trivial irreps in the twisted sector,
$\bGS = b_C$.  The T-moduli are massless
to the order of the approximation made
in \myref{mtm}, and the moduli problem
reappears with a vengence.  To see how $\bGS = b_C$
occurs in Patterns 1.1 and 1.2, it is only necessary
to note a few simple facts.  In Appendix \ref{cma}
I use well-known results to demonstrate that,
for the class of models studied here,
the Green-Schwarz coefficient is given by
\beq
\bGS = \tbeta{a} - 2 \sum_{\rho \in \stxt{tw}} X_a(R^\rho),
\qquad \forall \; G_a \in \GNA,
\label{gsc}
\eeq
where $\tbeta{a}$ is the $\beta$ function
coefficient (given by \myref{bcd} with
$G_C \to G_a$) calculated from the entire
pseudo-massless spectrum of a given
model, and the index $\rho$ runs only over
twisted matter chiral supermultiplet irreps.
In Table~\ref{tbgs} I show $\bGS$ for each of the
twenty patterns; the value is universal to all
models in a given pattern.
From \myref{gsc} it is clear that
$\bGS = \tbeta{a}$ for $G_a$ with
only trivial irreps in the twisted sector.
This occurs for $SO(10)$ in Patterns 1.1 and 1.2,
so that one has $\bGS = \tbeta{10}$;
we also recall $G_C = SO(10)$ in these patterns;
this leads to vanishing T-moduli masses
in \myref{mtm} if $b_C = \tbeta{10}$.
One might hope to get around this
by giving some of the $SO(10)$ charged matter
$\ordnt{\LamX}$ vector mass couplings so that $b_C$, the
effective coefficient which appears in
the theory below the scale $\LamX$,
is different from $\tbeta{10}$.
Pattern 1.1 does not contain $SO(10)$ charged
matter so this is fruitless.  In Pattern 1.2,
the $SO(10)$ matter is in $16$s, which have
as their lowest dimensional invariant
$(16)^4$.  To have effective vector masses for these
states from superpotential terms would require
breaking $SO(10)$.
I leave these issues to further research.
Another way resolve the light moduli problem
in Patterns 1.1 and 1.2 would involve
alternative inflation scenarios.  For example,
light moduli could be diluted via
the thermal inflation 
of Lyth and Stewart \cite{LS96}.
Lastly, I note that the BGW result \myref{mtm}
is obtained in an effective theory which does not
account for a \ux; until it is understood how the
BGW effective theory is modified in the presence of
a \ux\ factor \cite{Gai01}, firm conclusions about the Pattern
1.2 models cannot be drawn.  (Recall that
Pattern 1.1 has no \ux\ factor.)

\begin{table}
\begin{center}
\begin{tabular}{ccccc}
Pattern & $\bGS$ & \hspace{50pt} &
Pattern & $\bGS$ \\ \hline 
1.1 & -24 & & 1.2, 2.1 & -18 \\
2.2-5, 3.1 & -15 & & 3.2-4, 4.1-4, 4.6 & -12 \\
2.6, 4.5, 4.7-8 & -9 & & & \\ 
\hline
\end{tabular}
\caption{Green-Schwarz coefficients. \label{tbgs}}
\end{center}
\end{table}

The values for $\bGS$ are problematic for
more than just the Pattern 1.1 and 1.2 models.
For example, in the $G_C=SU(5)$ Patterns 2.2-5,
the Green-Schwarz coefficient is $\bGS = -15$ and
we can constrain $-15 \leq b_C \leq -6$.
The bound $-15$ comes from a scenario
of pure $SU(5)$; i.e., no matter.  Pattern
2.2 for instance allows for the possibility
that the vector-like
$3(5 + \bar 5)$ matter acquires mass at $\LamX$,
so that effectively there is no $SU(5)$ charged
matter in the running which dynamically generates the
condensation scale.  The bound $-6$ comes from
the ``marginal'' case of very low $\LamC$
discussed in Section \ref{ssb}.  For this
range of $b_C$ we have
from \myref{mtm}
\beq
0 \leq m_T \leq 3 m_{\tilde G}.
\eeq
From the arguments of \cite{BGW,GLM99},
the T-moduli mass appears to be too light
even in the marginal case $b_C= -6$,
which gives the upper bound for $m_T$.
Taking the $b_C = -6$ limit for each of the
values of $\bGS$ (except $\bGS= -24$ which
corresponds to Pattern 1.1 discussed above---where
it seems $m_T \approx 0$ is unavoidable),
we find upper bounds of $m_T^{\stxt{max}}/m_{\tilde G}
\approx 4,3,2,1$ for $\bGS = -18,-15,-12,-9$ respectively.
Thus, the light T-moduli mass problem is a
general feature of the \bsa\ models.\myfoot{However,
I need to incorporate MK's results of DTERM analysis.}

Most of the 20 patterns contain $(3 + \bar 3,1)$
representations under $SU(3)_C \times SU(2)_L$.  
It is necessary to find a
vacuum solution which gives these fields
vector mass couplings at a high enough scale.
The greater the number of such pairs,
the more difficult this is to achieve,
since one must simultaneously
avoid high scale supersymmetry
breaking;  more and more
fields must be identified as Xiggses in order
to give all of the required effective
supersymmetric mass couplings.  As each new
Xiggs is introduced, it is harder to avoid
nonzero F-terms at the scale $\LamX$.
Similarly, large vector masses are generally
required for the many additional
$(1,2)$ and $(1,1)$ representations
present in all of the models.  The electroweak
hypercharges of these representations
depend on how the several {\uone}s
are broken in choosing a D-flat direction.
States with exotic electric charge (i.e.,
leptons with fractional charges and quarks
which may form fractionally charged color
singlet bound states) typically occur.
I will address constraints on the presence
of such matter in Chapter~\ref{app} below.

The distinction between observable and hidden
sectors is blurred by twisted states
in nontrivial representations of both
$G_O$ and $G_H$.  
Gauge interactions
communicating with both sectors
are a well-known effect in orbifold
models.  Communication via {\uone}s
was for example noted in Refs.~\cite{IMNQ88,CKM89,FINQ88a},
while the occurence of states in nontrivial
representations of both observable and
hidden nonabelian factors has been
noted in other orbifold constructions,
for example in a $Z_3 \times Z_3$ model
in Ref.~\cite{BL92}.
Cases 2 through 4 (cf. Table \ref{gcs}) have
at least one hidden $SU(2)$ factor (which
I denote $SU(2)'$),
and $(1,2,2)$ representations
under $SU(3)_C \times SU(2)_L \times SU(2)'$
occur in several of the patterns.  No 
$(\bar 3, 1, 2)$ representations occur, so
it is not possible to use $SU(2)'$
to construct a left-right symmetric model
in any of the 175 models studied here.
(Left-right symmetric models would place
the $u^c$- and $d^c$-type quarks
in $(\bar 3, 1, 2)$ representations.)
All 175 models
contain twisted states 
in nontrivial irreps of $SU(3)_C \times SU(2)_L$
charged under
{\uone}s contained in $G_H$.  

It is an interesting question to what degree
these features might communicate
supersymmetry breaking to the observable
sector.  While such
communications may be suppressed by large masses,
they are likely competitive with gravity mediation,
which is suppressed by inverse powers of
the Planck mass.
A similar scenario has
been considered by
Antoniadis and Benakli \cite{AB92}.
Specifically, they examined hidden
sector matter
with supersymmetric
masses $M$ and a soft mass $\delta M$
splitting the matter scalars from fermions,
gauginos from vector gauge bosons,
with the assumption $\delta M \ll M$;
this ``hidden'' matter was also assumed
to be in nontrivial irreps of $\GSM$.
They found significant
contributions to the soft terms
which break supersymmetry in the
MSSM.  To evaluate the implications
of such gauge mediation of supersymmetry
breaking in the 175 models at hand
requires a significant extension of
their results, given the strong dynamics
of the hidden sector in a gaugino
condensation scenario; much of
the hidden sector matter now consists
of bound states of $G_C$ which
are $\GSM$ neutral (certainly
the case for those condensates which
acquire the supersymmetry
breaking nonvanishing vevs) yet contain
particles in nontrivial $\GSM$ irreps.
I leave these matters to future research.

The generic presence of an anomalous \ux\
has implications for low energy 
supersymmetric models which aim
to be ``string-inspired'' or ``string-derived.''
The effective theory in 
the low energy limit is obtained
by integrating out states which get
large masses due to the \ux\ FI term.
The surviving spectrum
of states will generally contain superpositions
of the original states, mixing
the various sectors.  Thus, assigning
each state in the MSSM to a {\it definite}
sector (i.e., the untwisted sector or one
of the 27 $(n_1,n_3,n_5)$ twisted sectors)
is in many cases inconsistent with 
the mixing which occurs in the presence
of a \ux, as was for instance remarked recently
in Ref.~\cite{Den01}.  Mixings of sectors
was considered for quarks, for example,
in the FIQS model and in the toy model
of Ref.~\cite{Gie01a}.
In addition to modified properties for
the spectrum, integrating out the
massive states will modify the interactions
of the light fields and create
threshold effects for running couplings.
These threshold effects can be
large due to the large number of
extra states, and need to be
considered in any analysis of gauge
coupling unification, for example.

%

\chapter{Hypercharge}
\label{hyc}
\index{hypercharge}
In this chapter I discuss the important issue of
hypercharge normalization.  As has been seen above,
the unbroken rank sixteen gauge group $G$ has
many \uone\ factors for each of the 175 models
in the \bsa\ class.  It is not at all clear which
combination of these generators should be taken
as the electroweak hypercharge generator in a
given model.  Indeed, the possibilities are diverse
and depend on the flat direction chosen in canceling
the FI term discussed in Section \ref{au1}.  Furthermore,
we shall see below that the normalization relative
to other generators of the Standard Model gauge
group is not necessarily the same as the one which
appears in conventional Grand Unified Theories (GUTs).
Because unification of the running gauge couplings
near the string scale is required by the underlying
theory, nonstandard hypercharge normalization leads
to nonstandard scenarios for matter intermediate in
mass between the electroweak scale and the unification
scale.  Alternatively, if one demands standard hypercharge
normalization, many of the 175 models are excluded.
Those models which can accomodate standard hypercharge
normalization do so at the cost of a significant
restriction on the flat directions which can be chosen,
and thus on the effective couplings in the low energy
theory.  Of course, as will be discussed in more
detail in Chapter \ref{app}, a problem exists for
unification of couplings at the string scale
in the standard MSSM based scheme.  Regardless,
the resolution of this discrepancy depends on
the hypercharge normalization and this therefore
remains a crucial issue in the phenomenology of the
models studied here.

In Section \ref{guts} I review hypercharge normalization
in the case of GUTs.  This prepares the reader for the
discussion of Section \ref{strn}, where I address hypercharge
normalization in the context of string models.  Both of
these sections summarize and explain well-known facts
established by the work of others.  In Section \ref{su5s}
I define what I shall call $SU(5)$ {\it embeddings} of
hypercharge in the models under consideration.  These
embeddings lead naturally to the conventional
hypercharge normalization.  I describe my result
that none of these embeddings allows for the spectrum
of particles contained in the MSSM to be fit into
the spectrum of pseudomassless states in any of
the 175 models; I previously presented this analysis
in Ref.~\cite{Gie01c}.  More general, {\it extended
embeddings} are considered in Section \ref{ehye}.
I have conducted a detailed study of these embeddings
for the 175 models and have found minimum values for
the hypercharge normalization in each case.  My
results are presented and implications are drawn.
This final section of the chapter is also based on
work which I previously discussed in Ref.~\cite{Gie01c}.

\section{Normalization in GUTs}
\label{guts}
\glossary{grand unified theories}
An important feature of GUTs
is that the \uone\ generator corresponding to
electroweak hypercharge does not have arbitrary normalization.
This is because the hypercharge generator is embedded into
the Lie algebra of the GUT group.
That is, $\GGUT \supset \GSME$.
The {\it unified normalization}
is most clear when one identifies a Cartesian basis for the
GUT group generators $T^a$ for a given representation $R$:
\beq
\tra{R} T^a T^b = X(R) \; \delta^{ab}.
\eeq
The normalization prevalent in phenomenology has
$X(F)=1/2$ for an $SU(N)$
fundamental representation $F$.  Because of the GUT symmetry,
the interaction strength of a gauge particle with matter
is given by
\beq
g_U(\mu) \; T^a, \quad \forall \; a,
\label{4a}
\eeq
where $g_U(\mu)$ is the running coupling for the GUT
gauge group at the scale $\mu \geq \LamU$,
with $\LamU$ the unification scale.  One of the
$T^a$, say $T^1$, is then identified with the
electroweak hypercharge generator.  However,
to obtain the usual eigenvalues for MSSM particles
(e.g., $Y=1$ for $e^c$) we
generally must rescale the generator:
\beq
Y \equiv \sqrt{k_Y} \; T^1 .
\label{3b}
\eeq
The reason for writing the rescaling
constant in this way will become clear
below.  Because of (\ref{4a},\ref{3b}), 
the hypercharge coupling $g_Y(\mu)$ will
be related to $g_U(\mu)$ at the boundary scale $\LamU$.  More
precisely,
\beq
g_U(\LamU) \; T^1 = g_Y(\LamU) \; Y 
=  \sqrt{k_Y} \; g_Y(\LamU) \; T^1,
\eeq
since the interaction strength should not depend on normalization
conventions for the generators.
I maintain the GUT normalization for the generators $T^a$
which correspond to the unbroken $SU(2)$ and $SU(3)$ groups,
so that there are no rescalings analogous to \myref{3b} for
these two groups; their running couplings are denoted by
$g_2(\mu)$ and $g_3(\mu)$ respectively.  Because of \myref{4a},
they too must be matched to the boundary value $g_U(\LamU)$
when $\mu=\LamU$; thus, we obtain the well-known
GUT boundary conditions
\beq
g_3(\LamU) = g_2(\LamU) = \sqrt{k_Y} \; g_Y(\LamU) 
= g_U(\LamU).
\label{4c}
\eeq

For example, consider an SU(5) GUT \cite{GG74}.
The $SU(5)$ embedding of hypercharge,
which I write as $T^1$, can
be determined from the requirement
that $\tr (T^1)^2 = 1/2$ for a
fundamental or antifundamental irrep.  For example,
\beq
T^1 = {1 \over \sqrt{60}} \diag ( -3,
-3, 2, 2, 2), \qquad \mtxt{for} \qquad \bar 5 =
{L \choose d^c} .
\label{gyc}
\eeq
Here, $L$ is a $(1,2)$ lepton,
and $d^c$ is a $(\bar 3,1)$ down-type quark,
where I denote \nasm\ quantum numbers.
On the other hand, the electroweak normalization
has by convention
\beq
Y = {1 \over 6} \diag ( -3,
-3, 2, 2, 2)
\eeq
for the same set of states.
Since $Y = \sqrt{5/3} \; T^1$,
we see from \myref{3b} that
\beq
k_Y = 5 / 3 .
\label{s5y}
\eeq
It is this value which, when assumed in \myref{4c},
yields the amazingly successful
gauge coupling unification in the MSSM,
detailed for example in Refs.~\cite{MSSMu,LP93}.

\section{Normalization in String Theory}
\label{strn}
As in GUTs, the normalization of \uone\
generators in string-derived field theories
requires care.  Above,
I have alluded to the fact that
gauge coupling unification at the 
heterotic string scale $\LamH$ is
a prediction of the underlying theory \cite{Gin87}.
Just as
in GUTs, unification of the hypercharge coupling with the
couplings of other factors of the gauge symmetry
group $G$ corresponds to a particular normalization.
However, the unified normalization of hypercharge
is often different than the one which appears in $SU(5)$
or $SO(10)$ GUTs; in fact it is often difficult
or impossible
to obtain \myref{s5y}.  Examples
of this hypercharge normalization ``difficulty''
will be examined below.
I will show how the unified normalization can be identified
from simple arguments.
In the process
I will make it clear why, in the class of
orbifold models considered here, nonstandard hypercharge
normalization is generic and fractionally charged exotic matter
is abundant.

It was noted in Chapter \ref{mss} that the
basis \myref{2g} is larger by a factor
of two than the phenomenological normalization.
Thus, $\tr (T^a)^2 = 2$ for
an $SU(N)$ fundamental representation.
For instance, consider an untwisted $SU(2)_L$ doublet
with respect to $\alpha_{1,1}$ in \myref{2m} above,
for CMM 2 observable sector embeddings.
(The embedding label here corresponds to Table \ref{tabcmm}.)
The lowest and highest weight states are respectively
\beqa
K_1 & = & (0,1,0,0,0,1,0,0;0,\ldots,0), \nnn
K_2 & = & K_1 + \alpha_{1,1} = (1,0,0,0,0,1,0,0;0,\ldots,0).
\label{aad}
\eeqa
Using Eqs.~(\ref{2m},\ref{wtd}),
the corresponding weights are $\mp 1$; this gives
$\tr (H_1^1)^2 = 2$,
where $H_1^1 = T^3$,
the isospin operator of
$SU(2)_L$.  To get to the phenomenological
normalization, we should rescale generators
by $1/2$.  Thus, instead of \myref{2g},
I define properly normalized Cartan generators
$\hat H_a^i$ according to
\beq
\hat H_a^i = \sum_{I=1}^{16} \hat h_a^{iI} H^I
\equiv \sum_{I=1}^{16} \half \alpha_{ai}^I H^I .
\label{4d}
\eeq
In this case, the sixteen-vectors $\hat h_a^i$
satisfy 
\beq
(\hat h_a^i)^2 = 1/2.
\label{4i}
\eeq
It is hardly surprising that the 
properly normalized generator $\hat Q_a$
of $U(1)_a$ must also satisfy 
$(\hat q_a)^2 = 1/2$,
where $\hat q_a$ is the sixteen-vector
appearing in \myref{Qdf},
but now with a special normalization.  After all,
the generator of $U(1)_a$ just corresponds
to a different linear combination of the
\eetee\ Cartan generators $H^I$,
and taking a linear combination of
the same norm is the logical choice.
If, on the other hand, we work with
a generator $Q_a = \sqrt{k_a} \hat Q_a$,
then it follows that $q_a^2 = k_a/2$.
This is one way
of motivating the ``affine level''
of a \uone\ factor:
\beq
k_a = 2 \sum_{I=1}^{16} (q_a^I)^2 .
\label{kad}
\eeq
(This relation also follows from a
consideration of the double-pole
Schwinger term which occurs in the
operator product of \uone\ currents 
in the underlying conformal field
theory \cite{FIQS90,DFMR96,KN97,Die97},
details which I have purposely
avoided here.)
The unified normalization,
where nonabelian Cartan generators $\hat H_b^i$
and \uone\ generators $\hat Q_a$ have in
common $(\hat h_b^i)^2 = \hat q_a^2 = 1/2$,
corresponds to $k_a = 1$.

\section[$SU(5)$ Embeddings]{${\bf SU(5)}$ Embeddings}
\label{su5s}
Note that the generator
\beq
Y_1 = \sum_{I=1}^{16} y_1^I H^I, \qquad
y_1 = {1 \over 6} (-3,-3,2,2,2,0,0,0;0,\ldots,0),
\label{aac}
\eeq
satisfies $k_{Y_1} = 5/3$, is orthogonal to the
$SU(3)_C \times SU(2)_L$ roots in \myref{nz32}, and
has nonzero entries only in the subspace where
the $SU(3)_C \times SU(2)_L$ roots have nonzero
entries.  Furthermore, it gives 
$Y_1 = y_1 \cdot K_{1,2} = -1/2$ to
the doublet in \myref{aad}, corresponding to the
lepton doublets $L$ or the $H_d$ Higgs
doublet of the MSSM.
The \bsa\ models with observable sector
embedding CMM 2
also include $(\bar 3, 1)$ states
in the untwisted sector with weights
\beq
K_{3,4,5} = (0,0,\underline{1,0,0},-1,0,0;0,\ldots,0).
\label{aaf}
\eeq
These have $Y_1=y_1 \cdot K_{3,4,5}= 1/3$,
corresponding to the $d^c$ states.  Finally,
the untwisted sector contains $(3,2)$ states
with weights
\beq
K_{6,\ldots,11} =
(\underline{-1,0},\underline{-1,0,0},0,0,0;0,\ldots,0)
\label{4z}
\eeq
which have $Y_1=y_1 \cdot K_{6,\ldots,11}=1/6$, corresponding to the
quark doublets $Q$.
Thus, the untwisted sector
contains a $\bar 5$ and an incomplete $10$ under
the ``would-be'' $SU(5)$ into which we
wish to embed $SU(3)_C \times SU(2)_L \times U(1)_{Y_1}$,
taking \myref{aac} to be
the hypercharge generator.
The fact that the
$e^c$ and $u^c$ representations needed to
fill out the $10$ irrep are not present in
the untwisted sector is a troubling feature
which is generic to the 175 models studied
here.

In Table~\ref{unpat} I display the
irreps present in the untwisted sector for
each of the twenty patterns.  In no case do
we have the required irreps to build
a $10$ of $SU(5)$.  In those cases where
one finds $(3,2) + (\bar 3,1)$, the states
which are singlets of the observable
$SU(3) \times SU(2)$ are in nontrivial
irreps of the hidden sector group.  One
could imagine breaking the hidden sector
group and using a singlet of the surviving
group to give the necessary $(1,1)$ irrep
to fill out a $10$.  For instance, in
Pattern 2.2, the $(1,1,1,2)$ irrep,
a $2$ of the hidden $SU(2)'$,
would give two singlets if we break
$SU(2)'$ with nonvanishing vevs
for a pair of twisted
sector $(1,1,1,2)$ irreps along a
D-flat direction.  (A pair is required
to have vanishing D-terms for $SU(2)'$.)
We would thereby obtain three
generations of two $(1,1,1)$ irreps
with respect to the surviving
nonabelian gauge symmetry
$SU(3) \times SU(2) \times SU(5)$,
where the $SU(5)$ shown here is the
hidden condensing gauge group.  However,
the untwisted $(1,1,1,2)$ irrep which
gives these states has an \eetee\ weight vector
$K$ of the form $K=(0;\beta), \; \beta \in \elat$,
since it is an untwisted state
charged under the hidden sector gauge
group.  Then it has vanishing charge
with respect to the generator $Y_1$
according to \myref{aac}, rather than
the required $Y_1=1$.
We could overcome this by modifying $y_1$
to have nonzero entries in the hidden
sector portion, represented by $0,\ldots,0$
in \myref{aac}.  However, according to
\myref{kad}, this would increase $k_Y$
over the value of $5/3$ which $y_1$
gives.  Moreover,
it can be seen that one never has enough
untwisted $(\bar 3,1)$ irreps to give three
generations of both $u^c$- and $d^c$-type
quarks, and that untwisted $(1,2)$ irreps always occur
when an untwisted $(\bar 3,1)$ is present.  Thus,
even if we break the hidden gauge
group, use a singlet to complete the
$10$, are willing to consider $k_Y > 5/3$,
and find the $(\bar 3,1)$ has $Y_1 = -2/3$
so that it fits into a $10$,
the $(1,2)$ would
stand for an incomplete $\bar 5$.
It is inevitable that we use states from
the twisted sectors to fill out the MSSM;
as I have already alluded to in Section~\ref{rec},
twisted states have unusual \uone\ charges 
(partly) because
the \eetee\ weights are shifted by the
embedding vectors $E(n_1,n_3,n_5)$.

Let us now examine the relationship
of \myref{aac} to $SU(5)$.
To begin with we
relabel the $SU(3) \times SU(2)$ simple roots in
(\ref{2m},\ref{2.39}) as
\beq
\alpha_1 \equiv \alpha_{1,1}, \qquad
\alpha_2 \equiv \alpha_{2,1}, \qquad
\alpha_3 \equiv \alpha_{2,2}.
\label{4e}
\eeq
These may be supplemented by a fourth \eetee\ root
\beq
\alpha_4 = (0, 1, -1, 0,0,0,0,0;0,\ldots,0)
\label{aaa}
\eeq
to give the correct Cartan matrix for $SU(5)$,
according to \myref{2c}.  In
this way we embed $SU(3) \times SU(2)$ into
a would-be $SU(5)$ subgroup of the observable $E_8$
factor of \eetee.
A (properly normalized)
basis $\hat H_1, \dots, \hat H_4$ for the Cartan
subalgebra of the would-be $SU(5)$
is given in terms of the \eetee\ Cartan generators
$H^I$ according to the methods described
in Section~\ref{rec}, supplemented by the
normalization considerations which led to
\myref{4d}.  That is, we take linear combinations
described by sixteen-vectors $\hat h^i = \alpha_i/2$,
so that
\beq
\hat H^i = \sum_{I=1}^{16} \half \alpha_i^I H^I.
\label{4g}
\eeq
However, when we decompose
$SU(5) \supset SU(3) \times SU(2) \times U(1)$
we want to take the \uone\ generator to be
orthogonal to the generators $\hat H^{1,2,3}$
associated with the
simple roots \myref{4e},
unlike $\hat H^4$.
(This is the analogue of \myref{2s}.)
We thus
make a change of basis, keeping $\hat h^i = \alpha_i/2$
for $i=1,2,3$ while taking the fourth vector
to be an orthogonal linear combination of the
four simple roots:
\beq
y = \sum_{i=1}^4 r_i \alpha_i, \qquad
\mtxt{where}
\qquad y \cdot \alpha_i = 0, \quad i=1,2,3.
\label{aab}
\eeq
The orthogonality constraint in \myref{aab}
and the fact that $\alpha_i^I=0$ for 
$I=6,\ldots,16$ requires
\beq
y = (a,a,b,b,b,0,0,0;0,\ldots,0),
\label{4m}
\eeq
while $\sum_I \alpha_i^I =0$ requires
$2a= -3b$.  From here it is easy to check
that with normalization $k_Y=5/3$, we have
$y=y_1$, Eq.~\myref{aac}.
Thus we see that $y_1$
corresponds to a natural completion of the $SU(3) \times SU(2)$
roots \myref{4e} into a would-be $SU(5)$ subgroup
of the observable $E_8$.  I note that
\myref{4m} has the form of a {\it minimal
embedding} of hypercharge, in the spirit
of the analysis carried out in \cite{DFMR96}.

\begin{figure}
\begin{center}
\unitlength=1mm
\begin{picture}(100,30)
\put(10,20){\makebox(0,0){$\alpha_1$}}
\put(20,20){\makebox(0,0){$\alpha_4$}}
\put(30,20){\makebox(0,0){$\alpha_2$}}
\put(40,20){\makebox(0,0){$\alpha_3$}}
\put(10,25){\circle*{1}}
\put(20,25){\circle*{1}}
\put(30,25){\circle*{1}}
\put(40,25){\circle*{1}}
\put(10,25){\line(1,0){10}}
\put(20,25){\line(1,0){10}}
\put(30,25){\line(1,0){10}}
\put(25,10){\makebox(0,0){Case 1}}
\put(60,20){\makebox(0,0){$\alpha_1$}}
\put(70,20){\makebox(0,0){$\alpha_4$}}
\put(80,20){\makebox(0,0){$\alpha_3$}}
\put(90,20){\makebox(0,0){$\alpha_2$}}
\put(60,25){\circle*{1}}
\put(70,25){\circle*{1}}
\put(80,25){\circle*{1}}
\put(90,25){\circle*{1}}
\put(60,25){\line(1,0){10}}
\put(70,25){\line(1,0){10}}
\put(80,25){\line(1,0){10}}
\put(75,10){\makebox(0,0){Case 2}}
\end{picture}
\end{center}
\caption{Would-be $SU(5)$ Dynkin diagrams.}
\label{wb5}
\end{figure}
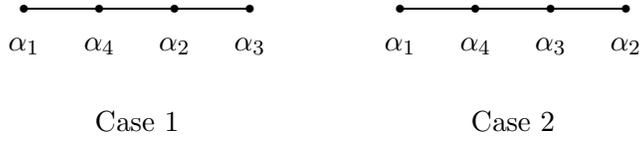

Now I come to the origin of the subscript in \myref{aac}.
It turns out that \myref{aaa} is not the unique $E_8$
root which may be appended to  $\alpha_1, \alpha_2, \alpha_3$
to obtain the simple roots of an $SU(5)$ subalgebra
of the observable $E_8$.  The two ways that a
supposed $\alpha_4$ could be related to the
roots $\alpha_1, \alpha_2, \alpha_3$ are
shown in the Dynkin diagrams of Figure \ref{wb5}.
A line connecting $\alpha_i$ to $\alpha_j$
indicates $\alpha_i \cdot \alpha_j = -1$;
if not connected by a line, 
$\alpha_i \cdot \alpha_j = 0$.

I define $y$ as in \myref{aab}, except that
now I allow $\alpha_4$ to be any observable
$E_8$ root (i.e., $\alpha_4 = (\beta;0)$,
$\beta \in \elat$, $\beta^2=2$)
consistent with Figure \ref{wb5}.
I simultaneously demand $2y^2=5/3$, corresponding
to $k_Y=5/3$ from \myref{kad}.  This gives
solutions:
\beqa
y & = & \pm {1 \over 6} (3 \alpha_1 + 4 \alpha_2
+ 2 \alpha_3 + 6 \alpha_4) \qquad \mtxt{Case 1}, 
\label{4y} \\
y & = & \pm {1 \over 6} (3 \alpha_1 + 2 \alpha_2
+ 4 \alpha_3 + 6 \alpha_4) \qquad \mtxt{\rm Case 2}.
\label{4x}
\eeqa

In each of the 175 models I consider here, 
the only $(3,2)$ representations
under the observable $SU(3) \times SU(2)$ are contained
in the untwisted sector, and they all take the form
\myref{4z}.  To accomodate the MSSM
we require that this representation have 
$Y = y \cdot K_{6,\ldots,11} =1/6$.
It suffices to demand this for any of the
six $K_i$ since by \myref{aab}
\beq
(K_i + \alpha_j)\cdot y = K_i \cdot y,
\qquad \forall \quad i=6,\ldots,11, \quad j=1,2,3.
\eeq
(Recall from the discussion in Chapter \ref{mss}
that the weights $K_{6,\ldots,11}$ are related
to each other by the addition
of $SU(3) \times SU(2)$ roots.)
I choose to employ 
\beq
K_6=(-1,0,-1,0,0,0,0,0;0,\ldots,0).
\eeq
It is easy to check that for Eq.~\myref{4y},
$K_6 \cdot y = 1/6$ imposes  
\beq
K_6 \cdot \alpha_4 = \left\{
\begin{array}{cc}
4/3 & (+), \\
1 & (-).
\end{array}
\right.
\eeq
Since $\alpha_4$ can only have integral or
half-integral entries, we must take the
negative sign in \myref{4y} 
and $K_6 \cdot \alpha_4=1$.
For Eq.~\myref{4x},
$K_6 \cdot y = 1/6$ imposes 
\beq
K_6 \cdot \alpha_4 = \left\{
\begin{array}{cc}
1 & (+), \\
2/3 & (-).
\end{array}
\right.
\eeq
Now we must take the
positive sign in \myref{4x}.
To summarize, imposing that the quark doublet have
$Y=1/6$ constrains $\alpha_4$ to satisfy
the additional constraint
\beq
K_6 \cdot \alpha_4 = 1
\eeq
and determines the signs in (\ref{4y},\ref{4x}):
\beqa
y & = & - {1 \over 6} (3 \alpha_1 + 4 \alpha_2
+ 2 \alpha_3 + 6 \alpha_4) \qquad \mtxt{Case 1}, 
\\
y & = & {1 \over 6} (3 \alpha_1 + 2 \alpha_2
+ 4 \alpha_3 + 6 \alpha_4) \qquad \mtxt{Case 2}.
\eeqa

As noted briefly in Section~\ref{eer}, the ordering
by which nonzero \eetee\ roots are determined to
be positive is arbitrary.  A particular 
{\it lexicographic ordering} for the first
$E_8$ can be specified by an eight-tuple
$(n_1,n_2,\ldots,n_8)$.  Here, $n_1$ tells us
which entry should be checked first, $n_2$
tells us which entry should be checked second,
etc.  For example, $(8,7,6,5,4,3,2,1)$ would
instruct us to determine positivity by
reading the entries of a given $E_8$ root
vector backwards, right to left.
It is easy to see that several
{\it lexicographic orderings} are consistent with
$\alpha_1, \alpha_2, \alpha_3$ being regarded
as positive; in fact, the number of such
orderings is 3360.  
Our final restriction on $\alpha_4$ is
that for one of these 3360
orderings, $\alpha_4$ is also positive.
This is necessary if it is to be regarded
as a simple root of a would-be $SU(5)$.

When all of the conditions described above are taken
into account, the complete list
of observable $E_8$ roots $\alpha_4$ and the corresponding
vectors $y$ which result can be
determined by straightforward analysis
of the 240 nonzero $E_8$ roots.  The
results are given in Table \ref{su5e}.
I label the four additional
$y$ solutions according to:
\beqa
y_2 & = & {1 \over 6} (0,0,-1,-1,-1,-3,-3,-3;0,\ldots,0), \nnn
y_{3,4,5} & = & 
{1 \over 6} (0,0,-1,-1,-1,\underline{-3,3,3};0,\ldots,0).
\eeqa
In what follows, I refer to $Y_i$, $i=1,\ldots,5$, as the
five possible $SU(5)$ {\it embeddings} of the hypercharge
in the \bsa\ models.

\begin{table}[ht!]
\begin{center}
$$
\begin{array}{cc}
\alpha_4 & y \\ \hline
(0,1,-1,0,0,0,0,0;0,\ldots,0) & 
{1 \over 6} (-3,-3,2,2,2,0,0,0;0,\ldots,0) \\
(-\half,\half,-\half,\half,\half,\half,\half,\half;0,\ldots,0) &
{1 \over 6} (0,0,-1,-1,-1,-3,-3,-3;0,\ldots,0) \\
(-\half,\half,-\half,\half,\half,
	\underline{\half,-\half,-\half};0,\ldots,0) &
{1 \over 6} (0,0,-1,-1,-1,\underline{-3,3,3};0,\ldots,0) \\
(-1,0,0,0,1,0,0,0;0,\ldots,0) &
{1 \over 6} (-3,-3,2,2,2,0,0,0;0,\ldots,0) \\
(-\half,\half,-\half,-\half,\half,
	\underline{-\half,\half,\half};0,\ldots,0) &
{1 \over 6} (0,0,-1,-1,-1,\underline{-3,3,3};0,\ldots,0) 
\vspace{5pt} \\
\hline
\end{array}
$$
\end{center}
\caption{Observable $E_8$ roots which embed
$SU(3)_C \times SU(2)_L$ into a would-be
$SU(5)$. \label{su5e}}
\end{table}

A model must also have $Y$ non-anomalous
for it to survive unmixed with other \uone\ factors below
$\LamX$.  Many models have a trace anomaly for one or
more of the five $Y_i$.  This would not
occur if complete $SU(5)$ irreps
were present.
We have already seen that the untwisted
sector does not contain complete would-be
$SU(5)$ irreps for any of the 175 models
(cf. Table~\ref{unpat}).
Of course, whether or not $Y_1$ is
anomalous in those models also depends on the matter content
of the twisted sectors.  This in turn depends on the
hidden sector embedding through \myref{tps};
consequently, each of the 175 models must be
studied separately.

I have determined the charges of all matter
irreps with respect to $Y_i \; (i=1,\ldots,5)$
for all of models.
In those models where a given $Y_i$
is not anomalous, the MSSM particle spectrum is never
accomodated.  That is not to say that we do not have
enough $(3,2)$s, $(\bar 3,1)$s, $(1,2)$s and $(1,1)$s;
in fact, we typically have too many of the latter
three types, as can be seen from 
Tables~\ref{pt1}-\ref{pt4}.
The difficulty comes in their
hypercharge assignments when we take $Y$ to be one of
the five $Y_i$.  Although there are always a few
irreps with the right hypercharges, there are never enough.

As suggested by the discussion in Section~\ref{rec},
the origin of bizzare
hypercharges with respect to the $SU(5)$ embeddings
$Y_i$ is due to the fact that twisted states
generically have \eetee\ weights on a shifted
lattice, as is apparent in \myref{tps}.
To further understand these matters, 
I now discuss the decomposition
of the two lowest lying $E_8$ representations,
of dimension 248 and 3875 respectively.
The decomposition of these irreps under
$E_8 \supset SU(5)$ is tabulated,
for instance, in the review by Slansky~\cite{Sla81}.
I identify this $SU(5)$ as the
subgroup of $E_8$ in which
irreps of $\GSM$ are embedded.  The decompositions
are (numbers in parentheses denote $SU(5)$ irreps)
\beqa
248 & = & 24(1) + (24) + 10(5+\bbar{5}) + 5(10 + \bbar{10}), \nnn
3875 & = & 100(1) + 65(5 + \bbar{5}) + 50(10 + \bbar{10})
+ 5(15 + \bbar{15}) + 25(24) + 5(40+\bbar{40}) \nnn
& & \quad {} + 10(45 + \bbar{45}) + (75).
\label{edc}
\eeqa
Although these are real representations, a chiral
four-dimensional theory is obtained by compactification
on a quotient manifold (i.e., the $Z_3$ orbifold),
a mechanism pointed out some time ago \cite{CGS}.
Also from Slansky, I take the decomposition of
the $SU(5)$ irreps shown in \myref{edc} with
respect to $SU(5) \supset SU(3) \times SU(2) \times U(1)$,
with the standard electroweak normalization for the
\uone\ charge given in the last entry:
\beqa
1 & = & (1,1,0) \nnn
5 & = & (1,2,1/2) + (3,1,-1/3) \nnn
10 & = & (1,1,1) + (\bar 3,1,-2/3) + (3,2,1/6) \nnn
15 & = & (1,3,1) + (3,2,1/6) + (6,1,-2/3) \nnn
24 & = & (1,1,0) + (1,3,0) + (3,2,-5/6)
+ (\bar 3, 2, 5/6) + (8,1,0) \nnn
40 & = & (1,2,-3/2) + (3,2,1/6) + (\bar 3,1,-2/3)
+ (\bar 3,3,-2/3) + (8,1,1) + (\bar 6,2,1/6) \nnn
45 & = & (1,2,1/2) + (3,1,-1/3) + (3,3,-1/3)
+ (\bar 3,1,4/3) + (\bar 3,2,-7/6) \nnn & & \quad {} + (\bar 6,1,-1/3)
+ (8,2,1/2) \nnn
75 & = & (1,1,0) + (3,1,5/3) + (\bar 3,1,-5/3)
+ (3,2,-5/6) + (\bar 3,2,5/6)  \nnn & & \quad {} + (6,2,5/6)
+ (\bar 6,2,-5/6) + (8,1,0) + (8,3,0)
\label{4k}
\eeqa
While the higher dimensional $SU(5)$ irreps certainly
contain states with unusual hypercharge
(e.g., the $(1,2)$ irrep in the $40$ of $SU(5)$ with
$Y=-3/2$),
given the number of $5$, $\bar 5$ and $10$ representations
present in \myref{edc} it is perhaps
surprising that we do not obtain the
$SU(3) \times SU(2) \times U(1)$ irreps to fill out
the MSSM for any of the 175 models.

Beside the projections (\ref{2i},\ref{ums})
in the untwisted sector---which lead to incomplete
would-be $SU(5)$ irreps as discussed in
detail above---the problem, 
of course, is that in the twisted sectors
the \eetee\ weights do not correspond to the decomposition
of $E_8$ representations described by
(\ref{edc},\ref{4k}).  The weights
are of the form $\tK = K + E(n_1,n_3,n_5)$;
whereas $K \in \eelat$,
for any twisted sector with
solutions to \myref{tps}
the embedding vector is a 
strict fraction of a lattice vector:
\beq
3 E(n_1,n_3,n_5) \in \eelat,
\qquad
E(n_1,n_3,n_5) \not\in \eelat.
\eeq
Specializing \myref{2t},
the hypercharge for
any of the $Y_i$ is given by
\beq
Y_i(\tK;n_1,n_3,n_5)  =  
y_i \cdot K + \delta_{y_i}(n_1,n_3,n_5) , \qquad
\delta_{y_i}(n_1,n_3,n_5) =  y_i \cdot E(n_1,n_3,n_5).
\eeq
For a massless state, the value of $y_i \cdot K$
will take values corresponding 
to the decompositions \myref{4k};
$y_i \cdot K$ values from the 3875 of $E_8$
occur because $K^2 > 2$ is possible,
as discussed in Section~\ref{rec}.
The second term on the right-hand side
is the Wen-Witten defect,
briefly discussed above in Section~\ref{rec}.
Since each $y_i$ is nonzero
only in the first eight entries, the Wen-Witten defect
only depends on the observable sector embeddings
enumerated by CMM.  It is easy to 
check that
for each of the $y_i$ the defect in each twisted
sector is a multiple of $1/3$.  This is
consistent with general arguments 
\cite{AADF88,Ant90} which show that
fractionally charged color
singlet (bound) states in $Z_N$ orbifolds
have electric charges which are quantized
in units of $1/N$.

\section{Extended Embeddings}
\label{ehye}
Having failed to accomodate the MSSM with
any of the five $Y_i$, I envision
the most general hypercharge consistent
with leaving at least a hidden $SU(3)'$
unbroken to serve as the condensing
group $G_C$.  (Such a $Y$ is of the
{\it extended} hypercharge embedding
variety, studied for example in Ref.~\cite{CHL96}.)
That is, I include the
possibility that Cartan generators of
the nonabelian hidden sector group
mix into $Y$ under a Higgs effect,
perhaps induced by the FI term.  
(A well-known example of the mixing of 
a nonabelian Cartan generator into
a surviving \uone\ is the
electroweak symmetry breaking 
$SU(2)_L \times U(1)_Y
\to U(1)_E$.)  Thus, I assume
a hypercharge generator of the form
\beq
6 Y = \sum_{a \not= X} c_a Q_a + \sum_{a,i} c_a^i H_a^i.
\label{gym}
\eeq
A factor of six has
been included for later convenience.
The Cartan generators written here
are {\it not} those of \myref{2g}
or \myref{4g}.  Rather, I choose
a basis where the $H_a^i$ are mutually
orthogonal (i.e., $\tra{R} H_a^i H_a^j
= 0$ for $i \not= j$, any irrep $R$ of
$G_a$).

Nontrivial irreps of the hidden sector
gauge group $G_H$ may decompose
under the partial breaking of $G_H$ implied
by \myref{gym} to give some of
the $(1,2)$ and $(1,1)$ irreps of the MSSM.
For instance, if the pattern of
gauge symmetry breaking
in an irrep Pattern 2.5 model is
\beq
SU(3)_C \times SU(2)_L \times SU(5) \times SU(2)' \times U(1)^8
\to SU(3)_C \times SU(2)_L \times SU(3)' \times U(1)_Y,
\eeq
then we have the following decompositions
of nontrivial irreps of $G_H$ onto the
surviving gauge symmetry group:
\beqa
(1,1,5,1) & \to & (1,1,3) + 2(1,1,1), \nnn
(1,1,10,1) & \to & 2(1,1,3) + (1,1,\bar 3) + (1,1,1), \nnn
(1,2,1,2) & \to & 2(1,2,1) .
\label{4f}
\eeqa
Thus, we get many candidates for $e^c$
as well as candidates for $L,H_d$ or $H_u$.
The Cartan generator of $SU(2)'$ is allowed
to mix into $Y$; this is also true of the
two Cartan generators of $SU(5)$ which commute
with all of the generators of the surviving
$G_C=SU(3)'$.  The weights of the 
$(1,2,1)$ and $(1,1,1)$ states in
\myref{4f} with respect to these generators
then contribute to the hypercharges of
these states.

Corresponding to \myref{gym} is an
assumption for the sixteen-vector $y$
which describes the linear combination
of \eetee\ Cartan generators $H^I$ which
give $Y$:
\beq
6 y = \sum_{a \not= X} c_a q_a + \sum_{a,i} c_a^i h_a^i.
\label{4h}
\eeq
To calculate $k_Y$, we use Eq.~\myref{kad}
and the orthogonality of the
sixteen-vectors appearing in \myref{4h}:
\beq
k_Y = {1 \over 36} \left(
\sum_{a \not= X} c_a^2 k_a 
+ \sum_{a,i} 2 (c_a^i h_a^i)^2 \right).
\eeq
I define, as above, $\hat H_a^i$ to be the
generator $H_a^i$ rescaled to the unified
normalization (e.g., $\tr (\hat H_a^i)^2 = 1/2$
for an $SU(N)$ fundamental irrep).
We express the rescaling by $H_a^i = \sqrt{k_a^i}
\hat H_a^i$.  Then in terms of the sixteen-vectors
associated with these generators, using
Eq.~\myref{4i},
\beq
2 (h_a^i)^2  = 2 k_a^i (\hat h_a^i)^2 = k_a^i.
\eeq
Thus, the hypercharge normalization
may be expressed as
\beq
k_Y = {1 \over 36} \left(
\sum_{a \not= X} c_a^2 k_a 
+ \sum_{a,i} (c_a^i)^2 k_a^i \right) .
\label{4j}
\eeq

Eq.~\myref{4j} gives $k_Y$ as a quadratic form
of the real coefficients $c_a$ and $c_a^i$,
a function which is easy to minimize subject
to the linear constraints imposed by demanding
that the seven types of chiral supermultiplets
in the MSSM ($Q,u^c,d^c,L,H_d,H_u,e^c$)
be accomodated, including hypercharges.
(For instance, I used standard
routines available on the math package Maple.)
I have performed an automated analysis
to determine the minimum $\dkY \equiv k_Y- 5/3$ 
values allowed by each model,
for each possible assignment of the
MSSM to the full pseudo-massless
spectrum.  My results are shown in Table~\ref{mkytab}.

\begin{table}[ht!]
\begin{center}
\begin{tabular}{cccccccc}
Pattern & $\dkYm$ & Pattern & $\dkYm$
   & Pattern & $\dkYm$ & Pattern & $\dkYm$ \\
\hline
1.1 & 0       & 2.4 & 8/29    & 3.3 & -4/61  & 4.4 & 16/61 \\
1.2 & 1/5     & 2.5 & 11/73   & 3.4 & 16/59  & 4.5 & -1/31 \\
2.1 & 4/29    & 2.6 & 4/11    & 4.1 & -8/113  & 4.6 & 11/73 \\
2.2 & -8/167  & 3.1 & 1/7     & 4.2 & -8/113  & 4.7 & -1/31 \\
2.3 & 0       & 3.2 & -8/119  & 4.3 & 8/81  & 4.8 & 14/5 \\
\hline
\end{tabular}
\caption{Minimum values of $\dkY = k_Y - 5/3$. \label{mkytab}}
\end{center}
\end{table}

It can be seen from the table
that $k_Y=5/3$ is possible
in some patterns.  I remark, however,
that this value has lost most of
its motivation in the present context.
Whereas in a GUT the normalization
$k_Y=5/3$ came out
naturally, we now obtain this value
by artifice, choosing a ``just so''
linear combination of observable
and hidden sector generators.  Perhaps
this is to be expected, since \nasm\
was obtained from
the start at the string scale, without
ever being---properly speaking---embedded
into a GUT.

For some of the assignments of
$Q,u^c,d^c,L,H_d,H_u,e^c$ to the pseudo-massless
spectrum, other states in the spectrum
may have the right charges with
respect to $SU(3)_C \times SU(2)_L \times U(1)_Y$
to also be candidates for some of these MSSM states.
In this case, the MSSM states will generally
be a mixture of all the candidate states from
the pseudo-massless spectrum,
as described above in Section \ref{mds}.
An example of this will be seen in the
following section.  This, however,
does not alter our conclusions for the
coefficients $c_a$ and $c_a^i$, as
well as the hypercharge normalization $k_Y$.

%

\chapter[Example: ${\rm {BSL}}_{A}$ 6.5]
{Example: ${\rm {\bf BSL}}_{\bf A}$ 6.5}
\label{app}
In this chapter I illustrate more detailed
examination of the \bsa\ models, by focusing
on one specific example from within the class.
I can only scratch the surface of the phenomenological
issues which arise as one looks more carefully
at any one of the 175 models.  The discussion
here should nevertheless serve to illustrate
the fact that the further we push our analysis
toward confrontation with experimental facts,
the more constrained is the effective theory.
That is, the flat direction which must
be chosen---for the model to be phenomenologically
viable---is selected; in fact, a viable flat
direction may not exist once all of the
relevant requirements are imposed!

Section \ref{agen} presents general features
of \bsa\ 6.5 and places it in context vis-\`a-vis
the preceding chapters.  Section \ref{acm}
addresses implications of accomodating the
MSSM into the pseudo-massless spectrum.
In Section \ref{gcu}, unification of the
running gauge couplings is examined.  It
is shown that the exotic states must take
masses intermediate between the electroweak
scale and the Planck scale.  The amount of
fine-tuning on these mass scales is also
studied.  The intermediate masses required
by gauge coupling unification impose requirements
on the choice of flat direction in any subsequent
analysis contemplated for the model.


\section{General}
\label{agen}
The model labeling here is the same as described in Section \ref{mds}:
the observable embedding is
CMM~6 from Table \ref{tabcmm}
and the hidden sector embedding is No.~5 from Table \ref{tab6}.
Thus, the model has embedding
\beqa
3 V & = & (-1,-1,0,0,0,2,0,0;2,1,1,0,0,0,0,0), \nnn
3 a_1 & = & (1,1,-1,-1,-1,2,1,0;-1,0,0,1,0,0,0,0), \nnn
3 a_3 & = & (0,0,0,0,0,1,1,2;2,0,0,-1,1,0,0,0).
\label{5e}
\eeqa
(Recall that $a_5 \equiv 0$ in the class of models
studied here.)
Using the recipes of Section~\ref{rec}, it is easy to
determine the simple roots and to
check that the unbroken gauge group is
\beq
G = SU(3)_C \times SU(2)_L \times SU(5) \times SU(2)'
\times U(1)^8.
\label{5d}
\eeq
The untwisted sector pseudo-massless
matter states are also obtained by simple
calculations; the twisted sectors
are somewhat tedious because of the
large number of states involved.  The full
spectrum of pseudo-massless matter states
is given in Table~\ref{tb5}.
Each entry corresponds
to a {\it species} of chiral matter multiplets,
with three families to each species.  I have
assigned labels 1 through 51 to the species
for convenience of reference in the discussion
which follows.  The irrep of each species with
respect to the nonabelian factors of $G$
is given in the second column of Table~\ref{tb5},
with the order of entries corresponding to
the order of the nonabelian factors in \myref{5d}.
It is not hard to check that
the model falls into Pattern~2.6 of Table~\ref{pt2}.
This pattern has the attractive feature
that it contains only three
extra $(3 + \bar 3,1)$ representations.
Thus, we can expect less finagling with flat
directions to arrange masses
for these exotic isosinglet quarks.
The subscript on the Irrep column data denotes
the sector to which a species belongs:  ``U''
is for untwisted, while for the twisted
species, $n_1,n_3$ pairs of fixed
point labels are given.  The $n_5$
fixed point label now serves as a family
index, so that for each twisted species, it takes
on all three values $n_5=0,\pm 1$.
Twisted oscillator matter states do
not occur in the pseudo-massless spectrum
of this model.  The
remainder of the columns in Table~\ref{tb5}
provide information about \uone\ charges.

As discussed in Section~\ref{rec},
the eight \uone\ generators correspond to
sixteen-dimensional vectors $q_a$ which
are orthogonal to the simple roots and
to each other.  It is not hard to determine
a set of eight $q_a$s.  However, once
the pseudo-massless spectrum of matter
states has been calculated using the recipes
of Section~\ref{rec}, one finds that a naive choice
of the $q_a$s
does not isolate the trace anomaly to a single
\uone.  Using the redefinition technique
described in Section~\ref{rec}, I have isolated
the anomaly to the eighth generator, which
I denote $Q_X$.  Unfortunately,
the redefinitions required to do this, while
maintaining orthogonality of the $q_a$s,
lead to large entries for many of the $q_a$s
when the charges of states are kept integral.
I display my choice of $q_a$s
in Table~\ref{tb6}, along
with $k_a$ (determined by Eq.~\myref{kad}) and
$\tr Q_a$ (determined from the pseudo-massless
spectrum).
I note that $q_1/6=y_1$ of \myref{aac}.
States 27 and 42 would be
electrically neutral exotic isoscalar quarks
if we took $Q_1/6$ as hypercharge.
This provides an explicit example of
the effects of charge fractionalization;
in the low energy theory these states
would bind with ordinary quarks
to form fractionally charged color
singlet composite states.

\begin{table}[ht!]
\begin{center}
$$
\begin{array}{cccc}
a & q_a & \tr Q_a & k_a/4 \\ \hline
1 & (-3,-3,2,2,2,0,0,0;0,0,0,0,0,0,0,0) & 0 & 15 \\
2 & 3(-1,-1,-1,-1,-1,15,0,0;0,0,0,0,0,0,0,0) & 0 & 1035 \\
3 & 3(3,3,3,3,3,1,-46,0;0,0,0,0,0,0,0,0) & 0 & 9729 \\
4 & {3 \over 2}
	(-3,-3,-3,-3,-3,-1,-1,-47;0,0,0,0,0,0,0,0) & 0 & 2538 \\
5 & {3 \over 2}
	(-15,-15,-15,-15,-15,-5,-5,5;12,-12,-12,-48,-12,0,0,0)
   & 0 & 4590 \\
6 & \half (-15,-15,-15,-15,-15,-5,-5,5;-22,-12,-12,20,22,0,0,0)
   & 0 & 357 \\
7 & 3(0,0,0,0,0,0,0,0;1,0,0,0,1,0,0,0) & 0 & 9 \\
X & \half (-3,-3,-3,-3,-3,-1,-1,1;4,6,6,4,-4,0,0,0)
   & 504 & 21 \\ \hline
\end{array}
$$
\caption{Charge generators of \bsa\ 6.5 (cf.~\myref{Qdf}).
\label{tb6}}
\end{center}
\end{table}

For fields which are not $Q_X$ neutral,
we see from Table~\ref{tb5} that $|Q_X|$
has minimum value 1 and maximum value 6.
On the other hand, from Table~\ref{tb6}
we see that $k_X = 84$.  Then
the generator with unified normalization is
$\hat Q_X = Q_X / \sqrt{84}$ and
for fields which are not $\hat Q_X$ neutral,
$|\hat Q_X|$ has minimum value $1/\sqrt{84}
\approx 0.11$ and maximum value $6/\sqrt{84}
\approx 0.65$.  I appealed to this range
in Section~\ref{mds} above.

Note that the $SU(5)$ charged
states in the model consist of
\beq
3 \; [ \; (1,1,5,1) + 3(1,1,\bar 5,1)
+ (1,1,10,2) \; ] .
\label{4n}
\eeq
Using $C(SU(5))=5$, $X(5)=X(\bar 5)=1/2$,
and $X(10)=3/2$ (apparent from \myref{4k}
taking $\tr T^a T^a$ with respect to
a generator of an $SU(3)$ subgroup of $SU(5)$),
we find that 
\beq
\tbeta{5}= - 3 \cdot 5 + 3(4 \cdot 1/2 + 2 \cdot 3/2) = 0 .
\eeq
Thus, in order to have supersymmetry
broken by gaugino condensation in
the hidden sector,
it is necessary that vector masses be
given to some of the states in \myref{4n}.
If we can arrange to give large masses to
the $3(5 + \bar 5)$ vector pairs, then
the effective $\beta$ function coefficient
is only $b_5=-3$.  This gives a lower
$\LamC$ than the pure $G_C=SU(2)$ 
case ($b_C = -6$) which was regarded
as ``marginal'' in Section \ref{ssb}.
Consequently, the
hidden $SU(5)$ must be broken to a
subgroup so that vevs can be given
to components of the $(\bar 5 \cdot \bar 5 \cdot 10)$
and $(5 \cdot 10 \cdot 10)$ invariants,
allowing more states to
get large masses.  (For the $SU(5)$ invariant
$(\bar 5 \cdot \bar 5 \cdot 10)$ to generate an
effective mass term, the hidden $SU(2)'$ would
also have to be broken since the 10s belong to
doublet representations of $SU(2)'$, as is
evident from Eq.~\myref{4n}.)

As an example, consider
breaking $SU(5) \to SU(4)$.  For many choices
of the hypercharge generator, some (but generally
not all) of the $5$ and $\bar 5$ irreps are hypercharge
neutral.  Decomposing these onto $SU(4)$
irreps, we have $5 = 4 + 1$ and $\bar 5 = \bar 4 + 1$.
The breaking can be achieved by
giving vevs to the $SU(4)$ singlets in
these decompositions, though one should
be careful to avoid generating
non-vanishing F- or D-terms in the process.
The $10$ of $SU(5)$ decomposes according
to $10 = 4 + 6$.  
The invariants mentioned above may generate
masses for many of the nontrivial $SU(4)$
irreps, since under the $SU(5) \supset SU(4)$ decomposition
\beq
(5 \cdot \bar 5) \ni (4 \cdot \bar 4), \qquad
(\bar 5 \cdot \bar 5 \cdot 10) \ni (1 \cdot \bar 4 \cdot 4), \qquad
(5 \cdot 10 \cdot 10) \ni (1 \cdot 6 \cdot 6).
\eeq
It is conceivable that {\it all} of the
$SU(4)$ charged matter may be given
$\ordnt{\LamX}$ masses in this way,
yielding $b_4 = -12$.  If some matter remains
light and $SU(4)$ is identified as the
condensing group $G_C$, values in the range $-12 < b_C \leq -6$
could be obtained.  To say whether or not these
arrangements can actually be made requires
an analysis of D- and F-flat directions which
is beyond the scope of the present work.
However, as promised in the introductory
remarks to this chapter, examination of
a phenomenological issue (supersymmetry breaking
scale) has placed broad constraints on the
flat directions which may be chosen for the
model to be viable (the hidden $SU(5)$ must be
broken).  Further examples of this ``tightening''
of the allowed effective theory will be seen
in the following sections.

\section{Accomodating the MSSM}
\label{acm}
Inspection of Table~\ref{tb5} shows that
while appropriate \nasm\ charged multiplets are
present to accomodate the MSSM spectrum, the ``obvious''
choice for hypercharge, $Y_1=Q_1/6$,
does not provide for the
three $e^c$ supermultiplets nor does
it provide enough $(1,2)$ representations with
hypercharge $-1/2$ to accomodate three $L$s
and an $H_d$.
As discussed above, one problem is that
most of the twisted states have bizarre
$Y_1$ charges due to the Wen-Witten defect.
We also have the problem that
$\tK^2=4/3$ for twisted (non-oscillator)
states (versus $K^2=2$ for untwisted),
so that the \eetee\ weights are
``smaller'' and it is harder to obtain
the ``large'' $e^c$ hypercharge; this
explains why $k_Y > 5/3$ is generically
required.
Note that the $Y_1$ charges
are ordinary in the untwisted sector:
the hidden irreps $(1,1,10,2)$ and $(1,1,5,1)$
are $Y_1$ neutral while the observable irreps
$(3,2,1,1)$, $(1,2,1,1)$ and $(\bar 3,1,1,1)$
have $Y_1$ charges $1/6$, $1/2$ and $-2/3$
respectively.
Furthermore, if we subtract off the Wen-Witten
defect, we expect $Y_1$ charges which would
appear in the decompositions \myref{4k}
for twisted states.
With this in mind, I define {\it Z charge}
to be $Z=Y_1$ for untwisted states while
for twisted states
\beq
Z(n_1,n_3,n_5) \equiv Y_1 - y_1 \cdot E(n_1,n_3,n_5)
= {Q_1 \over 6} - \third + n_1 \twthird,
\eeq
where the last equality is easy to check using
the embedding vectors \myref{5e}.
The Z charges are given in
Table~\ref{tb5}.  To see that these
charges are ordinary,
one should compare to the decompositions
(\ref{edc},\ref{4k}).
Checking the Z charges and $SU(3) \times SU(2)$ irrep
labels from Table~\ref{tb5}, it can be seen
that all are in correspondence to some irrep
contained in a decomposition of the 248 and 3875
irreps of $E_8$.
An example of the role of the 3875 irrep
can be seen in state 11 of Table \ref{tb5},
which is a $(1,2)$ irrep of \nasm\
with Z charge $-3/2$; from
\myref{4k} we see that this occurs in
the 40 of $SU(5)$, which itself occurs in
the 3875 but not the 248 of $E_8$.
This shows how it is precisely
the peculiar role of higher dimensional
\eetee\ irreps and the shift $E(n_1,n_3,n_5)$ 
that is responsible for
the bizarre $Y_1$ charges
in the twisted sectors.

Thus, we are forced to assume hypercharge of
the more general form \myref{gym}, which in the present case
I write as
\beq
6 Y = c_1 Q_1 + \cdots + c_7 Q_7 + c_8 H_{(2')}
+ c_{9} H_{(5)}^1 + c_{10} H_{(5)}^2 .
\label{mhc}
\eeq
The generator $H_{(2')}$ is the Cartan
element for the hidden $SU(2)'$,
which I take to be
\beq
H_{(2')} = \diag (1,-1)
\label{5b}
\eeq
in the fundamental irrep.  The
generators $H_{(5)}^1, H_{(5)}^2$ are
the two Cartan elements for the hidden
$SU(5)$ which could combine into hypercharge
while still leaving unbroken a hidden
$SU(3)'$ for the condensing group $G_C$,
as explained in Section~\ref{ehye}.
I take them to be given by
\beq
H_{(5)}^1 = \diag (4,-1,-1,-1,-1),
\qquad
H_{(5)}^2 = \diag (0,3,-1,-1,-1),
\label{5c}
\eeq
for the fundamental representation.
We seek solutions
$c_1, \ldots,  c_{10}$ which
allow for the accomodation of the MSSM.
As mentioned in Section~\ref{ehye},
assigning the MSSM amounts to the
imposition of seven linear constraints
on the coefficients $c_i$,
one for each of the
species $Q,u^c,d^c,L,H_d,H_u,e^c$.
Because of the enormous number of species
to which $L,H_d,H_u$ and $e^c$ could be assigned,
a very large number of assignments
accomodate the MSSM.
However, it is also important to consider
the hypercharge normalization $k_Y$.
From the discussion given in
Section~\ref{ehye}, we know that
\beq
k_Y = {1 \over 36} (c_1^2 k_1 + \cdots + c_{10}^2 k_{10}),
\label{mky}
\eeq
with $k_1, \ldots, k_7$ given in Table~\ref{tb6},
and where $k_8, k_{9}, k_{10}$ depend on the normalization
of the hidden $SU(2)' \times SU(5)$
Cartan generators (\ref{5b},\ref{5c}).
It is easy to see that the generators
(\ref{5b},\ref{5c}) have been
rescaled from the unified normalization
according to
$$
H_{(2')} = \sqrt{k_8} \hat H_{(2')}, \qquad
H_{(5)}^1 = \sqrt{k_{9}} \hat H_{(5)}^1, \qquad
H_{(5)}^2 = \sqrt{k_{10}} \hat H_{(5)}^2,
$$
\beq
k_8 = 4, \qquad k_{9} = 40, \qquad k_{10} = 24.
\eeq

I have investigated the range of $k_Y$ that
is allowed in \bsa\ 6.5, consistent with assignment of
the MSSM spectrum to the model.  This is
not a difficult exercise.  We first obtain
seven linear constraint equations on the $c_i$s
from a given assignment of
the seven types of fields in the MSSM.
We use these constraint equations to rewrite
\myref{mky} in terms of a set of independent
$c_i$s.  The result is a quadratic form
$k_Y$ depending on the independent $c_i$s.
I minimize this quadratic form subject to
the constraint of real $c_i$ using
a standard algorithm provided with the
math package Maple.  I have verified
the automated results by hand in a few sample cases
and find agreement.  An exhaustive
analysis of all possible assignments of the MSSM to the
\bsa\ 6.5 spectrum shows that in every case $k_Y > 5/3$,
consistent with Table \ref{mkytab} (Pattern 2.6).
As above, it is convenient to define $\dkY = k_Y - 5/3$.
I find that constraining $\dkY \leq 2$
still gives 274 possible assignments.
A manageable set is obtained if we impose the limit
$\dkY \leq 1$.
The only possible assignments in this case are
given in Table~\ref{mkya}.  I also give
the minimum value $\dkYm$ for each of
the assignments.  For the cases where
$\dkYm=4/11$ or $\dkYm=1/2$, some of the
MSSM states have been assigned to
$(1,2,1,2)$ irreps, which are each effectively
two $(1,2,1)$ irreps when the hidden $SU(2)'$ is broken
to give an effective nonabelian gauge symmetry
group $SU(3) \times SU(2) \times SU(5)$.
None of the assignments in Table~\ref{mkya}
require breaking the hidden $SU(5)$ to provide
the $e^c$ species or $SU(5)$ Cartan generators
contributing to $Y$; that is, each of these
assignments has $c_{9}=c_{10}=0$ for the
minimum value $\dkYm$.  These two coefficients
are independent parameters for any of the
assignments in Table~\ref{mkya} and {\it could}
be made nonzero without affecting the $Y$
values of the MSSM spectrum; however, this
would alter the $Y$ charges of $SU(5)$ charged
states and would increase $\dkY$ above the
minimum value $\dkYm$.  In principle, $k_Y$
could be made arbitrarily large!
Subscripts on species labels in
Table~\ref{mkya} denote which of the two
$H_{(2')}$ eigenstates the MSSM state
has been assigned to.
For instance, in the $\dkYm=1/2$
assignments, $30_1$ and $30_2$ are states of opposite
$SU(2)'$ isospin.

\begin{table}[ht!]
\begin{center}
$$
\begin{array}{cccccc}
{\rm No.} & Q,u^c,d^c,\underline{L,H_d},H_u,e^c & \dkYm &
{\rm No.} & Q,u^c,d^c,\underline{L,H_d},H_u,e^c & \dkYm \\
\hline
1 & 1,3,10,11,30_1,2,48_2 & 4/11 & 10 & 1,3,42,30_1,30_2,2,29 & 1/2 \\
2 & 1,3,10,25,30_1,2,33_2 & 4/11 & 11 & 1,3,10,11,25,2,43 & 4/5 \\
3 & 1,3,10,30_1,31,2,28_2 & 4/11 & 12 & 1,3,10,11,31,2,49 & 4/5 \\
4 & 1,3,10,30_1,44,2,16_2 & 4/11 & 13 & 1,3,10,25,44,2,34 & 4/5 \\
5 & 1,3,42,11,30_1,2,48_2 & 4/11 & 14 & 1,3,10,31,44,2,23 & 4/5 \\
6 & 1,3,42,25,30_1,2,33_2 & 4/11 & 15 & 1,3,42,11,25,2,35 & 4/5 \\
7 & 1,3,42,30_1,31,2,28_2 & 4/11 & 16 & 1,3,42,11,31,2,24 & 4/5 \\
8 & 1,3,42,30_1,44,2,16_2 & 4/11 & 17 & 1,3,42,25,44,2,9  & 4/5 \\
9 & 1,3,10,30_1,30_2,2,29 & 1/2  & 18 & 1,3,42,31,44,2,17 & 4/5 \\
\hline
\end{array}
$$
\caption{Assignments satisfying $\dkY \leq 1$ 
in \bsa\ 6.5.  Underlining on $H_d$ and $L$
indicates that either permutation may be assigned to
the fourth and fifth entries.
Where applicable, the subscript on a
state label denotes which of the two $H_{(2')}$
eigenstates of a $(1,2,1,2)$ irrep is used in
an assignment.
\label{mkya}}
\end{center}
\end{table}

With these assignments and $\dkY$ set to its
minimum value $\dkYm$, the coefficients $c_i$ in
\myref{mhc} are uniquely determined for each
case; examples are:
\beq
\begin{array}{cccl}
{\rm Assign.~1:} &
(c_1,\ldots,c_{10})
& = & (1,3/253,1/11891,-4/517,0,0,2/11,-18/11,0,0), \\
{\rm Assign.~9:} &
(c_1,\ldots,c_{10})
& = & (2/5,1/10,0,0,1/68,-3/68,3/4,0,0,0), \\
{\rm Assign.~11:} &
(c_1,\ldots,c_{10})
& = & (1,-6/115,-2/5405,8/235,0,0,2/5,0,0,0).
\end{array}
\eeq
From these one can calculate the hypercharges
of the pseudo-massless spectrum, using
the $Q_a$ values and $SU(2)'$ irrep
data provided in Table~\ref{tb5}.
As an example, I have calculated
the hypercharges of the spectrum for
Assignment~11.  These are tabulated
in the last column of Table~\ref{tb5}.

For all of the $\dkYm=4/5$ cases, the \nasm\
charged exotic matter is
\beq
3 \; [ \; (3,1,1/15) + (\bar 3,1,-1/15)
+ 2(1,2,1/10) + 2(1,2,-1/10) \; ]
+ 2 \; [ \; (1,2,1/2)+(1,2,-1/2) \; ].
\eeq
The last number in each term 
gives the hypercharge of the corresponding
state.  I refer to the $SU(3)_C$ charged
states as {\it exoquarks} and to the $SU(2)_L$
charged states as {\it exoleptons.}
The last four exolepton states correspond to
the two extra families of $H_u$-like
and $H_d$-like states which are an artifact of the
three generation construction.
However, the other exoleptons have $Y = \pm 1/10$,
a rather bizarre value, and certainly not one
that appears in GUT scenarios,
as can be seen by comparison to \myref{4k}.
Here again we see the effect of
charge fractionalization.
Similar comments apply to the
exoquarks which have $Y= \pm 1/15$.

For all of the $\dkYm=1/2$ assignments, 
the \nasm\ charged exotic matter is
\beq
3 \; [ \; (3,1,-1/3) + (\bar 3,1,1/3) + 4(1,2,0) \; ]
+ 2 \; [ \; (1,2,1/2)+(1,2,-1/2) \; ].
\eeq
The exoquarks in these assignments have
SM charges of the colored Higgs fields
in an $SU(5)$ GUT.  Whether or not their
masses are similarly constrained 
by proton decay depends
on a detailed study of
the allowed effective superpotential couplings
along a given flat direction,
since we do not have the $SU(5)$ symmetry
to relate Yukawa couplings.
Since altogether we have six
$(\bar 3,1,1/3)$ representations, each of the
three $d^c$-type quarks and their
three exoquark relatives will generally be a mixture
of States 10 and 42, corresponding to
a cross between Assignments 9 and 10.
Some consequences of 
such mixing was discussed above in Section~\ref{mds}.

For all of the $\dkYm=4/11$ assignments, 
the \nasm\ charged exotic matter is
\beqa
&& 3 \; [ \; (3,1,-2/33) + (\bar 3,1,2/33) 
+ (1,2,1/22) + 2(1,2,-3/22)
+ (1,2,5/22) \; ] \nnn
&& \hspace{1in} + 2 \; [ \; (1,2,1/2)+(1,2,-1/2) \; ].
\eeqa
Note that a portion of the exolepton spectrum is
chiral and would lead to a massless states
if the usual electroweak symmetry
breaking is assumed.  For this reason
the Assignments 1-8 are not viable.

\section{Gauge Coupling Unification}
\label{gcu}
Gauge coupling unification in semi-realistic
four-dimensional string models has been a
topic of intense research for several years.
The situation in the heterotic theory has
been reviewed by Dienes in Ref.~\cite{Die97},
which contains a thorough discussion
and extensive references to the original
articles.  I will only present a brief
overview; the interested reader is recommended
to Dienes' review for further details.

It has been known since the
earliest attempts \cite{SS74} to use closed string
theories as unified theories of all fundamental
interactions that
\beq
g^2 \sim \kappa^2/\alpha',
\label{u01}
\eeq
where $g$ is the gauge coupling, $\kappa$ is the gravitational
coupling and $\alpha'$ is the {\it Regge slope,}
related to the string scale
by $\Lambda_{\stxt{string}} \approx 1/\sqrt{\alpha'}$.
In particular, this relation holds
for the heterotic string~\cite{GHMR85}.
However, $g$ and $\kappa$ in \myref{u01} are the 
ten-dimensional couplings.  By {\it dimensional reduction}
\glossary{dimensional reduction}
of the ten-dimensional effective field theory
obtained from the ten-dimensional heterotic string
(cf. Section \ref{rot})
in the zero slope limit,
the relation \myref{u01} may be translated into
a constraint relating the heterotic 
string scale $\LamH$ to the 
four-dimensional Planck mass $m_P$.  
One finds, as expected on dimensional
grounds, $m_P \sim 1/\sqrt{\kappa}$,
where the coefficients which
have been supressed depend on the
size of the six compact dimensions;
similarly, the four-dimensional gauge coupling
satisfies $g_H \sim g$; for details
see Ref.~\cite{DS85}.
Then \myref{u01} gives
\beq
\LamH \sim g_H m_P.
\eeq
Kaplunovsky has made this relation more precise,
including one loop effects from
heavy string states \cite{Kap88}.
Subject to various conventions described
in \cite{Kap88}, including a choice
of the $\overline{\mbox{DR}}$ 
renormalization scheme in
the effective field theory, 
the result is:
\beq
\LamH \approx 0.216 \times g_H m_P
= g_H \times 5.27 \times 10^{17} \mtxt{GeV}.
\label{ssc}
\eeq

In \myref{ssc}, a single gauge coupling, $g_H$, is shown.
However, in the heterotic orbifolds under
consideration the gauge group $G$ 
has several factors, each of which will
have its own running gauge coupling.
One may ask how these running couplings are related to
$g_H$.  This question was studied by
Ginsparg \cite{Gin87}, with the result
that the running couplings unify to a common value $g_H$
at the string scale $\LamH$, up to string threshold effects
and affine levels (discussed below).
(In the case of {\uone}s, normalization conventions
must be accounted for, as I have described in
detail in Section~\ref{strn}.)
Specifically, unification in four-dimensional string
models makes the following requirements
on the running gauge couplings $g_a(\mu)$:
\beq
k_a g_a^2(\LamH) = g_H^2, \quad \forall \; a.
\label{hbc}
\eeq
Here, $k_a$ for a nonabelian factor $G_a$ is
the {\it affine} or {\it Kac-Moody level} of
the current algebra---in the underlying
theory---which is responsible for the gauge symmetry in
the effective field theory.  I will
not trouble the reader with a detailed
explanation of this quantity or its string theoretic
origins, since $k_a = 1$ for any nonabelian
factor in the heterotic orbifolds we are considering.
For this reason, these heterotic orbifolds are referred
to as affine level one constructions.
In the case of $G_a$ a \uone\ factor, $k_a$
carries information about the normalization
of the corresponding current in the underlying theory, and
hence the normalization of the charge generator
in the effective field theory.  We saw explicit
examples of this 
in the previous section.

The important point, which has been emphasized many
times before, is that a gauge coupling
unification prediction is made by the underlying
string theory.  The SM gauge couplings
are known (to varying levels of accuracy), say,
at the Z scale (approximately 91 GeV).  Given the
particle content and mass spectrum 
of the theory between the Z scale
and the string scale, one can easily check at the one loop
level whether or not the unification prediction
is approximately consistent with the Z scale boundary values.
To go beyond one loop requires some knowledge of
the other couplings in the theory, and the analysis
becomes much more complicated.  However,
the one loop success is not typically spoiled by
two loop corrections, but rather requires a slight
adjustment of flexible parameters (such as superpartner
masses) which enter the one
loop analysis.

In what follows I briefly discuss
the one loop running of SM gauge couplings in \bsa\ 6.5,
Assignment 11 of Table~\ref{mkya}, estimating two loop
effects using previous studies of the MSSM.
Due to the presence of exotic matter,
I am able to achieve string scale unification.
This sort of unification scenario
has been studied
many times before, for example in 
Refs.~\cite{GX92,Far93,MR95,AK96}.
However, in contrast to the Refs.~\cite{GX92,MR95,AK96},
\bsa\ 6.5 has states which would not appear in
decompositions of standard GUT groups,
such as \myref{4k}.  Indeed, it was found
by Gaillard and Xiu in Ref.~\cite{GX92} that $(3 + \bar 3,2)$
representations with hypercharge
$Y=\pm 1/6$ were necessary to string
scale unification, while Faraggi
achieved string unification in Ref.~\cite{Far93}
in a model where
the only colored exotics were $(3 + \bar 3,1)$
states.  The resolution of this apparent
conflict is that the unification scenario of Faraggi
contains $(1,2)$ exoleptons with vanishing hypercharge
and $(3 + \bar 3,1)$ exoquarks with hypercharge
$Y=\pm 1/6$;
such states have exotic electric charge
and {\it do not} appear in \myref{4k}.
The appearance of these states is due to the
Wen-Witten defect in the free fermionic
construction used in the model discussed
by Faraggi,
which has a $Z_2 \times Z_2$ orbifold underlying
it, leading to shifts in
hypercharge values by integer multiples of
$1/2$.  Because $SU(3)_C \times SU(2)_L$ charged
representations with small hypercharge
values---much like the $(3 + \bar 3,2)$
representations used by Gaillard and
Xiu---appear in the model employed by Faraggi,
the $SU(3) \times SU(2)$ running can be
altered to unify at the string scale without
having an overwhelming modification on
the running of the $U(1)_Y$ coupling.

Similar to the unification scenario
of Faraggi, in the model studied here
exotic representations with small hypercharges
are present; these exotics allow us to unify
at the string scale without the presence
of extra quark doublets.  However, we
also have nonstandard hypercharge normalization:
for Assignment~11 the minimum value is $k_Y=37/15>5/3$.
Nonstandard
hypercharge normalization has been
studied previously, for example in
Refs.~\cite{Iba93,DFMR96}.  In these analyses,
it was found that {\it lower} values
$k_Y < 5/3$ were preferred if only the
MSSM spectrum is present up to the unification
scale; the preferred values were between $1.4$
to $1.5$.  Unfortunately, we are faced with
the opposite effect---a larger than normal
$k_Y=37/15$.
This larger value requires a larger correction
to the running from the exotic states,
and has the effect of pushing down the
required mass scale of the exotics
from what was found in Faraggi's analysis---particularly
in the case of the exoquarks.  

Standard evolution of the gauge couplings from
the Z scale (i.e., the solution to \myref{bfdef}
for groups other than $G_C$),
together with the unification
prediction \myref{hbc}, 
leads to three constraint equations:
\beqa
4 \pi \icoup{H} & = & {1 \over k_Y} \left[
   4 \pi \icoup{Y}(m_Z) - b_Y \ln {\LamHsq \over m_Z^2}
   - \Delta_Y \right], \label{rgeY} \\
4 \pi \icoup{H} & = & 4 \pi \icoup{a}(m_Z)
   - b_a \ln {\LamHsq \over m_Z^2} - \Delta_a,
   \quad a=2,3. \label{rge23}
\eeqa
The notation is conventional, with 
$\alpha_a = g_a^2/ 4 \pi \; (a=H,Y,2,3)$.
Corrections are captured by the quantities $\Delta_a$,
and will be discussed below.
The quantities $b_a, \; a=Y,2,3$ are the
$\beta$ function coefficients
\beq
b_{a} = -3 C(G_a) + \sum_R X_a(R)
\label{5a}
\eeq
evaluated for the MSSM spectrum.  
Here, $C(SU(N))=N$ while $C(U(1))=0$.  For
a fundamental or antifundamental
representation of $SU(N)$ we have $X_a=1/2$
while for hypercharge $X_Y(R) = Y^2(R)$.
This gives
\beq
b_{Y} = 11, \qquad b_{2} = 1,
\qquad b_{3} = -3.
\label{bcm}
\eeq

Throughout, I use Z scale boundary
values from the Particle Data Group 2000
review \cite{PDG00}, which are given in the $\bbar{\rm MS}$
scheme.  For a supersymmetric running,
these boundary values should be converted 
to the $\bbar{\rm DR}$ scheme, so that the
supersymmetry
algebra is kept four-dimensional \cite{DRB,BAM}.
These scheme conversion effects are included
in the corrections $\Delta_a$.
Due to very small
errors (relative to other uncertainties in
the analysis), I take as precise
\beq
m_Z = 91.19 \mtxt{GeV}, \qquad \icoup{e}(m_Z) = 127.9 .
\eeq
For the other couplings I utilize global fits to
experimental data \cite{PDG00}:
\beq
\sin^2 \theta_W(m_Z) = 0.23117 \pm 0.00016, \qquad
\alpha_3(m_Z) = 0.1192 \pm 0.0028 .
\label{4b}
\eeq
Using
\beq
\icoup{2} = \icoup{e} \sin^2 \theta_W, \qquad
\icoup{Y} = \icoup{e} \cos^2 \theta_W,
\eeq
we obtain the boundary values
\beq
\icoup{Y}(m_Z) = 98.333 \pm 0.020 , \qquad
\icoup{2}(m_Z) = 29.567 \pm 0.020 , \qquad
\icoup{3}(m_Z) = 8.39 \pm 0.20 .
\label{Zin}
\eeq

I now discuss the various corrections contributing
to $\Delta_a \; (a=Y,2,3)$.  Each may be written as the
sum of six terms:
\beq
\Delta_a = \Delta_a^{\stxt{conv}} + \Delta_a^{\stxt{HL}} 
+ \Delta_a^{\stxt{string}} + \Delta_a^{\stxt{light}}
+ \Delta_a^{\stxt{exotic}} + \Delta_a^{\stxt{heavy}}.
\eeq

The quantities $\Delta_a^{\stxt{conv}}$ convert the
$\bbar{MS}$ renormalization scheme input values \myref{Zin}
to the $\bbar{DR}$ scheme \cite{BAM,LP93}.
They are given by:
\beq
\Delta_a^{\stxt{conv}} = \third C(G_a) \quad \Rightarrow \quad
\Delta_Y^{\stxt{conv}} = 0, \quad
\Delta_2^{\stxt{conv}} = 2/3, \quad
\Delta_3^{\stxt{conv}} = 1.
\eeq
As will be seen below, these corrections are negligible
in comparison to the other terms in $\Delta_a$,
and we could ignore them without changing our results
in a meaningful way.

The quantities $\Delta_a^{\stxt{HL}}$ represent corrections
from higher loop orders, which are sensitive to Yukawa
couplings for the MSSM spectrum and the exotic states.
If either the top or bottom Yukawa
coupling evolves to nonperturbative
values somewhere between Z scale and the string scale
(as can happen for small or very large
values of the ratio of MSSM Higgs vevs, $\tan \beta$),
the $\Delta_a^{\stxt{HL}}$ correction is out of control.  However,
if the Yukawa couplings arise from a weakly coupled heterotic
string theory, as we assume, then this does not occur;
$\Delta_a^{\stxt{HL}}$ will take more reasonable
values.  For example, Dienes, Faraggi and
March-Russell \cite{DFMR96} have studied the range of MSSM two loop
corrections with the Yukawa couplings taking
values $\lambda_t(m_Z) \approx 1.1$ and
$\lambda_b(m_Z) \approx 0.175$.  (Using $m_b(m_Z) \approx 3.0$ GeV
from Ref.~\cite{FK98} and $m_t(m_Z) \approx 174$ GeV from
\cite{PDG00}, these Yukawa couplings correspond to
$\tan \beta \approx 9.2$.)
These authors found that the two loop (TL) correction
terms took approximate values
\beq
\Delta_Y^{\stxt{TL}} \approx 11.6,
\qquad
\Delta_2^{\stxt{TL}} \approx 12.3,
\qquad
\Delta_3^{\stxt{TL}} \approx 6.0.
\label{5f}
\eeq
These should dominate $\Delta_a^{\stxt{HL}}$,
so we assume that to the same level of
approximation $\Delta_a^{\stxt{HL}}
\approx \Delta_a^{\stxt{TL}}, \; \forall \; a=Y,2,3.$
Relative to the boundary values
for $4 \pi \icoup{a}$, these
are 0.9\%, 3.3\% and 5.7\% corrections,
respectively.
By comparison, the largest experimental
error is 2.4\% for $\icoup{3}$.

The third type of correction is peculiar
to unified theories with large numbers of
gauge-charged states above or near the unification
scale.  These effects have been extensively
studied \cite{GUTth} in GUTs.  In attempts
to bring unification predictions into good
agreement with precision data
these corrections
play an important role \cite{LP93}.
When very large GUT group representations are
introduced near the unification scale, these
corrections can be considerable \cite{HY93}.
With the standard-like string constructions which
we study here, a GUT symmetry group and
heavy states which complete GUT multiplets
are not
restored at the unification scale.  Rather,
the chief concern is with threshold effects
due to the enormous towers of massive string states.
These may be computed from one loop diagrams
in the
underlying string theory, using background
field methods quite similar to those
exploited in ordinary field theory \cite{Kap88}.
As noted above,
in some four-dimensional heterotic theories,
string threshold corrections
exist which grow
in size as the T-moduli
vevs increase \cite{Nothr}.
This corresponds to the large volume limit for the compact
dimensions; the potentially large contribution
in this limit can also be understood from the
fact that the compactification scale drops below
the string scale and entire excited mass levels of
the string enter the running below the string scale.
In any event, such T-moduli dependent string
threshold effects are irrelevant for
the 175 models studied here, as they do not occur in $Z_3$
orbifold compactifications of the heterotic string
\cite{Nothr}.

However, threshold corrections which do not
increase with the vevs of T-moduli must
also be considered.  These threshold effects have
been calculated by Mayr, Nilles and Stieberger \cite{MNS93}
for an example model which is equivalent to one of the
175 studied here.  They find that the string threshold
effects are given by
\beq
\Delta^{\stxt{string}}_a = 0.079 \; \tbeta{a} + 4.41 \; k_a .
\label{stef}
\eeq
(Actually, Ref.~\cite{MNS93} states that \myref{stef}
is valid with $k_a=1$.  However, starting with
the hypercharge coupling in the
unified normalization $\icoup{1} = \icoup{Y} / k_Y$,
it can be seen from \myref{rgeY} that
by our conventions $\tbeta{1} = \tbeta{Y} / k_Y$
and $\Delta_1 = \Delta_Y / k_Y$.  Substituting these
expressions into \myref{stef} for $a=1$, and then
solving for $\Delta^{\stxt{string}}_Y$, one
finds that the formula is also valid for $a=Y$
where $k_Y \not= 1$.)
It is important to keep in mind that $\tbeta{a}$
is the $\beta$ function coefficient for $G_a$ with
the full spectrum of pseudo-massless states.
This includes
those states which get $\LamX \approx \LamH$
scale masses when the vacuum shifts
to cancel the FI term.  Because of the large
number of states with charge under a given
\uone\ factor, the hypercharge correction
$\Delta^{\stxt{string}}_Y$ is usually much larger
than $\Delta^{\stxt{string}}_2$
or $\Delta^{\stxt{string}}_3$.
The precise values of the coefficients in \myref{stef}
will vary from model to model; these must be worked
out by the numerical evaluation of a rather complicated
integral, as explained in \cite{MNS93}.  However,
Mayr, Nilles and Stieberger
analyzed a few other $Z_3$ orbifold
models, which do not fall into the class of models
considered here, and found that the threshold corrections
differed only slightly from \myref{stef}.  This
was found to be due to the fact that the leading
term in the integrand did not depend on the embedding.
From this we conclude that Eq.~\myref{stef}
gives a fair estimate of the string
threshold corrections in all 175 models
which we study here.

The hypercharge values of the 51 species
must be calculated in order to compute
$\tbeta{Y}$ for the example model.  
This of course depends on what linear
combination \myref{mhc} of generators
we take to be the hypercharge generator $Y$.
As an example we take Assignment 11
from Table~\ref{mkya}, which has
(for $\dkY=\dkYm$) hypercharge
normalization $k_Y=37/15$
and hypercharges $Y$ given in Table~\ref{tb5}.
It is easy to check that
\beq
\tbeta{Y} = \tr Y^2 = 171/5,
\qquad
\tbeta{2} = 9, \qquad \tbeta{3} = 0.
\eeq
Applying \myref{stef}, one finds
\beq
\Delta^{\stxt{string}}_Y \approx 13.6,
\qquad
\Delta^{\stxt{string}}_2 \approx 5.1,
\qquad
\Delta^{\stxt{string}}_3 \approx 4.4,
\eeq
which are comparable to the two loop corrections
in \myref{5f}.

Next I discuss one loop threshold corrections for
pseudo-massless states which have masses greater than the Z mass
but less than the string scale $\LamH$.
Heuristically, these corrections
may be understood as follows.  At a running
scale $\mu$, only states with masses less
than this scale contribute significantly to
the running of the gauge couplings.  Then the
more accurate one loop $\beta$ function coefficients
in this regime are calculated using the spectrum of
states with masses less than $\mu$.  If some of
the superpartner states are more massive than
$\mu$, the $\beta$ function coefficients will
not take the MSSM values given in \myref{bcm}.
Non-MSSM values for the coefficients will
also be obtained if exotic states with masses
less than $\mu$ are present.  In
(\ref{rgeY},\ref{rge23}) we assumed the MSSM
values for the $\beta$ function coefficients.
The threshold corrections we now discuss
account for the non-MSSM $\beta$ function coefficients
which ``should'' have been used over regimes
where the MSSM was not the spectrum of states
with masses less than $\mu$.  This simple
picture is valid in the $\bbar{DR}$ renormalization
scheme; in other schemes there are modifications
to the one loop threshold corrections
presented below, as has been recounted
for example in \cite{LP93}.

The first correction is
due to MSSM superpartners to the SM.
In the coefficients \myref{bcm}, we have implicitly
included these particles in the running all the
way from the Z scale; however, if they are more
massive than the Z scale, this is not quite right.
We introduce ``light'' threshold corrections
which subtract
out the running which should never have been there
in the first place:
\beq
\Delta_a^{\stxt{light}}
= - \sum_{m_i > m_Z} b_{a,i} \ln {m_i^2 \over m_Z^2} ,
\label{ltc}
\eeq
where $b_{a,i}$ is the contribution to the MSSM
$b_{a}$ coming from the state $i$ of mass $m_i$.
Properly speaking, the top quark and the light scalar Higgs doublet
threshold corrections should also be
included in $\Delta_a^{\stxt{light}}$.  The top mass is
near enough to the Z mass that the correction is
negligibly small for our purposes; we assume
that this is likewise true for the light scalar Higgs
doublet.  Following Langacker and Polonsky \cite{LP93},
one often defines effective thresholds $T_a \; (a=Y,2,3)$
which give the same $\Delta_a^{\stxt{light}}$
as \myref{ltc}:
\beq
\Delta_a^{\stxt{light}}
\equiv -(b_a - b_{a}^{\stxt{SM}}) \ln { T_a^2 \over m_Z^2}.
\label{5s}
\eeq
Here, $b_{a}^{\stxt{SM}}$ are the $\beta$
function coefficients in the SM (which we
take to include a light Higgs doublet and
the top quark):
\beq
b_{Y}^{\stxt{SM}} = 7, \quad b_{2}^{\stxt{SM}} = -3,
\quad b_{3}^{\stxt{SM}} = -7, \qquad
\Rightarrow \qquad b_a - b_{a}^{\stxt{SM}} = 4, \quad
a=Y,2,3,
\label{bsm}
\eeq
where we make use of \myref{bcm}.
Eq.~\myref{5s} has the interpretation that
it gives the equivalent threshold correction to
$\icoup{a}$ if all superpartners contributing
to $b_a$ had a uniform mass scale $T_a$.
One may study how the 
prediction for $\alpha_3(m_Z)$ in terms of
$\sin^2 \theta_W (m_Z)$ depends on $T_a$
and determine a combination of
the three effective thresholds which
would give the same effect as a uniform
superpartner mass threshold $\LamSB$ \cite{LP93}:
\beqa
\lefteqn{ \hspace{-15pt} (b_Y - b_3 k_Y)(b_2 - b_2^{\stxt{SM}}) \, \ln {T_2 \over m_Z}
- (b_2 - b_3)(b_Y - b_Y^{\stxt{SM}}) \, \ln {T_Y \over m_Z}
- (b_Y - b_2 k_Y)(b_3 - b_3^{\stxt{SM}}) \, \ln {T_3 \over m_Z}} 
& & \nnn
& {\hspace{-25pt} \equiv} & \hspace{-10pt} \[ (b_Y - b_3 k_Y)(b_2 - b_2^{\stxt{SM}})
- (b_2 - b_3)(b_Y - b_Y^{\stxt{SM}}) 
- (b_Y - b_2 k_Y)(b_3 - b_3^{\stxt{SM}}) \]
\, \ln {\LamSB \over m_Z}.
\label{5r}
\eeqa
From this one can define the
single effective threshold $\LamSB$ in terms
of a geometric average of superpartner masses \cite{CPW93}.
Because of terms of opposite sign
in \myref{5r}, it should be clear that $\LamSB$ can
be much lower than the typical superpartner
mass, which we denoted $\MSB$ in the
Introduction; $\LamSB \lappeq m_Z$ is not at all unreasonable,
even with the typical superpartner mass $\MSB$ several
hundred GeV.  Furthermore, it should be noted
that the formulae for $\LamSB$ given in Refs.~\cite{LP93,CPW93}
are modified in the present context due to the
nonstandard hypercharge normalization,
as has been accounted for in \myref{5r},
which holds generally.
(Our $b_a^{\stxt{SM}}$, as given in \myref{bsm},
also differ slightly due to the inclusion of a light scalar Higgs
doublet; however, Eq.~\myref{5r} has been written
such that it is valid in either case.)
Lastly, the effective threshold $\LamSB$
completely encodes the effects of split thresholds on the 
$\alpha_3(m_Z)$ versus $\sin^2 \theta_W (m_Z)$
prediction, but for other unification
predictions, such as the unified coupling and
scale of unification, a fixed value of
$\LamSB$ corresponds to many different
outcomes \cite{CPW93}; this is because other unification
predictions depend on
combinations of the $T_a$ other than
\myref{5r}.  In the present context,
simply using $\LamSB$ would
not cover the full range of
$g_H$, $\LamH$ and the predictions for
intermediate scales where exotic
matter thresholds alter the running.
An exhaustive analysis would require
scanning over the parameters $T_a \; (a=Y,2,3)$
independently, or subject to model constraints
on the generation of soft masses
by supersymmetry breaking.  Our
purpose here is simply to demonstrate the
possibility of string scale unification with
nonstandard hypercharge normalization and to
estimate the order of magnitude required
for the exotic scales.
For these purposes it is therefore
sufficient to take $\LamSB \approx T_a \; (a=Y,2,3)$.
Within this universal scale $\LamSB$ approximation,
\beq
\Delta_a^{\stxt{light}}
= - 4 \ln {\LamSB^2 \over m_Z^2}, \quad a = Y,2,3.
\eeq
If we limit $m_Z \lappeq \LamSB \lappeq 1$ TeV, then
\beq
0 \gappeq \Delta_a^{\stxt{light}} \gappeq -19.2,
\quad a=Y,2,3.
\eeq

The second set of mass threshold corrections
comes from exotic matter at intermediate
scales.  For the sake of
simplicity, we assume that exoleptons with mass
much less than the string scale enter the running
at a {\it single} scale $\Lambda_2$.  We assume that
{\it all} the exoquarks enter at a single scale $\Lambda_3$.
(Introducing only {\it some} of the exoquarks forces
$\Lambda_3$ to even lower values than we will
find below, which are already a bit of a problem
given the exotic hypercharges that these
exoquarks have.)
The exotic matter threshold corrections 
can be thought of as due to shifts
in the total $\beta$ function coefficients
between $\Lambda_{2,3}$ and the string scale.
Since we introduce $3(3 + \bar 3,1)$ chiral multiplets
$q_i$ and $q_i^c$ at $\Lambda_3$, we have
\beq
\Delta_3^{\stxt{exotic}}
= 3 \ln {\LamHsq \over \Lambda_3^2} .
\eeq
The shift in the $\beta$ function coefficient
for $SU(2)_L$ due to extra $(1,2)$ representations---the
exolepton chiral multiplets $\ell_i$ and
$\ell_i^c$ introduced at $\Lambda_2$---is given by
\beq
\dbtw = \sum_{\ell_i,\ell_i^c} \half .
\eeq
That is, $\dbtw$ is just the number of exolepton
pairs $\ell_i + \ell_i^c$.  The threshold corrections are
\beq
\Delta_2^{\stxt{exotic}}
= \dbtw \ln {\LamHsq \over \Lambda_2^2} .
\eeq
The exoquark and exolepton chiral multiplets
also carry hypercharge.  We denote the shifts in
the $\beta$ function coefficient for $U(1)_Y$ by
\beq
\dbY = \sum_{q_i,q_i^c} (Y_i)^2,
\qquad
\dbYp = \sum_{\ell_i,\ell_i^c} (Y_i)^2 .
\eeq
In this notation the threshold corrections are
\beq
\Delta_Y^{\stxt{exotic}}
= \dbY \ln {\LamHsq \over \Lambda_3^2} 
+ \dbYp \ln {\LamHsq \over \Lambda_2^2} .
\eeq

Let $m,n$ denote the numbers of
exolepton pairs entering the running at
$\Lambda_2$, where $m$ is the number of $Y=\pm 1/2$ exolepton
pairs and $n$ is the number of $Y=\pm 1/10$ exolepton
pairs.  We then have
\beq
\dbY = {2 \over 25}, \qquad \dbYp = m + {n \over 25},
\qquad \dbtw = m + n .
\eeq
For purposes of illustration below,
we will study only the case $(m,n)=(0,6)$, for which
\beq
\dbY = {2 \over 25}, \qquad \dbYp = {6 \over 25},
\qquad \dbtw = 6.
\label{5g}
\eeq
It is not difficult to generalize our results to
other $(m,n)$ values.

Finally, there is the spectrum of particles which
get masses of order $\LamX$ when the vacuum shifts
to cancel the FI term.  Since $\LamX < \LamH$
in \bsa\ 6.5 (cf.~Table~\ref{fir}, Pattern 2.6),
these can give an appreciable heavy threshold correction.
Corrections of this type have been noted
previously; for example, in Ref.~\cite{BL92}.
We assume that all pseudo-massless
states other than the MSSM spectrum
plus exotics associates with
$\Lambda_{2,3}$ enter the running at
$\LamX$, which is convenient because the ratio
\beq
\ln{\Lambda_H^2  \over \Lambda_X^2}
= 2 \ln {0.216 \times g_H m_P \over 0.170 \times g_H m_P} = 0.479
\eeq
is independent of $g_H$ (both $\LamX$ and $\LamH$
are proportional to $g_H$); here we use the value
for Pattern 2.6 from Table~\ref{fir}.  Taking into account the
exotic matter assumed at intermediate scales 
$\Lambda_{2,3}$ and
the total $\beta$ function coefficients mentioned
above, we have
\beq
\Delta_Y^{\stxt{heavy}} = (\tbeta{Y} - b_Y - \dbY - \dbYp)
\; \ln {\Lambda_H^2 \over \Lambda_X^2} = 10.3,
\eeq
\beq
\Delta_2^{\stxt{heavy}} = (\tbeta{2} - b_2 - \dbtw)
\; \ln {\Lambda_H^2 \over \Lambda_X^2} = 1.0,
\qquad
\Delta_3^{\stxt{heavy}} = 0 .
\eeq
The hypercharge threshold correction is
comparable to the larger corrections discussed
above.  On the other hand, we could just
as well ignore $\Delta_{2,3}^{\stxt{heavy}}$
at the level of approximation made here.

As we tune $\Lambda_{2,3}$ to satisfy
the unification constraints,
it is convenient to define the 
sum of all the corrections
{\it except} $\Delta_a^{\stxt{exotic}}$:
\beq
\Delta_a^0 \equiv
\Delta_a - \Delta_a^{\stxt{exotic}} = \Delta_a^{\stxt{conv}} + \Delta_a^{\stxt{HL}} 
+ \Delta_a^{\stxt{string}} + \Delta_a^{\stxt{light}}
+ \Delta_a^{\stxt{heavy}}.
\label{5u}
\eeq
Using the above estimates for each of the terms, we find
for the case of $\LamSB = m_Z$
\beq
\Delta_Y^0 \approx 35.5,
\qquad
\Delta_2^0 \approx 19.1,
\qquad
\Delta_3^0 \approx 11.4.
\label{5h}
\eeq
For the case of $\LamSB=1$ TeV, the estimate is
\beq
\Delta_Y^0 \approx 16.3,
\qquad
\Delta_2^0 \approx -0.1,
\qquad
\Delta_3^0 \approx -7.8.
\label{5i}
\eeq

We now proceed to study the unification
constraint in \bsa\ 6.5, Assignment 11, 
subject to the
assumptions described above.
For convenience, we define
\beq
a_H = 4 \pi \icoup{H}; \qquad
d_a = 4 \pi \icoup{a}(m_Z) - \Delta_a^0, \quad a=Y,2,3;
\eeq
\beq
t_2 = \ln { \Lambda_2^2 \over m_Z^2 }, \qquad
t_3 = \ln { \Lambda_3^2 \over m_Z^2 }.
\eeq
Because the string
scale $\LamH$ contains a dependence on $g_H$
through \myref{ssc}, it will prove convenient to write
\beq
\ln \left({\LamH \over m_Z}\right)^2 = t_P - \ln (4 \pi \icoup{H}),
\eeq
\beq
t_P \equiv 2 \ln \left( {4 \pi \LamH \over g_H m_Z} \right)
= 2 \ln \left( {4 \pi \times \zeta \times 5.27 \times 10^{17}
\over 91.19 } \right) = 77.6 + 2 \, \ln \zeta.
\label{5j}
\eeq
Here we introduce a coefficient $\zeta$
which expresses uncertainty in \myref{ssc}
described in \cite{Kap88};
we study 10\% deviations by
taking $0.9 \leq \zeta \leq 1.1$,
leading to $t_P = 77.6 \pm 0.2$.
Eqs.~(\ref{rgeY},\ref{rge23}) give the following equations
which must be simultaneously satisfied:
\beqa
a_H & = & d_3 + 3 t_3 \label{u22} \\
a_H & = & d_2 + \dbtw \, t_2
- (1 + \dbtw)(t_P- \ln a_H) \label{u23} \\
k_Y a_H & = & d_Y + \dbY \, t_3
+ \dbYp \, t_2
- (11 + \dbY + \dbYp)(t_P - \ln a_H) \label{u24}
\eeqa
The first equation shows the nice feature that since
the $SU(3)_C$ coupling becomes conformal above $\Lambda_3$,
the $\ln a_H$ dependence is gone and we can solve for
$a_H$ explicitly.  Since this equation does not depend
at all on $t_2$, we obtain $a_H = a_H(d_3,t_3)$.  Substituting
this into the second equation allows us to solve for
$t_2$ explicitly, yielding $t_2 = t_2(d_2,d_3,t_3)$.  Thus,
the last equation becomes the only nontrivial constraint,
which is transcendental and must be solved numerically.
Through it we can determine $t_3=t_3(d_Y,d_2,d_3)$ after having
substituted the expressions for $a_H$ and $t_2$ from
the first two equations.  Taking the values \myref{5g}
for the $(m,n)=(0,6)$ example, the implicit
equation for $t_3$ is
\beq
t_3 = {1 \over 540} \left[ 75 d_Y - 182 d_3 - 3 d_2 \right]
- {23 \over 15} \left[ t_P - \ln(d_3 + 3 t_3) \right],
\label{5q}
\eeq
which can easily be solved iteratively.  Once
$t_3$ is determined, $a_H$ is easily
obtained from \myref{u22} and
\beq
t_2 = {1 \over 6} \left[ a_H - d_2
+ 7 (t_P - \ln a_H) \right].
\eeq

Note that if $g_H$ and $\LamH$
were independent, as in the GUT case, we would have one
more degree of freedom and we could not uniquely determine
$t_2, t_3, g_H, \LamH$ in terms of $d_Y,d_2,d_3$.
Related to this is an alternative,
but equivalent, method of
solution to that employed above.  We could
treat $\LamH$ and $g_H$ as independent
and solve (\ref{rgeY},\ref{rge23}) keeping
$t_3$ as the extra free parameter.  Then
solutions to (\ref{rgeY},\ref{rge23})
would have $\LamH=\LamH(t_3)$ and
$g_H=g_H(t_3)$.  We could then determine
the range of $t_3$ which allow the fourth
constraint \myref{ssc} to be satisfied to within,
say, 10\%.  Instead we impose \myref{ssc}
from the start and
address uncertainty of $\pm$10\% with the
parameter $\xi$.  The results are
of course the same by either method.

In the case of $\LamSB=m_Z$, we find
\beqa
\Lambda_2 & =&  \( 2.25 \mp 0.07 \mp 0.006 \pm 0.09 \)
\times 10^{13} \; {\rm GeV}, \nnn
\Lambda_3 & = & \( 5 \mp 0.1 \mp 3 \mp 1 \)
\times 10^{6} \; {\rm GeV}, \nnn
g_H & = & 0.995 \pm 0.0004 \pm 0.0001 \pm 0.003, \nnn
\LamH & = & \( 5.1 \pm 0.002 \pm 0.0005 \pm 0.6 \)
\times 10^{17} \; {\rm GeV}.
\label{5k}
\eeqa
The first two uncertainties for each quantity
give the modified estimates if
$\swZ$ and $\icoup{3}(m_Z)$
are taken at the ends of
the $1\s$ ranges given in \myref{4b} and \myref{Zin}
respectively.  Upper signs in \myref{5k} correspond
to the upper limits of the $1\s$ ranges;
asymmetric uncertainties (due to logarithms)
have been rounded up to the larger of the two.
The third uncertainty gives
the modified estimates if the ``fudge parameter''
$\zeta$ in \myref{5j} is taken at the
ends of the range $0.9 \leq \zeta \leq 1.1$.
Again, the upper signs in \myref{5k}
correspond to the upper limit of the range
for $\zeta$.
Sensitivities are logical:
the exoquark scale $\Lambda_3$ is most
sensitive to $\icoup{3}(m_Z)$, while the
sensitivity to $\swZ$ is below significance.
Only the exolepton scale $\Lambda_2$ is
has significant sensitivity to $\swZ$;
$\Lambda_2, \; \LamH$ and $g_H$,
quantities more closely related
to the high scale physics, are sensitive
the high scale uncertainty $\zeta$.
For the case of $\LamSB=1$ TeV, we find
\beqa
\Lambda_2 & =&  \( 8.4 \mp 0.3 \mp 0.02 \pm 0.4 \)
\times 10^{12} \; {\rm GeV}, \nnn
\Lambda_3 & = & \( 7 \mp 0.1 \mp 4 \mp 1 \)
\times 10^{5} \; {\rm GeV}, \nnn
g_H & = & 0.972 \pm 0.0003 \pm 0.0001 \pm 0.003, \nnn
\LamH & = & \( 5.0 \pm 0.002 \pm 0.0004 \pm 0.5 \)
\times 10^{17} \; {\rm GeV}.
\label{5m}
\eeqa

\begin{figure}[p]
\begin{center}
\includegraphics[height=6.0in,width=5.0in,angle=90]{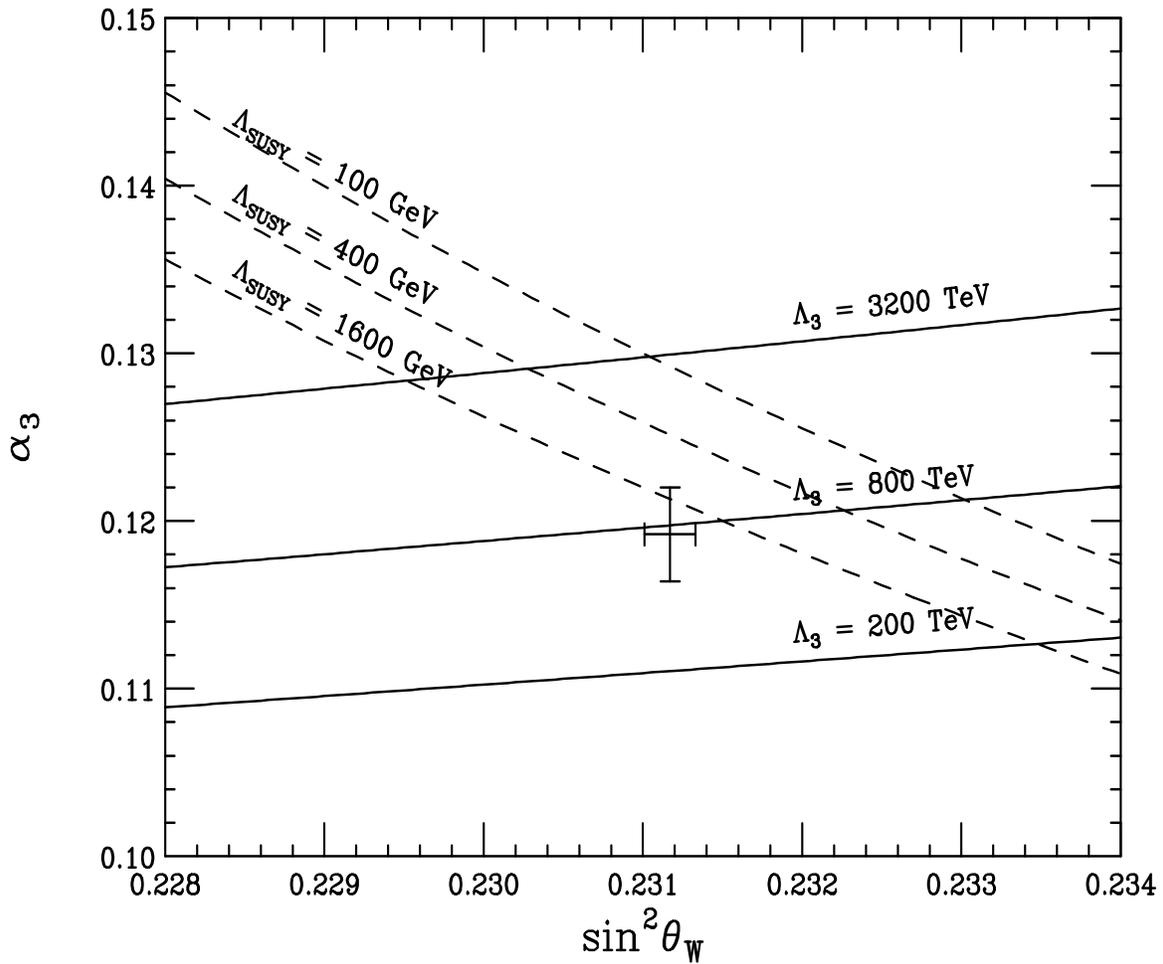}
\end{center}
\caption{Predicted Z scale values per the string
unification scenario (solid), for values of $\Lambda_3$
stepping by factors of four, with $\LamSB=1$ TeV.
For comparison, the MSSM unification prediction
is shown (dashed), with $\LamSB$ stepping by factors
of four.  Experimental values are show with error
bars.}
\label{fg1}
\end{figure}

I next address concerns over fine-tuning
in the unification scenario considered here.
Ghilencea and Ross have recently argued
that a realistic string model should not
disturb the ``significance of the prediction
for the gauge couplings'' which
occurs in the MSSM \cite{GR01}.
They note that for reasonable values of
$\LamSB$, the portion of the
$\alpha_3(m_Z)$ versus $\sin^2 \theta_W (m_Z)$
plane allowed by conventional MSSM unification
is a very small strip.  We can rewrite
Eq.~\myref{5q} as an implicit equation
$d_3=d_3(d_Y,d_2,t_3)$, so that for fixed
value of the exoquark scale, and thereby
$t_3$, we can predict $\alpha_3(m_Z)$ 
as a function of $\sin^2 \theta_W (m_Z)$.
In Figure~\ref{fg1} I show my results
for values of $\Lambda_3$ which step by
a factor of four; I assume $\LamSB=1$ TeV
for these (solid) curves.
For comparison I also show the MSSM
unification predictions (dashed) 
with $\LamSB$ stepping
by factors of four; in the
MSSM case I take $k_Y=5/3$
and assume threshold corrections
\beq
\Delta_a^{\stxt{MSSM}} \approx
\Delta_a^{\stxt{conv}} +
\Delta_a^{\stxt{HL}} +
\Delta_a^{\stxt{light}},
\qquad a=Y,2,3,
\eeq
where each of the quantities 
on the right-hand side are assumed
as above.  I also show with error bars
the experimental values \myref{4b}.
The experimental error bars define the
major and minor axes of an ``error ellipse.''
In any give direction, the distance from
the center of this ellipse to its edge
gives a measure which is independent
of how we scale the axes of the graph. 
We compare the widths of strips to those
of the MSSM in these units.
It can be seen
that the sensitivity to
$\Lambda_3$ is only a factor of
approximately three greater
than the sensitivity to
$\LamSB$ in the MSSM.  Roughly
speaking, the tuning is not much worse than
in the MSSM.  Another way to see that the
tuning is not ``fine''
is that deviations of up to roughly 60\%
in $\Lambda_3$
from the central value given in \myref{5m}
can be accomodated by the uncertainty in
$\icoup{3}(m_Z)$.
It is also interesting to note that
setting the scale $\Lambda_3$ is equivalent
to predicting $\icoup{3}(m_Z)$, since the
(solid) curves
in Figure~\ref{fg1}
are nearly horizontal; this is
reflected in that fact that uncertainty
in $\swZ$ had no
appreciable effects on the estimates
of $\Lambda_3$ in Eqs.~(\ref{5k},\ref{5m}).

\begin{figure}[p]
\begin{center}
\includegraphics[height=6.0in,width=5.0in,angle=90]{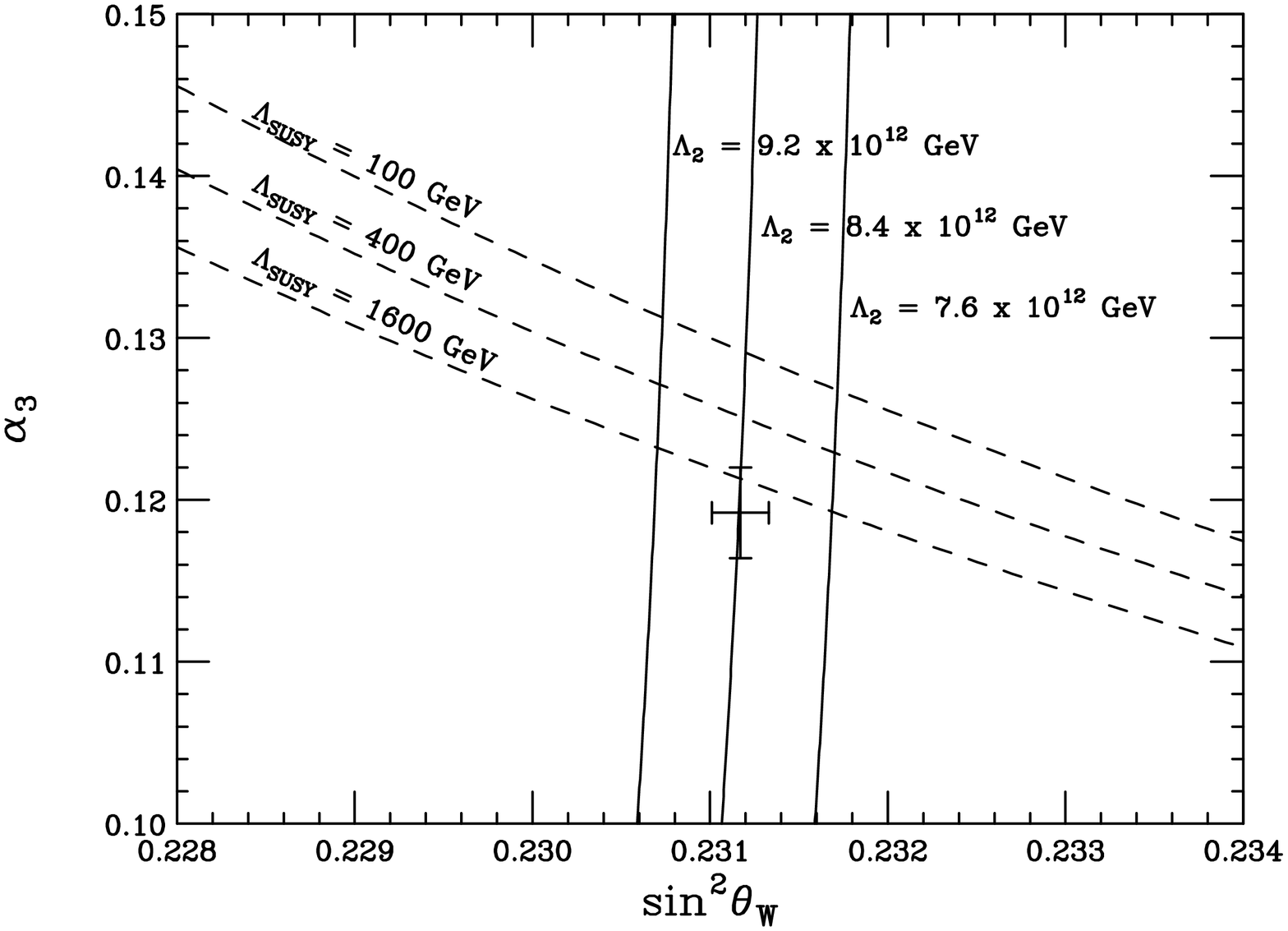}
\end{center}
\caption{Predicted Z scale values per the string
unification scenario (solid), for values of $\Lambda_2$
stepping by $\pm$10\% from the best fit value, with $\LamSB=1$ TeV.
For comparison, the MSSM unification prediction
is shown (dashed), with $\LamSB$ stepping by factors
of four.  Experimental values are show with error
bars.}
\label{fg2}
\end{figure}

In Figure~\ref{fg2} I present a similar analysis
for $\Lambda_2$, the exolepton scale.  
I fix $t_2$ and solve
Eqs.~(\ref{u22}-\ref{u24}) numerically 
eliminating $t_3$ and $a_H$ to obtain
$d_3=d_3(d_Y,d_2,t_2)$.  
For a given value of $t_2$ we
obtain a curve for $\alpha_3(m_Z)$ 
as a function of $\sin^2 \theta_W (m_Z)$;
I take $\LamSB=1$ TeV.
The sensitivity to the exolepton
scale is {\it much} higher,
so I only step by $\pm$10\% from
$\Lambda_2 = 8.4 \times 10^{12}$ GeV,
the approximate central value of \myref{5m}.
I compare
the widths of the strips to those of
the MSSM unification as describe
above.  It can be seen that
they are roughly three times wider, implying that a
10\% variation of $\Lambda_3$
in the string unification scenario
studied here is on a par with a 1200\% variation
of $\LamSB$ in the MSSM unification scenario.
That is, sensitivity to the exolepton scale is
roughly 120 times worse than the $\LamSB$
sensitivity of the MSSM.
From \myref{5m}
we note that deviations of up to
3.5\% for $\Lambda_2$ from the central
value can be accomodated by the uncertainty
in $\swZ$.  Although this tuning
is ``fine,'' it is not horrendous.
The vertical (solid) curves in Figure~\ref{fg2}
demonstrate that choosing $\Lambda_2$ is
essentially equivalent to predicting $\swZ$;
this is reflected in \myref{5m} by the fact
that $\Lambda_2$ has no significant
sensitivity to the uncertainty in $\icoup{3}(m_Z)$.

To summarize, relative to the tuning of superpartner
thresholds in the MSSM unification scenario, the
the exoquark scale is {\it not} finely-tuned, but the
exolepton scale {\it is} finely-tuned; however, the
fine-tuning of the exolepton scale is not astronomical
and is perhaps acceptable.  If one is prepared to
accept a tuning 120 times worse than the tuning of
$\LamSB$ in the MSSM, then one still must explain
{\it why} the exotic scales have the order of
magnitudes that they do.  Presumably, this would be
determined by a detailed study of the flat directions
which produce Xiggs vevs and the selection rules which
restrict couplings in the effective theory.  If the
leading couplings giving exoquarks mass were of
high enough dimension, a natural explanation of the low
exoquark scale may be possible; the exolepton scale
may be easier to explain because it is near the
condensation scale.

Using our results for the scales $\Lambda_{2,3}$, we can extract the
range of exotic thresholds corrections $\Delta_a^{\stxt{exotic}}$
which are required:
\beq
9 \lappeq \Delta_Y^{\stxt{exotic}} \lappeq 10, \qquad
120 \lappeq \Delta_2^{\stxt{exotic}} \lappeq 130, \qquad
150 \lappeq \Delta_3^{\stxt{exotic}} \lappeq 160.
\eeq
Comparing to (\ref{5h},\ref{5i}), it can be seen that the
exotic threshold corrections for $\icoup{2}$ and $\icoup{3}$ are
quite large compared to other effects; they represent
roughly 35\% and 150\% corrections to $4 \pi \icoup{2}$
and $4 \pi \icoup{3}$ respectively!  However, the
hypercharge correction is fairly modest (0.8\%).  
(To a good approximation, we could have neglected
the $\Delta^0_a$ of Eq.~\myref{5u} and solved for the
order of magnitude of the exotic
threshold corrections.)
This can
be traced to the fact that the exoquarks and exoleptons
which we have introduced at $\Lambda_3$ and $\Lambda_2$
have very small hypercharges.  This is precisely what
is needed to overcome the nonstandard hypercharge normalization.
It can be seen from \myref{rgeY} that as $k_Y$ is increased
above its standard value, the prediction for $\icoup{H}$
will tend to decrease, all other quantities being held constant
and ignoring the constraints (\ref{ssc},\ref{rge23}).
We can correct for this tendancy by making
$\Delta_2$ and $\Delta_3$ significantly larger
than what is typical in the MSSM, so long as we
do not greatly change $\Delta_Y$.  This is possible
because we have exoquarks and exoleptons with very
small hypercharge.

The bizarre hypercharges of the exotic particles
lead to fractionally charged particles;
the most problematic are the exoquarks, given
the rather low value of $\Lambda_3$.
Thermal production of exoquarks or exoleptons
at an early stage of the universe would violate
relic abundance bounds on fractionally charged
particles (FECs) by several orders of
magnitude, as discussed for example in
Refs.~\cite{AADF88,CCF96,Per01}.  
Thus, viabilty of this unification
scenario requires inflation, to dilute the abundances
of FECs, with a reheating temperature $T_R$ which
is sufficiently low that the FECs will not
be appreciably produced following inflation;
such scenarios have been examined
for example in free fermionic models \cite{CCF96}.
Chung, Kolb and Riotto \cite{CKR98} have
recently pointed out that the dilution
of heavy particle abundances
by inflation imposes
a much stronger limit than was initially
imagined:  to avoid thermal production of
heavy particles with $G_{SM}$ gauge quantum
numbers, the masses of these heavy particles
must be greater than $T_R$ by a factor
of roughly $10^3$.  Then to escape conflict with the
relic density data for fractionally charged
particles, we require inflation with
\beq
T_R \lappeq 10^{-3} \Lambda_3 \lappeq  5 \mtxt{TeV}.
\eeq
While inflationary scenarios with such low
reheating temperatures have certainly been
proposed (see for example Ref.~\cite{GRS00}),
it is not at all clear that
such scenarios can be achieved in the present
context.  I will not address this question
here, leaving it to further investigation.

\chapter*{Conclusions}
\typeout{Conclusions}
\addcontentsline{toc}{chapter}{Conclusions}
The first two chapters were entirely review
and require little comment, except that it is
my hope that they prove useful to the orbifold
or string novice.  In particular, the detailed
descriptions of the classical-to-quantum transition
and Hilbert space construction were unlike that which
usually appears in the literature; I presented
this material in a manner much closer to elementary
discussions of quantum mechanical systems, and spelled
out projections and boundary conditions which are usually phrased only in
a classical language.  The three chapters which
followed represent the bulk of the original
work which I performed for this thesis.  

Allow me to summarize the tasks which I accomplished,
as described in this report.

\bfe{Enumeration.}  In a significant extension of
initial efforts by CMM \cite{CMM89}, I have enumerated
a complete set of embedings for the \bsa\ class.
This list was greatly reduced through the
exploitation of equivalence relations,
in an automated analysis.

\bfe{Tabulation.}  This has been systematic
and detailed, focused on properties of
\bsa\ class models.  In my dissertation
research, the calculation and
tabulation aspect involved the most effort;
I composed over thirty thousand lines 
of computer code for the automated analysis.
As a result of this work, I can legitimately say what
is typical in this class of models.  
For each model, I have determined a number of
quantities which are useful for phenomenological
studies.

I have determined the hidden sector
gauge group $G_H$ and the representations of matter
charged under the nonabelian part of
$G_H$.  These details are key to predicting
the superpartner spectrum and couplings
from supergravity effective lagrangians
with hidden sector dynamical supersymmetry
breaking.  Superpartner masses and
couplings are the aspects of supersymmetric
extensions to the SM which are constrained
by existing experimental data and have the
potential to be measured at forthcoming experiments.

I have listed all of the patterns of irreps
under the nonabelian factors of $G$.  Using these
results, one can easily select a model from
the \bsa\ class on the basis of exotic matter
content.
The tables of irreps also suggest topics for
further study, such as gauge mediation of
hidden sector supersymmetry breaking, due to
mixed representations of the observable
and hidden sector gauge groups.

The FI terms in Table~\ref{fir} allow one to determine
the scale of initial gauge symmetry breaking.
Because many of the
low energy effective operators have coefficients
which at leading order depend on large powers of
the Xiggs vevs, $\ordnt{1}$ variations in the FI
term can be greatly enhanced.  For this reason,
an accurate determination of the FI term is
of practical interest.  Indeed, such operators
typically yield the leading order Yukawa interactions
for SM particles; ample experimental data on
the masses and mixings of most SM particles
exists, placing strong constraints on
these effective Yukawa couplings.

Table~\ref{tbgs} gives
the Green-Schwarz coefficient $\bGS$ for
each model.
I have applied these
results to the T-moduli mass formula
\myref{mtm} as an example of how this
quantity plays a prominent role
in the BGW effective theory.
It was found that this
may imply a problem of too light T-moduli masses
in the \bsa\ class.\myfoot{However, I need to address
the correction the MK has found in her D-term notes.}

The minimum hypercharge normalization $k_Y$
(consistent with accomodation of the MSSM and
at least $SU(3)'$ surviving in the hidden
sector---to provide for gaugino condensation) was
determined for each model.  If one is determined
to obtain the standard normalization
$k_Y=5/3$, Table~\ref{mkytab}
spares effort on fruitless models where this
is not possible---over half of the 175 studied here.
I am able to conclude that extended hypercharge
embeddings allow for $k_Y < 5/3$ in some of
the models, similar to what was found for
free fermionic models in Ref.~\cite{CHL96}.
However, it is not possible
to obtain small enough $k_Y$, in the range of
$1.4$ to $1.5$, to achieve string scale
unification with only the
MSSM field content---a string
unification scenario studied in Refs.~\cite{Iba93,DFMR96}
and reviewed in \cite{Die97}.

All of the quantities tabulated here are necessary to
detailed model-building in the effective
supergravity theory and have implications for
soft terms in the MSSM and the unification
of running gauge couplings.  To my knowledge,
this is the first complete and systematic survey of
a large class of
three generation standard-like bosonic
heterotic orbifold models performed {\it at
this level of detail.}

\bfe{Organization.}  By organizing the models into twenty patterns
of irreps and enumerating various other
properties which are universal to models
within a given pattern, I allow the
phenomenologist to quickly select a subset
of the models within the \bsa\ class which
have the desired properties (or perhaps to
conclude that no model in this class
suits her or his tastes).  It is an interesting
result that so many of the features of the various
models within an irrep pattern are universal.

\bfe{Cross-referencing.}  The embeddings enumerated
in Appendix \ref{emt} are each identified with one of the
twenty patterns, using Table \ref{tb1}.
One can employ the recipes provided in Section \ref{rec}
to quickly generate the matter spectra
for a given model, without a detailed understanding
of the underlying theory; alternatively,
full tables of all 175 models are available
from the author upon request.  It is hoped
that through these efforts the \bsa\ class
of string-derived models has been rendered
amenable to further study by a wider audience.


The unusual features of string-derived models,
charge fractionalization and nonstandard hypercharge
normalization, have been discussed in the simplest
of terms.  I have endeavored to make clear
as is possible how it is that states occur
which would not be discovered through
straightforward dimensional reduction
and irrep decompositions of the original
ten-dimensional \eetee\ theory.  I have
discussed at length the problems which
these features present for the construction
of a phenomenologically viable model.
I have described the size of Xiggs vevs in
general terms, and have found that large
T-moduli vevs would seem to spoil perturbativity
of the $\s$ model expansion of the effective
theory.

In an example model
where nonstandard hypercharge 
normalization cannot be avoided,
I have described the
lengths to which one must go in order to
achieve unification at the string scale.
Exotic matter states with very small
hypercharges were introduced at intermediate
scales to obtain agreement with Z scale
data for the gauge couplings.  The
exoquark scale was found to be rather
low.  The exotic hypercharges of the 
exotic matter in turn implied a low
reheating temperature to avoid problems
with FEC relic abundance constraints.
Fine-tuning of the intermediate scales
was examined and was shown to be,
in my opinion, rather mild.  However, I did not
study flat directions and superpotential couplings
in the example model,
and for this reason the intermediate scales
and intermediate field content remain to
be justified.

To defend the unification scenario
presented in Section~\ref{gcu}, 
one must be willing to take the position that
the apparent unification at roughly $2 \times
10^{16}$ GeV in the MSSM with $k_Y=5/3$ is
purely accidental; I find this
point of view difficult to accept.
On the other hand, the unification scenario
I have studied serves as
an illustration of how ugly things really
are when one attempts to refine many
of the models into a realistic theory.
Though I have studied only one example,
it can be seen from Table~\ref{mkytab}
that a good fraction of the \bsa\ class
models have $k_Y > 5/3$ and the unification
constraint in these models leads inevitably
to the contortions exhibited in our example.

In conclusion, the more promising models
will be those with $k_Y \leq 5/3$.  One
might invoke M-theory \cite{Mth}
to explain unification
at $2 \times 10^{16}$, as was done in Refs.~\cite{CFN99};
or, one might introduce {\it many} exotics
at a intermediate scale with a ``just so''
arrangement of irreps and charges in the hope that
with enough exotics the intermediate scale would
quite near the unification scale of the MSSM
and the apparent approximate unification
at $2 \times 10^{16}$ would not be an accident.
In either case, the enumeration and classification performed
here has moved the effort further along 
for the \bsa\ class and has
narrowed down the number of ``attractive
models.''

\appendix

%

\chapter{Proof of Three Equivalences}
\label{erp}
The equivalences \myref{cmmiss} were uncovered using the
automated routines developed for the analysis of hidden
sector embeddings; any further equivalences
between the observable sector embeddings of CMM would require
four or more half-integral Weyl reflections, transformations
which were not studied for reasons explained
above.  Because the equivalences
\myref{cmmiss} are a significant revision to the results
of Ref.~\cite{CMM89}, I have chosen to explicitly
demonstrate them in this appendix.
In addition to operations \opI\ and \opII\ used in the main text,
I make use of two redefinitions of the Wilson lines which give
equivalent embeddings (cf.~Ref.~\cite{CMM89}):
\beq
a_1 \to a_1' = -a_1 - a_3, \qquad
a_3 \to a_3' = a_1 - a_3 ;
\label{B1}
\eeq
\beq
a_1 \to a_1' = a_1 - a_3, \qquad
a_3 \to a_3' = a_1 + a_3 .
\label{B2}
\eeq
In what follows I will ignore the hidden sector embedding
vectors, since in the end I complete the observable
sector embeddings with all consistent choices.

First consider CMM 3, as given in Table \ref{tabcmm}.  We Weyl
reflect (operation \opII) by
$e=\half (1,1,-1,-1,-1,1,1,-1)$ to obtain
\beqa
3V_A & \to & 3V_A' = (-1,-1,0,0,0,2,0,0), \nnn
3a_{1A} & \to & 3a_{1A}' = \half (-1,-1,1,1,1,-3,-1,3), \nnn
3a_{3A} & \to & 3a_{3A}' = \half (-1,-1,1,1,1,3,1,3).
\label{B3}
\eeqa
Application of \myref{B1} yields
\beqa
3a_{1A}' & \to & 3a_{1A}'' = -3a_{1A}' - 3a_{3A}'
= (1,1,-1,-1,-1,0,0,-3), \nnn
3a_{3A}' & \to & 3a_{3A}'' = 3a_{1A}' - 3a_{3A}'
= (0,0,0,0,0,-3,-1,0).
\eeqa
Finally, we employ operation \opI\ to shift
\beq
3a_{1A}'' \to 3a_{1A}''' = 3a_{1A}'' + 3 \ell_1,
\qquad
3a_{3A}'' \to 3a_{3A}''' = 3a_{3A}'' + 3 \ell_3,
\label{B5}
\eeq
where
\beq
3 \ell_1 = (0,0,0,0,3,0,0,3), \qquad
3 \ell_3 = (0,0,0,0,0,3,3,0),
\eeq
to obtain
\beq
3 a_{1A}''' = (1,1,-1,-1,2,0,0,0), \qquad
3 a_{3A}''' = (0,0,0,0,0,0,2,0).
\eeq
With $V_A'$ as given in \myref{B3}, one can see by
comparison to Table \ref{tabcmm} that
$\{ V_A', a_{1A}''', a_{3A}''' \}$ is precisely
the observable sector embedding of CMM 1;
thus, we have shown the first equivalence of
\myref{cmmiss}.

Next consider CMM 5.  We Weyl reflect by
$e = \half (1,1,-1,-1,-1,1,-1,1)$ to obtain
\beqa
3V_A & \to & 3V_A' = (-1,-1,0,0,0,2,0,0), \nnn
3a_{1A} & \to & 3a_{1A}' = \half (1,1,-1,-1,-1,-3,3,-3), \nnn
3a_{3A} & \to & 3a_{3A}' = \half (-1,-1,1,1,1,3,3,1).
\eeqa
Application of \myref{B2} yields
\beqa
3a_{1A}' & \to & 3a_{1A}'' = 3a_{1A}' - 3a_{3A}'
= (1,1,-1,-1,-1,-3,0,-2), \nnn
3a_{3A}' & \to & 3a_{3A}'' = 3a_{1A}' + 3a_{3A}'
= (0,0,0,0,0,0,3,-1).
\eeqa
Shifting as in \myref{B5}, but with
\beq
3 \ell_1 = (0,0,0,0,0,3,0,3), \qquad
3 \ell_3 = (0,0,0,0,0,0,-3,3),
\eeq
we obtain
\beq
3 a_{1A}''' = (1,1,-1,-1,-1,0,0,1), \qquad
3 a_{3A}''' = (0,0,0,0,0,0,0,2).
\eeq
Performing a Weyl reflection of
$\{ V_A', a_{1A}''', a_{3A}''' \}$ by
the root $e' = (0,0,0,0,0,0,1,-1)$ interchanges
entries seven and eight of each embedding
vector:
\beqa
3V_A' & \to & 3V_A'' = (-1,-1,0,0,0,2,0,0), \nnn
3a_{1A}''' & \to & 3a_{1A}'''' = (1,1,-1,-1,-1,0,1,0), \nnn
3a_{3A}''' & \to & 3a_{3A}'''' = (0,0,0,0,0,0,2,0).
\eeqa
Comparing to Table \ref{tabcmm}, we see that
$\{ V_A'', a_{1A}'''', a_{3A}'''' \}$ is the
observable sector embedding of CMM~4; this
proves the second equivalence of \myref{cmmiss}.

Finally consider CMM 7.  Weyl reflection by
$e = \half (1,1,-1,-1,-1,1,1,-1)$ yields
\beqa
3V_A & \to & 3V_A' = (-1,-1,0,0,0,2,0,0), \nnn
3a_{1A} & \to & 3a_{1A}' = (-1,-1,1,1,1,0,-1,2), \nnn
3a_{3A} & \to & 3a_{3A}' = \half (-1,-1,1,1,1,-1,3,1).
\eeqa
Application of \myref{B2} gives
\beqa
3a_{1A}' & \to & 3a_{1A}'' = 3a_{1A}' - 3a_{3A}'
= \half (-1,-1,1,1,1,1,-5,3), \nnn
3a_{3A}' & \to & 3a_{3A}'' = 3a_{1A}' + 3a_{3A}'
= \half (-3,-3,3,3,3,-1,1,5).
\eeqa
We shift as in \myref{B5}, but with
\beq
3 \ell_1 = \half (3,3,-3,-3,-3,3,3,-3), \qquad
3 \ell_3 = \half (3,3,-3,-3,-3,3,-3,-9),
\eeq
to obtain
\beq
3 a_{1A}''' = (1,1,-1,-1,-1,2,-1,0), \qquad
3 a_{3A}''' = (0,0,0,0,0,1,-1,-2).
\eeq
Weyl reflection of
$\{ V_A', a_{1A}''', a_{3A}''' \}$ by
$e' = (0,0,0,0,0,0,1,-1)$ then
$e'' = (0,0,0,0,0,0,1,1)$ flips the signs of
entries seven and eight of each embedding
vector, yielding
\beqa
3V_A' & \to & 3V_A'' = (-1,-1,0,0,0,2,0,0), \nnn
3a_{1A}''' & \to & 3a_{1A}'''' = (1,1,-1,-1,-1,2,1,0), \nnn
3a_{3A}''' & \to & 3a_{3A}'''' = (0,0,0,0,0,1,1,2).
\eeqa
Comparing to Table \ref{tabcmm}, we see that
$\{ V_A'', a_{1A}'''', a_{3A}'''' \}$ is the
observable sector embedding of CMM~6; this
demonstrates the third equivalence of \myref{cmmiss}.

\chapter{Anomaly Cancelation}
\section{Cancellation of the Modular Anomaly}
\label{cma}
For the $Z_3$ orbifold, $SU(3,3,\Zbf)$
reparameterizations of the nine T-moduli $T^{ij}$
are symmetries \cite{z3dual}
of the underlying perturbative string theory,
at least to one loop in string perturbation theory
\cite{TDTAF,Nothr}.
These are referred to as {\it target space modular transformations}
or {\it duality transformations} of the T-moduli.
Most commonly, projective $SL(2,\Zbf)$ subgroups acting
on the diagonal moduli are studied:
\beq
T^i \to {a^i T^i-i b^i \over ic^i T^i+d^i}, \qquad
a^i d^i - b^i c^i=1, \qquad \forall \; i=1,3,5,
\label{mtr}
\eeq
with $a^i,b^i,c^i,d^i$ all integers.  The indices on
these integers indicate that each of the three $T^i$
may transform with its own set.  In addition to
transformations on the T-moduli, accompanying T-dependent
reparameterizations of chiral matter superfields
must be made:
\beq
\Phi^A \to {\sum_B M^A_{\spc B} \Phi^B \over
\prod_{i=1,3,5} (ic^i T^i + d^i)^{q_i^A} } .
\label{6a}
\eeq
Here, $q_i^B$ is the modular weight of the field
$\Phi^B$; these quantities were
given in Section~\ref{mds}.
The matrix $M^A_{\spc B}$ is identity
for untwisted fields while it mixes 
subsets of twisted fields 
with the same modular weight \cite{TWMIX}
in a way which depends on the parameters
$a^i,b^i,c^i,d^i$.

Transformations (\ref{mtr},\ref{6a}) are symmetries of the
effective supergravity action at the classical 
level---isometries of the nonlinear $\s$ model.
However, at the quantum level there is a
$\s$ model anomaly \cite{NLSA} associated with the
duality tranformations, as originally
pointed in Refs.~\cite{DFKZ91,LCO92}.
To study this {\it modular anomaly}, 
one calculates the quantum corrections to the supergravity
lagrangian, in particular triangle diagrams involving
the composite $\s$ model connections of T-moduli to other
fields at one vertex and gauge boson currents at
the other two vertices.  Various calculations of the
modular anomaly have been performed.  Most often, supergravity
interactions have been studied at the component
level and then the anomaly written as a globally
supersymmetric
superspace integral, which is an approximation to
the true supergravity anomaly~\cite{DFKZ91,LCO92,KL945,Lou91,GT92}.
The supergravity one loop effective lagrangian
and its transformation properties has been studied
in great detail by Gaillard and collaborators, using Pauli-Villars
regularization techniques \cite{PVREG}.  These calculations
were recently used to infer a
locally supersymmetric superspace
expression for the anomaly at one loop \cite{GNW99}.
Equivalent expressions have also been obtained
in Ref.~\cite{BMP00}.
Keeping only the leading term important to the
present analysis, the quantum part of the one loop effective
supergravity lagrangian transforms under \myref{mtr} as
\beq
\delta {\cal L}_Q
= \sum_{a,j} {\alpha_a^j \over 64 \pi^2}
\int d^4\theta {E \over R} \ln (ic^j T^j + d^j) 
\sum_i(\W^\alpha \W_\alpha)_a^i + \mtxt{h.c.}
\label{man}
\eeq
The expression on the right-hand side is a superspace
integral in the K\"ahler $U(1)$ formulation of supergravity
\cite{BGGM87,BGG91,BGG01}.  The quantity $E$ is the superdeterminant
of the {\it vielbein;} it generalizes the tensor density
$e = \sqrt{g}$ which appears in the Einstein-Hilbert action
to a superfield.  The superfield $R$ is
chiral and has as its lowest component the
scalar auxilliary field of supergravity.  The chiral
spinor superfield $\W_{\alpha,a}^i$ is the superfield-strength
corresponding to the generator $T_a^i$
of the factor $G_a$ of the gauge group $G$ and has
as its lowest component the gaugino $\lambda_{\alpha,a}^i$.
The coefficient $\alpha_a^j$ reflects particles 
in the triangle loop which
contribute to the anomalous transformation, 
and is given by \cite{Lou91,GT92}
\beq
\alpha_a^j = - C(G_a) + \sum_A (1 - 2 q_j^A) X_a(R^A) ,
\label{A6}
\eeq
where the sum runs over matter
irreps $R^A$ of $G_a$ and $q_j^A$ is
the modular weight appearing in \myref{6a}.

Since the transformations (\ref{mtr},\ref{6a})
are known to be anomaly free in 
the underlying four-dimensional
string theory, we must add effective
terms to cancel the anomaly.
One possible cancellation is from the
shift in the T-moduli dependent threshold
corrections alluded to in Sections~\ref{mds} and~\ref{gcu}.
As mentioned there, however, such
threshold corrections are absent in $Z_3$ orbifold
compactifications \cite{Nothr}.
Thus, the entire modular anomaly given by \myref{man}
must be canceled by the Green-Schwarz mechanism.
That is, we include in the effective supergravity
lagrangian a term which will have an anomalous
transformation under (\ref{mtr},\ref{6a}),
just such as to cancel \myref{man}.
The overall coefficient 
$\bGS$ of the Green-Schwarz term is determined
by this matching.  

I now describe this
term in the BGW effective theory.
However, I note that
in expressions below,
I use a slightly different normalization
for the Green-Schwarz coefficient $\bGS$ than
BGW; rather, I adopt the more common
convention of Refs.~\cite{IL92,BIM}.
In the BGW notation, the Green-Schwarz
coefficient is written as $b$, which is related to
$\bGS$ by the equation $b = - \bGS / 24 \pi^2$.
In addition, in my formulae I do not use the BGW
conventions for the $\beta$ function coefficients of the
gauge groups.  The two
conventions are related by
$b_a^{\stxt{BGW}} = - b_a^{\stxt{here}}/ 24 \pi^2$.

In addition to the supergravity
multiplet, gauge multiplets, and matter multiplets,
string theory predicts the existence of other supermultiplets
of dynamic states.  One particularly important set of fields
is the following:  a real scalar field $\ell$ called
the ${\it dilaton}$, an antisymmetric tensor $B_{mn}$
whose field strength is dual to the universal
axion, and a Majorana spinor 
$\varphi$ which is referred to as the
${\it dilatino}$.  This is on-shell content of the superfield
$L$, which is a {\it linear} multiplet.  It satisfies
the modified linearity condition \cite{BGG91,ABGG93,GG99,BGG01}
\beq
(\Db^2 + 8 R) L = - \sum_{a,i}(\W^\alpha \W_\alpha)_a^i.
\label{6b}
\eeq
Following BGW,
I write the Green-Schwarz counterterm for the modular anomaly as
\beq
{\cal L}_{GS} = {\bGS \over 24 \pi^2} \int d^4\theta \;
E L \sum_j \ln (T^j + \bar T^j) .
\label{gst}
\eeq
Using \myref{mtr},
integration by parts in superspace \cite{Zum79}, 
chirality of $T^j$
and the modified linearity condition \myref{6b}, 
\beqa
\delta {\cal L}_{GS} & = & {- \bGS \over 24 \pi^2}
\int d^4\theta E L \sum_j \ln (ic^j T^j + d^j) + \mtxt{h.c.} \nnn
& = & { \bGS \over 8 \cdot 24 \pi^2}
\int d^4\theta {E \over R}
(\Db^2 + 8 R)[ L \sum_j \ln (ic^j T^j + d^j)] + \mtxt{h.c.} \nnn
& = & { - \bGS \over 192 \pi^2} \sum_{j,a}
\int d^4\theta {E \over R}
\ln (ic^j T^j + d^j) \sum_i (\W^\alpha \W_\alpha)_a^i + \mtxt{h.c.}
\eeqa
Comparing to \myref{man},
it is easy to see that in the present context (i.e.,
in the absence of T-moduli dependent string threshold corrections),
\beq
\bGS = 3 \alpha_a^j \qquad \forall \; a,j.
\label{gsc2}
\eeq
A generic spectrum of massless states which is
free of chiral gauge anomalies will not satisfy
\myref{gsc2}, since it requires that we get the
same result, $\bGS$, for each factor $G_a$ in the gauge
group $G$.  Thus, \myref{gsc2} is a highly nontrivial
constraint on the matter spectrum.  This was
exploited by Ib\'a\~nez and L\"ust to draw a
number of phenomenological conclusions
for $Z_3$ orbifold models \cite {IL92}.

As discussed in Section~\ref{mss},
untwisted states come in families of three;
we make explicit the family index
$i=1,3,5$ by taking $A \to (\alpha,i)$ for
untwisted fields, so that $\alpha$ denotes the
species of untwisted field.  For the twisted
fields we take $A \to \rho$ to distinguish them,
but do not separate out the family label.
For nonabelian factors $G_a$ in the models considered
here, a nice simplification can be made.  
As mentioned in Section~\ref{mds}, none of
the pseudo-massless twisted 
fields which are in nontrivial representations
of $G_a$ are oscillator states.  Consequently,
it follows from the discussion of Section~\ref{mds} that
they all have modular weights $q_j^\rho = 2/3$.
Also from Section~\ref{mds}, we have for the
untwisted states $q^{(\alpha,i)}_j = \delta_j^i$.
With these
facts, it is easy to show that
Eqs.~(\ref{A6},\ref{gsc2}) can be rewritten
\beq
\bGS = - 3 C_a 
+ \sum_{(\alpha,i) \in \stxt{untw}} X_a(R^{(\alpha,i)})
- \sum_{\rho\in \stxt{tw}} X_a(R^\rho)
= \tbeta{a} - 2 \sum_{\rho\in \stxt{tw}} X_a(R^\rho),
\label{A2}
\eeq
where the last equality follows from 
\myref{5a},
only now it is the total $\beta$ function coefficient
which appears, since all pseudo-massless
states contribute.
In the absence of twisted states in nontrivial
irreps of $G_a$, the last term on the right-hand
side vanishes.  This occurs for $SO(10)$ in
Patterns 1.1 and 1.2.  But then for a $G_a$ with
only trivial irreps in the twisted sector $\bGS = \tbeta{a}$.
This is the source of the (approximately vanishing)
T-moduli mass problem
discussed in Section~\ref{mds} and Ref.~\cite{GG00}.

As an example of the surprising matching
of \myref{A2} for different $G_a$, 
we examine Pattern~1.1.
The $SO(10)$ factor of $G$ has no nontrivial matter representations,
as can be seen from Table~\ref{pt1}, which gives
\beq
\bGS = \tbeta{10} = -3 C(SO(10)) = -24 .
\label{A3}
\eeq
For the $SU(3)$ factor, we have $15(3 + \bar 3,1,1)$ beyond
the MSSM which gives 
$\dbth = \tbeta{3} - b_3 = 15$, and consequently
$\tbeta{3} = 12$.  Comparison
of Table~\ref{pt1} to Table~\ref{unpat} shows that
the twisted sector irreps are $15(3,1,1) + 21(\bar 3,1,1)$, which
gives
\beq
\bGS = \tbeta{3} - 2 \sum_{\rho \in \stxt{tw}} X_3(R^\rho)
= 12 - 36 = -24.
\label{A4}
\eeq
Finally, the $SU(2)$ factor has $40(1,2,1)$ beyond the
MSSM, so that $\dbtw = \tbeta{2} - b_2 = 20$ 
and $\tbeta{2}= 21$.
Again comparing Table~\ref{pt1} to Table~\ref{unpat},
we find that the $SU(2)$ charged twisted matter
is $45(1,2,1)$ and so
\beq
\bGS = \tbeta{2} - 2 \sum_{\rho \in \stxt{tw}} X_2(R^\rho)
= 21 - 45 = -24.
\label{A5}
\eeq
It is reassuring that the groups $SO(10)$, $SU(3)$ and $SU(2)$
give the same answer for $\bGS$, as they must for universal
cancellation of the modular anomaly \cite{IL92}.
As a nontrivial check on
my results, I have verified that 
this matching holds among the nonabelian
factors in each of the twenty patterns.

\section{Green-Schwarz Mechanism for ${\bf U(1)_X}$}
\label{gsx}
Under a gauge transformation,
the \ux\ vector superfield $V_X$ is shifted
by $\delta V_X = (1/2)(\Lambda + \bar \Lambda)$,
with $\Lambda$ a chiral superfield
and $\bar \Lambda$ the corresponding
anti-chiral superfield.  The one-loop
effective lagrangian has an anomalous transformation
which contains the terms
\beq
\delta {\cal L} = {1 \over 16 \pi^2} \sum_a \tr (T^a T^a \hat Q_X)
\left[ \Real \lambda F^a \cdot F^a + \Imag \lambda F^a \cdot \tilde F^a
\right] + \cdots
\eeq
where $\lambda = \Lambda |$.  We introduce
our counterterm as
\beq
{\cal L}_{GS,V_X} = \delta_X \int E L V_X
\label{gsvx}
\eeq
from which it follows that under the shift in $V_X$
\beq
\delta {\cal L}_{GS,V_X} = {\delta_X \over 2} \int E L (\Lambda + \bar \Lambda)
\eeq
which when we go to components yields
\beq
\delta {\cal L}_{GS,V_X} = - {\delta_X \over 8}
\sum_a \left( \Real \lambda F^a \cdot F^a
+ \Imag \lambda F^a \cdot \tilde F^a \right) + \cdots
\eeq
The anomaly is cancelled if we choose
\beq
\delta_X = {1 \over 2 \pi^2} \tr T^a T^a \hat Q_X.
\eeq
When combined with other terms in the lagrangian,
the component form of \myref{gsvx} gives
\beq
D_X = \sum_i K_i \hat q^X_i \phi^i + {\delta_X \over 2} \ell
\equiv \sum_i K_i \hat q^X_i \phi^i + \xi.
\eeq
From this, we see that the FI term $\xi$ is given
by
\beq
\xi = (2 \ell) {\delta_X \over 4} = {2 \ell \over 8 \pi^2}
\tr T^a T^a \hat Q_X.
\eeq
Further study of the anomalous transformation
of the one-loop lagrangian leads to the anomaly
matching condition (in the unified normalization)
\beq
24 \tr (T^a T^a \hat Q_X) = 8 \tr {\hat Q_X}^3 = \tr \hat Q_X .
\eeq
This gives
\beq
\xi = {2 \ell \over 192 \pi^2} \tr \hat Q_X ,
\eeq
which may be recognized as the form typically quoted
in the literature once it is realized that if we neglect
nonperturbative corrections to the K\"ahler potential
of the dilaton $\ell$, the universal coupling constant
at the string scale is given by $g_H^2 = 2 \ell$.

\chapter{Lengthy Tables}
\section{Embedding Tables}
\label{emt}
To construct the full sixteen-dimensional
embedding vectors $V,a_1$, and $a_3$,
simply take the direct
sum of a CMM observable sector embedding (labeled by
subscript $A$) and a hidden sector embedding (labeled
by subscript $B$) from a corresponding table:
\beq
V = (V_A;V_B), \qquad
a_1 = (a_{1A};a_{1B}), \qquad
a_3 = (a_{3A};a_{3B}).
\eeq
For instance, the observable sector embedding CMM~1
from Table~\ref{tabcmm} may be completed by any of the
embeddings in Table~\ref{tab1}.  Any other hidden
sector embedding which is consistent with CMM~1
will be equivalent to one of the choices given in
Table~\ref{tab1}.
It should be noted that CMM~8 and CMM~9 each allow
two inequivalent hidden sector twist embeddings
$V_B$; as a consequence, two hidden sector embedding
tables are given for each.  I have abbreviated
$G_H$ by the cases defined in Table \ref{gcs}.

{ \scriptsize
\begin{longtable}{|r|l|l|r||r|l|l|r|}
\caption{CMM 1, $3V_{B}$ = (2,1,1,0,0,0,0,0).\label{tab1}} \\
\hline
\# & \hspace{25pt} $3a_{1B}$ & \hspace{25pt} $3a_{3B}$ & $G_H$ &
\# & \hspace{25pt} $3a_{1B}$ & \hspace{25pt} $3a_{3B}$ & $G_H$ \\
\hline \hline \endfirsthead
\caption{(continued) CMM 1, $3V_{B}$ = (2,1,1,0,0,0,0,0).} \\
\hline
\# & \hspace{25pt} $3a_{1B}$ & \hspace{25pt} $3a_{3B}$ & $G_H$ &
\# & \hspace{25pt} $3a_{1B}$ & \hspace{25pt} $3a_{3B}$ & $G_H$ \\
\hline \hline \endhead
1 & (-2,0,0,0,0,0,0,0) & (0,1,-1,0,0,0,0,0) & 1 & 2 & (0,2,0,0,0,0,0,0) & (-1,0,-1,0,0,0,0,0) & 1 \\ \hline
3 & (0,2,0,0,0,0,0,0) & (1,0,1,0,0,0,0,0) & 1 & 4 & (-2,0,0,0,0,0,0,0) & (0,0,0,2,1,1,1,-1) & 2 \\ \hline
5 & (-2,0,0,0,0,0,0,0) & (0,0,0,2,1,1,1,1) & 2 & 6 & (-1,1,0,1,1,0,0,0) & (-1,-1,0,0,0,0,0,0) & 2 \\ \hline
7 & (-1,1,0,1,1,0,0,0) & (0,0,0,1,-1,0,0,0) & 2 & 8 & (-1,1,0,1,1,0,0,0) & (1,1,0,0,0,0,0,0) & 2 \\ \hline
9 & (0,1,1,1,1,0,0,0) & (0,0,0,1,-1,0,0,0) & 2 & 10 & (0,1,1,1,1,0,0,0) & (0,1,-1,0,0,0,0,0) & 2 \\ \hline
11 & (0,2,0,0,0,0,0,0) & (0,0,0,2,1,1,1,-1) & 2 & 12 & (0,2,0,0,0,0,0,0) & (0,0,0,2,1,1,1,1) & 2 \\ \hline
13 & (-2,0,0,0,0,0,0,0) & (0,0,0,1,1,0,0,0) & 3 & 14 & (-1,1,0,1,1,0,0,0) & (-2,-1,-1,1,1,0,0,0) & 3 \\ \hline
15 & (-1,1,0,1,1,0,0,0) & (2,1,1,-1,-1,0,0,0) & 3 & 16 & (0,1,1,1,1,0,0,0) & (-2,-1,-1,1,1,0,0,0) & 3 \\ \hline
17 & (0,1,1,1,1,0,0,0) & (2,1,1,-1,-1,0,0,0) & 3 & 18 & (0,2,0,0,0,0,0,0) & (0,0,0,1,1,0,0,0) & 3 \\ \hline
19 & (-2,0,0,0,0,0,0,0) & (0,-1,-2,1,1,1,0,0) & 4 & 20 & (-1,1,0,1,1,0,0,0) & (-2,-1,-1,0,-1,1,0,0) & 4 \\ \hline
21 & (-1,1,0,1,1,0,0,0) & (-2,0,1,-1,-1,1,0,0) & 4 & 22 & (-1,1,0,1,1,0,0,0) & (-2,1,0,0,0,1,1,-1) & 4 \\ \hline
23 & (-1,1,0,1,1,0,0,0) & (-2,1,0,0,0,1,1,1) & 4 & 24 & (-1,1,0,1,1,0,0,0) & (2,1,1,1,0,1,0,0) & 4 \\ \hline
25 & (-1,1,0,1,1,0,0,0) & (2,-1,0,0,0,1,1,1) & 4 & 26 & (-1,1,0,1,1,0,0,0) & (-1,1,1,-1,-1,1,1,-1) & 4 \\ \hline
27 & (0,1,1,1,1,0,0,0) & (-2,-1,-1,0,-1,1,0,0) & 4 & 28 & (0,1,1,1,1,0,0,0) & (-2,1,0,1,1,1,0,0) & 4 \\ \hline
29 & (0,1,1,1,1,0,0,0) & (2,0,-1,-1,-1,1,0,0) & 4 & 30 & (0,1,1,1,1,0,0,0) & (2,1,1,1,0,1,0,0) & 4 \\ \hline
31 & (0,1,1,1,1,0,0,0) & (-1,1,1,-1,-1,1,1,-1) & 4 & 32 & (0,2,0,0,0,0,0,0) & (-2,0,1,1,1,1,0,0) & 4 \\ \hline
33 & (0,2,0,0,0,0,0,0) & (2,0,-1,1,1,1,0,0) & 4 &  & & & \\ \hline
\end{longtable}

\begin{longtable}{|r|l|l|r||r|l|l|r|}
\caption{CMM 2, $3V_{B}$ = (2,1,1,0,0,0,0,0).\label{tab2}} \\
\hline
\# & \hspace{25pt} $3a_{1B}$ & \hspace{25pt} $3a_{3B}$ & $G_H$ &
\# & \hspace{25pt} $3a_{1B}$ & \hspace{25pt} $3a_{3B}$ & $G_H$ \\
\hline \hline \endfirsthead
\caption{(continued) CMM 2, $3V_{B}$ = (2,1,1,0,0,0,0,0).} \\
\hline
\# & \hspace{25pt} $3a_{1B}$ & \hspace{25pt} $3a_{3B}$ & $G_H$ &
\# & \hspace{25pt} $3a_{1B}$ & \hspace{25pt} $3a_{3B}$ & $G_H$ \\
\hline \hline \endhead
1 & (-2,0,-1,1,0,0,0,0) & (-1,0,-1,0,0,0,0,0) & 1 & 2 & (-2,0,-1,1,0,0,0,0) & (1,0,1,0,0,0,0,0) & 1 \\ \hline
3 & (-2,1,1,0,0,0,0,0) & (0,1,-1,0,0,0,0,0) & 1 & 4 & (-2,0,-1,1,0,0,0,0) & (-2,-1,-1,1,1,0,0,0) & 2 \\ \hline
5 & (-2,0,-1,1,0,0,0,0) & (2,1,1,-1,1,0,0,0) & 2 & 6 & (-2,1,1,0,0,0,0,0) & (0,0,0,2,1,1,1,-1) & 2 \\ \hline
7 & (-2,1,1,0,0,0,0,0) & (0,0,0,2,1,1,1,1) & 2 & 8 & (-1,0,0,1,1,1,1,-1) & (-1,1,1,1,1,1,1,-1) & 2 \\ \hline
9 & (-1,0,0,1,1,1,1,-1) & (0,1,-1,0,0,0,0,0) & 2 & 10 & (-1,1,-1,1,1,1,0,0) & (-1,-1,0,0,0,0,0,0) & 2 \\ \hline
11 & (-1,1,-1,1,1,1,0,0) & (0,0,0,0,0,0,1,1) & 2 & 12 & (-1,1,-1,1,1,1,0,0) & (1,1,0,0,0,0,0,0) & 2 \\ \hline
13 & (-2,0,-1,1,0,0,0,0) & (-1,1,1,-1,1,1,1,1) & 3 & 14 & (-2,1,1,0,0,0,0,0) & (0,0,0,1,1,0,0,0) & 3 \\ \hline
15 & (-1,0,0,1,1,1,1,-1) & (0,0,0,2,1,-1,-1,1) & 3 & 16 & (-1,0,0,1,1,1,1,-1) & (0,0,0,1,1,-1,-2,-1) & 3 \\ \hline
17 & (-1,1,-1,1,1,1,0,0) & (-2,-1,-1,0,-1,-1,0,0) & 3 & 18 & (-1,1,-1,1,1,1,0,0) & (2,1,1,1,1,0,0,0) & 3 \\ \hline
19 & (-2,0,-1,1,0,0,0,0) & (-2,0,1,0,1,1,1,0) & 4 & 20 & (-2,0,-1,1,0,0,0,0) & (-2,1,0,-1,1,1,0,0) & 4 \\ \hline
21 & (-2,1,1,0,0,0,0,0) & (0,-1,-2,1,1,1,0,0) & 4 & 22 & (-1,0,0,1,1,1,1,-1) & (-2,-1,-1,0,0,-1,-1,0) & 4 \\ \hline
23 & (-1,0,0,1,1,1,1,-1) & (2,1,1,1,0,0,0,-1) & 4 & 24 & (-1,0,0,1,1,1,1,-1) & (0,-1,-2,0,-1,-1,-1,0) & 4 \\ \hline
25 & (-1,0,0,1,1,1,1,-1) & (0,-1,-2,1,1,0,0,-1) & 4 & 26 & (-1,0,0,1,1,1,1,-1) & (0,0,0,0,0,0,-1,-1) & 4 \\ \hline
27 & (-1,1,-1,1,1,1,0,0) & (-2,-1,-1,1,0,0,1,0) & 4 & 28 & (-1,1,-1,1,1,1,0,0) & (-2,0,1,0,0,-1,1,-1) & 4 \\ \hline
29 & (-1,1,-1,1,1,1,0,0) & (-2,0,1,1,-1,-1,0,0) & 4 & 30 & (-1,1,-1,1,1,1,0,0) & (-2,1,0,-1,-1,-1,0,0) & 4 \\ \hline
31 & (-1,1,-1,1,1,1,0,0) & (-2,1,0,1,1,1,0,0) & 4 & 32 & (-1,1,-1,1,1,1,0,0) & (2,1,1,0,0,-1,1,0) & 4 \\ \hline
33 & (-1,1,-1,1,1,1,0,0) & (-1,1,1,1,-1,-1,1,-1) & 4 &  & & & \\ \hline
\end{longtable}

\begin{longtable}{|r|l|l|r||r|l|l|r|}
\caption{CMM 4, $3V_{B}$ = (2,1,1,0,0,0,0,0).\label{tab4}} \\
\hline
\# & \hspace{25pt} $3a_{1B}$ & \hspace{25pt} $3a_{3B}$ & $G_H$ &
\# & \hspace{25pt} $3a_{1B}$ & \hspace{25pt} $3a_{3B}$ & $G_H$ \\
\hline \hline \endfirsthead
\caption{(continued) CMM 4, $3V_{B}$ = (2,1,1,0,0,0,0,0).} \\
\hline
\# & \hspace{25pt} $3a_{1B}$ & \hspace{25pt} $3a_{3B}$ & $G_H$ &
\# & \hspace{25pt} $3a_{1B}$ & \hspace{25pt} $3a_{3B}$ & $G_H$ \\
\hline \hline \endhead
1 & (-2,1,-1,0,0,0,0,0) & (-1,-1,0,0,0,0,0,0) & 1 & 2 & (-2,1,-1,0,0,0,0,0) & (0,-1,1,0,0,0,0,0) & 1 \\ \hline
3 & (2,-1,-1,0,0,0,0,0) & (1,1,0,0,0,0,0,0) & 1 & 4 & (-2,0,0,1,1,0,0,0) & (-2,-1,-1,1,-1,0,0,0) & 2 \\ \hline
5 & (-2,0,0,1,1,0,0,0) & (1,1,0,0,0,0,0,0) & 2 & 6 & (-2,1,-1,0,0,0,0,0) & (1,-1,-1,1,1,1,1,-1) & 2 \\ \hline
7 & (-2,1,-1,0,0,0,0,0) & (1,-1,-1,1,1,1,1,1) & 2 & 8 & (-1,1,0,1,1,1,1,0) & (-1,0,-1,0,0,0,0,0) & 2 \\ \hline
9 & (-1,1,0,1,1,1,1,0) & (-1,1,1,-1,-1,-1,-1,1) & 2 & 10 & (-1,1,0,1,1,1,1,0) & (0,1,-1,0,0,0,0,0) & 2 \\ \hline
11 & (2,-1,-1,0,0,0,0,0) & (1,-1,-1,1,1,1,1,-1) & 2 & 12 & (2,-1,-1,0,0,0,0,0) & (1,-1,-1,1,1,1,1,1) & 2 \\ \hline
13 & (-2,0,0,1,1,0,0,0) & (2,1,1,1,1,0,0,0) & 3 & 14 & (-2,0,0,1,1,0,0,0) & (-1,1,1,1,1,1,1,-1) & 3 \\ \hline
15 & (-2,1,-1,0,0,0,0,0) & (-2,-1,-1,1,1,0,0,0) & 3 & 16 & (-1,1,0,1,1,1,1,0) & (2,1,1,0,0,0,-1,-1) & 3 \\ \hline
17 & (-1,1,0,1,1,1,1,0) & (-1,1,1,-1,-1,-1,-1,-1) & 3 & 18 & (2,-1,-1,0,0,0,0,0) & (-2,-1,-1,1,1,0,0,0) & 3 \\ \hline
19 & (-2,0,0,1,1,0,0,0) & (-2,-1,-1,0,0,1,1,0) & 4 & 20 & (-2,0,0,1,1,0,0,0) & (-2,1,0,0,0,1,1,-1) & 4 \\ \hline
21 & (-2,0,0,1,1,0,0,0) & (-2,1,0,0,0,1,1,1) & 4 & 22 & (-2,0,0,1,1,0,0,0) & (2,1,1,0,-1,1,0,0) & 4 \\ \hline
23 & (-2,0,0,1,1,0,0,0) & (0,0,0,1,0,1,0,0) & 4 & 24 & (-2,1,-1,0,0,0,0,0) & (2,-1,0,1,1,1,0,0) & 4 \\ \hline
25 & (-2,1,-1,0,0,0,0,0) & (0,-1,-2,1,1,1,0,0) & 4 & 26 & (-1,1,0,1,1,1,1,0) & (-2,-1,-1,1,0,0,-1,0) & 4 \\ \hline
27 & (-1,1,0,1,1,1,1,0) & (-2,0,1,1,1,0,0,-1) & 4 & 28 & (-1,1,0,1,1,1,1,0) & (-2,1,0,0,0,-1,-1,1) & 4 \\ \hline
29 & (-1,1,0,1,1,1,1,0) & (2,1,1,0,0,0,-1,1) & 4 & 30 & (-1,1,0,1,1,1,1,0) & (2,-1,0,0,0,-1,-1,1) & 4 \\ \hline
31 & (-1,1,0,1,1,1,1,0) & (2,0,-1,0,-1,-1,-1,0) & 4 & 32 & (-1,1,0,1,1,1,1,0) & (-1,1,1,1,1,1,-1,1) & 4 \\ \hline
33 & (2,-1,-1,0,0,0,0,0) & (-2,1,0,1,1,1,0,0) & 4 &  & & & \\ \hline
\end{longtable}

\begin{longtable}{|r|l|l|r||r|l|l|r|}
\caption{CMM 6, $3V_{B}$ = (2,1,1,0,0,0,0,0).\label{tab6}} \\
\hline
\# & \hspace{25pt} $3a_{1B}$ & \hspace{25pt} $3a_{3B}$ & $G_H$ &
\# & \hspace{25pt} $3a_{1B}$ & \hspace{25pt} $3a_{3B}$ & $G_H$ \\
\hline \hline \endfirsthead
\caption{(continued) CMM 6, $3V_{B}$ = (2,1,1,0,0,0,0,0).} \\
\hline
\# & \hspace{25pt} $3a_{1B}$ & \hspace{25pt} $3a_{3B}$ & $G_H$ &
\# & \hspace{25pt} $3a_{1B}$ & \hspace{25pt} $3a_{3B}$ & $G_H$ \\
\hline \hline \endhead
1 & (-1,0,0,1,0,0,0,0) & (-2,0,-1,1,0,0,0,0) & 1 & 2 & (0,1,0,1,0,0,0,0) & (-2,-1,0,1,0,0,0,0) & 1 \\ \hline
3 & (0,1,0,1,0,0,0,0) & (0,-1,2,1,0,0,0,0) & 1 & 4 & (-1,0,0,1,0,0,0,0) & (-1,0,0,-1,1,1,1,-1) & 2 \\ \hline
5 & (-1,0,0,1,0,0,0,0) & (2,0,0,-1,1,0,0,0) & 2 & 6 & (0,1,0,1,0,0,0,0) & (-2,0,-1,0,1,0,0,0) & 2 \\ \hline
7 & (0,1,0,1,0,0,0,0) & (0,-2,0,-1,1,0,0,0) & 2 & 8 & (-2,0,-1,1,1,1,0,0) & (-1,1,-1,1,1,1,0,0) & 2 \\ \hline
9 & (-2,0,-1,1,1,1,0,0) & (0,1,0,1,1,1,1,1) & 2 & 10 & (-2,0,-1,1,1,1,0,0) & (1,1,1,1,1,1,0,0) & 2 \\ \hline
11 & (-2,1,1,1,1,0,0,0) & (0,2,-1,0,-1,0,0,0) & 2 & 12 & (-2,1,1,1,1,0,0,0) & (0,0,-2,1,1,0,0,0) & 2 \\ \hline
13 & (-1,0,0,1,0,0,0,0) & (-1,0,0,-1,1,1,1,1) & 3 & 14 & (0,1,0,1,0,0,0,0) & (-1,-1,1,1,1,1,0,0) & 3 \\ \hline
15 & (-2,0,-1,1,1,1,0,0) & (-2,0,-1,1,0,0,0,0) & 3 & 16 & (-2,0,-1,1,1,1,0,0) & (-1,-1,1,1,-1,-1,0,0) & 3 \\ \hline
17 & (-2,1,1,1,1,0,0,0) & (-2,1,1,0,0,0,0,0) & 3 & 18 & (-2,1,1,1,1,0,0,0) & (2,0,0,-1,-1,0,0,0) & 3 \\ \hline
19 & (-1,0,0,1,0,0,0,0) & (-1,1,-1,-1,1,1,0,0) & 4 & 20 & (0,1,0,1,0,0,0,0) & (2,0,0,0,1,1,0,0) & 4 \\ \hline
21 & (0,1,0,1,0,0,0,0) & (-1,1,-1,-1,1,1,0,0) & 4 & 22 & (-2,0,-1,1,1,1,0,0) & (-2,-1,0,0,0,-1,0,0) & 4 \\ \hline
23 & (-2,0,-1,1,1,1,0,0) & (-2,1,1,0,0,0,0,0) & 4 & 24 & (-2,0,-1,1,1,1,0,0) & (-1,-1,1,0,0,-1,1,-1) & 4 \\ \hline
25 & (-2,0,-1,1,1,1,0,0) & (-1,-1,1,0,0,-1,1,1) & 4 & 26 & (-2,0,-1,1,1,1,0,0) & (-1,-1,1,1,1,0,1,0) & 4 \\ \hline
27 & (-2,0,-1,1,1,1,0,0) & (2,0,0,0,-1,-1,0,0) & 4 & 28 & (-2,0,-1,1,1,1,0,0) & (-1,1,-1,-1,-1,-1,0,0) & 4 \\ \hline
29 & (-2,1,1,1,1,0,0,0) & (-2,0,-1,0,0,1,0,0) & 4 & 30 & (-2,1,1,1,1,0,0,0) & (-1,0,0,-1,-1,1,1,-1) & 4 \\ \hline
31 & (-2,1,1,1,1,0,0,0) & (2,0,0,1,0,1,0,0) & 4 & 32 & (-2,1,1,1,1,0,0,0) & (-1,1,-1,-1,-1,1,0,0) & 4 \\ \hline
33 & (-2,1,1,1,1,0,0,0) & (0,-1,-1,0,-1,1,1,1) & 4 &  & & & \\ \hline
\end{longtable}

\begin{longtable}{|r|l|l|r||r|l|l|r|}
\caption{CMM 8, $3V_{B}$ = (1,1,0,0,0,0,0,0).\label{tab8a}} \\
\hline
\# & \hspace{25pt} $3a_{1B}$ & \hspace{25pt} $3a_{3B}$ & $G_H$ &
\# & \hspace{25pt} $3a_{1B}$ & \hspace{25pt} $3a_{3B}$ & $G_H$ \\
\hline \hline \endfirsthead
\caption{(continued) CMM 8, $3V_{B}$ = (1,1,0,0,0,0,0,0).} \\
\hline
\# & \hspace{25pt} $3a_{1B}$ & \hspace{25pt} $3a_{3B}$ & $G_H$ &
\# & \hspace{25pt} $3a_{1B}$ & \hspace{25pt} $3a_{3B}$ & $G_H$ \\
\hline \hline \endhead
1 & (0,0,1,1,1,1,0,0) & (0,0,-1,-1,-1,-1,1,1) & 1 & 2 & (0,0,1,1,1,1,0,0) & (-1,-2,0,0,0,-1,0,0) & 2 \\ \hline
3 & (0,0,1,1,1,1,0,0) & (0,0,2,0,0,0,1,-1) & 2 & 4 & (0,0,1,1,1,1,0,0) & (0,0,-1,-1,-1,-1,1,-1) & 3 \\ \hline
5 & (0,0,1,1,1,1,0,0) & (0,0,2,0,0,0,1,1) & 4 &  & & & \\ \hline
\end{longtable}

\begin{longtable}{|r|l|l|r||r|l|l|r|}
\caption{CMM 8, $3V_{B}$ = (2,1,1,1,1,0,0,0).\label{tab8b}} \\
\hline
\# & \hspace{25pt} $3a_{1B}$ & \hspace{25pt} $3a_{3B}$ & $G_H$ &
\# & \hspace{25pt} $3a_{1B}$ & \hspace{25pt} $3a_{3B}$ & $G_H$ \\
\hline \hline \endfirsthead
\caption{(continued) CMM 8, $3V_{B}$ = (2,1,1,1,1,0,0,0).} \\
\hline
\# & \hspace{25pt} $3a_{1B}$ & \hspace{25pt} $3a_{3B}$ & $G_H$ &
\# & \hspace{25pt} $3a_{1B}$ & \hspace{25pt} $3a_{3B}$ & $G_H$ \\
\hline \hline \endhead
1 & (-1,0,0,0,-1,1,1,0) & (-1,-1,-1,-1,-1,0,0,-1) & 2 & 2 & (-1,0,0,0,-1,1,1,0) & (-1,1,1,1,-1,0,0,1) & 2 \\ \hline
3 & (0,1,1,1,0,1,0,0) & (1,-1,-1,-1,1,-1,0,0) & 2 & 4 & (-1,0,0,0,-1,1,1,0) & (-1,1,1,0,0,-1,-1,1) & 3 \\ \hline
5 & (-1,0,0,0,-1,1,1,0) & (0,0,-1,-1,-1,-1,-1,-1) & 3 & 6 & (-1,1,1,0,0,1,0,0) & (-2,0,0,-1,-1,0,0,0) & 3 \\ \hline
7 & (-1,1,1,0,0,1,0,0) & (1,1,1,1,1,1,0,0) & 3 & 8 & (-1,0,0,0,-1,1,1,0) & (-2,0,-1,-1,0,0,0,0) & 4 \\ \hline
9 & (-1,0,0,0,-1,1,1,0) & (-2,0,0,0,1,1,0,0) & 4 & 10 & (-1,0,0,0,-1,1,1,0) & (-2,1,0,0,0,0,0,1) & 4 \\ \hline
11 & (-1,0,0,0,-1,1,1,0) & (-1,0,-1,-1,1,0,-1,1) & 4 & 12 & (-1,0,0,0,-1,1,1,0) & (-1,1,1,-1,1,0,-1,0) & 4 \\ \hline
13 & (-1,1,1,0,0,1,0,0) & (-2,1,0,0,0,-1,0,0) & 4 & 14 & (-1,1,1,0,0,1,0,0) & (-1,-1,-1,-1,-1,0,1,0) & 4 \\ \hline
15 & (-1,1,1,0,0,1,0,0) & (-1,-1,-1,1,0,0,1,-1) & 4 & 16 & (0,1,1,1,0,1,0,0) & (-2,0,0,0,1,-1,0,0) & 4 \\ \hline
17 & (0,1,1,1,0,1,0,0) & (0,1,0,-2,1,0,0,0) & 4 &  & & & \\ \hline
\end{longtable}

\begin{longtable}{|r|l|l|r||r|l|l|r|}
\caption{CMM 9, $3V_{B}$ = (1,1,0,0,0,0,0,0).\label{tab9a}} \\
\hline
\# & \hspace{25pt} $3a_{1B}$ & \hspace{25pt} $3a_{3B}$ & $G_H$ &
\# & \hspace{25pt} $3a_{1B}$ & \hspace{25pt} $3a_{3B}$ & $G_H$ \\
\hline \hline \endfirsthead
\caption{(continued) CMM 9, $3V_{B}$ = (1,1,0,0,0,0,0,0).} \\
\hline
\# & \hspace{25pt} $3a_{1B}$ & \hspace{25pt} $3a_{3B}$ & $G_H$ &
\# & \hspace{25pt} $3a_{1B}$ & \hspace{25pt} $3a_{3B}$ & $G_H$ \\
\hline \hline \endhead
1 & (1,0,1,0,0,0,0,0) & (-1,1,-1,1,1,1,1,1) & 1 & 2 & (1,0,1,0,0,0,0,0) & (1,2,0,1,1,1,0,0) & 2 \\ \hline
3 & (-1,-1,1,1,1,1,1,-1) & (0,0,2,0,-1,-1,-1,1) & 2 & 4 & (-1,-1,1,1,1,1,1,-1) & (0,0,0,0,0,-1,-1,0) & 3 \\ \hline
5 & (-1,-1,1,1,1,1,1,-1) & (-1,-2,1,0,0,0,-1,-1) & 4 &  & & & \\ \hline
\end{longtable}

\begin{longtable}{|r|l|l|r||r|l|l|r|}
\caption{CMM 9, $3V_{B}$ = (2,1,1,1,1,0,0,0).\label{tab9b}} \\
\hline
\# & \hspace{25pt} $3a_{1B}$ & \hspace{25pt} $3a_{3B}$ & $G_H$ &
\# & \hspace{25pt} $3a_{1B}$ & \hspace{25pt} $3a_{3B}$ & $G_H$ \\
\hline \hline \endfirsthead
\caption{(continued) CMM 9, $3V_{B}$ = (2,1,1,1,1,0,0,0).} \\
\hline
\# & \hspace{25pt} $3a_{1B}$ & \hspace{25pt} $3a_{3B}$ & $G_H$ &
\# & \hspace{25pt} $3a_{1B}$ & \hspace{25pt} $3a_{3B}$ & $G_H$ \\
\hline \hline \endhead
1 & (-1,0,0,0,0,1,0,0) & (-1,0,0,0,-1,0,0,0) & 2 & 2 & (0,1,0,0,0,1,0,0) & (-1,-1,-1,-1,-1,-1,1,1) & 2 \\ \hline
3 & (-2,0,0,0,-1,1,1,1) & (-1,-1,-1,-1,-1,1,1,-1) & 2 & 4 & (-1,0,0,0,0,1,0,0) & (-2,1,0,0,0,-1,1,-1) & 3 \\ \hline
5 & (-2,0,0,0,-1,1,1,1) & (-2,0,0,-1,-1,1,1,0) & 3 & 6 & (-1,0,0,0,0,1,0,0) & (-2,0,0,-1,-1,-1,1,0) & 4 \\ \hline
7 & (0,1,0,0,0,1,0,0) & (-2,-1,0,0,-1,-1,1,0) & 4 & 8 & (-2,0,0,0,-1,1,1,-1) & (-2,0,0,-1,-1,1,0,-1) & 4 \\ \hline
9 & (-2,0,0,0,-1,1,1,-1) & (-2,1,0,0,0,-1,-1,1) & 4 & 10 & (-2,0,0,0,-1,1,1,-1) & (-2,1,0,0,0,1,1,-1) & 4 \\ \hline
11 & (-2,0,0,0,-1,1,1,-1) & (-1,-1,-1,-1,-1,1,-1,-1) & 4 & 12 & (-2,0,0,0,-1,1,1,-1) & (-1,1,1,-1,1,-1,-1,1) & 4 \\ \hline
13 & (-2,0,0,0,-1,1,1,-1) & (-1,1,1,1,-1,1,-1,-1) & 4 & 14 & (-2,0,0,0,-1,1,1,-1) & (1,-1,-1,-1,1,1,-1,-1) & 4 \\ \hline
15 & (-2,0,0,0,-1,1,1,-1) & (1,1,1,1,1,1,-1,-1) & 4 & 16 & (-2,0,0,0,-1,1,1,1) & (-2,0,-1,-1,0,1,0,-1) & 4 \\ \hline
\end{longtable}
}

\newpage

\section{Pattern Tables}
\label{tbs}

\setcounter{mytab}{\value{table}}

{ \footnotesize
\begin{center}
\begin{tabular}{ll}
Pattern & \multicolumn{1}{c}{
   $SU(3) \times SU(2) \times SO(10)$ \hspace{3pt} Irreps}
   \\ \hline
1.1  &  $ 3[(3,2,1) + 5(3,1,1) + 7(\bar 3,1,1) + 15(1,2,1)
   + 48(1,1,1)_0 + 15(1,1,1)_1] $ \\
1.2  &  $ 3[(3,2,1) + 4(3,1,1) + 6(\bar 3,1,1) + 13(1,2,1)
   + (1,1,16) + 48(1,1,1)_0$ \\ & $+ 9(1,1,1)_1] $ \\
\hline
\end{tabular}
\mycap{pt1}{20pt}{Patterns of irreps in Case 1 models.}
\end{center}

\vspace{25pt}

\begin{center}
\begin{tabular}{ll}
Pattern & \multicolumn{1}{c}{ $SU(3) \times SU(2) \times SU(5)
   \times SU(2)$ \hspace{3pt} Irreps} \\ \hline
2.1 &  $ 3[(3,2,1,1) + 3(3,1,1,1) + 5(\bar 3,1,1,1) + 9(1,2,1,1) + (1,1,5,1) $ \\
   & $  + (1,1,\bar 5,1) + 6(1,1,1,2) + (1,2,1,2) + 34(1,1,1,1)_0 + 9(1,1,1,1)_1] $ \\
2.2 &  $ 3[(3,2,1,1) + 3(3,1,1,1) + 5(\bar 3,1,1,1) + 9(1,2,1,1) + (1,1,5,1) $ \\
   & $  + (1,1,\bar 5,1) + 6(1,1,1,2) + (1,2,1,2) + 37(1,1,1,1)_0 + 6(1,1,1,1)_1] $ \\
2.3 &  $ 3[(3,2,1,1) + 3(3,1,1,1) + 5(\bar 3,1,1,1) + 11(1,2,1,1) + (1,1,5,1) $ \\
   & $  + (1,1,\bar 5,1) + 8(1,1,1,2) + 33(1,1,1,1)_0 + 6(1,1,1,1)_1] $ \\
2.4 &  $ 3[(3,2,1,1) + 2(3,1,1,1) + 4(\bar 3,1,1,1) + 9(1,2,1,1) + (1,1,5,1) $ \\
   & $  + 2(1,1,\bar 5,1) + (1,1,10,1) + 6(1,1,1,2) + 32(1,1,1,1)_0 + 6(1,1,1,1)_1] $ \\
2.5 &  $ 3[(3,2,1,1) + 2(3,1,1,1) + 4(\bar 3,1,1,1) + 7(1,2,1,1) + (1,1,5,1) $ \\
   & $  + 2(1,1,\bar 5,1) + (1,1,10,1) + 4(1,1,1,2) + (1,2,1,2) + 36(1,1,1,1)_0 $ \\
   & $  + 6(1,1,1,1)_1] $ \\
2.6 &  $ 3[(3,2,1,1) + (3,1,1,1) + 3(\bar 3,1,1,1) + 5(1,2,1,1) + (1,1,5,1) $ \\
   & $  + 3(1,1,\bar 5,1) + (1,1,10,2) + 10(1,1,1,2) + (1,2,1,2) + 25(1,1,1,1)_0] $ \\
\hline
\end{tabular}
\mycap{pt2}{20pt}{Patterns of irreps in Case 2 models.}
\end{center}

\vspace{25pt}

\begin{center}
\begin{tabular}{ll}
Pattern & \multicolumn{1}{c}{ $SU(3) \times SU(2) \times SU(4)
   \times SU(2)^2$ \hspace{3pt} Irreps} \\ \hline
3.1 &  $ 3[(3,2,1,1,1) + 2(3,1,1,1,1) + 4(\bar 3,1,1,1,1) + 7(1,2,1,1,1) + 2(1,1,4,1,1) $ \\
   & $  + 2(1,1,\bar 4,1,1) + 6(1,1,1,2,1) + 4(1,1,1,1,2) + (1,2,1,1,2) + 27(1,1,1,1,1)_0 $ \\
   & $  + 6(1,1,1,1,1)_1] $ \\
3.2 &  $ 3[(3,2,1,1,1) + 2(3,1,1,1,1) + 4(\bar 3,1,1,1,1) + 7(1,2,1,1,1) + 2(1,1,\bar 4,1,1) $ \\
   & $  + 8(1,1,1,2,1) + 4(1,1,1,1,2) + (1,1,4,2,1) + (1,2,1,1,2) + 26(1,1,1,1,1)_0 $ \\
   & $  + 3(1,1,1,1,1)_1] $ \\
3.3 &  $ 3[(3,2,1,1,1) + 2(3,1,1,1,1) + 4(\bar 3,1,1,1,1) + 7(1,2,1,1,1) + 2(1,1,\bar 4,1,1) $ \\
   & $  + 6(1,1,1,2,1) + 6(1,1,1,1,2) + (1,1,4,2,1) + (1,2,1,2,1) + 26(1,1,1,1,1)_0 $ \\
   & $  + 3(1,1,1,1,1)_1] $ \\
3.4 &  $ 3[(3,2,1,1,1) + (3,1,1,1,1) + 3(\bar 3,1,1,1,1) + 5(1,2,1,1,1) + 2(1,1,4,1,1) $ \\
   & $  + 2(1,1,\bar 4,1,1) + 8(1,1,1,2,1) + 4(1,1,1,1,2) + (1,1,6,2,1) + (1,2,1,2,1) $ \\
   & $  + 24(1,1,1,1,1)_0 + 3(1,1,1,1,1)_1] $ \\
\hline
\end{tabular}
\mycap{pt3}{20pt}{Patterns of irreps in Case 3 models.}
\end{center}

\begin{center}
\begin{tabular}{ll}
Pattern & \multicolumn{1}{c}{ $SU(3) \times SU(2) \times SU(3)
   \times SU(2)^2$ \hspace{3pt} Irreps} \\ \hline
4.1 &  $ 3[(3,2,1,1,1) + 2(3,1,1,1,1) + 4(\bar 3,1,1,1,1)
   + 9(1,2,1,1,1) + (1,1,3,1,1) $ \\
   & $  + (1,1,\bar 3,1,1) + 6(1,1,1,2,1) + 6(1,1,1,1,2)
   + 30(1,1,1,1,1)_0 + 3(1,1,1,1,1)_1] $ \\
4.2 &  $ 3[(3,2,1,1,1) + 2(3,1,1,1,1) + 4(\bar 3,1,1,1,1)
   + 7(1,2,1,1,1) + (1,1,3,1,1) $ \\
   & $  + (1,1,\bar 3,1,1) + 4(1,1,1,2,1) + 6(1,1,1,1,2)
   + (1,2,1,2,1) + 34(1,1,1,1,1)_0 $ \\
   & $  + 3(1,1,1,1,1)_1] $ \\
4.3 &  $ 3[(3,2,1,1,1) + (3,1,1,1,1) + 3(\bar 3,1,1,1,1)
   + 7(1,2,1,1,1) + 3(1,1,3,1,1) $ \\
   & $  + 3(1,1,\bar 3,1,1) + 4(1,1,1,2,1) + 4(1,1,1,1,2)
   + 36(1,1,1,1,1)_0 $ \\
   & $  + 3(1,1,1,1,1)_1] $ \\
4.4 &  $ 3[(3,2,1,1,1) + (3,1,1,1,1) + 3(\bar 3,1,1,1,1)
   + 7(1,2,1,1,1) + (1,1,3,1,1) $ \\
   & $  + 3(1,1,\bar 3,1,1) + 4(1,1,1,2,1) + 7(1,1,1,1,2)
   + (1,1,3,1,2) + 30(1,1,1,1,1)_0 $ \\
   & $  + 3(1,1,1,1,1)_1] $ \\
4.5 &  $ 3[(3,2,1,1,1) + (3,1,1,1,1) + 3(\bar 3,1,1,1,1)
   + 7(1,2,1,1,1) + (1,1,3,1,1) $ \\
   & $  + 3(1,1,\bar 3,1,1) + 4(1,1,1,2,1) + 7(1,1,1,1,2)
   + (1,1,3,1,2) + 33(1,1,1,1,1)_0] $ \\
4.6 &  $ 3[(3,2,1,1,1) + (3,1,1,1,1) + 3(\bar 3,1,1,1,1)
   + 5(1,2,1,1,1) + (1,1,3,1,1) $ \\
   & $  + 3(1,1,\bar 3,1,1) + 4(1,1,1,2,1) + 5(1,1,1,1,2)
   + (1,1,3,1,2) + (1,2,1,1,2) $ \\
   & $  + 34(1,1,1,1,1)_0 + 3(1,1,1,1,1)_1] $ \\
4.7 &  $ 3[(3,2,1,1,1) + (3,1,1,1,1) + 3(\bar 3,1,1,1,1)
   + 5(1,2,1,1,1) + 3(1,1,3,1,1) $ \\
   & $  + (1,1,\bar 3,1,1) + 4(1,1,1,2,1) + 5(1,1,1,1,2)
   + (1,2,1,1,2) + (1,1,\bar 3,1,2) $ \\
   & $  + 37(1,1,1,1,1)_0] $ \\
4.8 &  $ 3[(3,2,1,1,1) + 2(\bar 3,1,1,1,1) + 3(1,2,1,1,1)
   + (1,1,3,1,1) + 5(1,1,\bar 3,1,1) $ \\
   & $  + 8(1,1,1,2,1) + 6(1,1,1,1,2) + (1,2,1,1,2)
   + (1,1,3,2,2) + 25(1,1,1,1,1)_0] $ \\
\hline
\end{tabular}
\mycap{pt4}{20pt}{Patterns of irreps in Case 4 models.} 
\end{center}

\begin{center}
\begin{tabular}{cl}
Patterns & Untwisted Irreps \\
\hline
1.1 & $ 3[(3,2,1) + 3(1,1,1)_0] $ \\
1.2 & $ 3[(3,2,1) + (\bar 3,1,1) + (1,2,1) + (1,1,16)] $ \\
2.1 & $ 3[(3,2,1,1) + 3(1,1,1,1)_0] $ \\
2.2, 2.3 &
   $ 3[(3,2,1,1) + (\bar 3,1,1,1) + (1,2,1,1) + (1,1,5,1) + (1,1,1,2)] $ \\
2.4, 2.5 & $ 3[(3,2,1,1) + (1,1,10,1) + 2(1,1,1,1)_0] $ \\
2.6 &
   $ 3[(3,2,1,1) + (\bar 3,1,1,1) + (1,2,1,1) + (1,1,5,1) + (1,1,10,2)] $ \\
3.1 & $ 3[(3,2,1,1,1) + (1,1,4,1,1) + 2(1,1,1,1,1)_0] $ \\
3.2, 3.3 & $ 3[(3,2,1,1,1) + (\bar 3,1,1,1,1) + (1,2,1,1,1)
   + (1,1,1,1,2) + (1,1,4,2,1)] $ \\
3.4 & $ 3[(3,2,1,1,1) + (1,1,6,2,1) + 3(1,1,1,1,1)_0] $ \\
4.1, 4.2 & $ 3[(3,2,1,1,1) + (\bar 3,1,1,1,1) + (1,2,1,1,1)
   + (1,1,1,2,1) + (1,1,1,1,2)] $ \\
4.3 & 
   $ 3[(3,2,1,1,1) + (1,1,3,1,1) + (1,1,\bar 3,1,1) + 3(1,1,1,1,1)_0] $ \\
4.4, 4.6 & $ 3[(3,2,1,1,1) + (1,1,3,1,2) + 3(1,1,1,1,1)_0] $ \\
4.5, 4.7 & $ 3[(3,2,1,1,1) + (\bar 3,1,1,1,1) + (1,2,1,1,1)
   + (1,1,\bar 3,1,1) + (1,1,1,2,1) $ \\*
   & $  + (1,1,1,1,2) + (1,1,3,1,2)] $ \\
4.8 &
   $ 3[(3,2,1,1,1) + (1,1,3,1,1) + (1,1,3,2,2) + 3(1,1,1,1,1)_0] $ \\
\hline
\end{tabular}
\mycap{unpat}{20pt}{Irreps of the 
untwisted sectors for each pattern of
total irreps.}
\end{center}

\begin{center}
\begin{tabular}{ll}
Pattern & \multicolumn{1}{c}{Models} \\ \hline
1.1 & 1.1, 1.2, 1.3, 4.1, 4.2, 4.3, 8.1 \\
1.2 & 2.1, 2.2, 2.3, 6.1, 6.2, 6.3, 9.1 \\
2.1 & 1.4, 1.5, 1.11, 1.12, 4.4, 4.6, 4.9, 4.11, 8.2, 8.3 \\
2.2 & 2.4, 2.5, 2.6, 2.7, 6.4, 6.6, 6.9, 6.11, 9.2, 11.3 \\
2.3 & 2.9, 2.10, 2.12, 6.8, 6.10, 6.12, 9.3 \\
2.4 & 1.6, 1.8, 1.10, 4.5, 4.8, 4.10, 10.2 \\
2.5 & 1.7, 1.9, 4.7, 4.12, 10.1, 10.3 \\
2.6 & 2.8, 2.11, 6.5, 6.7, 11.1, 11.2 \\
3.1 & 1.14, 1.15, 1.16, 1.17, 4.13, 4.15,
   4.16, 4.18, 10.4, 10.5, 10.6, 10.7 \\
3.2 & 2.13, 2.14, 6.15, 6.17, 11.5 \\
3.3 & 2.15, 2.16, 2.17, 2.18, 6.13, 6.14, 6.16, 6.18, 9.4, 11.4 \\
3.4 & 1.13, 1.18, 4.14, 4.17, 8.4 \\
4.1 & 2.19, 2.20, 2.21, 6.22, 6.23, 6.29, 11.16 \\
4.2 &  2.22, 2.23, 2.27, 2.32, 6.24, 6.26, 6.30,
   6.31, 9.5, 11.9, 11.11, 11.14 \\
4.3 & 1.19, 1.32, 1.33, 4.20, 4.27, 4.31, 8.5 \\
4.4 & 1.21, 1.22, 1.23, 1.25, 1.28, 1.29, 4.21,
   4.24, 4.25, 4.28, 4.30, 4.33, 10.8, 10.11, \\
   & 10.14 \\
4.5 & 2.24, 2.25, 2.28, 2.29, 2.30, 2.31, 6.19,
   6.20, 6.21, 6.27, 6.28, 6.32, 11.6, 11.7, \\
   & 11.12, 11.13, 11.15 \\
4.6 & 1.20, 1.24, 1.27, 1.30, 4.19, 4.22, 4.26,
   4.29, 10.10, 10.12, 10.15, 10.16, 10.17 \\
4.7 & 2.26, 2.33, 6.25, 6.33, 11.8, 11.10 \\
4.8 & 1.26, 1.31, 4.23, 4.32, 10.9, 10.13 \\
\hline
\end{tabular}
\mycapw{tb1}{20pt}{Irrep patterns versus the models
enumerated in \cite{Gie01b}.  See explanation
of model labeling in Section~\ref{mds}.}
\end{center}
}

\section{Example Spectrum Table}

\setcounter{table}{\value{mytab}}

{\footnotesize
\begin{longtable}{cccccccccccc}
\caption{\bsa\ 6.5 Pseudo-Massless Spectrum \label{tb5}} \\
No. & Irrep 
 & $Q_1$ & $Q_2$ & $Q_3$ & $Q_4$
 & $Q_5$ & $Q_6$
 & $Q_7$ & $Q_X$ & $Z$ & $Y$ \\ \hline \endfirsthead
\caption{\bsa\ 6.5 Pseudo-Massless Spectrum (Cont.)} \\
No. & Irrep 
 & $Q_1$ & $Q_2$ & $Q_3$ & $Q_4$
 & $Q_5$ & $Q_6$
 & $Q_7$ & $Q_X$ & $Z$ & $Y$ \\ \hline \endhead
1 & $(3, 2, 1, 1)_U$ & $1$ & $6$ & $-18$ & $9$ & $45$ & $15$ & $0$ & $3$ & $1/6$ & $1/6$ \\ 
2 & $(1, 2, 1, 1)_U$ & $3$ & $18$ & $-54$ & $27$ & $-45$ & $-15$ & $0$ & $-3$ & $1/2$ & $1/2$ \\
3 & $(\bar 3, 1, 1, 1)_U$ & $-4$ & $-24$ & $72$ & $-36$ & $0$ & $0$ & $0$ & $0$ & $-2/3$ & $-2/3$ \\
4 & $(1, 1, 10, 2)_U$ & $0$ & $0$ & $0$ & $0$ & $-18$ & $-6$ & $0$ & $3$ & $0$ & $0$ \\
5 & $(1, 1, 5, 1)_U$ & $0$ & $0$ & $0$ & $0$ & $36$ & $12$ & $0$ & $-6$ & $0$ & $0$ \\
6 & $(1, 1, 1, 1)_{-1,-1}$ & $0$ & $-20$ & $-32$ & $-31$ & $-35$ & $-23$ & $0$ & $1$ & $-1$ & $0$ \\
7 & $(1, 1, 1, 1)_{-1,-1}$ & $0$ & $-35$ & $13$ & $17$ & $25$ & $-3$ & $0$ & $5$ & $-1$ & $2/5$ \\
8 & $(1, 1, 1, 1)_{-1,-1}$ & $0$ & $10$ & $16$ & $-55$ & $25$ & $-3$ & $0$ & $5$ & $-1$ & $-2/5$ \\
9 & $(1, 1, 1, 1)_{-1,-1}$ & $0$ & $10$ & $-122$ & $14$ & $10$ & $-8$ & $0$ & $4$ & $-1$ & $0$ \\
10 & $(\bar 3, 1, 1, 1)_{-1,-1}$ & $2$ & $7$ & $25$ & $11$ & $-5$ & $-13$ & $0$ & $3$ & $-2/3$ & $1/3$ \\
11 & $(1, 2, 1, 1)_{-1,-1}$ & $-3$ & $7$ & $25$ & $11$ & $-5$ & $-13$ & $0$ & $3$ & $-3/2$ & $-1/2$ \\
12 & $(1, 1, 1, 2)_{-1,0}$ & $0$ & $-5$ & $61$ & $-7$ & $-5$ & $-13$ & $0$ & $-4$ & $-1$ & $0$ \\
13 & $(1, 1, 1, 1)_{-1,0}$ & $0$ & $-5$ & $61$ & $-7$ & $-95$ & $-9$ & $0$ & $1$ & $-1$ & $0$ \\
14 & $(1, 1, 1, 2)_{-1,0}$ & $0$ & $-20$ & $-32$ & $-31$ & $55$ & $7$ & $0$ & $0$ & $-1$ & $0$ \\
15 & $(1, 1, 1, 1)_{-1,0}$ & $0$ & $-20$ & $-32$ & $-31$ & $-35$ & $11$ & $0$ & $5$ & $-1$ & $0$ \\
16 & $(1, 1, 1, 2)_{-1,0}$ & $0$ & $25$ & $-29$ & $38$ & $40$ & $2$ & $0$ & $-1$ & $-1$ & $0$ \\
17 & $(1, 1, 1, 1)_{-1,0}$ & $0$ & $25$ & $-29$ & $38$ & $-50$ & $6$ & $0$ & $4$ & $-1$ & $0$ \\
18 & $(1, 1, 1, 2)_{-1,1}$ & $0$ & $-5$ & $61$ & $-7$ & $-5$ & $21$ & $0$ & $0$ & $-1$ & $0$ \\
19 & $(1, 1, \bar 5, 1)_{-1,1}$ & $0$ & $-5$ & $61$ & $-7$ & $49$ & $5$ & $0$ & $1$ & $-1$ & $0$ \\
20 & $(1, 1, 1, 1)_{-1,1}$ & $0$ & $-5$ & $61$ & $-7$ & $-5$ & $4$ & $-3$ & $5$ & $-1$ & $-1/5$ \\
21 & $(1, 1, 1, 1)_{-1,1}$ & $0$ & $-5$ & $61$ & $-7$ & $-5$ & $4$ & $3$ & $5$ & $-1$ & $1/5$ \\
22 & $(1, 1, 1, 1)_{0,-1}$ & $2$ & $-28$ & $38$ & $-19$ & $-5$ & $4$ & $-1$ & $5$ & $0$ & $2/5$ \\
23 & $(1, 1, 1, 1)_{0,-1}$ & $2$ & $17$ & $41$ & $50$ & $-20$ & $-1$ & $-1$ & $4$ & $0$ & $2/5$ \\
24 & $(1, 1, 1, 1)_{0,-1}$ & $2$ & $17$ & $-97$ & $-22$ & $-20$ & $-1$ & $-1$ & $4$ & $0$ & $0$ \\
25 & $(1, 2, 1, 1)_{0,-1}$ & $-1$ & $14$ & $50$ & $-25$ & $-35$ & $-6$ & $-1$ & $3$ & $-1/2$ & $-1/2$ \\
26 & $(1, 1, 1, 1)_{0,-1}$ & $-4$ & $-4$ & $-34$ & $17$ & $-5$ & $4$ & $-1$ & $5$ & $-1$ & $-3/5$ \\
27 & $(3, 1, 1, 1)_{0,-1}$ & $0$ & $-10$ & $-16$ & $8$ & $-50$ & $-11$ & $-1$ & $2$ & $-1/3$ & $1/15$ \\
28 & $(1, 1, 1, 2)_{0,0}$ & $2$ & $32$ & $-4$ & $2$ & $10$ & $9$ & $-1$ & $-1$ & $0$ & $0$ \\
29 & $(1, 1, 1, 1)_{0,0}$ & $2$ & $32$ & $-4$ & $2$ & $10$ & $-8$ & $2$ & $4$ & $0$ & $1/5$ \\
30 & $(1, 2, 1, 2)_{0,0}$ & $-1$ & $-16$ & $2$ & $-1$ & $-5$ & $4$ & $-1$ & $-2$ & $-1/2$ & $-1/10$ \\
31 & $(1, 2, 1, 1)_{0,0}$ & $-1$ & $-16$ & $2$ & $-1$ & $-5$ & $-13$ & $2$ & $3$ & $-1/2$ & $1/10$ \\
32 & $(1, 1, \bar 5, 1)_{0,1}$ & $2$ & $2$ & $-52$ & $26$ & $4$ & $7$ & $-1$ & $0$ & $0$ & $2/5$ \\
33 & $(1, 1, 1, 2)_{0,1}$ & $2$ & $2$ & $-52$ & $26$ & $40$ & $2$ & $2$ & $-1$ & $0$ & $3/5$ \\
34 & $(1, 1, 1, 1)_{0,1}$ & $2$ & $2$ & $-52$ & $26$ & $40$ & $-15$ & $-1$ & $4$ & $0$ & $2/5$ \\
35 & $(1, 1, 1, 1)_{0,1}$ & $2$ & $2$ & $-52$ & $26$ & $-50$ & $6$ & $2$ & $4$ & $0$ & $3/5$ \\
36 & $(1, 1, 1, 2)_{1,-1}$ & $-2$ & $3$ & $-9$ & $-19$ & $55$ & $7$ & $-2$ & $0$ & $0$ & $-3/5$ \\
37 & $(1, 1, \bar 5, 1)_{1,-1}$ & $-2$ & $3$ & $-9$ & $-19$ & $19$ & $12$ & $1$ & $1$ & $0$ & $-2/5$ \\
38 & $(1, 1, 1, 1)_{1,-1}$ & $-2$ & $3$ & $-9$ & $-19$ & $-35$ & $11$ & $-2$ & $5$ & $0$ & $-3/5$ \\
39 & $(1, 1, 1, 1)_{1,-1}$ & $-2$ & $3$ & $-9$ & $-19$ & $55$ & $-10$ & $1$ & $5$ & $0$ & $-2/5$ \\
40 & $(1, 1, 1, 1)_{1,0}$ & $-2$ & $-27$ & $-57$ & $5$ & $-5$ & $4$ & $1$ & $5$ & $0$ & $0$ \\
41 & $(1, 1, 1, 1)_{1,0}$ & $-2$ & $18$ & $84$ & $5$ & $-5$ & $4$ & $1$ & $5$ & $0$ & $-2/5$ \\
42 & $(\bar 3, 1, 1, 1)_{1,0}$ & $0$ & $15$ & $-45$ & $-1$ & $-35$ & $-6$ & $1$ & $3$ & $1/3$ & $-1/15$ \\
43 & $(1, 1, 1, 1)_{1,0}$ & $4$ & $-6$ & $18$ & $38$ & $-20$ & $-1$ & $1$ & $4$ & $1$ & $1$ \\
44 & $(1, 2, 1, 1)_{1,0}$ & $1$ & $-9$ & $27$ & $-37$ & $-35$ & $-6$ & $1$ & $3$ & $1/2$ & $1/10$ \\
45 & $(1, 1, 1, 1)_{1,0}$ & $-2$ & $-12$ & $36$ & $29$ & $-65$ & $-16$ & $1$ & $1$ & $0$ & $0$ \\
46 & $(1, 1, 1, 2)_{1,1}$ & $-2$ & $-12$ & $36$ & $29$ & $25$ & $14$ & $1$ & $0$ & $0$ & $0$ \\
47 & $(1, 1, 1, 1)_{1,1}$ & $-2$ & $-12$ & $36$ & $29$ & $25$ & $-3$ & $-2$ & $5$ & $0$ & $-1/5$ \\
48 & $(1, 1, 1, 2)_{1,1}$ & $4$ & $9$ & $-27$ & $-10$ & $10$ & $9$ & $1$ & $-1$ & $1$ & $3/5$ \\
49 & $(1, 1, 1, 1)_{1,1}$ & $4$ & $9$ & $-27$ & $-10$ & $10$ & $-8$ & $-2$ & $4$ & $1$ & $2/5$ \\
50 & $(1, 1, 1, 2)_{1,1}$ & $-2$ & $3$ & $-9$ & $-19$ & $-35$ & $-6$ & $1$ & $-4$ & $0$ & $-2/5$ \\
51 & $(1, 1, 1, 1)_{1,1}$ & $-2$ & $3$ & $-9$ & $-19$ & $-35$ & $-23$ & $-2$ & $1$ & $0$ & $-3/5$ \\
\hline
\end{longtable}
}

\chapter{Supplementary References}
\label{mysg}

\bfe{Quantum field theory and particle physics.}
A standard modern text is that of Peskin and
Schroeder \cite{PS95}.  The annotated bibliography
provided at the end of the book gives ample
references to other texts which the reader might
find helpful.  The Standard Model of particle
physics has been reviewed, for example,
in \cite{GGS99}.

\bfe{Supersymmetric field theory.}  The monograph
of Wess and Bagger \cite{WB92} summarizes the mathematical
features of this class of theories, but leaves
much to be desired with respect to applications.
These weaknesses are best compensated by reference
to reviews on the MSSM \cite{mssmr}
and its origins in minimal supergravity
theories \cite{BL94}.

\bfe{Grand unified theories.}  The non-supersymmetric
versions have been reviewed in \cite{gutrvw}.
The supersymmetric GUTs have been reviewed
in \cite{sgtrvw}.

\bfe{String theory.}  In my opinion, because of
its completeness with respect to the topics
covered, the best text remains Green, Schwarz
and Witten \cite{GSW87a,GSW87b}.  More modern
topics are discussed in \cite{Pol98b}.
Aspects of string theory are also discussed
in \cite{BL94}.  Heterotic orbifolds have
been reviewed in \cite{orbrvw,BL94,BL99}.

\bfe{Lie algebras and groups.}
Numerous references have evolved over the
years.  For example, the reader
might consult Refs.~\cite{Cah84,Cor84,Geo99}.

\addtocounter{page}{1}
\typeout{Bibliography}
\addcontentsline{toc}{chapter}{Bibliography}
\addtocounter{page}{-1}

\bibliography{dis}

\begin{thebibliography}{100}

\bibitem{Gie01b}
J. Giedt, {\it Ann. of Phys. (N.Y.)} {\bf 289} (2001) 251.

\bibitem{Gie01c}
J. Giedt, {\it Ann. of Phys. (N.Y.)} {\bf 297} (2002) 67\xxx{hep-th/0108244}.

\bibitem{GSW87a}
M. B. Green, J. H. Schwarz and E. Witten, ``Superstring Theory,'' Vol.~1,
  Cambridge University Press, Cambridge, UK, 1987.

\bibitem{GGS99}
M. K. Gaillard, P. D. Grannis and F. J. Sciulli, {\it Rev. Mod. Phys.} {\bf 71}
  (1999) S96.

\bibitem{GHMR85}
D. J. Gross, J. A. Harvey, E. Martinec and R. Rohm, {\it Phys. Rev. Lett.} {\bf
  54} (1985) 502. \bibann{The OR for the heterotic string.}

\bibitem{KKth}
O. Klein, {\it Z. Phys.} {\bf 37} (1926) 895; T. Kaluza, Sitzunber, Preuss.
  Akad. Wiss. Berlin, Math-Phys. {\bf K1} (1921) 966.

\bibitem{CGS}
G. Chapline and R. Slansky, {\it Nucl. Phys.} {\bf B 209} (1982) 461; G. F.
  Chapline and B. Grossman, {\it Phys. Lett.} {\bf B 135} (1984) 109.

\bibitem{CHSW85}
P. Candelas, G. Horowitz, A. Strominger, E. Witten, {\it Nucl. Phys.} {\bf B
  355} (1985) 46.

\bibitem{DHVW85}
L. Dixon, J. Harvey, C. Vafa and E. Witten, {\it Nucl. Phys.} {\bf B 261}
  (1985) 678.

\bibitem{DHVW86}
L. Dixon, J. Harvey, C. Vafa and E. Witten, {\it Nucl. Phys.} {\bf B 274}
  (1986) 285.

\bibitem{MP77}
R. S. Millman and G. D. Parker, ``Elements of Differential Geometry,''
  Prentice-Hall Inc., Englewood Cliffs, New Jersey, 1977.

\bibitem{GSO}
F. Gliozzi, J. Scherk and D. Olive, {\it Phys. Lett.} {\bf B 65} (1976) 282;
  {\it Nucl. Phys.} {\bf B 122} (1977) 253.

\bibitem{DeW92}
B. S. DeWitt, ``Supermanifolds,'' 2nd Edition, Cambridge University Press, New
  York, 1992\xxx{UCB QC 20.7 .M24 D47 1992 Physics}.

\bibitem{GSW87b}
M. B. Green, J. H. Schwarz and E. Witten, ``Superstring Theory,'' Vol.~2,
  Cambridge University Press, Cambridge, UK, 1987.

\bibitem{GT90}
D. M. Gitman and I. V. Tyutin, ``Quantization of Fields with Constraints,''
  Springer-Verlag, New York, 1990\xxx{UCB QC 793.3 .F5 .G5813 1990 Phys}.

\bibitem{AG93}
N. I. Akhiezer and I. M. Glazman, ``Theory of Linear Operators in Hilbert
  Space,'' 2 Volumes, Translated by M. Nestell, Frederick Ungar Publishing Co.,
  New York, 1961 and 1963 (Reprint, Dover, New York, 1993).

\bibitem{Hos83b}
Y. Hosotani, Phys. Lett. B 129 (1983) 193.

\bibitem{IMNQ88}
L. E. {Ib\'a\~nez}, J. Mas, H.-P. Nilles and F. Quevedo, {\it Nucl. Phys.} {\bf
  B 301} (1988) 157.

\bibitem{BLT88}
D. Bailin, A. Love and S. Thomas, {\it Mod. Phys. Lett.} {\bf A 3} (1988) 167.

\bibitem{Cor84}
J. F. Cornwell, ``Group Theory in Physics,'' Vol.~1, Academic Press, New York,
  1984.

\bibitem{GM94}
W. Greiner and B. {M\"uller}, ``Quantum Mechanics: Symmetries,'' 2nd
  Rev.~Edition, Springer-Verlag, New York, 1994.

\bibitem{Gou82}
M. Gourdin, ``Basics of Lie Groups,'' Editions {Fronti\`eres}, Gif sur Yvette,
  France, 1982.

\bibitem{Geo99}
H. Georgi, ``Lie Algebras in Particle Physics,'' 2nd Edition, Perseus Books,
  Reading, Mass., 1999.

\bibitem{Cah84}
R. N. Cahn, ``Semi-Simple Lie Algebras and Their Representations,''
  Ben\-ja\-min/Cum\-mings, Men\-lo Park, 1984.

\bibitem{Sla81}
R. Slansky, {\it Phys. Rep.} {\bf 79} (1981) 1.

\bibitem{BL94}
D. Bailin and A. Love, ``Supersymmetric Gauge Field Theory and String Theory,''
  Institute of Physics Publishing, Philadelphia, 1994.

\bibitem{Pol98b}
J. Polchinski, ``String Theory, Vol. 2: Superstring Theory and Beyond,''
  Cambridge University Press, Cambridge, UK, 1998.

\bibitem{Har86}
J. A. Harvey, {\it in} ``Unified String Theories'' (M. Green and D. Gross,
  Eds.) World Scientific, Singapore, 1986.

\bibitem{orbrvw}
J. E. Kim, {\it in} ``Superstrings, Proceedings, Boulder, CO, Jul 27 - Aug 1,
  1987'' (P. G. O. Freund and K. T. Mahanthappa, Eds.) NATO Advanced Study
  Institute, Series B: Physics, Vol. 175, Plenum Press, N.Y., 1988; F. Quevedo,
  {\it in } ``Summer Workshop on High Energy Physics and Cosmology, Trieste,
  Italy, Jun 29 - Aug 7, 1987'' (G. Furlan et al., Eds.) ICTP Ser. Theor. Phys.
  Vol. 4.; L.~E.~Ibanez, {\it in} ``Strings and Superstrings: XVIII
  International GIFT Seminar on Theoretical Physics, el Escorial, Spain, 1-6
  Jun 1987'' (J. R. Mittelbrunn, M. {Ram\'on-Medrano} and G. S. Rodero, Eds.)
  World Scientific, Singapore, 1988;.

\bibitem{BL99}
D. Bailin and A. Love, {\it Phys. Rep.} {\bf 315} (1999) 285.

\bibitem{WW85}
X.-G. Wen and E. Witten, {\it Nucl. Phys.} {\bf B 261} (1985) 651.

\bibitem{CMM89}
J. A. Casas, M. Mondragon and C. {Mu\~noz}, {\it Phys. Lett.} {\bf B 230}
  (1989) 63.

\bibitem{Sch71}
J. Scherk, {\it Nucl. Phys.} {\bf B 31} (1971) 222.

\bibitem{HV87}
S. Hamidi and C. Vafa, {\it Nucl. Phys.} {\bf B 279} (1987) 465.

\bibitem{FIQS90}
A. Font, L. E. {Ib\'a\~nez}, F. Quevedo and A. Sierra, {\it Nucl. Phys.} {\bf B
  331} (1990) 421.

\bibitem{Gre97}
B. R. Greene, {\it in} ``Fields, Strings and Duality (TASI 1996)'' (C.
  Efthimiou and B. Greene, Eds.) Singapore, World Scientific, 1997
  [hep-th/9702155].

\bibitem{GG00}
M. K. Gaillard and J. Giedt, {\it Phys. Lett.} {\bf B 479} (2000) 308.

\bibitem{ABK87}
I. Antoniadis, C. Bachas and C. Kounnas, {\it Nucl. Phys.} {\bf B 289} (1987)
  87.

\bibitem{AEHN88}
I. Antoniadis, J. Ellis, J. S. Hagelin and D. V. Nanopoulos, {\it Phys. Lett.}
  {\bf B 205} (1988) 459; {\bf B 208} (1988) 209.

\bibitem{CFN99}
G. B. Cleaver, A. E. Faraggi and D. V. Nanopoulos, {\it Phys. Lett.} {\bf B
  455} (1999) 135.

\bibitem{CFNW00}
G. B. Cleaver, A. E. Faraggi, D. V. Nanopoulos, and J. W. Walker, {\it Mod.
  Phys. Lett.} {\bf A 15} (2000) 1191.

\bibitem{CFNW01a}
G. B. Cleaver, A. E. Faraggi, D. V. Nanopoulos, and J. W. Walker, {\it Nucl.
  Phys.} {\bf B 593} (2001) 471.

\bibitem{CFNW01b}
G. B. Cleaver, A. E. Faraggi, D. V. Nanopoulos, and J. W. Walker,
  hep-ph/0104091.

\bibitem{FNY90}
A. E. Faraggi, D. V. Nanopoulos and K. Yuan, {\it Nucl. Phys.} {\bf B 335}
  (1990) 347.

\bibitem{mssmr}
H. E. Haber and G. L. Kane, {\it Phys. Rep.} {\bf 117} (1985) 75; H. E. Haber
  {\it in} ``Recent Directions in Particle Theory: From Superstrings and Black
  Holes to the Standard Model, TASI Proceedings, 3-28 Jun 1992, Boulder,
  Colorado'' (J. Harvey and J. Polchinski, Eds.) World Scientific, River Edge,
  N. J., 1993; C. {Cs\'aki}, {\it Mod. Phys. Lett.} {\bf A 11} (1996) 599.

\bibitem{INQ87}
L. E. {Ib\'a\~nez}, H.-P. Nilles and F. Quevedo, {\it Phys. Lett.} {\bf B 187}
  (1987) 25.

\bibitem{IKNQ87}
L. E. {Ib\'a\~nez}, J. E. Kim, H.-P. Nilles and F. Quevedo, {\it Phys. Lett.}
  {\bf B 191} (1987) 282.

\bibitem{FINQ88b}
A. Font, L. E. {Ib\'a\~nez}, H.-P. Nilles and F. Quevedo, {\it Phys. Lett.}
  {\bf B 210} (1988) 101; {\bf B 213} (1988) 564.

\bibitem{CM88}
J. A. Casas and C. {Mu\~noz}, {\it Phys. Lett.} {\bf B 209} (1988) 214; {\bf B
  214} (1988) 63.

\bibitem{Gie01a}
J. Giedt, {\it Nucl. Phys.} {\bf B 595} (2001) 3.

\bibitem{gutrvw}
M. Gell-Mann, P. Ramond and R. Slansky, {\it Rev. Mod. Phys.} {\bf 50} (1978)
  721; P. Langacker, {\it Phys. Rep.} {\bf 72} (1981) 185; M. Srednicki, {\it
  in} ``From the Planck Scale to the Weak Scale: Toward a Theory of the
  Universe, TASI Proceedings, U.C. Santa Cruz, 1986'' (H. Haber, Ed.) World
  Scientific, Singapore, 1987.

\bibitem{sgtrvw}
H.-P. Nilles, {\it Phys. Rep.} {\bf 110} (1984) 1; C. Kounnas, A. Masiero, D.
  V. Nanopoulos and K. A. Olive, ``Grand Unification With and Without
  Supersymmetry and Cosmological Implications,'' World Scientific, Singapore,
  1984; S. Raby, {\it in} ``From the Planck Scale to the Weak Scale: Toward a
  Theory of the Universe, TASI Proceedings, U.C. Santa Cruz, 1986'' (H. Haber,
  Ed.) World Scientific, Singapore, 1987; R. N. Mohapatra, Lectures at Trieste
  Summer School, 1999, hep-ph/9911272.

\bibitem{HM88}
T. J. Hollowood and R. G. Myhill, {\bf Int. J. Mod. Phys. A 3} (1988) 899.

\bibitem{PDG00}
Particle Data Group, D. E. Groom et al., {\it Eur. Phys. J.} {\bf C 15} (2000)
  1.

\bibitem{BGW}
P. {Bin\'etruy}, M. K. Gaillard and Y.-Y. Wu, {\it Nucl. Phys.} {\bf B 481}
  (1996) 109; {\bf B 493} (1997) 27; {\it Phys. Lett.} {\bf B 412} (1997) 288.

\bibitem{GN00}
M. K. Gaillard and B. Nelson, {\it Nucl. Phys.} {\bf B 571} (2000) 3.

\bibitem{GS84}
M. B. Green and J. H. Schwarz, {\it Phys. Lett.} {\bf B 149} (1984) 117.

\bibitem{DIS87}
M. Dine, I. Ichinose and N. Seiberg, {\it Nucl. Phys.} {\bf B 293} (1987) 253.

\bibitem{DSW87}
M. Dine, N. Seiberg and E. Witten, {\it Nucl. Phys.} {\bf B 289} (1987) 585.

\bibitem{ADS87}
J. J. Atick, L. Dixon and A. Sen, {\it Nucl. Phys.} {\bf B 292} (1987) 109.

\bibitem{Gai01}
M. K. Gaillard et al., in preparation.

\bibitem{BDFS82}
F. Buccella, J. P. Derendinger, S. Ferrara and C. A. Savoy, {\it Phys. Lett.}
  {\bf B 115} (1982) 375.

\bibitem{CCF01}
G. B. Cleaver, D. J. Clements and A. E. Faraggi, hep-ph/0106060.

\bibitem{Kap88}
V. S. Kaplunovsky, {\it Nucl. Phys.} {\bf B 307} (1988) 145; Erratum {\bf B
  382} (1992) 436.

\bibitem{twk}
D. Bailin, S. K. Gandhi and A. Love, {\it Phys. Lett.} {\bf B 275} (1992) 55;
  D. Bailin and A. Love, {\it Phys. Lett.} {\bf B 288} (1992) 263.

\bibitem{Cas90}
J. A. Casas, Z. Lalak, C. Munoz and G. G. Ross, {\it Nucl. Phys.} {\bf B 347}
  (1990) 243.

\bibitem{ILR91}
L. E. {Ib\'a\~nez}, D. {L\"ust} and G. G. Ross, {\it Phys. Lett.} {\bf B 272}
  (1991) 251.

\bibitem{BL92}
D. Bailin and A. Love, {\it Phys. Lett.} {\bf B 278} (1992) 125.

\bibitem{DKL91}
L. Dixon, V. Kaplunovsky and J. Louis, {\it Nucl. Phys.} {\bf B 355} (1991)
  649.

\bibitem{ANT91}
I. Antoniadis, K. S. Narain and T. R. Taylor, {\it Phys. Lett.} {\bf B 267}
  (1991) 37.

\bibitem{CFKRR83}
G. D. Coughlan et al., {\it Phys. Lett.} {\bf B 131} (1983) 59.

\bibitem{GLM99}
M. K. Gaillard, D. H. Lyth and H. Murayama, {\it Phys. Rev.} {\bf D 58} (1998)
  123505.

\bibitem{LS96}
D. H. Lyth and E. D. Stewart, {\it Phys. Rev.} {\bf D53} (1996) 1784.

\bibitem{CKM89}
J. A. Casas, E. K. Katehou and C. {Mu\~noz}, {\it Nucl. Phys.} {\bf B 317}
  (1989) 171.

\bibitem{FINQ88a}
A. Font, L. E. {Ib\'a\~nez}, H.-P. Nilles and F. Quevedo, {\it Nucl. Phys.}
  {\bf B 307} (1988) 109.

\bibitem{AB92}
I. Antoniadis and K. Benakli, Phys. Lett. B 295 (1992) 219; Erratum B 407
  (1997) 449.

\bibitem{Den01}
T. Dent, {\it Phys. Rev.} {\bf D 64} (2001) 056005.

\bibitem{GG74}
H. Georgi and S. L. Glashow, {\it Phys. Rev. Lett.} {\bf 32} (1974) 438.

\bibitem{MSSMu}
U. Amaldi, W. de Boer and H. {F\"urstenau}, {\it Phys. Lett.} {\bf B 260}
  (1991) 447; J. Ellis, S. Kelly and D. V. Nanopoulos, {\it Phys. Lett.} {\bf B
  249} (1990) 441. \bibann{ORs on MSSM unification predictions and LEP data.}

\bibitem{LP93}
P. Langacker and N. Polonsky, {\it Phys. Rev.} {\bf D 47} (1993) 4028.

\bibitem{Gin87}
P. Ginsparg, {\it Phys. Lett.} {\bf B 197} (1987) 139.

\bibitem{DFMR96}
K. R. Dienes, A. E. Faraggi and J. March-Russell, {\it Nucl. Phys.} {\bf B 467}
  (1996) 44.

\bibitem{KN97}
T. Kobayashi and H. Nakano, {\it Nucl. Phys.} {\bf B 496} (1997) 103.

\bibitem{Die97}
K. Dienes, {\it Phys. Rep.} {\bf 287} (1997) 447.

\bibitem{AADF88}
G. G. Athanasiu, J. J. Atick, M. Dine and W. Fischler, {\it Phys. Lett.} {\bf B
  214} (1988) 55.

\bibitem{Ant90}
I. Antoniadis, {\it Nucl. Phys. Proc. Suppl.} {\bf A 22} (1991) 73.

\bibitem{CHL96}
S. Chaudhuri, G. Hockney and J. D. Lykken, {\it Nucl. Phys.} {\bf B 469} (1996)
  357.

\bibitem{SS74}
J. Scherk and J. H. Schwarz, {\it Nucl. Phys.} {\bf B 81} (1974) 118.

\bibitem{DS85}
M. Dine and N. Seiberg, {\it Phys. Rev. Lett.} {\bf 55} (1985) 366.

\bibitem{GX92}
M. K. Gaillard and R. Xiu, {\bf Phys. Lett.} {\bf B 296} (1992) 71.

\bibitem{Far93}
A. E. Faraggi, {\it Phys. Lett.} {\bf B 302} (1993) 202.

\bibitem{MR95}
S. P. Martin and P. Ramond, {\it Phys. Rev.} {\bf D 51} (1995) 6515.

\bibitem{AK96}
B. C. Allanach and S. F. King, {\it Nucl. Phys.} {\bf B 473} (1996) 3.

\bibitem{Iba93}
L. E. {Ib\'a\~nez}, {\it Phys. Lett.} {\bf B 318} (1993) 73.

\bibitem{DRB}
W. Siegel, {\it Phys. Lett.} {\bf B 84} (1979) 193; D.M. Capper, D.R.T. Jones
  and P. van Nieuwenhuizen, {\it Nucl. Phys.} {\bf B 167} (1980) 479.

\bibitem{BAM}
P. {Bin\'etruy} and T. {Sch\"ucker}, {\it Nucl. Phys.} {\bf B 178} (1981) 301;
  I. Antoniadis, C. Kounnas and K. Tamvakis, { \it Phys. Lett.} {\bf B 119}
  (1982) 377; S. P. Martin and M. T. Vaughn, {\it Phys. Lett.} {\bf B 318}
  (1993) 331.

\bibitem{FK98}
H. Fusaoka and Y. Koide, {\it Phys. Rev.} {\bf D 57} (1998) 3986.

\bibitem{GUTth}
D. Ross, {\it Nucl. Phys.} {\bf B 140} (1978) 1; W. J. Marciano, {\it Phys.
  Rev.} {\bf D 20} (1979) 274; T. J. Goldman and D. A. Ross, {\it Phys. Lett.}
  {\bf B 84} (1979) 208. S. Weinberg, {\it Phys. Lett.} {\bf B 91} (1980) 51;
  L. J. Hall, {\it Nucl. Phys.} {\bf B 178} (1981) 75; P. {Bin\'etruy} and T.
  {Sch\"ucker}, {\it Nucl. Phys.} {\bf B 178} (1981) 293.

\bibitem{HY93}
K. Hagiwara and Y. Yamada, {\it Phys. Rev. Lett.} {\bf 70} (1993) 709.

\bibitem{Nothr}
L. Dixon, V. Kaplunovsky and J. Louis, {\it Nucl. Phys.} {\bf B 355} (1991)
  649; I. Antoniadis, K. S. Narain and T. R. Taylor, {\it Phys. Lett.} {\bf B
  267} (1991) 37.

\bibitem{MNS93}
P. Mayr, H.-P. Nilles and S. Stieberger, {\it Phys. Lett.} {\bf B 317} (1993)
  53.

\bibitem{CPW93}
M. Carena, S. Pokorski and C. E. M. Wagner, {\it Nucl. Phys.} {\bf B 406}
  (1993) 59.

\bibitem{GR01}
D. M. Ghilencea and G. G. Ross, hep-ph/0102306.

\bibitem{CCF96}
S. Chang, C. {Corian\`o} and A. E. Faraggi, {\it Nucl. Phys.} {\bf B 477}
  (1996) 65.

\bibitem{Per01}
M. L. Perl et al., hep-ex/0102033.

\bibitem{CKR98}
D. J. H. Chung, E. W. Kolb and A. Riotto, {\it Phys. Rev.} {\bf D 60} (1999)
  063504.

\bibitem{GRS00}
G. {Germ\'an}, G. Ross and S. Sarkar, {\it Phys. Rev. Lett.} {\bf 84} (2000)
  4284.

\bibitem{Mth}
E. Witten, {\it Nucl. Phys.} {\bf B 471} (1996) 135; D. V. Nanopoulos,
  Unpublished, hep-th/9711080.

\bibitem{z3dual}
S. Ferrara, C. Kounnas and M. Porrati, {\it Phys. Lett.} {\bf B 181} (1986)
  263; M. {Cveti\v c}, J. Louis and B. A. Ovrut, {\it Phys. Lett.} {\bf B 206}
  (1988) 227.

\bibitem{TDTAF}
R. Dijkgraaf, E. Verlinde and H. Verlinde, {\it Comm. Math. Phys.} {\bf 115}
  (1988) 649; {\it in} ``Proceedings, Perspectives in String Theory,
  Copenhagen, 1987'' (P. Di Vecchia and J. L. Petersen, Eds.) Singapore, World
  Scientific, 1988; A. Shapere and F. Wilczek, {\it Nucl. Phys.} {\bf B 320}
  (1989) 669; S. Ferrara, D. {L\"ust}, A. Shapere and S. Thiesen, {\it Phys.
  Lett.} {\bf B 225} (1989) 363; J. Lauer, J. Mas and H.-P. Nilles, {\it Phys.
  Lett.} {\bf B 226} (1989) 251; {\it Nucl. Phys.} {\bf B 351} (1991) 353; E.
  J. Chun, J. Mas, J. Lauer and H.-P. Nilles, {\it Phys. Lett.} {\bf B 233}
  (1989) 141; S. Ferrara, D. {L\"ust} and S. Thiesen, {\it Phys. Lett.} {\bf B
  233} (1989) 147.

\bibitem{TWMIX}
E. J. Chun, J. Mas, J. Lauer and H.-P. Nilles, {\it Phys. Lett.} {\bf B 233}
  (1989) 141; J. Lauer, J. Mas and H.-P. Nilles, {\it Nucl. Phys.} {\bf B 351}
  (1991) 353.

\bibitem{NLSA}
G. Moore and P. Nelson, {\it Phys. Rev. Lett.} {\bf 53} (1984) 1519; L.
  {Alvarez-Gaum\'e} and P. Ginsparg, {\it Nucl. Phys.} {\bf B 262} (1985) 439;
  J. Bagger, D. Nemeschansky and S. Yankielowicz, {\it Nucl. Phys.} {\bf B 262}
  (1985) 478; A. Manohar, G. Moore and P. Nelson, {\it Phys. Lett.} {\bf B 152}
  (1985) 68; W. {Buchm\"uller} and W. Lerche, {\it Ann. Phys. (N.Y.)} {\bf 175}
  (1987) 159.

\bibitem{DFKZ91}
J. P. Derendinger, S. Ferrara, C. Kounnas and F. Zwirner, {\it Phys. Lett.}
  {\bf B 271} (1991) 307.

\bibitem{LCO92}
G. Lopes Cardoso and B. Ovrut, {\it Nucl. Phys.} {\bf B 369} (1992) 351.

\bibitem{KL945}
V. Kaplunovsky and J. Louis, {\it Nucl. Phys.} {\bf B 422} (1994) 57; {\bf B
  444} (1995) 191.

\bibitem{Lou91}
J. Louis, {\it in} ``2nd International Symposium on Particles, Strings and
  Cosmology, Boston, March 25-30, 1991'' (P. Nath and S. Reucroft, Eds.) River
  Edge, N.J., World Scientific, 1992;.

\bibitem{GT92}
M. K. Gaillard and T. R. Taylor, {\it Nucl. Phys.} {\bf B 381} (1992) 577.

\bibitem{PVREG}
J. Burton, M. K. Gaillard and V. Jain, {\it Phys. Rev.} {\bf D 41} (1990) 3118;
  M. K. Gaillard, {\it Phys. Lett.} {\bf B 342} (1995) 125; {\it Phys. Rev.}
  {\bf D 58} (1998) 105027; {\bf D 61} (2000) 084028.

\bibitem{GNW99}
M. K. Gaillard, B. Nelson and Y.-Y. Wu, {\it Phys. Lett.} {\bf B 459} (1999)
  549.

\bibitem{BMP00}
J. A. Bagger, T. Moroi and E. Poppitz, {\it JHEP} {\bf 0004} (2000) 009.

\bibitem{BGGM87}
P. {Bin\'etruy}, G. Girardi, R. Grimm and M. {M\"uller}, {\it Phys. Lett.} {\bf
  B 189} (1887) 83; {\bf B 195} (1987) 389.

\bibitem{BGG91}
P. {Bin\'etruy}, G. Girardi and R. Grimm, {\it Phys. Lett.} {\bf B 265} (1991)
  111.

\bibitem{BGG01}
P. {Bin\'etruy}, G. Girardi and R. Grimm, {\it Phys. Rep.} {\bf 343} (2001)
  255.

\bibitem{IL92}
L. E. {Ib\'a\~nez} and D. {L\"ust}, {\it Nucl. Phys.} {\bf B 382} (1992) 305.

\bibitem{BIM}
A. Brignole, C. E. {Ib\'a\~nez} and C. {Mu\~noz}, {\it Nucl. Phys.} {\bf B 422}
  (1994) 125; {\bf B (E) 436} (1995) 747; hep-ph/9707209; A. Brignole, C. E.
  {Ib\'a\~nez}, C. {Mu\~noz} and C. Scheich, {\it Z. Phys.} {\bf C 74} (1997)
  157.

\bibitem{ABGG93}
P. Adamietz, P. {Bin\'etruy}, G. Girardi and R. Grimm, {\it Nucl. Phys.} {\bf B
  401} (1993) 257.

\bibitem{GG99}
G. Girardi and R. Grimm, {\it Ann. Phys. (N.Y.)} {\bf 272} (1999) 49.

\bibitem{Zum79}
B. Zumino, {\it in} ``Recent Developments in Gravitation, {Carg\`ese} 1978''
  (M. Levy and S. Deser, Eds.) NATO ASI Series B44, Plenum Press, New York,
  1979.

\bibitem{PS95}
M. E. Peskin and D. V. Schroeder, ``An Introduction to Quantum Field Theory,''
  Addison-Wesley Publishing Company, Menlo Park, California, 1995.

\bibitem{WB92}
J. Wess and J. Bagger, ``Supersymmetry and Supergravity,'' Princeton University
  Press, Princeton, New Jersey, 1992.

\end{thebibliography}

\end{document}